# THE INTERNATIONAL LINEAR COLLIDER

Technical Design Report | Volume 2: Physics

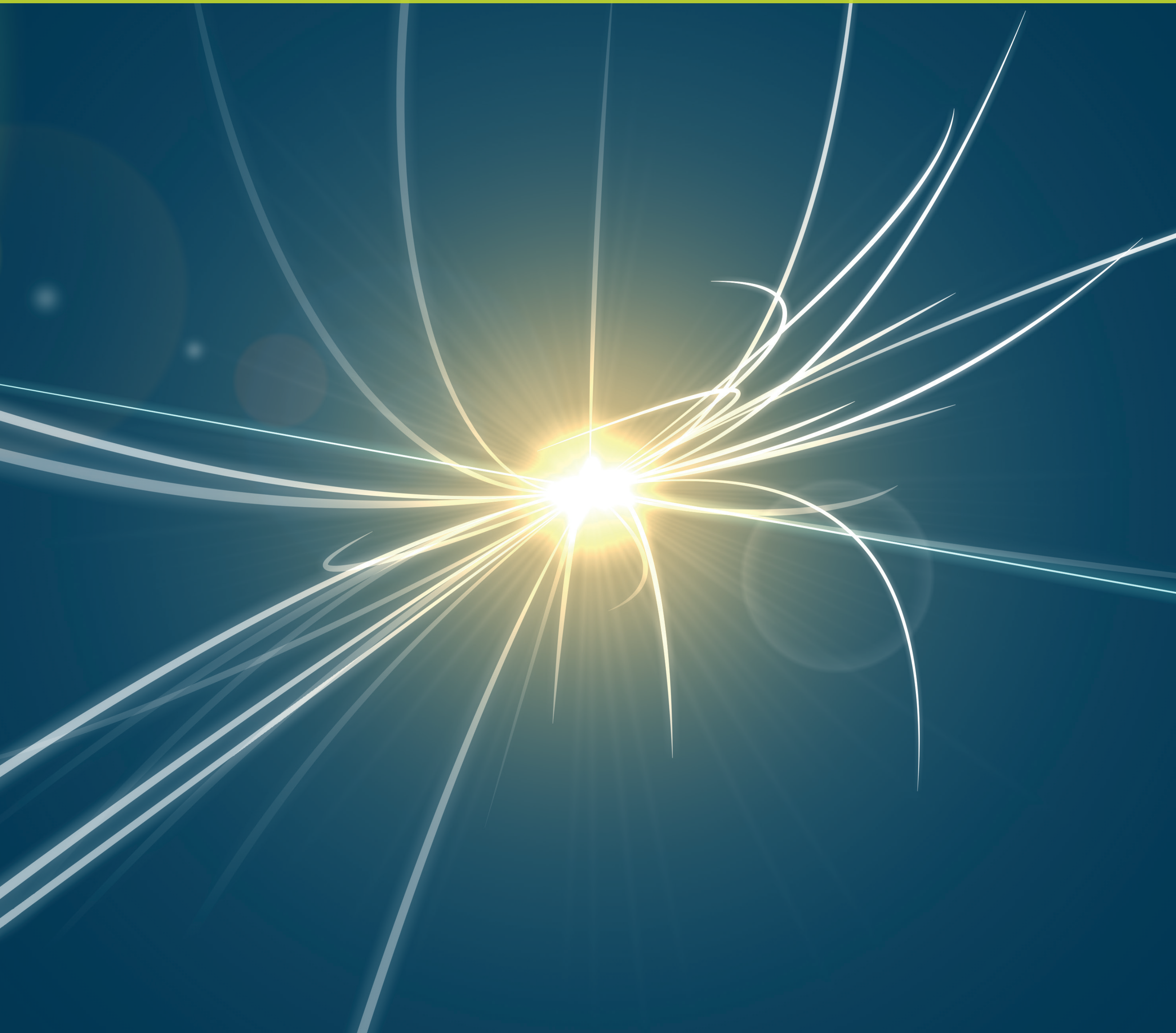

# The International Linear Collider

# Technical Design Report

2013

international linear collider

Volume 1: Executive Summary

Volume 2: Physics

Volume 3: Accelerator

Part I: R&D in the Technical Design Phase

Volume 3: Accelerator

Part II: Baseline Design

Volume 4: Detectors

# Volume 2

# Physics

Editors

Howard Baer, Tim Barklow, Keisuke Fujii, Yuanning Gao, Andre Hoang, Shinya Kanemura, Jenny List, Heather E. Logan, Andrei Nomerotski, Maxim Perelstein, Michael E. Peskin, Roman Pöschl, Jürgen Reuter, Sabine Riemann, Aurore Savoy-Navarro, Geraldine Servant, Tim M. P. Tait, Jaehoon Yu

# Acknowledgements

We acknowledge the support of BMWF, Austria; MinObr, Belarus; FNRS and FWO, Belgium; NSERC, Canada; NSFC, China; MPO CR and VSC CR, Czech Republic; Commission of the European Communities; HIP, Finland; IN2P3/CNRS, CEA-DSM/IRFU, France; BMBF, DFG, Helmholtz Association, MPG and AvH Foundation, Germany; DAE and DST, India; ISF, Israel; INFN, Italy; MEXT and JSPS, Japan; CRI(MST) and MOST/KOSEF, Korea; FOM and NWO, The Netherlands; NFR, Norway; MNSW, Poland; ANCS, Romania; MES of Russia and ROSATOM, Russian Federation; MON, Serbia and Montenegro; MSSR, Slovakia; MICINN-MINECO and CPAN, Spain; SRC, Sweden; STFC, United Kingdom; DOE and NSF, United States of America.

# Contents

# Chapter 1
# Introduction

## 1.1 Physics at the ILC

For more than twenty years, an advanced electron-positron collider has been put forward as a key component of the future program of elementary particle physics. We have a well-established Standard Model of particle physics, but it is known to be incomplete. Among the many questions that this model leaves open, there are two — the origin of the masses of elementary particles and the particle identity of cosmic dark matter – that should be addressed at energy scales below 1 TeV. It has been appreciated for a long time that a next-generation electron-positron collider would give us the ability to make precision measurements that would shed light on these mysteries.

Now the technology to build this electron-positron collider has come of age. This report is a volume of the Technical Design Report for the International Linear Collider (ILC). The accompanying volumes of this report lay out the technical design of a high-luminosity $e^+e^-$ collider at 500 GeV in the center of mass and of detectors that could make use of the collisions to perform high-precision measurements. In this volume, we summarize the physics arguments for building this collider and their appropriate relation to the situation of particle physics as of the fall of 2012. The discussion in this volume supplements the presentation of the physics opportunities for a 500 GeV $e^+e^-$ collider given in the review articles [1–3], the 2001 regional study reports [4–6], the 2006 study of ILC/LHC complementarity [7], and the 2007 ILC Reference Design Report [8].

There are two important reasons to review the physics arguments for the ILC now. First, the Large Hadron Collider (LHC) has now begun to explore the energy region up to 1 TeV in proton-proton collisions. The LHC experiments have discovered a resonance that is a strong candidate for a Higgs boson similar to that of the Standard Model and have measured the mass of this resonance to be about 125 GeV [9, 10] It has been understood for a long time that there are intrinsic limitations to the ability of hadron colliders to study color-singlet scalar particles, and that precision measurements, to the few percent level, are needed to place a new scalar particle correctly within our model of particle physics. The ILC is an ideal machine to address this question. In this report, we will describe the system of measurements that will be needed to probe the identity of the Higgs boson and present new estimates of the capability of the ILC to make those measurements.

We will also describe many other opportunities that the ILC provides to probe for and study new physics, both through the production of new particle predicted by models of physics beyond the Standard Model and through the study of indirect effects of new physics on the $W$ and $Z$ bosons, the top quark, and other systems that can be studied with precision at the ILC. It is important to re-evaluate the merits of these experiments in view of new constraints from the LHC, and we will do that in this report.

The experience of operating the LHC and its detectors also allows us to make more concrete projections of the long-term capabilities of the LHC experiments and the complementarity of the measurements from the ILC experiments. We have tried to incorporate the best available information into this report.

A second reason to revisit to physics case for the ILC is that the studies for the technical design and benchmarking of the ILC detectors have given us a more precise understanding of their eventual capabilities. In many cases, the performance of the detectors found in full-simulation studies exceeds the capabilities claimed from studies done at earlier stages of the conceptual detector design process. Our estimates here will be based on these new results.

To support a major accelerator project such as the ILC, it should be a criterion that this project will advance our knowledge of particle physics *qualitatively* beyond the information that will be available from currently operating accelerators, including the results expected from the future stages of the LHC. In this report, we will address this question. We will demonstrate the profound advances that the ILC will make in our understanding of fundamental physics.

## 1.2 Advantages of $e^+e^-$ colliders

Over the past forty years, experiments at proton and electron colliders have played complementary roles in illuminating the properties of elementary particles. For example, the bottom quark was first discovered in 1977 through the observation of the $\Upsilon$ resonances in proton-proton collisions. However, many of the most revealing properties of the $b$ quark, from its unexpectedly long lifetime to its decays with time-dependent CP violation, were discovered at $e^+e^-$ colliders.

Today, the LHC offers obvious advantages for experimenters in providing very high energy and very high rates in typical reactions. The advantages of the ILC are different and perhaps more subtle to appreciate. In this section, we will review these advantages in general terms. We will revisit these points again and again in our discussions of specific processes that will be studied at the ILC.

### 1.2.1 Cleanliness

An elementary difference between hadron and electron collisions is apparent in the design of detectors: The environment for electron-positron collisions is much more benign. At LHC energies, the proton-proton total cross section is roughly 100 mb. In the current scheme for running the LHC, proton-proton bunch collisions occur every 50 nsec, each bunch crossing leads to about 30 proton-proton collisions, and each of these produces hundreds of energetic particles. At the ILC, the most important chronic background source comes from photon-photon collisions, for which the cross section is hundreds of nb. Bunch crossings are spaced by about 300 nsec; at each bunch crossing we expect about 1 photon-photon collision, producing a few hadrons in the final state. Each $e^+e^-$ bunch crossing does produce a large number of secondary electron-positron pairs, but these are mainly confined to a small volume within 1 cm of the beam.

The difference between hadron-hadron and $e^+e^-$ collisions has profound implications for the detectors and for experimentation. The LHC detectors must be made of radiation-hard materials to handle a high occupancy rate. They must have thick calorimeters to contain particles with a wide range of energies, requiring also the placement of solenoids inside the calorimeter volume. They must have complex trigger systems that cut down rates to focus on the most interesting events. At the ILC, tracking detectors can be made as thin as technically feasible. All elements, from the vertex detector to the calorimeter, can be brought much closer to the interaction point and contained inside the solenoid. The ILC detectors are projected to improve the momentum resolution from tracking by a factor of 10 and the jet energy resolution of the detector by a factor of 3 or better. The very close placement of the innermost pixel vertex detector layer leads to excellent $b$, charm and $\tau$ tagging capabilities. In addition, the complications in analyzing LHC events due to hadrons from the underlying-event and pileup from multiple collisions in each beam crossing are essentially removed at the ILC. The $e^+e^-$ environment thus provides a setting in which the basic high-energy collision can be measured with high precision.

### 1.2.2 Democracy

The elementary coupling $e$ of the photon is the same for all species of quarks and leptons, and the same also for new particles from beyond the Standard Model. Thus, $e^+e^-$ annihilation produces pairs of all species, new and exotic, at similar rates.

At the LHC, the gluon couples equally to all quarks and to new colored particles. However, here, this democracy is hardly evident experimentally. Soft, non-perturbative strong interactions are the dominant mechanism for particle production and involve only the light quarks and gluons. Further, because the proton is a composite object with parton distributions that fall steeply, the production cross sections are much lower for heavy particles than for light ones. At the LHC, the cross section for producing bottom quarks is of the order of 1 mb, already much lower than the total inelastic cross section. The cross section for top quark pair production at the 14 TeV LHC is expected to be about 1 nb. Production cross sections for new particles will be 1 pb or smaller. Thus, interesting events occur at rates of $10^{-7}$ to $10^{-13}$ of the total event rate. This implies, first, that a trigger system is needed to exclude all events but 1 in $10^6$ before any data analysis is possible. Beyond this, only events with unusual and striking properties can be recognized in the much larger sample of background QCD events. A new particle or process can be studied only if its signals can be clearly discriminated from those of QCD reactions.

At the ILC, the cross sections for light quark and lepton pair production are much smaller, but also more comparable to the cross sections for interesting new physics processes. The main Standard Model processes in $e^+e^-$ annihilation — annihilation to quark and lepton pairs, annihilation to $W^+W^-$, and single $W$ and $Z$ production – all have cross sections at the pb level at 500 GeV. New particle production processes typically have cross sections of order 10–100 fb and result in events clearly distinguishable from the basic Standard Model reactions.

This has a number of important implications for $e^+e^-$ experimentation. First, no trigger is needed. The ILC detectors can record all bunch crossings and performed any needed event reduction offline. Second, no special selection is needed in classifying events. That is, all final states of a decaying particle, not only the most characteristic ones, can be used for physics analyses. At the LHC, it is not possible to measure absolute branching ratios or total widths; at the ILC, these quantitites are directly accessible. Third and perhaps most importantly, at the ILC, it is much easier to recognize $W$ and $Z$ bosons in their hadronic decay modes than at the LHC. Since most $W$ and $Z$ decays are to hadronic modes, this is a tremendous advantage in the systematic study of heavy particles whose decay products typically include the weak bosons. We will see that this advantage applies not only to exotic particles but also in the study of the top quark and the Higgs boson.

The Higgs boson is produced in roughly one in one billion $pp$ collisions at LHC energies. The modes actually used in the Higgs discovery occur at the rate of one in a trillion $pp$ collisions. At the ILC, Higgs events occur at about 1% of all $e^+e^-$ annihilations, and the resulting events are quite characteristic. They can be picked out and analyzed by eye. Figure 1.1 shows typical simulated events of $e^+e^- \to Zh$.

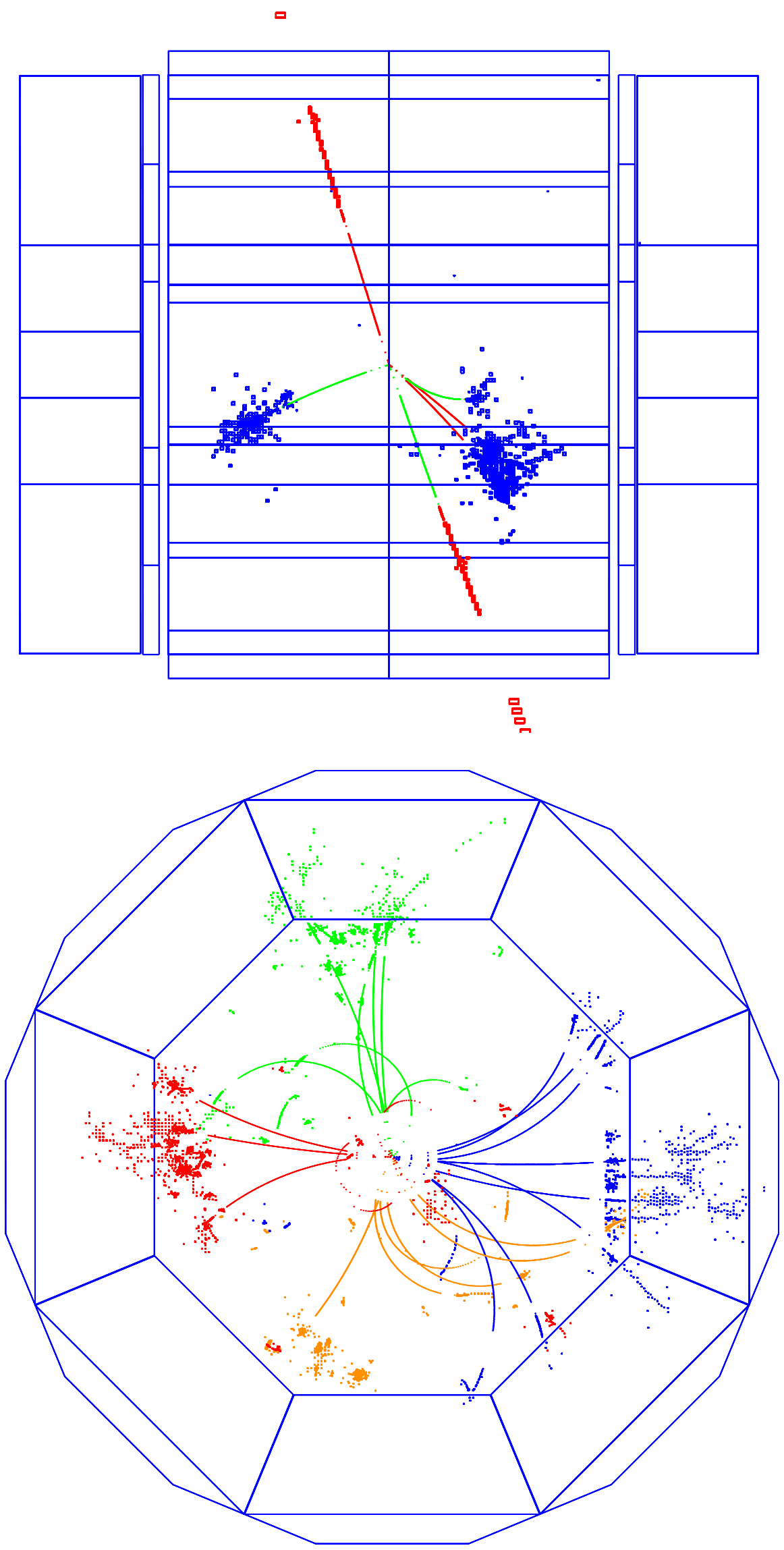

**Figure 1.1**
Simulated $e^+e^- \to Zh$ events: Top: $e^+e^- \to Zh \to \mu^-\mu^-\tau^+\tau^-$; Bottom: $e^+e^- \to Zh \to b\bar{b}b\bar{b}$ [11].

### 1.2.3 Calculability

At the LHC, all cross section calculations rely on QCD. Any theoretical calculation of signal or background has systematic uncertainties from the proton structure functions, from unknown higher-order perturbative QCD corrections, and from nonperturbative QCD effects. NLO QCD corrections to cross section calculations are typically at the 30-50% level. For the Higgs boson cross section, the first correction is +100%. To achieve theoretical errors smaller than 10% requires computations to NNLO or beyond, a level that is not feasible now except for the simplest reactions.

At the ILC, the initial-state $e^-$ and $e^+$ are pointlike elementary particles, coupling only to the electroweak interactions. The first radiative corrections to cross sections are at the few-percent level. With effort, one can reach the part-per-mil level of theoretical precision, a level already achieved in the theoretical calculations for the LEP program.

Thus, it is possible to study heavy particles through their effects in perturbing the Standard Model at lower energies. For example, the LHC will be able to detect $Z'$ bosons up to 4-5 TeV by searches for production of high-mass $\mu^+\mu^-$ pairs. The ILC at 500 GeV is sensitive to the presence of bosons with comparably high masses by searching for deviations from the precise Standard Model predictions for $e^+e^- \to f\bar{f}$ cross sections. By studying the dependence of these deviations on flavor and polarization, the ILC experiments can reconstruct the complete phenomenological profile of the heavy boson. Similar precision measurements can give new information about heavy particles that couple to the top quark and the Higgs boson.

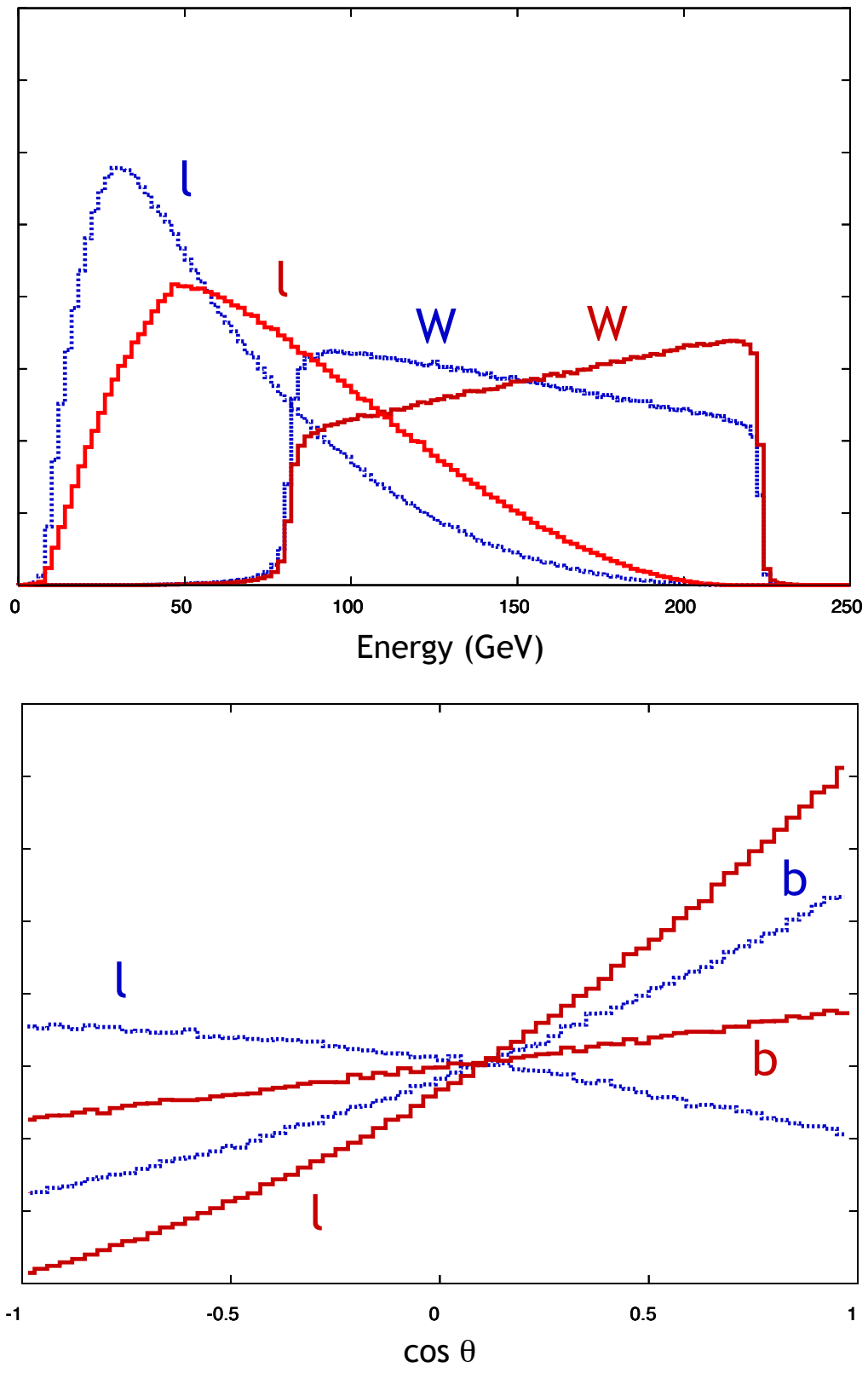

**Figure 1.2** Spin asymmetries in $e^+e^- \to t\bar{t}$: For the two fully polarized beam initial state ($e^-_R e^+_L$ - red, solid, $e^-_L e^+_R$ - blue, dotted), the figures show: Top: the energy distributions of the $W$ and $\ell$ (or $d, s$ quark). Bottom: the $\cos\theta$ distributions of the $b$ and $\ell$.

Beyond this, the high precision theoretical understanding of Standard Model signal and background processes available at the ILC can make it possible to find elusive new physics interactions, and to characterize these interactions fully.

### 1.2.4 Detail

Because of the simplicity of event selections at the ILC and the absence of a complicating underlying event, physics analyses at the ILC can be done by reconstructing complete events and determining quark and lepton momenta by kinematic fitting. Such an analysis reveals the spin-dependence of production and decay processes. The ILC will also provide polarized electron and positron beams, and so the processes studied there can be completely characterized for each initial and final polarization state.

We are used to thinking of quarks and leptons at low energy as single massive objects. However, at energies above the $Z^0$ mass, the left- and right-handed components of quarks and leptons behave as distinct particles with different $SU(2) \times U(1)$ quantum numbers. The weak-interaction decays of heavy particles, including the top quark and the $W$ and $Z$ bosons, have order-1 spin asymmetries. These spin effects are difficult to observe at hadron colliders. Most typically, they are used as inputs to perform signal/background discrimination using the matrix element method of multivariate event selection. At the ILC, they are obvious aspects of the physics. That is, we do not rely on the correctness of the predicted Standard Model distribution but instead observe these distributions in full detail. In Fig. 1.2, we present an array of nontrivial energy and angular distributions generated by the spin asymmetries in the process $e^+e^- \to t\bar{t}$. In every process studied at the ILC, spin effects provide a crucial new handle on the physics, allowing us to make interpretations at the basic level of the underlying weak-interaction quantum numbers.

**Table 1.1.** Major physics processes to be studied by the ILC at various energies. The table indicates the various Standard Model reactions that will be accessed at increasing collider energies, and the major physics goals of the study of these reactions. A reaction listed at a given energy will of course be studied at all higher energies. The last column gives the motivation for the use of polarized beams. Polarization is always an important component of the ILC program, but for different reasons in different reactions. The codes **A**, **H**, **L**, and **B** are explained in the text.

| Energy | Reaction | Physics Goal | Polarization |
|---|---|---|---|
| 91 GeV | $e^+e^- \to Z$ | ultra-precision electroweak | **A** |
| 160 GeV | $e^+e^- \to WW$ | ultra-precision $W$ mass | **H** |
| 250 GeV | $e^+e^- \to Zh$ | precision Higgs couplings | **H** |
| 350–400 GeV | $e^+e^- \to t\bar{t}$ | top quark mass and couplings | **A** |
| | $e^+e^- \to WW$ | precision $W$ couplings | **H** |
| | $e^+e^- \to \nu\bar{\nu}h$ | precision Higgs couplings | **L** |
| 500 GeV | $e^+e^- \to f\bar{f}$ | precision search for $Z'$ | **A** |
| | $e^+e^- \to t\bar{t}h$ | Higgs coupling to top | **H** |
| | $e^+e^- \to Zhh$ | Higgs self-coupling | **H** |
| | $e^+e^- \to \tilde{\chi}\tilde{\chi}$ | search for supersymmetry | **B** |
| | $e^+e^- \to AH, H^+H^-$ | search for extended Higgs states | **B** |
| 700–1000 GeV | $e^+e^- \to \nu\bar{\nu}hh$ | Higgs self-coupling | **L** |
| | $e^+e^- \to \nu\bar{\nu}VV$ | composite Higgs sector | **L** |
| | $e^+e^- \to \nu\bar{\nu}t\bar{t}$ | composite Higgs and top | **L** |
| | $e^+e^- \to \tilde{t}\tilde{t}^*$ | search for supersymmetry | **B** |

## 1.3 Modes of operation of the ILC

At a proton-proton collider, one creates collisions at a fixed center of mass energy, relying on the energy distribution of partons in the proton to sample a range of collisions energies for elementary processes. At a circular $e^+e^-$ collider, the maximum energy is preset by the size of the ring, and typically the performance of the accelerator is best just near this maximum energy. An $e^+e^-$ linear collider is more forgiving in terms of operating at different energies and in different running conditions. In principle, it is possible to run at any energy up to the energy set by the length of the machine, with a penalty in luminosity roughly proportional to the reduction in the energy. Increasing the length of the machine of course requires the purchase of more components, but in principle a linear collider can also be lengthened to smoothly raise its maximum collision energy if physics discoveries call for this.

This flexibility has let the designers of the ILC to envision an experimental programs at series of energies well adapted to individual physics goals. In Table 1.1, we list possible center of mass energies at which the ILC could be run. These encompass the following:

- **91 GeV and 160 GeV**: These energies correspond to the $Z$ resonance and the threshold for $e^+e^- \to W^+W^-$. The ILC is capable of achieving a luminosity much higher than that of the LEP program of the 1990's. This motivates a *Giga-Z* program, to improve the precision electroweak measurements of $Z$ asymmetries and couplings by an order of magnitude, and a *Mega-W* program to measure the $W$ mass with MeV precision.
- **250 GeV**: This energy is the peak of the cross section for the reaction $e^+e^- \to Zh$, for $h$ the new boson resonance discovered near 125 GeV. Whether or not $h$ is a Higgs boson, these experiments will begin the precision study of the nature and couplings of this particle. The $h$ production events are tagged, allowing study of invisible and unexpected decay modes.
- **350-400 GeV**: Within a few GeV of 350 GeV, the $e^+e^-$ annihilation cross section is expected to show a prominent rise associated with the threshold for top quark pair production. Because of its short lifetime, the top quark has no stable bound states. Instead, it has a threshold structure whose shape is precisely predicted by *perturbative* QCD. Measurement of this threshold shape will yield the top quark mass to an accuracy of 100 MeV for input to grand unification and other fundamental physics predictions. Measurements of the full details of the $t\bar{t}$ final states near threshold and in the continuum will provide a new program of precision measurements constraining electroweak symmetry breaking.

The Higgs boson reaction $e^+e^- \to \nu\overline{\nu}h$ turns on in this energy region. The study of this reaction gives a measurement of the $hWW$ coupling, an essential ingredient in a program of precision Higgs boson studies. The cross section for this reaction grows at higher energies roughly as $\log(E_{\text{CM}}/m_h)$, providing large statistics for the study of rare Higgs decays.

Finally, the reaction $e^+e^- \to W^+W^-$ becomes exceptionally sensitive to possible modification of the Standard Model couplings at high energy, with the effect of modified couplings growing as $(E/m_w)^2$. In this energy region and above, precision $W$ coupling measurements provide a third powerful probe for new physics.

- **500 GeV**: Running at the full energy and highest luminosity of the ILC increases the power of the precision experiments just described. In addition, precision studies of two-fermion reactions $e^+e^- \to f\overline{f}$ can probe sensitively for vector resonances at high energy, new fermion interactions, and quark and lepton compositeness. This program also allows us to search for new particles such as color-singlet supersymmetry particles and states of an extended Higgs sector in parameter regions that are very difficult for the LHC experiments to explore.
- **up to 1000 GeV**: Running at even high energies, which is envisioned in upgrades of the ILC, allows a number of new measurements sensitive to the Higgs boson coupling to the top quark and the Higgs boson self-coupling, to additional probes of strongly-interacting or composite models of the Higgs boson, and to searches for new exotic particles.

The exact run plan that will be carried out at the ILC will depend on the situation in particle physics at the time of the ILC operation, taking into account new discoveries and measurements from the LHC in its running at 14 TeV. For definiteness in our projections for the ILC capabilities for Higgs boson couplings, we will discuss here a canonical program with stages at 250 GeV, 500 GeV, and 1000 GeV, with integrated luminosity 250 fb$^{-1}$, 500 fb$^{-1}$, and 1000 fb$^{-1}$, respectively, at these stages.

Both the electron and the positron beams at the ILC will be polarized. As we have emphasized already in the previous section, ILC cross sections depend on beam polarization in order 1. Thus, polarization is an essential ingredient in the experimental program.

In Table 1.1, we have devoted the last column to the role of polarization in the study of each of the major physics reactions. Beam polarization always plays an important role, but this role differs from one reaction to the next [12]. Going down the Table, we see four distinct modes in which beam polarization is used. These are:

**A:** At the $Z$ resonance, in the precision measurement of the electroweak couplings of the top quark, and in precision measurement of $e^+e^- \to f\overline{f}$, the beam polarization asymmetry is a crucially informative observable.

**H:** In $e^+e^-$ annihilation, an electron annihilates a positron of the opposite helicity. Thus, opposite polarization of electrons and positrons ($e^-_L e^+_R$ and $e^-_R e^+_L$) enhances the luminosity.

**L:** Certain Standard Model processes, especially at high energy, occur dominantly from the $e^-_L e^+_R$ polarization state. Polarizing to this state greatly enhances the rates and suppresses background.

**B:** Conversely, new physics searches at high energy benefit from use of the polarization state $e^-_R e^+_L$ to suppress Standard Model background.

In this volume, we will discuss only $e^+e^-$ experiments, but the ILC also has the possibility of hosting experimental programs with $\gamma\gamma$ and $e^-e^-$ collisions. A review of these options and more detailed discussion of the role of polarization can be found in Part 4 of [4].

## 1.4 Key physics explorations at the ILC

In the following sections of this volume, we will present the major aspects of the physics program of the ILC in detail. We will see explicitly how the features of $e^+e^-$ experimentation and the specific reactions outlined above translate into measurements with direct and illuminating physical interpretation.

We begin by discussing the ILC program on the Higgs boson. There is now great excitement over the discovery of a bosonic resonance at the LHC whose properties are consistent with those of the Higgs boson [9, 10]. This particle might indeed be the Higgs boson predicted by the Standard Model, a similar particle arising from a different model of electroweak symmetry breaking, or a particle of totally different origin that happens to be a scalar resonance. To choose among these options, detailed precision measurements of this particle are needed.

In Section 2, we will present the program of precision measurements of the properties of this new boson that would be made by the ILC experiments. Since the new boson is observed to decay to $WW$ and $ZZ$ at rates comparable to the predictions for the Standard Model Higgs boson, we already know that its production cross section at the ILC will be sufficient to carry out this program. We will first set out the requirements for an experimental program that has sufficient sensitivity to distinguish the various hypotheses for the nature of the new scalar. Very high precision—at the level of percent accuracy in the new coupling constants—is needed. It is unlikely that the LHC experiments will reach this level of performance. We will then describe the variety of measurements that the ILC experiment would be expected to carry out for this particle at the various stage of ILC operation. As we have already emphasized, the ILC program on the Higgs boson includes experiments at 250 GeV, the peak of the cross section for $e^+e^- \to Zh$, and at higher energies to access the process $e^+e^- \to \nu\bar{\nu}h$ with $WW$ fusion production of the Higgs boson. We will show that these measurements will be extremely powerful probes. They will definitively settle the question of the nature of the new boson and will give insight into any larger theory of which it might be a part.

The LHC has not yet provided evidence for signals of new physics beyond the Standard Model from its early running at 7 and 8 TeV. There are two distinct attitudes to take toward the current situation. The first is that it is premature to draw any conclusions at the present time. The LHC experimental program is still in its early stages. The accelerator has not yet reached its design energy and has so far accumulated only 1% of its eventual data set. The second is that the discovery of the new scalar boson—especially if turns out to have the properties similar to the Standard Model Higgs boson—and the deep exclusions already made for supersymmetry and other new physics models have already changed our ideas about new physics at the TeV energy scale. Our information from the LHC is certainly incomplete. We look forward to new information and new discoveries in the LHC run at 14 TeV that will take place in the latter years of this decade. And, yet, we must take seriously the implications of what we have already learned.

Though the Standard Model of particle physics is internally consistent and, so far, is not significantly challenged experimentally, it is incomplete in many respects. Most challenging is the lack, in the Standard Model, of any explanation for the spontaneous breaking of electroweak symmetry that leads to the masses of all quarks, leptons, and gauge bosons and provides the qualitative structure of their fundamental interactions.

The problem of electroweak symmetric breaking has motivated a large number of proposals for new physics at the TeV energy scale. These proposals fall into three classes. The first postulates that electroweak symmetry is broken by new strong interactions at the TeV energy scale. In these models, the key observables are the parameters of weak vector boson scattering at TeV energies. The discovery of a new light scalar, especially if its couplings to $W$ and $Z$ are seen to be those characteristic of a Higgs boson with a nonzero vacuum expectation value, deals a signficant blow to this whole set of models.

The second class of models posulates that electroweak symmetry breaking is due to the expectation value of an effective Higgs field that is composite at a higher mass scale. Little Higgs models, in which the Higgs boson is a Goldstone boson of a higher energy theory, and Randall-Sundrum models and other theories with new dimensions of space, are examples of theories in this class. These theories predict new particles with the quantum numbers of the top quark and the $W$ and $Z$ bosons, with TeV masses. These particles should eventually be discovered at the LHC in its 14 TeV program. The other crucial predictions of these models are modifications of the couplings of the heaviest particles of the Standard Model, the $W$, $Z$, and top quark. The ILC is ideally suited to observe these effects through precision measurement of the properties of $W$, $Z$, and $t$. Extreme energies are not required; the ILC design center of mass energy of 500 GeV is quite sufficient.

The third class of models postulates the Higgs field as an elementary scalar field, requiring supersymmetry to tame its ultraviolet divergences. The LHC has now excluded the constrained supersymmetric models that were considered paradigmatic in the period up to 2009 for masses low enough that supersymmetry dynamics naturally drives electroweak symmetry breaking. However, supersymmetry has a large parameter space, and compelling regions are still consistent with the LHC exclusions. The typical property of these regions is that the lightest supersymmetric particles are the fermionic partners of the Higgs bosons. These particles are very difficult to discover or study at the LHC but are expected to be readily accessible to the ILC at 500 GeV. Models of this type are also likely to contain additional Higgs bosons at relatively low masses that would be targets of study at the ILC.

Thus, we argue, the exclusion of new physics at this early stage of the LHC program, combined with the observation of a new boson resembling the Standard Model Higgs boson, strengthens the case for the ILC as probe of new physics beyond the Standard Model.

In Sections 3–7 of this report, we will explain this viewpoint in full detail. We will begin in Section 3 with a review of the ILC program on $e^+e^- \to f\overline{f}$ processes, where $f$ is a light quark or lepton. The precision study of these processes is sensitive to new heavy gauge bosons. These reactions also probe models with extra space dimensions, and models in which the electron is composite with a very small size. We will explain how experiments at 500 GeV can reveal the nature of any such boson or composite structure, qualitatively improving on the information that we will obtain from the LHC.

In Sections 4–5, we will describe the ILC program relevant to models with a light Higgs boson that is composite at a higher energy scale. In Section 4, we will review the ILC program on the $W$ and $Z$ bosons. We will describe the capabilities of the ILC for the measurement of $W$ boson couplings and $W$ boson scattering. We will show that how these measurements are capable of revealing new terms in the couplings of $W$ and $Z$ induced by Higgs composite structure.

In Section 5, we will review the ILC program on the top quark. We will describe the study of top quark production at threshold and at higher energies near the maximum of the cross section for $e^+e^- \to t\bar{t}$. This study gives new, nontrivial, tests of QCD and also gives access to couplings of the top quark that are extremely difficult to study at the LHC. In models in which the top quark couples to a composite Higgs boson or a strongly interacting Higgs sector, the couplings of the top quark to the $Z$ boson provide crucial tests not available at the LHC. We will describe the beautiful probes of these couplings availabe at the 500 GeV ILC.

In Sections 6–7, we will discuss the ILC program in searching for and measuring the properties of new particles predicted by supersymmetry and other models in which the Higgs boson is an elementary scalar field. We will discuss particles that, although they are within the energy range of the ILC, they would not be expected to be found at the LHC at the current stage of its program. These particles might be discovered at the LHC with higher energy or luminosity, or their discovery might have to wait for the ILC. In either case, the ILC will make measurements that will be key to understanding their role in models of new physics.

In Section 6, we will review ILC measurements on new bosons associated with the Higgs boson within a larger theory of electroweak symmetry breaking. We will note many aspects of these more complex theories that the ILC will be able to clarify, beyond the results anticipated from the LHC.

In Section 7, we will review the program of ILC measurements on supersymmetric particles that might be present in the ILC mass range. In this discussion, we will review the current constraints on supersymmetry. We will observe that many scenarios are still open in which new particles can found at the 500 GeV ILC. We will present the detailed program of measurements that the ILC can carry out on these particles. This discussion will also illustrate that broad capabilities that the ILC experiments provide to understand the nature of new particles discovered at the LHC, whatever their origin in terms of an underlying model.

As we have already noted, the current exclusions of new particles by the LHC experiments drive us, in models of supersymmetry, to models in which the lightest supersymmetric particles are the charged and neutral Higgsinos, which would naturally lie in the 100–200 GeV mass range. These particles are very difficult to identify at the LHC but would be easily seen and studied at the ILC. More generally, if supersymmetry is indeed realized in nature, the ILC can be expected to directly probe those parameters of supersymmetry most intimately connected to the mechanism of electroweak symmetry breaking. We will explain this point of view in detail in Section 7.

Finally, in Section 8, we will discuss the role of the ILC in understanding cosmology and, in particular, the unique experiments possible at the ILC that will shed light on the nature of the dark matter of the universe. Section 9 will give some general conclusions.

# Chapter 2
# Higgs Boson

Precision studies of the weak interactions at LEP, SLC, Tevatron, and LHC have shown that they are described by a spontaneously broken $\mathrm{SU}(2)_L \times \mathrm{U}(1)_Y$ gauge theory. The quantum numbers of all fermions are verified experimentally, and the properties of the heavy vector bosons $W$ and $Z$ predicted by the theory are in excellent accord with the theory at the level of one-loop electroweak corrections [1]. However, the basic $SU(2) \times U(1)$ symmetry of the model forbids the generation of mass for all quarks, leptons, and vector bosons. Thus, this symmetry must be spontaneously broken. The theory of weak interactions then requires a vacuum condensate that carries charge under the $\mathrm{SU}(2)_L \times \mathrm{U}(1)_Y$ gauge groups.

In local quantum field theory, it is not possible to simply postulate the existence of a uniform vacuum condensate. This condensate must be associated with a field that has dynamics and quantum excitations. To prove the correctness of our theory of weak interactions, it is essential to study this field directly and to prove through experiments that the field and its quantum excitations have the properties required to generate mass for all particles. We have little direct or indirect information about the nature of this field. The Standard Model postulates the simplest possibility, that the needed spontaneous symmetry breaking is generated by one $SU(2)$ doublet scalar field, the Higgs field, with one new physical particle, the Higgs boson. The true story of electroweak spontaneous symmetry breaking could be much more complex.

The Higgs field, or a more general Higgs sector, couples to every type of particle. It likely plays an important role in all of the unanswered questions of elementary particle physics, including the nature of new forces and underlying symmetries, CP violation and baryogenesis, and the nature and relation of quark and lepton flavors. To make progress on these problems, we must understand the Higgs sector in detail.

In July 2012, the ATLAS and CMS experiments presented very strong evidence for a new particle whose properties are consistent with those of the Standard Model Higgs boson [2, 3]. Additional evidence for this particle was provided by the CDF and DO experiments [4]. This gives us a definite point of entry into the exploration of the Higgs sector. It would be ideal to produce this particle in a well-controlled setting and measure its mass, quantum numbers, and couplings with high precision. The particle is at a mass, 125 GeV, that is readily accessible to a next-generation $e^+e^-$ collider. It has been observed to couple to $ZZ$ and $WW$, insuring that the major production reactions in $e^+e^-$ collisions are present. The ILC is precisely the right accelerator to make these experiments available.

Though there is no reason to believe that the simple picture given by the Standard Model is correct, the minimal theory of electroweak symmetry breaking given by the Standard Model is a convenient place to begin in describing the capabilities of any experimental facility. This is especially true because, as we will discuss in Section 2.2, most models with larger and more complex Higgs sectors contain a particle that strongly resembles the Standard Model Higgs boson.

In this chapter, then, we will describe the capabilities of the ILC to obtain a comprehensive understanding of the Standard Model Higgs boson. In Section 2.1, we will review the Higgs mechanism and write its basic formulae. In Section 2.2, we will discuss the relation of the Standard Model Higgs boson to similar particles in more general theories of elementary particles. We will review the Decoupling Theorem that requires a boson similar to the Standard Model Higgs boson in a wide variety of models, and we will review the expected sizes of deviations from the simplest Standard Model expectations. In Section 2.3, we will review the prospects for measurements on the Higgs boson at the LHC. In Sections 2.4–2.6, we will discuss the capabilities of the ILC to measure properties of the Higgs boson in stages of center of mass energy—250 GeV, 500 GeV, and 1 TeV.

The prospects for the ILC to investigate other possible states of the Higgs sector will be discussed separately in Chapter 6 of this report.

## 2.1 The Higgs mechanism in the Standard Model

We begin by briefly reviewing the Higgs mechanism in the Standard Model (SM). In the SM, electroweak symmetry is broken by an SU(2)-doublet scalar field,

$$\Phi = \begin{pmatrix} G^+ \\ (h+v)/\sqrt{2} + iG^0/\sqrt{2} \end{pmatrix}. \tag{2.1}$$

Here $h$ is the physical SM Higgs boson and $G^+$ and $G^0$ are the Goldstone bosons eaten by the $W^+$ and $Z$. Electroweak symmetry breaking is caused by the Higgs potential, the most general gauge-invariant renormalizable form of which is,

$$V = \mu^2 \Phi^\dagger \Phi + \lambda \left(\Phi^\dagger \Phi\right)^2. \tag{2.2}$$

A negative value of $\mu^2$ leads to a minimum away from zero field value, causing electroweak symmetry breaking. Minimizing the potential, the Higgs vacuum expectation value (vev) and the physical Higgs mass are

$$v^2 = -\mu^2/\lambda \simeq (246\ \mathrm{GeV})^2, \quad m_h^2 = 2\lambda v^2 = 2|\mu^2|. \tag{2.3}$$

For $m_h \sim 125$ GeV, we have a weakly coupled theory with $\lambda \sim 1/8$ and $|\mu^2| \sim m_W^2$. The potential also gives rise to triple and quartic interactions of $h$, with Feynman rules given by

$$hhh: \ -6i\lambda v = -3i\frac{m_h^2}{v}, \quad hhhh: \ -6i\lambda = -3i\frac{m_h^2}{v^2}. \tag{2.4}$$

The couplings of the physical Higgs boson to other SM particles are predicted entirely in terms of $v$ and the known particle masses via the SM Higgs mass generation mechanism. The couplings of the $W$ and $Z$ bosons to the Higgs arise from the gauge-kinetic terms,

$$\mathcal{L} \supset (\mathcal{D}^\mu \Phi)^\dagger (\mathcal{D}_\mu \Phi), \quad \mathcal{D}_\mu = \partial_\mu - igA_\mu^a T^a - ig' B_\mu Y, \tag{2.5}$$

where $g$ and $g'$ are the SU(2)$_L$ and U(1)$_Y$ gauge couplings, respectively, and the hypercharge of the Higgs doublet is $Y = 1/2$. This gives rise to the $W$ and $Z$ masses,

$$m_W = g\,\frac{v}{2}, \quad m_Z = \sqrt{g^2 + g'^2}\,\frac{v}{2}, \tag{2.6}$$

and couplings to the Higgs given by

$$\begin{aligned} W_\mu^+ W_\nu^- h: \ i\frac{g^2 v}{2} g_{\mu\nu} &= 2i\frac{M_W^2}{v} g_{\mu\nu}, \qquad & W_\mu^+ W_\nu^- hh: \ i\frac{g^2}{2} g_{\mu\nu} &= 2i\frac{M_W^2}{v^2} g_{\mu\nu}, \\ Z_\mu Z_\nu h: \ i\frac{(g^2+g'^2)v}{2} g_{\mu\nu} &= 2i\frac{M_Z^2}{v} g_{\mu\nu}, \qquad & Z_\mu Z_\nu hh: \ i\frac{(g^2+g'^2)}{2} g_{\mu\nu} &= 2i\frac{M_Z^2}{v^2} g_{\mu\nu}. \end{aligned} \tag{2.7}$$

**Figure 2.1** Branching fractions of the Standard Model Higgs as a function of the Higgs mass.

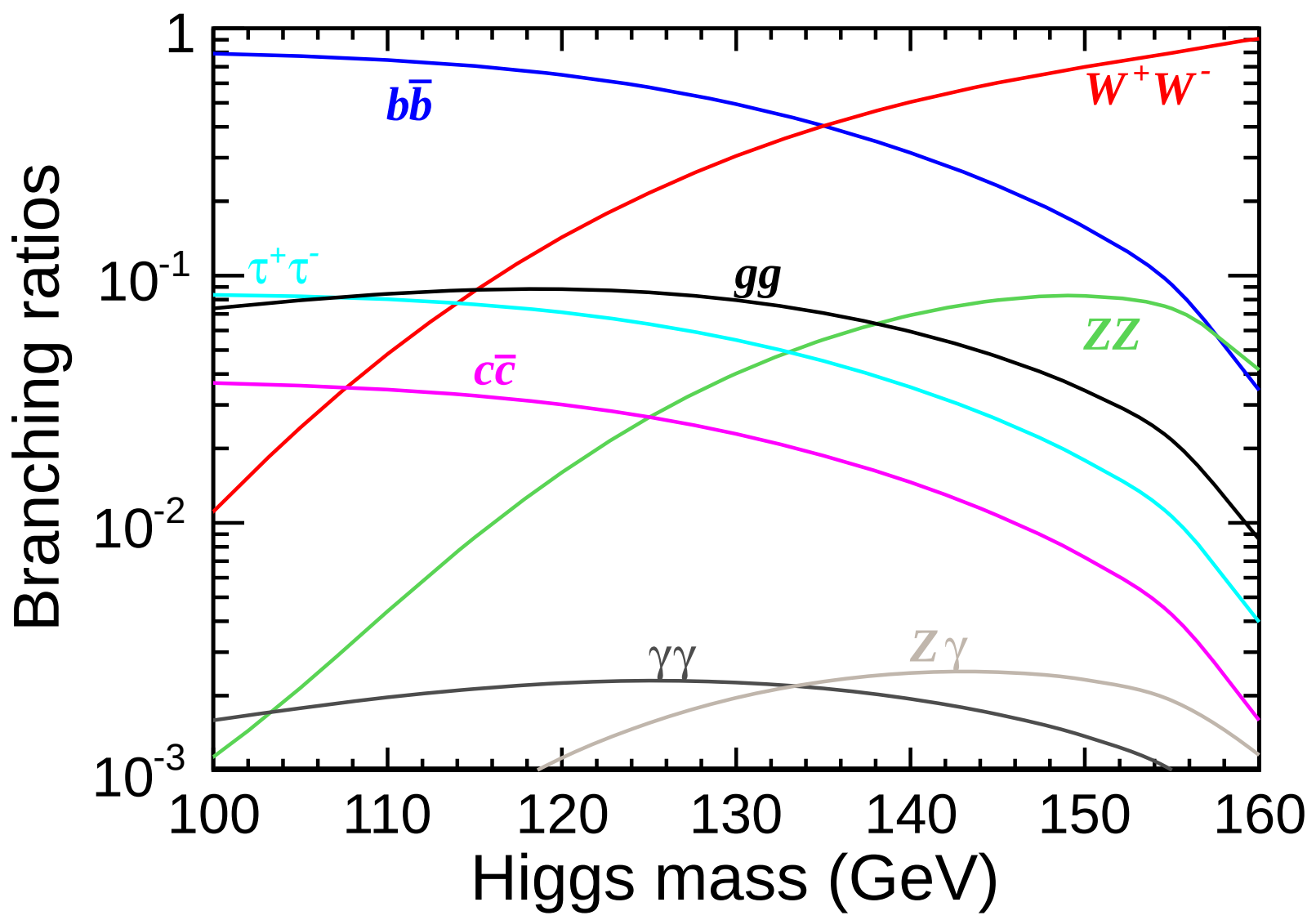

The photon remains massless and has no tree-level coupling to the Higgs.

The couplings of the quarks and charged leptons to the Higgs arise from the Yukawa terms,

$$\mathcal{L} \supset -y^u_{ij}\overline{u}_{Ri}\tilde{\Phi}^\dagger Q_{Lj} - y^d_{ij}\overline{d}_{Ri}\Phi^\dagger Q_{Lj} - y^\ell_{ij}\overline{\ell}_{Ri}\Phi^\dagger L_{Lj} + \text{h.c.}, \tag{2.8}$$

where $Q_L = (u_L, d_L)^T$, $L_L = (\nu_L, e_L)^T$, $\tilde{\Phi} = i\sigma^2\Phi^*$ is the conjugate Higgs doublet, and $y^u$, $y^d$, and $y^\ell$ are $3\times 3$ Yukawa coupling matrices for the up-type quarks, down-type quarks, and charged leptons, respectively. The Yukawa matrices can be eliminated in favor of the fermion masses, yielding Higgs couplings to fermions proportional to the fermion mass,

$$h\overline{f}f : \; -i\,\frac{y_f}{\sqrt{2}} = -i\,\frac{m_f}{v}, \tag{2.9}$$

where $y_f v/\sqrt{2} = m_f$ is the relevant fermion mass eigenvalue.

Thus we see that, in the SM, all the couplings of the Higgs are predicted with no free parameters once the Higgs mass is known. This allows the Higgs production cross sections and decay branching ratios to be unambiguously predicted. The key regularity is that each Higgs coupling is proportional to the mass of the corresponding particle. One-loop diagrams provide additional couplings and decay modes to $gg$, $\gamma\gamma$, and $\gamma Z$. In the SM, the Higgs coupling to $gg$ arises mainly from the one-loop diagram involving a top quark. The Higgs couplings to $\gamma\gamma$ and $\gamma Z$ arise at the one-loop level mainly from diagrams with $W$ bosons and top quarks in the loop.

The Higgs boson branching fractions in the Standard Model are currently predicted with relatively small errors, of the order of 5% in most cases. A current assessment is given in [5]. These errors may be improved to below the 1% level in the era in which the ILC experiments will run. A case of particular interest is the partial width for $h \to b\bar{b}$ The uncertainty estimated in [5] is 3%. The estimate is dominated by errors of order 1% on the measured values of $m_b$ and $\alpha_s$ and by errors from missing electroweak radiative corrections at NLO. The uncertainty from truncation of the QCD perturbation series is much smaller, 0.2%, since this series is known to N$^4$LO from a heroic calculation by Baikov, Chetyrkin, and Kühn [6]. The $\overline{MS}$ bottom quark mass is already known better than the estimate used in [5]. QCD sum rules [7] and lattice gauge theory calculations [8] give consistent estimates with errors below 0.6%. The same papers also give consistent estimates of the $\overline{MS}$ charm quark mass at 3 GeV, with an error on the lattice side of 0.6%. The lattice results, based on our precise knowledge of the heavy quark meson masses, are improvable. Electroweak radiative correction calculations to NNLO are within the state of the art. The value of $\alpha_s$ will be known with an error well below 1% from event shape measurments at the ILC. In all, we expect that the theoretical error on $\Gamma(h \to b\bar{b})$ will be below 1% in the era when the ILC measurements on the Higgs boson are ready for interpretation. Most of the considerations of this paragraph apply also to the partial width for $\Gamma(h \to c\bar{c})$. In particular, the same papers cited above also give consistent estimates of the $\overline{MS}$ charm quark mass at 3 GeV, with an error on the lattice side of 0.6%. We expect that the total theoretical error on this quantity can be brought down close to 1%.

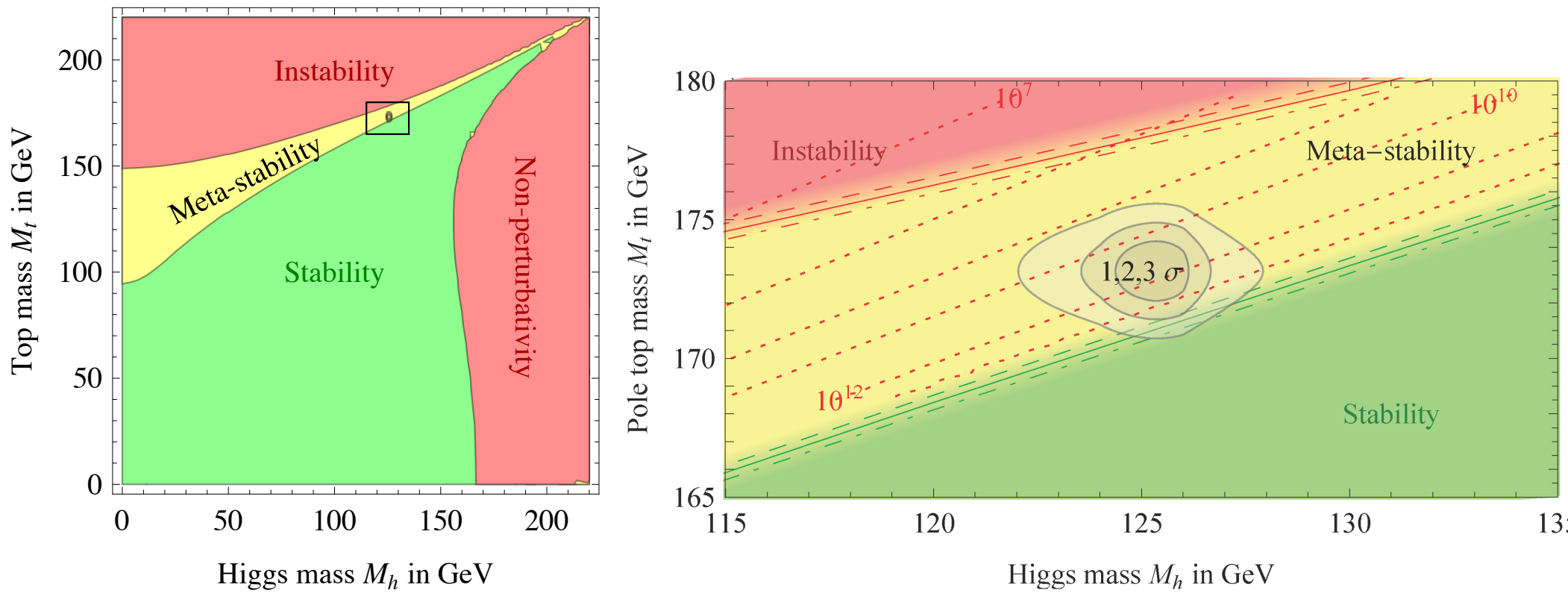

**Figure 2.2.** Regions of stability and instability for the Higgs potential of the Standard Model, in the plane of $m_h$ vs. $m_t$, from [14]. The right-hand figure show the 1, 2, and 3 $\sigma$ contours corresponding to the currently preferred values of the Higgs boson and top quark masses.

Figure 2.1 plots the branching fraction of the Standard Model Higgs boson as a function of the Higgs mass. The figure tells us that the Higgs boson mass $m_h \simeq 125$ GeV provides a very favorable situation in which a large number of decay modes have similar sizes and are accessible to experiments that provide a large Higgs event sample. The ILC, including its eventual 1 TeV stage, will allow measurement of the Higgs boson couplings to $W$, $Z$, $b$, $c$, $\tau$, and $\mu$, plus the loop-induced couplings to $gg$, $\gamma\gamma$, and $\gamma Z$. The regularity of the SM that the Higgs couplings are precisely proportional to mass can thus be verified or refuted through measurements of many couplings spanning a large dynamic range.

A deviation of any of the tree-level Higgs boson couplings to $WW$, $ZZ$, or SM fermions indicates that additional new physics—either additional Higgs bosons or electroweak symmetry-breaking strong dynamics—is needed to generate the full masses of these particles and to assure good behavior of the associated scattering amplitudes in the high-energy limit [9, 10].

The Higgs potential in the Standard Model has another very unusual feature to which we call attention. These remarks apply specifically to the situation in which there is no new physics close to the TeV energy scale. In that case, for large values of the top quark Yukawa coupling $y_t$, renormalization group running to small distance scales drives the Higgs coupling $\lambda$ negative and creates an instability of the Higgs potential [11]. If the low-energy value of $\lambda$ is large enough, the Higgs potential is stable for all values of $\langle\Phi\rangle$ below the Planck scale. However, it turns out that the currently measured value of the top mass is too high to guarantee stability for a Higgs boson mass value of 125 GeV. The minimum of the Higgs boson potential corresponding to the Standard Model might still be metastable for times longer than the age of the universe [12].

The stability region of the Standard Model, in relation to the current value of the top quark mass and the value near 125 GeV for the Higgs boson mass, is shown in Fig. 2.2 [13, 14]. There is a small sliver of the $(m_h, m_t)$ plane in which the Higgs potential is metastable, and the currently preferred values lie in that region. However, as shown in the inset, these values are plausibly consistent with eternal stability of the Standard Model. If the Standard Model turns out to be correct, we will need to know the value of the top quark mass very precisely to understand its eventual fate. As we will discuss in Section 5.1, it is not clear that hadron collider experiments will provide a substantially improved measurement of the top quark mass. Only the ILC, which will measure the top quark mass to an accuracy of about 100 MeV, will have the power to resolve the question of the ultimate fate of the Standard Model vacuum.

## 2.2 Higgs coupling deviations from new physics

### 2.2.1 The Decoupling Limit

In this section, we will discuss possible modifications of the Higgs boson couplings that might be searched for in precision Higgs experiments. It is a general property in many class of models of new physics beyond the Standard Model that they contain a light scalar field, elementary or effective, whose vacuum expectation value is the main source of electroweak symmetry breaking. It is possible that this particle can look very different from the Standard Model Higgs boson. At the moment, there is much interest in this question, stimulated by the values of the first measured Higgs production rates. Models predicting such large deviations can be found in [15–18] and other recent theoretical papers. If it turns out that the new boson has couplings very different from the Standard Model predictions, it will of course be important to measure those couplings as accurately as possible.

However, it is much more common that the lightest Higgs boson of new physics models has coupling that differ at most at the 5-10% level from the Standard Model expectations. This point was made recently through the study of a number of examples by Gupta, Rzehak, and Wells [19]; we will provide some additional examples here. A future program of Higgs physics must acknowledge this point and strive for the level of accuracy that is actually called for in these models.

The logic of this prediction is expressed by the Decoupling Limit of Higgs models described by Haber in [20]. Consider a model with many new particles, in which all of these new particles are heavy while an $SU(2)$ doublet of scalars has a relatively small mass parameter. There are many reasons why the mass parameter of the doublet might be smaller than the typical mass scale of new particles. It might be driven small by renormalization group running, as happens in supersymmetry; it might be suppressed because the scalar is a pseudo-Goldstone boson, as happens in Little Higgs models. In any event, if there is separation between the masses of other new particles and the mass parameter of the scalar doublet, we can integrate out the heavy particles and write an effective Lagrangian for the light doublet. The resulting effective theory is precisely the Standard Model, plus possible higher-dimension operators. If the light doublet acquires a vev, its physical degree of freedom is an effective Higgs particle, with precisely the properties of the Standard Model Higgs up to the effects of the higher-dimension operators. These effects are then required to be of the order of

$$m_h^2/M^2 \quad \text{or} \quad m_t^2/M^2 \ , \tag{2.10}$$

where $M$ is the mass scale of the new particles. The following sections will give quantitative examples of Higgs coupling deviations that follow this systematic dependence.

### 2.2.2 Additional Higgs bosons

If there is one doublet of Higgs field that breaks the electroweak gauge symmetry, there could well be more. Models that give mechanisms of electroweak symmetry breaking often require more than one Higgs field doublet. A prominent example is supersymmetry, which requires one Higgs doublet to give mass to the up-type fermions and a different Higgs doublet to give mass to the down-type fermions. Any enlargement of the Higgs sector has visible effects on the couplings of the lightest Higgs boson.

We can explore this in the case of the model with two Higgs doublets. Both doublets contribute to the $W$ and $Z$ masses. If fermions acquire masses from one or the other doublet, their couplings to the lightest Higgs boson are modified according to the Higgs sector mixing angles $\alpha$ and $\beta$. For the Higgs structure of the Minimal Supersymmetric Standard Model (MSSM), the couplings of the light SM-like Higgs boson $h$ are modified at tree level to

$$\begin{aligned} \frac{g_{hVV}}{g_{h_{\rm SM}VV}} &= \sin(\beta-\alpha) \\ \frac{g_{htt}}{g_{h_{\rm SM}tt}} = \frac{g_{hcc}}{g_{h_{\rm SM}cc}} &= \sin(\beta-\alpha)+\cot\beta\cos(\beta-\alpha) \\ \frac{g_{hbb}}{g_{h_{\rm SM}bb}} = \frac{g_{h\tau\tau}}{g_{h_{\rm SM}\tau\tau}} &= \sin(\beta-\alpha)-\tan\beta\cos(\beta-\alpha). \end{aligned} \tag{2.11}$$

The constrained form of the MSSM Higgs potential lets us express the couplings in terms of the mass $M_A$ of the CP-odd Higgs boson $A^0$ (for large $M_A$, the other Higgs states $H^0$ and $H^\pm$ are nearly degenerate with $A^0$). For $\tan\beta$ larger than a few, this yields [29]

$$\begin{aligned} \frac{g_{hVV}}{g_{h_{\rm SM}VV}} &\simeq 1-\frac{2c^2m_Z^4\cot^2\beta}{m_A^4} \\ \frac{g_{htt}}{g_{h_{\rm SM}tt}} = \frac{g_{hcc}}{g_{h_{\rm SM}cc}} &\simeq 1-\frac{2cm_Z^2\cot^2\beta}{m_A^2} \\ \frac{g_{hbb}}{g_{h_{\rm SM}bb}} = \frac{g_{h\tau\tau}}{g_{h_{\rm SM}\tau\tau}} &\simeq 1+\frac{2cm_Z^2}{m_A^2}, \end{aligned} \tag{2.12}$$

where the coefficient $c$ denotes the SUSY radiative corrections to the CP-even Higgs mass matrix.

We will review the LHC capabilities for detecting the heavy Higgs states in Section 6. The reach depends strongly on $\tan\beta$, but for moderate values of $\tan\beta$ it will be very difficult for the LHC to observe these states if their masses are 200 GeV. If we choose this value as a reference point, then, for $\tan\beta = 5$ and taking $c \simeq 1$, the $h^0$ couplings are approximately given by

$$\begin{aligned} \frac{g_{hVV}}{g_{h_{\rm SM}VV}} &\simeq 1-0.3\%\left(\frac{200\ \rm GeV}{m_A}\right)^4 \\ \frac{g_{htt}}{g_{h_{\rm SM}tt}} = \frac{g_{hcc}}{g_{h_{\rm SM}cc}} &\simeq 1-1.7\%\left(\frac{200\ \rm GeV}{m_A}\right)^2 \\ \frac{g_{hbb}}{g_{h_{\rm SM}bb}} = \frac{g_{h\tau\tau}}{g_{h_{\rm SM}\tau\tau}} &\simeq 1+40\%\left(\frac{200\ \rm GeV}{m_A}\right)^2. \end{aligned} \tag{2.13}$$

At the lower end of the range, the LHC experiments should see the deviation in the $hbb$ or $h\tau\tau$ coupling. However, the heavy MSSM Higgs bosons can easily be as heavy as 1 TeV without fine tuning of parameters. In this case, the deviations of the gauge and up-type fermion couplings are well below the percent level, while those of the Higgs couplings to $b$ and $\tau$ are at the percent level,

$$\frac{g_{hbb}}{g_{h_{\rm SM}bb}} = \frac{g_{h\tau\tau}}{g_{h_{\rm SM}\tau\tau}} \simeq 1+1.7\%\left(\frac{1\ \rm TeV}{m_A}\right)^2. \tag{2.14}$$

In this large-$m_A$ region of parameter space, vertex corrections from SUSY particles are typically also at the percent level.

More general two-Higgs-doublet models follow a similar pattern, with the largest deviation appearing in the Higgs coupling to fermions that get their mass from the Higgs doublet with the smaller vev. The decoupling with $m_A$ in fact follows the same quantitative pattern so long as the dimensionless couplings in the Higgs potential are not larger than $\mathcal{O}(g^2)$, where $g$ is the weak gauge coupling.

### 2.2.3 New states to solve the gauge hierarchy problem

Many models of new physics are proposed to solve the *gauge hierarchy problem* by removing the quadratic divergences in the loop corrections to the Higgs field mass term $\mu^2$. Supersymmetry and Little Higgs models provide examples. Such models require new scalar or fermionic particles with masses below a few TeV that cancel the divergent loop contributions to $\mu^2$ from the top quark. For this to work, the couplings of the new states to the Higgs must be tightly constrained in terms of the top quark Yukawa coupling. Usually the new states have the same electric and color charge as the top quark, which implies that they will contribute to the loop-induced $hgg$ and $h\gamma\gamma$ couplings. The new loop corrections contribute coherently with the Standard Model loop diagrams.

For new scalar particles (*e.g.*, the two top squarks in the MSSM), the resulting effective $hgg$ and $h\gamma\gamma$ couplings are given by

$$\begin{aligned} g_{hgg} &\propto \left| F_{1/2}(m_t) + \frac{2m_t^2}{m_T^2} F_0(m_T) \right| , \\ g_{h\gamma\gamma} &\propto \left| F_1(m_W) + \frac{4}{3} F_{1/2}(m_t) + \frac{4}{3}\frac{2m_t^2}{m_T^2} F_0(m_T) \right| . \end{aligned} \tag{2.15}$$

Here $F_1$, $F_{1/2}$, and $F_0$ are the loop factors defined in [21] for spin 1, spin 1/2, and spin 0 particles in the loop, and $m_T$ is the mass of the new particle(s) that cancels the top loop divergence. For application to the MSSM, we have set the two top squark masses equal for simplicity. For new fermionic particles (*e.g.*, the top-partner in Little Higgs models), the resulting effective couplings are

$$\begin{aligned} g_{hgg} &\propto \left| F_{1/2}(m_t) + \frac{m_t^2}{m_T^2} F_{1/2}(m_T) \right| , \\ g_{h\gamma\gamma} &\propto \left| F_1(m_W) + \frac{4}{3} F_{1/2}(m_t) + \frac{4}{3}\frac{m_t^2}{m_T^2} F_{1/2}(m_T) \right| . \end{aligned} \tag{2.16}$$

For simplicity, we have ignored the mixing between the top quark and its partner. For $m_h = 120$–130 GeV, the loop factors are given numerically by $F_1(m_W) = 8.2$–$8.5$ and $F_{1/2}(m_t) = -1.4$. For $m_T \gg m_h$, the loop factors tend to constant values, $F_{1/2}(m_T) \to -4/3$ and $F_0(m_T) \to -1/3$.

Very generally, then, such models predict deviations of the loop-induced Higgs couplings from top-partners of the decoupling form. Numerically, for a scalar top-partner,

$$\frac{g_{hgg}}{g_{h_{\rm SM}gg}} \simeq 1 + 1.4\% \left(\frac{1\ \mathrm{TeV}}{m_T}\right)^2 , \qquad \frac{g_{h\gamma\gamma}}{g_{h_{\rm SM}\gamma\gamma}} \simeq 1 - 0.4\% \left(\frac{1\ \mathrm{TeV}}{m_T}\right)^2 , \tag{2.17}$$

and for a fermionic top-partner,

$$\frac{g_{hgg}}{g_{h_{\rm SM}gg}} \simeq 1 + 2.9\% \left(\frac{1\ \mathrm{TeV}}{m_T}\right)^2 , \qquad \frac{g_{h\gamma\gamma}}{g_{h_{\rm SM}\gamma\gamma}} \simeq 1 - 0.8\% \left(\frac{1\ \mathrm{TeV}}{m_T}\right)^2 . \tag{2.18}$$

A "natural" solution to the hierarchy problem that avoids fine tuning of the Higgs mass parameter thus generically predicts deviations in the $hgg$ and $h\gamma\gamma$ couplings at the few percent level due solely to loop contributions from the top-partners. These effective couplings are typically also modified by shifts in the tree-level couplings of $h$ to $t\bar{t}$ and $WW$.

The Littlest Higgs model [22, 23] gives a concrete example. In this model, the one-loop Higgs mass quadratic divergences from top, gauge, and Higgs loops are cancelled by loop diagrams involving a new vector-like fermionic top-partner, new $W'$ and $Z'$ gauge bosons, and a triplet scalar. For a top-partner mass of 1 TeV, the new particles in the loop together with tree-level coupling modifications combine to give [24]

$$\begin{aligned} \frac{g_{hgg}}{g_{h_{\rm SM}gg}} &= 1-(5\% \sim 9\%) \\ \frac{g_{h\gamma\gamma}}{g_{h_{\rm SM}\gamma\gamma}} &= 1-(5\% \sim 6\%), \end{aligned} \tag{2.19}$$

where the ranges correspond to varying the gauge- and Higgs-sector model parameters. Note that the Higgs coupling to $\gamma\gamma$ is also affected by the heavy $W'$ and triplet scalars running in the loop. The tree-level Higgs couplings to $t\bar{t}$ and $WW$ are also modified by the higher-dimension operators arising from the nonlinear sigma model structure of the theory.

### 2.2.4 Composite Higgs

Another approach to solve the hierarchy problem makes the Higgs a composite bound state of fundamental fermions with a compositeness scale around the TeV scale. Such models generically predict deviations in the Higgs couplings compared to the SM due to higher-dimension operators involving the Higgs suppressed by the compositeness scale. This leads to Higgs couplings to gauge bosons and fermions of order

$$\frac{g_{hxx}}{g_{h_{\rm SM}xx}} \simeq 1 \pm \mathcal{O}(v^2/f^2), \tag{2.20}$$

where $f$ is the compositeness scale.

As an example, the Minimal Composite Higgs model [25] predicts [26]

$$\begin{aligned} a \equiv \frac{g_{hVV}}{g_{h_{\rm SM}VV}} &= \sqrt{1-\xi} \\ c \equiv \frac{g_{hff}}{g_{h_{\rm SM}ff}} &= \begin{cases} \sqrt{1-\xi} & \text{(MCHM4)} \\ (1-2\xi)/\sqrt{1-\xi} & \text{(MCHM5)}, \end{cases} \end{aligned} \tag{2.21}$$

with $\xi = v^2/f^2$. Here MCHM4 refers to the fermion content of the original model of Ref. [25], while MCHM5 refers to an alternate fermion embedding [27]. Again, naturalness favors $f \sim$ TeV, leading to

$$\begin{aligned} \frac{g_{hVV}}{g_{h_{\rm SM}VV}} &\simeq 1-3\%(1\ \text{TeV}/f)^2 \\ \frac{g_{hff}}{g_{h_{\rm SM}ff}} &\simeq \begin{cases} 1-3\%(1\ \text{TeV}/f)^2 & \text{(MCHM4)} \\ 1-9\%(1\ \text{TeV}/f)^2 & \text{(MCHM5)}. \end{cases} \end{aligned} \tag{2.22}$$

### 2.2.5 Mixing of the Higgs with an electroweak-singlet scalar

If the SM Higgs mixes with an electroweak-singlet scalar, all Higgs couplings become modified by the same factor,

$$\frac{g_{hVV}}{g_{h_{\rm SM}VV}} = \frac{g_{hff}}{g_{h_{\rm SM}ff}} = \cos\theta \simeq 1 - \frac{\delta^2}{2}, \tag{2.23}$$

where $h = h_{\rm SM}\cos\theta + S\sin\theta$, $S$ is the singlet, and the last approximation holds when $\delta \equiv \sin\theta \ll 1$. The orthogonal state, $H = -H_{\rm SM}\sin\theta + S\cos\theta$, has couplings to SM particles proportional to $-\sin\theta$.

When $H$ is heavy, the size of $\sin\theta$ is constrained by precision electroweak data (assuming no cancellations due to other BSM physics). At one loop, the contributions to the $T$ parameter from $h$ and $H$ are given by [19]

$$T = T_{\rm SM}(m_h)\cos^2\theta + T_{\rm SM}(m_H)\sin^2\theta, \tag{2.24}$$

where $T_{\rm SM}(m)$ refers to the SM $T$ parameter evaluated at a Higgs mass $m$. The same form holds for the $S$ parameter. Large $m_H$ is therefore only consistent with precision electroweak constraints for small $\sin\theta$; for example, for $m_H = 1$ TeV, Ref. [19] finds $\sin^2\theta \leq 0.12$, corresponding to $g_{hxx}/g_{H_{\rm SM}xx} \simeq 1 - 6\%$.

Similar effects follow from mixing of the SM Higgs with a radion in Randall-Sundrum models or a dilaton in models with conformally-invariant strong dynamics. The couplings of a radion or dilaton to SM particles are suppressed by a factor $v/f$ compared to those of the SM Higgs, where $f$ is the scale of the warped or conformal dynamics. The couplings of the mass eigenstate $h = H_{\rm SM}\cos\theta + \chi\sin\theta$ are modified according to

$$\frac{g_{hVV}}{g_{H_{\rm SM}VV}} = \frac{g_{hff}}{g_{H_{\rm SM}ff}} = \cos\theta + \frac{v}{f}\sin\theta \simeq 1 - \frac{\delta^2}{2} + \frac{v}{f}\delta. \tag{2.25}$$

For $f \simeq 1$ TeV and $\sin^2\theta$ as above, this corresponds to $g_{hxx}/g_{H_{\rm SM}xx} \simeq 1 - 6\% \pm 8.5\%$, where we allow for either sign of $\delta$.

### 2.2.6 The case of supersymmetry

The MSSM contains a mixture of effects discussed in the previous sections. It has an extended Higgs sector, affecting the tree level couplings to the lightest Higgs boson $h$, and it also introduces new particles, the top squarks, gauginos, and Higgsinos, whose loops cancel the quadratic divergences in the Higgs field mass term.

Supersymmetry is described by a large parameter space with many options for the form of the new particle spectrum. We will discuss this parameter space in some detail in Chapter 7. Here, we will give some examples of the effects that might be expected in the Higgs boson couplings.

We have already pointed out that the parameter space of the MSSM contains scenarios that give order 1 corrections to the Higgs boson couplings; examples are given in [15]. A more typical situation with heavy superparticle masses is given by the $m_h^{\rm max}$ benchmark scenario studied in [28, 29], with $m_A = 1$ TeV, $\tan\beta = 5$. This parameter set yields masses for the two top squarks of 857 GeV and 1200 GeV. We compute the Higgs couplings using `HDECAY4.43` [30]. The Higgs couplings to $gg$ and $\gamma\gamma$ are modified mainly by the loop effects from the new particles, to give

$$\begin{aligned} \frac{g_{hgg}}{g_{h_{\rm SM}gg}} &= 1 - 2.7\% \\ \frac{g_{h\gamma\gamma}}{g_{h_{\rm SM}\gamma\gamma}} &= 1 + 0.2\%, \end{aligned} \tag{2.26}$$

These estimates include the effect on the $\gamma\gamma$ coupling of charginos in the loop, since the lightest chargino mass is 201 GeV in this benchmark scenario, and the modification of the tree-level $ht\bar{t}$ coupling due to the presence of the second Higgs doublet. The couplings to massive vector bosons and to $c\bar{c}$ and $\tau^+\tau^-$ come mainly from the modification of the tree-level couplings. One finds

$$\begin{aligned} \frac{g_{hVV}}{g_{h_{\rm SM}VV}} &= 1 - \mathcal{O}(10^{-4}), \qquad \frac{g_{hcc}}{g_{h_{\rm SM}cc}} = 1 - 0.3\% \\ & \frac{g_{h\tau\tau}}{g_{h_{\rm SM}\tau\tau}} = 1 + 2.5\%. \end{aligned} \tag{2.27}$$

Finally, the $hb\bar{b}$ coupling receives corrections both from this source and from a loop effect involving

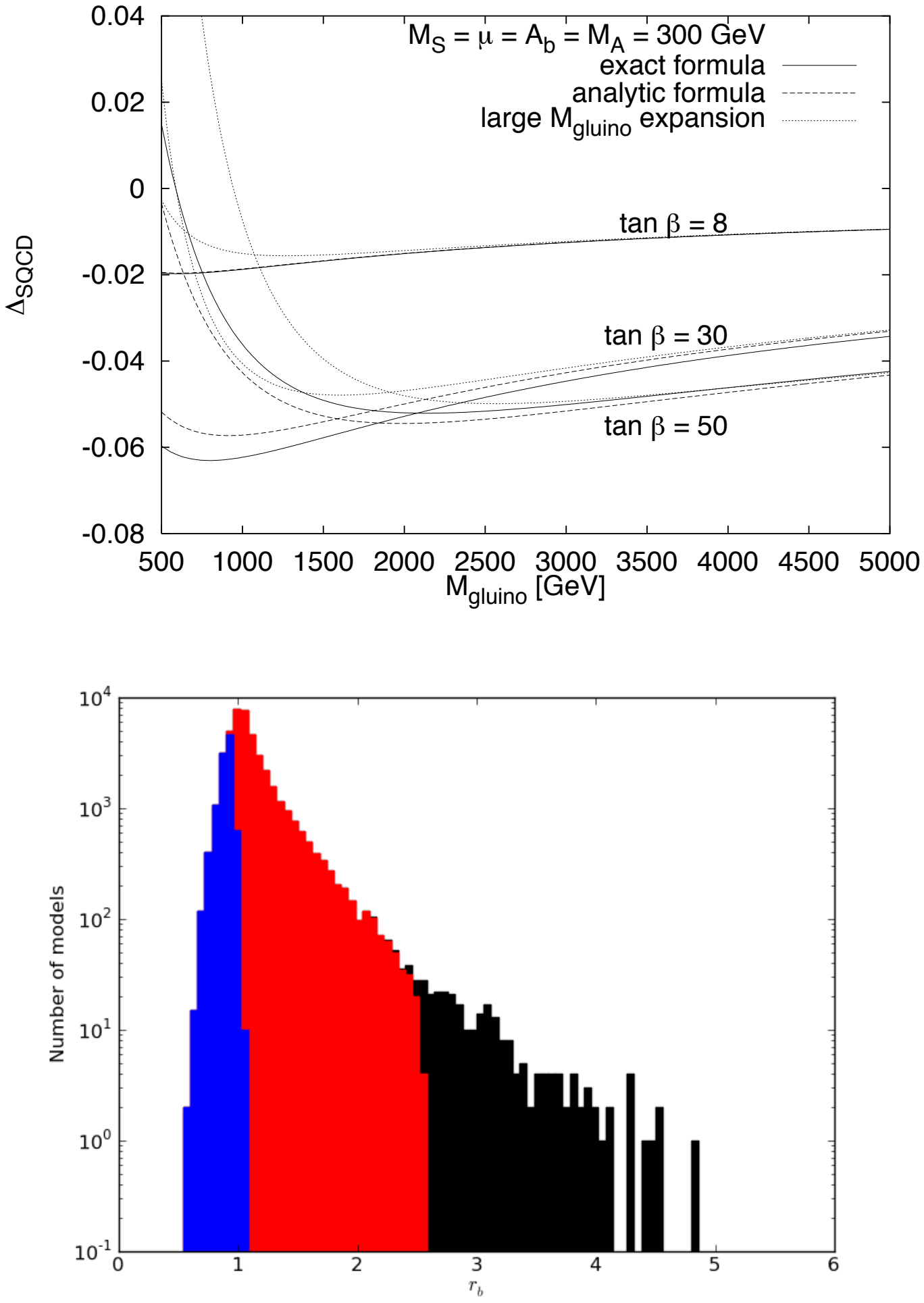

**Figure 2.3** (a) Fractional correction to the $hb\bar{b}$ coupling due to loop diagrams with supersymmetric particles in the MSSM, from [31], as a function of the mass of the gluino. (b) Values of $r_b$, the ratio of $\Gamma(h \to b\bar{b})$ to its Standard Model value, in a large set of MSSM models randomly generated in a 19-dimensional model space and then selected to satisfy all current experimental constraints, from [32].

the supersymmetry partners of the $b$ and $t$ quarks

$$\frac{g_{hbb}}{g_{h_{\rm SM}bb}} = 1 + 3.5\%. \tag{2.28}$$

It is dangerous, though, to view any particular model as typical in a model space as diverse as that of supersymmetry. As was shown already in [31], the loop corrections to the $hb\bar{b}$ vertex, though they formally follow the decoupling law, can be numerically large, especially for large values of $\tan\beta$. Fig. 2.3(a) illustrates this by showing the fractional correction to the $hb\bar{b}$ vertex for three values of $\tan\beta$. Fig. 2.3(b) shows the distribution of

$$r_b = \frac{g^2_{hbb}}{g^2_{h_{\rm SM}bb}} \tag{2.29}$$

in a very large sample of MSSM models satisfying current experimental constraints—including $m_h = 125.0 \pm 2.0$—generated in a 19-parameter supersymmetry parameter space [32]. Decoupling gives many models where $r_b$ is very close to 1, but there are also models with deviations from 1 of all magnitudes that are found as we explore the parameter space. The figure makes clear that $r_b$ is a useful discriminator of new physics models, both at the accuracy of a order-1 measurement and at any successive level of accuracy down to the percent level. Similar conclusions hold for all other coupling deviations, though it is the deviation in the $hb\bar{b}$ coupling that is most sensitive as the superpartner masses become large.

### 2.2.7 Conclusions

Though large deviations are possible in some models, the more general expectation in models of new physics is that a light Higgs boson has couplings to vector bosons, fermions, $gg$, and $\gamma\gamma$ similar to those of the Higgs boson of the Standard Model. Thus, the study of the Higgs boson couplings is likely to require precision measurements. Nevertheless, there are many models in which some of the Higgs couplings have 5-10% discrepancies from their Standard Model values. Discovery of these discrepancies would be an important clue to the nature of new physics at higher mass scales. To recognize these effects, it is important to be able to measure the Higgs boson couplings comprehensively and with high accuracy. We will now discuss how that can be done.

## 2.3 Status and prospects for Higgs measurements at LHC

The ATLAS and CMS experiments have now demonstrated that they have the capability to study the Standard Model Higgs boson. They have presented strong evidence for a scalar particle of mass about 125 GeV that is consistent with the profile of the Standard Model Higgs. The isolation of this signal in the LHC environment is extremely challenging. The strongest signal of the Higgs boson so far observed at the LHC comes in the Higgs decay to $\gamma\gamma$, a process that occurs less than once in $10^{12}$ proton-proton collisions. However, the Tevatron and LHC experiments have proven that they can make measurements of such rare events in the high background conditions of hadron colliders. In this section, we will review how far the LHC experiments are expected to go toward a comprehensive understanding of the Higgs boson in the case in which this particle has the couplings expected in the Standard Model.

### 2.3.1 The LHC Higgs discovery

As of July 2012, ATLAS and CMS presented Higgs results based on integrated luminosities up to 5.1 fb$^{-1}$ at 7 TeV plus 5.9 fb$^{-1}$ at 8 TeV [33, 34]. Each experiment observed an excess in $\gamma\gamma$ with local significance of 4.1–4.5$\sigma$ and an excess in $4\ell$ (consistent with being from $ZZ^*$) with local significance of 3.2–3.4$\sigma$. The signal strengths in these channels are consistent with SM expectations. The LHC experiments made a measurement of the resonance mass in these two final states with the result $125.3 \pm 0.4$ (stat) $\pm 0.5$ (syst) GeV (CMS) and $126.0 \pm 0.4$ (stat) $\pm 0.4$ (syst) GeV (ATLAS).

CMS also presented results including 8 TeV data for the final states $bb$, $\tau\tau$, and $WW$ [34]. ATLAS has presented results including 8 TeV data for the $WW$ final state [35]; results for the other channels are expected soon. These final states have poorer mass resolution than $\gamma\gamma$ and $ZZ^* \to 4\ell$. ATLAS observed an excess in the $WW$ channel at the 3.2$\sigma$ level. CMS saw a modest excess in $WW$ at the 1.5$\sigma$ level and no excess in the $bb$ and $\tau\tau$ channels. The rates in these channels are also broadly consistent with SM expectations.

A summary of the ATLAS and CMS results as of August 2012 have been published in [2, 3].

In addition to inclusive Higgs production, which is dominated in the SM by gluon fusion, the ATLAS and CMS analyses include event selections with enhanced sensitivity to vector boson fusion (VBF) and Higgs production in association with $W$, $Z$, or $t\bar{t}$. As of the fall of 2012, these subdominant production modes have not been conclusively observed.

The Tevatron experiments CDF and DO have also presented evidence for the presence of this particle [4]. The Tevatron search specifically targets the production reactions $q\overline{q} \to h + W, Z$ with the decay $h \to b\bar{b}$. The significance is 2.7 $\sigma$ assuming the resonance mass given by the LHC experiments.

Observation of the Higgs candidate in $\gamma\gamma$ excludes the possibility of the resonance being a spin-1 particle via the Landau-Yang theorem [36]. Observation of a signal in the $ZZ^*$ final state strongly disfavors the possibility that it is a pseudoscalar because in this case the $ZZ$ coupling must be loop-induced; most pseudoscalar models predict a ratio of rates in $ZZ^*$ versus $\gamma\gamma$ much smaller than observed. Prospects for direct LHC measurements of the spin and CP quantum numbers will be discussed below.

### 2.3.2 Prospects for measuring the Higgs mass and quantum numbers at LHC

The mass of the Higgs boson is an intrinsically important parameter of the Standard Model. Moreover, the Higgs mass must be known accurately in order to interpret other measurements in precision Higgs physics. In particular, because the Higgs decay widths to $WW$ and $ZZ$ depends sensitively on $m_h$ below the $WW$ threshold, a precise measurement of the Higgs mass is necessary in order to extract the Higgs couplings from branching ratio measurements. For $m_h = 115$–130 GeV, each 100 MeV of uncertainty in $m_h$ introduces 0.6–0.5% uncertainty in the ratio of the $hb\bar{b}$ and $hWW$ couplings, $g_b/g_W$.

The LHC is expected to make a precision measurement of the mass of the Higgs boson. As of this writing, the LHC experiments have already measured the Higgs mass with an uncertainty of 0.4 GeV (statistical) and 0.4–0.5 GeV (systematic) [33, 34]. Most of the sensitivity to the Higgs mass around 125 GeV comes from the $\gamma\gamma$ channel, with a subleading contribution from the $ZZ^* \to 4\ell$ channel. The ATLAS and CMS experiments estimate that, with large data samples $\sim 300$ fb$^{-1}$, they can determine the Higgs mass in absolute terms to an accuracy of 0.1 GeV [37–39]. Interference of the continuum $gg \to \gamma\gamma$ background with the diphoton signal shifts the peak downward by $\sim 150$ MeV or more [40] and must be taken into account at this level of precision.

The LHC also has excellent prospects to answer the question of the spin and parity of the Higgs boson. The SM Higgs coupling has the special form $hV_\mu V^\mu$, which arises specifically from the kinetic term of a scalar field with a vacumm expectation value that breaks $SU(2) \times U(1)$ symmetry. In contrast, generic loop-induced couplings for a neutral scalar take the form $\phi V_{\mu\nu} V^{\mu\nu}$ for a CP-even scalar, or $\phi V_{\mu\nu} \widetilde{V}^{\mu\nu}$ for a CP-odd scalar, with $\widetilde{V}^{\mu\nu} = \epsilon^{\mu\nu\rho\sigma} V_{\rho\sigma}$. These loop-induced couplings are typically suppressed in size by a factor $\alpha/4\pi$. So, already, the fact that the boson found by ATLAS and CMS is seen in its decay to $ZZ^*$ provides *prima facie* evidence that this boson is a CP even scalar with a vacuum expectation value. The true test of this hypothesis will come in the study of angular correlations in the boson's decays. The study of $h \to ZZ^* \to 4$ leptons is especially powerful [41–43]. The possible structures of couplings can also be distinguished experimentally using angular correlations of the forward tagging jets in weak boson fusion Higgs production or the four final-state fermions in $h \to VV$ decays. For example, the azimuthal angle $\Delta\phi_{jj}$ of the forward tagging jets in weak boson fusion has a fairly flat distribution for the SM $hV_\mu V^\mu$ coupling, while for the CP-even loop-induced vertex the distribution peaks at $\Delta\phi_{jj} \sim 0$, $\pi$ and for the CP-odd vertex it peaks at $\pi/2$, $3\pi/2$ [44–46]. Tests of the Higgs spin from $h \to \gamma\gamma$ decays are discussed in [47, 48].

### 2.3.3 Prospects for determining the Higgs couplings from LHC data

The LHC experiments are in principle sensitive to almost the full range of SM Higgs couplings. The decays to $\gamma\gamma$, $ZZ$ and $WW$ are already seen. The decay to $\tau^+\tau^-$ is expected to be straightforward to observe with luminosity samples of 30 fb$^{-1}$ at 14 TeV. The decay to $b\bar{b}$ and the process $pp \to t\bar{t}h$ should also be observed with similar luminosity samples, although that observation is much less straightfoward. We will discuss the observation of $h \to b\bar{b}$ further below. The LHC observations are sensitive to the $hgg$ coupling because $gg \to h$ is a primary channel for the production of the Higgs boson at the LHC. The only significant decay mode of the SM Higgs boson omitted from this list is $h \to c\bar{c}$, for which there is currently no strategy proposed. However, this is a relatively minor mode, with a branching ratio of about 3% for a Higgs boson of mass 125 GeV. In addition, it is possible to discover or bound invisible modes of Higgs decay by observing the $WW$ fusion production of a Higgs with two forward tagging jets [49].

The large number of measurements of $\sigma \cdot BR$ for the various modes of Higgs production and decay that are available at the LHC brings us very close to the situation in which LHC data can determine the Higgs couplings in a model-independent way. However, some problems remain. One is a genuine gap in the logic that needs to be filled by a model assumption. An observable $\sigma(A\overline{A} \to h) \cdot BR(h \to B\overline{B})$ depends on the Higgs boson couplings through the factor

$$\frac{g^2(hAA)g^2(hBB)}{\Gamma_T} . \tag{2.30}$$

where $\Gamma_T$ is the total width of the Higgs. For a Higgs boson of mass 125 GeV, the total width is expected to be about 4 MeV. Such a small value cannot be measured directly at either hadron-hadron or $e^+e^-$ colliders, so it must be determined by the fit to the collection of $\sigma \cdot BR$ measurements. However, there might always be decay modes of the Higgs boson that are unobservable in the LHC experimental environment. The presence of such modes would increase $\Gamma_T$. Thus, we need to impose a constraint that puts an upper limit on $\Gamma_T$.

A useful constraint comes from the fact that, under rather general conditions [50], that each scalar with a vev makes a positive contribution to the masses of the $W$ and $Z$. Since the Higgs couplings to the $W$ and $Z$ also arise from the vev, this implies that the coupling of any single Higgs field is bounded above by the coupling that would give the full mass of the vector bosons. This implies

$$g^2(hWW) \leq g^2(hWW)|_{SM} \quad \text{and } g^2(hZZ) \leq g^2(hZZ)|_{SM} \tag{2.31}$$

Then the measurement of the $\sigma \cdot BR$ for a process such as $WW$ fusion to $h$ with decay to $WW^*$, which is proportional to $g^4(hWW)/\Gamma_T$, puts an upper limit on $\Gamma_T$. This constraint was first applied to Higgs coupling fitting by Dührssen *et al.* [51]. In the literature, this constraint is sometimes applied together with the relation

$$g^2(hWW)/g^2(hZZ) = \cos^2\theta_w . \tag{2.32}$$

The relation (2.32), however, requires models in which the Higgs is a mixture of $SU(2)$ singlet and doublet fields only, while (2.31) is more general [52].

The application of this constraint solves the problem of principle for the determination of the the absolute strengths of Higgs boson couplings from LHC data. In practice, however, there is another important source of difficulty. A SM Higgs boson of mass 125 GeV has a 60% branching fraction to the final state $b\overline{b}$. Thus, measurements that involve the $b\overline{b}$ final state play a large role in determining the Higgs total width, and any errors in that determination feed back into all Higgs couplings. Unfortunately, it is very difficult to observe decays $h^0 \to b\overline{b}$ at the LHC. The simple argument for this is that the cross section for producing $h^0 \to b\overline{b}$ is of the order of pb while the cross section for producing a pair of $b$ jets at the Higgs boson mass is of the order of $\mu$b. The literature on Higgs boson measurements at the LHC has gone through cycles of optimism and pessimism about the possibility of overcoming this problem. Currently, we are in a state of optimism, due to the observation of Butterworth, Davison, Rubin, and Salam that highly boosted Higgs bosons can be distinguished by recognizing the Higgs as an exotic jet with special internal structure [53]. The Butterworth *et al.* paper discussed the observation of $h \to b\overline{b}$ in the reactions $pp \to W, Z + h$. Plehn, Salam, and Spannowsky have argued that an extension of this technique also allows the study of $pp \to t\overline{t} + h$ with $h \to b\overline{b}$ at the LHC [54]. However, it is one thing to observe these processes and quite another to use them to measure Higgs couplings with high precision. It is not yet understood how to calibrate these methods or what their ultimate systematic errors might be. Further, the selection of particular jet configurations potentially introduces large theoretical errors into the calculation of the relevant cross sections. The uncertainty in the extraction of couplings from these channels propagates back

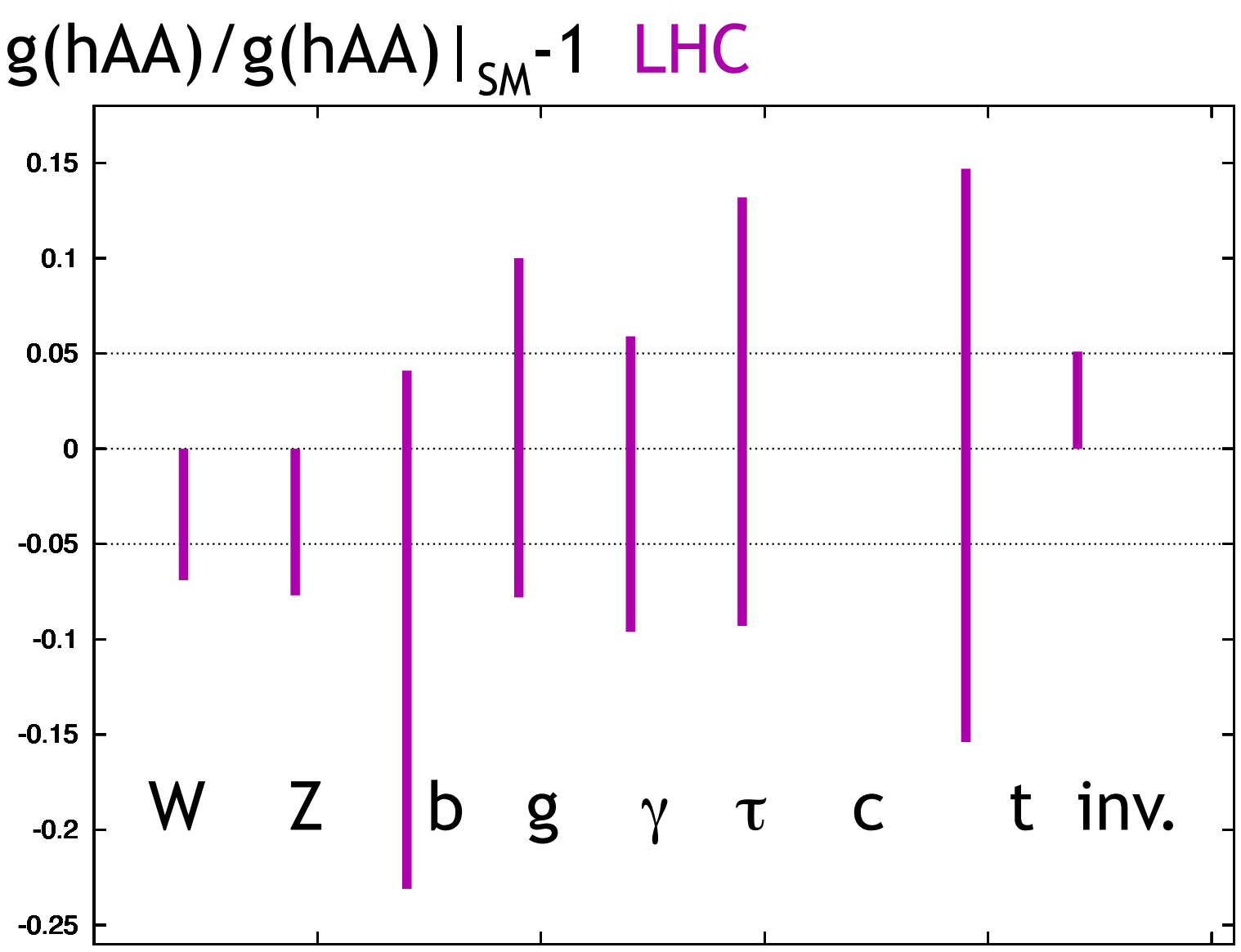

**Figure 2.4.** Estimate of the sensitivity of the LHC experiments to Higgs boson couplings in a model-independent analysis. The plot shows the 1 $\sigma$ confidence intervals for LHC at 14 TeV with 300 fb$^{-1}$ as they emerge from the fit described in the text. Deviation of the central values from zero indicates a bias, which can be corrected for. The upper limit on the $WW$ and $ZZ$ couplings arises from the constraints (2.31). No error is estimated for $g(hcc)$. The bar for the invisible channel gives the 1 $\sigma$ upper limit on the *branching ratio*. The analysis assumes a data set of 300 fb$^{-1}$ with one detector. The methodology leading to this figure is explained in [65].

**Table 2.1.** Expected Higgs self-coupling $1\sigma$ sensitivity limits for $m_h = 120$ GeV, from Refs. [71, 73]. Sensitivity is expressed in terms of $\Delta\lambda_{hhh} \equiv \lambda/\lambda_{\rm SM} - 1$. The $bb\tau\tau$ final state signal cross section is too small to be observed at the 300 fb$^{-1}$ LHC [71].

| | LHC (300 fb$^{-1}$) | SLHC (3000 fb$^{-1}$) |
|---|---|---|
| $4b$ [71] | $-6.8 < \Delta\lambda_{hhh} < 10.1$ | $-3.1 < \Delta\lambda_{hhh} < 6.0$ |
| $bb\tau\tau$ [71] | – | $-1.6 < \Delta\lambda_{hhh} < 3.1$ |
| | LHC (600 fb$^{-1}$) | SLHC (6000 fb$^{-1}$) |
| $bb\gamma\gamma$ [73] | $-0.74 < \Delta\lambda_{hhh} < +0.94$ | $-0.46 < \Delta\lambda_{hhh} < +0.52$ |

into the whole system of couplings determined from LHC data.

Over the years, there have been many attempts to estimate the ultimate sensitivity of the LHC experiments to the Higgs boson couplings. The most complete work on this subject to date is the 2003 Ph.D. thesis of Dührssen [55] and the subsequent analysis of this work with Heinemeyer, Logan, Rainwater, Weiglein, and Zeppenfeld [56]. This work has been updated by the SFitter group in [57, 58] and in the recent paper [59]. Other analysis using stronger model assumptions have been given in [60] and [61]. It is clear from the explanation given in the previous paragraph that any such analysis from before 2010 is excessively optimistic.

We have tried to make our own analysis of the model-independent LHC sensitivity to Higgs couplings, also bringing up to date the estimates in [55] and taking into account new results on the LHC capabilities for Higgs couplings presented by the ATLAS and CMS collaborations in [62–64]. The results are shown in Fig. 2.4. The details of the analysis are given in [65]. For comparison, the most recent estimates from the Sfitter group are shown in Fig. 2.5. The results differ in some details, but they are qualitatively similar. In [64], the CMS collaboration has presented a second scenario with more optimistic projections; however, these are based on the assumption, so far unsupported by simulation work, that systematic errors can be decreased with increasing data sets as $1/\sqrt{N}$, even in the high-luminosity LHC era.

This estimate leads to a quite definite conclusion. The LHC experiments will be able to simultaneously determine the Higgs couplings to Standard Model particle in a way that is, if not completely model-independent, at least relies onlly on the minimal theoretical assumptions described above. These determinations should be accurate enough to confirm or refute the hypothesis that the particle recently observed has the profile of the Standard Model Higgs boson. However, these experiments will not provide sufficient accuracy in the Higgs couplings to test for the deviations expected in new physics models in the Decoupling Limit. That is, they will not be able to access the deviation of Higgs couplings from the Standard Model for most of the effects described in Section 2.2. To reach the level required for this, a stronger tool is needed.

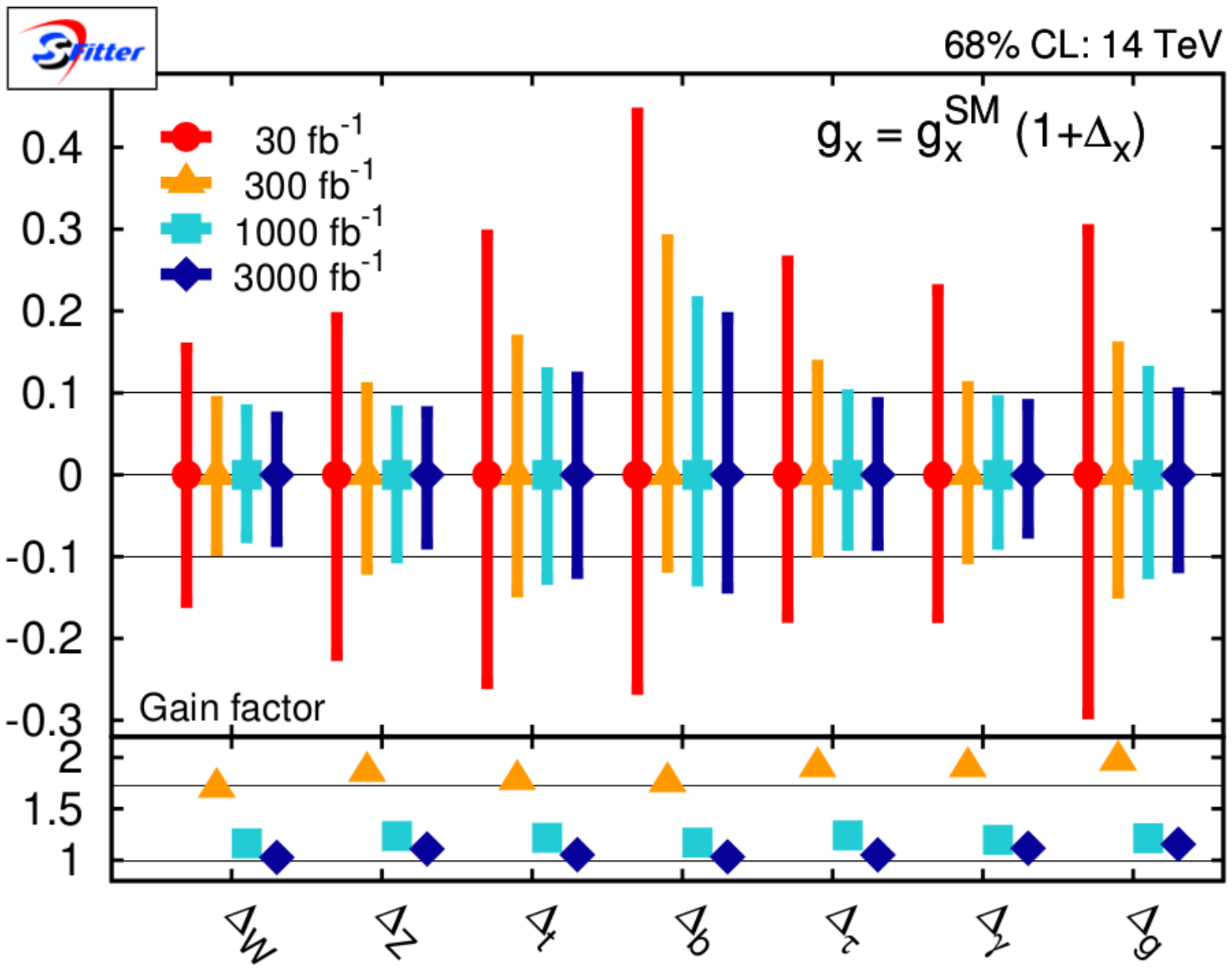

**Figure 2.5.** Estimate of the sensitivity of the LHC experiments to Higgs boson couplings in a model-independent analysis with one detector and varying luminosity sample, from the SFitter group [59].

### 2.3.4 Prospects for measurement of the triple Higgs coupling at the LHC

Measurement of the Higgs quartic coupling parameter $\lambda$ provides a test of the electroweak symmetry breaking mechanism through the structure of the Higgs potential. This coupling can be probed via a measurement of the triple-Higgs vertex, which contributes along with other diagrams to Higgs pair production. This coupling can be significantly modified in models with extended Higgs sectors, in particular in models that increase the strength of the electroweak phase transition to provide viable baryogenesis [66]. For Higgs pair production via $gg \to hh$, low-mass new physics in the loops can rather significantly affect the cross section even if it does not have a large effect on the $gg \to h$ cross section [67, 68].

Measuring the triple Higgs coupling at the LHC is very challenging for a 125 GeV Higgs boson. The largest production cross section is $gg \to hh$, with other potential production modes (VBF $qq \to qqhh$, $q\bar{q} \to Vhh$, and $gg, q\bar{q} \to t\bar{t}hh$) being severely rate-limited. The $4W$ final state has been studied for $m_h > 150$ GeV [69] and was found to be promising for $m_h \simeq 170$–$200$ GeV at the high-luminosity ($10^{35}$ cm$^{-2}$s$^{-1}$) LHC [70]; however, this final state is suppressed by the falling $h \to WW$ branching ratio at lower masses (a factor of $(0.22)^2 = 0.048$ at $m_h = 125$ GeV, compared to 0.92 (0.55) at $m_h = 170$ (200) GeV). This suppression will be compensated somewhat by an enhanced production cross section at lower masses, but no LHC study has been done in the $4W$ final state for a low-mass Higgs.

The $4b$ and $bb\tau\tau$ final states were studied for a 120 GeV Higgs in Ref. [71, 72] and the more promising $bb\gamma\gamma$ final state was studied in Ref. [73]. The expected triple-Higgs coupling sensitivity can be expressed as $\Delta\lambda_{hhh} \equiv \lambda/\lambda_{\rm SM} - 1$, assuming no new particles contribute to the $gg \to h$ and $gg \to hh$ loops. The results, summarized in Table 2.1, indicate that only order-1 sensitivity will be possible.

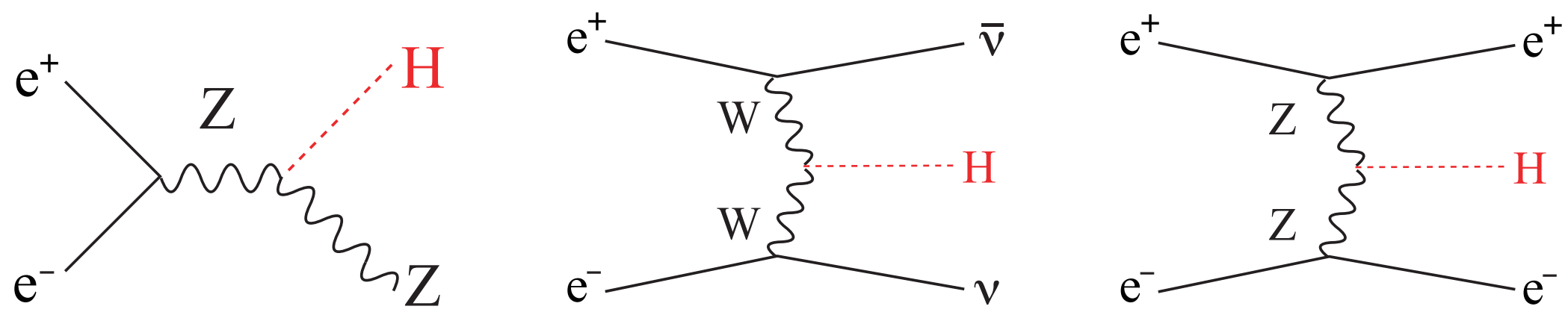

**Figure 2.6.** Feynman diagrams for the three major Higgs production processes at the ILC: $e^+e^- \to Zh$ (left), $e^+e^- \to \nu\bar{\nu}H$ (center), and $e^+e^- \to e^+e^-H$ (right).

The ATLAS submission to the European Strategy Study [62], gives some new results on the measurement of the triple Higgs coupling. The report estimates that, with 3000 fb$^{-1}$ and combining both LHC experiments, "a $\sim$ 30% measurement of $\lambda_{HHH}$ may be achieved". We look forward to the studies, not yet reported, that will support this conclusion.

## 2.4 Higgs measurements at ILC at 250 GeV

The physics program of the LHC should be contrasted with the physics program that becomes available at the ILC. The ILC, being an $e^+e^-$ collider, inherits traditional virtues of past $e^+e^-$ colliders such as LEP and SLC. We have described these in Chapter 1. The ILC offers well defined initial states, a clean environment, and reasonable signal-to-noise ratios even before any selection cuts. Thanks to the clean environment, it can be equipped with very high precision detectors. The experimental technique of Particle Flow Analysis (PFA), described in Volume 4 of this report, offers a qualitative improvement in calorimetry over the detectors of the LEP era and sufficient jet mass resolution to identify $W$ and $Z$ bosons in their hadronic decay modes. Thus, at the ILC, we can effectively reconstruct events in terms of fundamental particles — quarks, leptons, and gauge bosons. Essentially, we will be able to analyze events as viewing Feynman diagrams. By controlling beam polarization, we can even select the Feynman diagrams that participate a particular reaction under study. The Higgs boson can be observed in all important modes, including those with decay to hadronic jets. This is a great advantage over the experiments at the LHC and provides the opportunity to carry out a truly complete set of precision measurements of the properties of the Standard-Model-like Higgs boson candidate found at the LHC.

The precision Higgs program will start at $\sqrt{s} = 250$ GeV with the Higgs-strahlung process, $e^+e^- \to Zh$ (Fig. 2.6 (left)).The production cross section for this process is plotted in Fig. 2.7 as a function of $\sqrt{s}$ together with that for the weak boson fusion processes (Figs. 2.6-(center and right)). We can see that the Higgs-strahlung process attains its maximum at around $\sqrt{s} = 250$ GeV and dominates the fusion processes there. The cross section for the fusion processes increases with the energy and takes over that of the Higgs-strahlung process above $\sqrt{s} \gtrsim 400$ GeV.

The production cross section of the Higgs-strahlung process at $\sqrt{s} \simeq 250$ GeV is substantial for the low mass Standard-Model-like Higgs boson. Its discovery would require only a few fb$^{-1}$ of integrated luminosity. With 250 fb$^{-1}$, about $8. \times 10^4$ Higgs boson events can be collected. Note that, here and in the rest of our discussion, we take advantage of the ILC's positron polarization to increase the Higgs production rate over that expected for unpolarized beams.

The precise determination of the properties of the Higgs boson is one of the main goals of the ILC. Only after this study is completed can we settle the question of whether the new resonance is the Standard Model Higgs boson, a Higgs boson of a more general theory, or a particle of a different origin. Particular important for this question are the values of the Higgs boson mass, $m_h$, and the Higgs production cross sections and branching ratios.

**Figure 2.7** Production cross section for the $e^+e^- \to Zh$ process as a function of the center of mass energy for $m_h = 125$ GeV, plotted together with those for the $WW$ and $ZZ$ fusion processes: $e^+e^- \to \nu\bar{\nu}H$ and $e^+e^- \to e^+e^-H$.

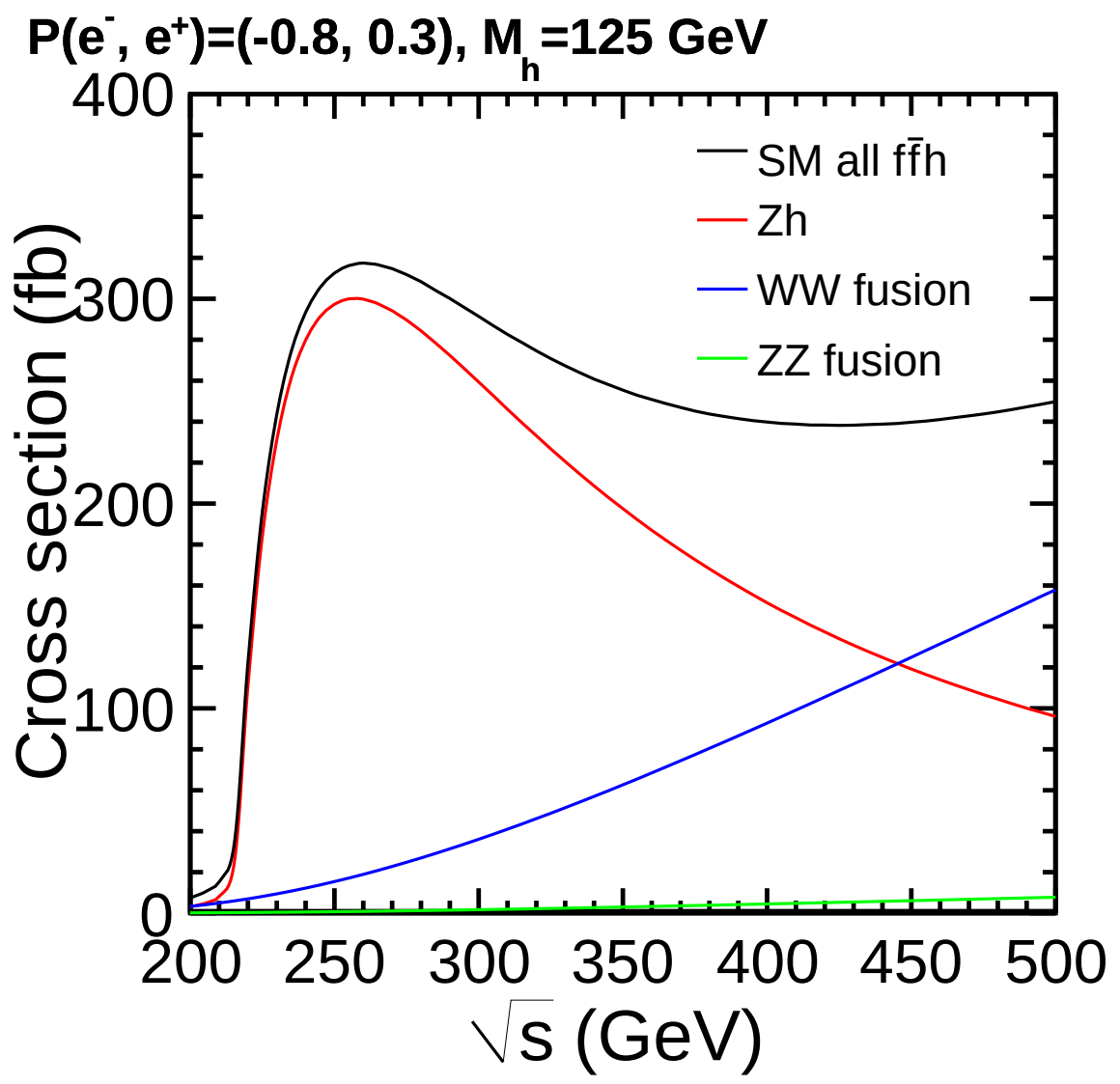

In this section and the following ones, we will present the measurement accuracies for the Higgs boson properties expected from the ILC experiments. These measurement accuracies are estimated from full simulation studies with the ILD and SiD detectors described in the Detector Volume, Volume 4 of this report. Because these full-simulation studies are complex and were begun long before the LHC discovery, the analyses assumed a Higgs boson of mass 120 GeV. In this section and the next two sections, then, all error estimates refer to 120 GeV Higgs boson. In Section 2.7, we will present a table in which our results are extrapolated to measurement accuracies for a 125 GeV Higgs boson, taking into appropriate account the changes in the signal and background levels in these measurements.

### 2.4.1 Mass and quantum numbers

We first turn our attention to the measurements of the mass and spin of the Higgs boson, which are necessary to confirm that the Higgs-like object found at the LHC has the properties expected for the Higgs boson. We have discussed in the previous section that the LHC already offers excellent capabilities to measure the mass and quantum numbers of the Higgs boson. However, the ILC offers new probes of these quantities that are very attractive experimentally. We will review them here.

We first discuss the precision mass measurement of the Higgs boson at the ILC. This measurement can be made particularly cleanly in the process $e^+e^- \to Zh$, with $Z \to \mu^+\mu^-$ and $Z \to e^+e^-$ decays. Here the distribution of the invariant mass recoiling against the reconstructed $Z$ provides a precise measurement of $m_h$, independently of the Higgs decay mode. In particular, the $\mu^+\mu^-X$ final state provides a particularly precise measurement as the $e^+e^-X$ channel suffers from larger experimental uncertainties due to bremsstrahlung. It should be noted that it is the capability to precisely reconstruct the recoil mass distribution from $Z \to \mu^+\mu^-$ that defines the momentum resolution requirement for an ILC detector.

The reconstructed recoil mass distributions, calculated assuming the $Zh$ is produced with four-momentum $(\sqrt{s}, 0)$, are shown in Fig.2.8. In the $e^+e^-X$ channel FSR and bremsstrahlung photons are identified and used in the calculation of the $e^+e^-(n\gamma)$ recoil mass. Fits to signal and background components are used to extract $m_h$. Based on this model-independent analysis of Higgs production in the ILD detector, it is shown that $m_h$ can be determined with a statistical precision of $40$ MeV $(80$ MeV) from the $\mu^+\mu^-X$ $(e^+e^-X)$ channel. When the two channels are combined an uncertainty of $32$ MeV is obtained [74, 75]. The corresponding model independent uncertainty on the Higgs production cross section is 2.5%. Similar results were obtained from SiD [76]. It should be emphasized that these measurements only used the information from the leptonic decay products of the $Z$ and are independent of the Higgs decay mode. As such this analysis technique could be applied even if the Higgs decayed invisibly and hence allows us to determine the absolute branching ratios including that of invisible Higgs decays. By combining the branching ratio to $ZZ$ with the production cross section, which involves the same $g_{hZZ}$ coupling, one can determine the total width and the absolute scale of partial widths with no need for the theoretical assumptions needed for the LHC case. We will return to this point later.

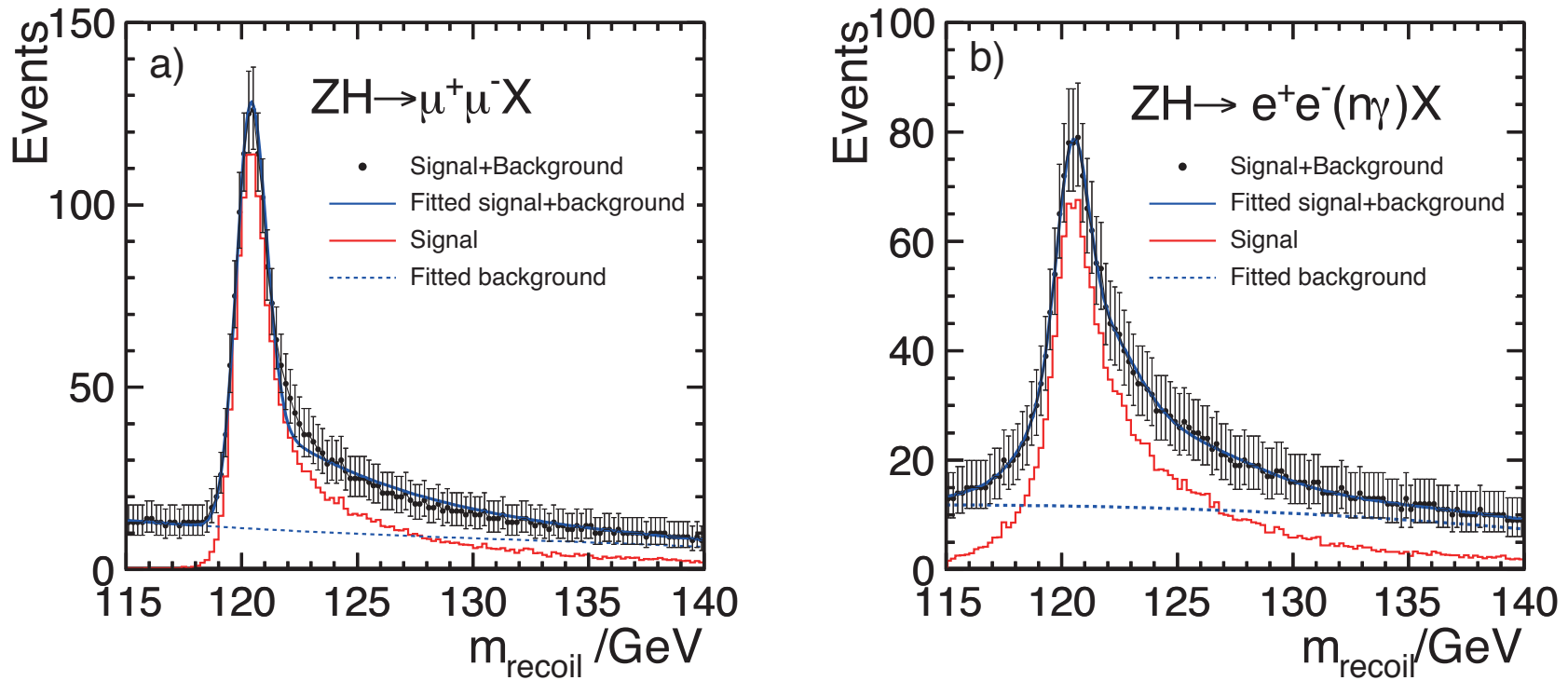

**Figure 2.8** Higgs recoil mass distribution in the Higgs-strahlung process $e^+e^- \to Zh$, with (a) $Z \to \mu^+\mu^-$ and (b) $Z \to e^+e^-(n\gamma)$. The results are shown for $P(e^+, e^-) = (+30\%, -80\%)$ beam polarization.

It is worth noting that, for the $\mu^+\mu^-X$ channel, the width of the recoil mass peak is dominated by the beam energy spread. In the above study Gaussian beam energy spreads of $0.28\,\%$ and $0.18\,\%$ are assumed for the incoming electron and positron beams respectively. For ILD the detector response leads to the broadening of the recoil mass peak from $560$ MeV to $650$ MeV. The contribution from momentum resolution is therefore estimated to be $330$ MeV. Although the effect of the detector resolution is not negligible, the dominant contribution to the observed width arises from the incoming beam energy spread rather than the detector response. This is no coincidence; the measurement of $m_h$ from the $\mu^+\mu^-X$ recoil mass distribution was one of the benchmarks used to determine the momentum resolution requirement for a detector at the ILC.

If there are additional Higgs fields with vacuum expectation values that contribute to the mass of the $Z$, the corresponding Higgs particles will also appear in reactions $e^+e^- \to Zh'$, and their masses can be determined in the same way.

We now turn to the determination of the spin and CP properties of the Higgs boson. The $h \to \gamma\gamma$ decay observed at the LHC rules out the possibility of spin 1 and restricts the charge conjugation C to be positive. We have already noted that the discrete choice between the CP even and CP odd charge assignments can be settled by the study of Higgs decay to $ZZ^*$ to 4 leptons at the LHC.

The ILC offers an additional, orthogonal, test of these assignments. The threshold behavior of the $Zh$ cross section has a characteristic shape for each spin and each possible CP parity. For spin 0, the cross section rises as $\beta$ near the threshold for a CP even state and as $\beta^3$ for a CP odd state. For spin 2, for the canonical form of the coupling to the energy-momentum tensor, the rise is also $\beta^3$. If the spin is higher than 2, the cross section will grow as a higher power of $\beta$. With a three-$20\,\mathrm{fb}^{-1}$-point threshold scan of the $e^+e^- \to Zh$ production cross section we can separate these possibilities as shown in Fig. 2.9 (left) [77]. The discrimination of more general forms of the coupling is possible by the use of angular correlations in the boson decay; this is discussed in detail in [78].

At energies well above the $Zh$ threshold, the $Zh$ process will be dominated by longitudinal $Z$ production as implied by the equivalence theorem. The reaction will then behave like a scalar pair production, showing the characteristic $\sim \sin^2\theta$ dependence if the $h$ particle's spin is zero. The measurement of the angular distribution will hence strongly corroborate that the $h$ is indeed a scalar particle.

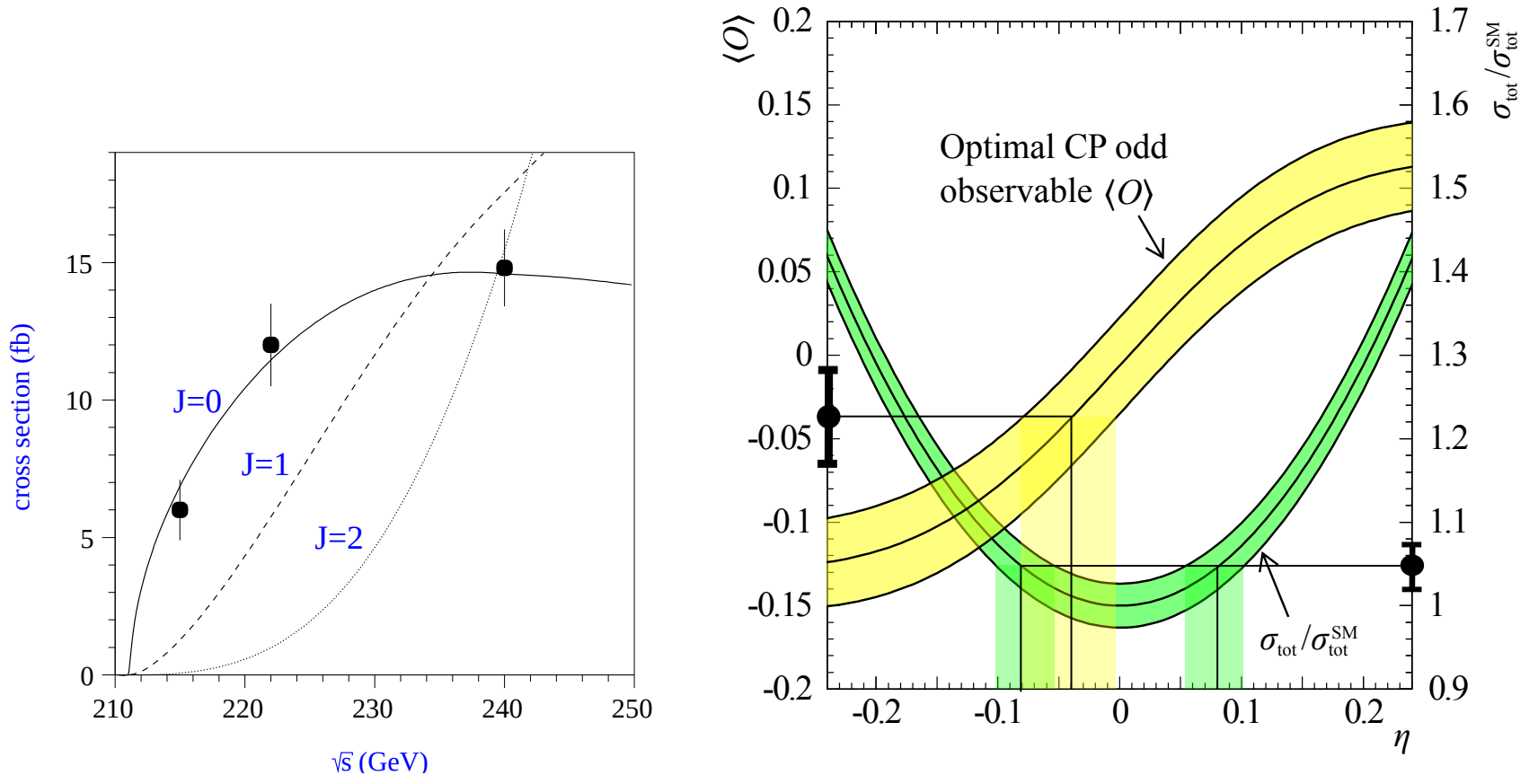

**Figure 2.9**
Left: Threshold scan of the $e^+e^- \to Zh$ process for $m_h = 120$ GeV, compared with theoretical predictions for $J^P = 0^+, 1^-$, and $2^+$ [77]. Right: Determination of $CP$-mixing with 1-$\sigma$ bands expected at $\sqrt{s} = 350$ GeV and $500\,\text{fb}^{-1}$ [79].

The analytic power of the ILC is emphasized when we consider more detailed questions. It is possible that the $h$ is not a CP eigenstate but rather a mixture of CP even and CP odd components. This occurs if there is CP violation in the Higgs sector. It is known that CP violation from the CKM matrix cannot explain the cosmological excess of baryons over antibaryons; thus, a second source of CP violation in nature is needed. One possibility is that this new CP violation comes from the Higgs sector and gives rise to net baryon number at the electroweak phase transitions, through mechanisms that we will discuss in Section 9.1 of this report. For these models, the $h$ mass eigenstates can be mainly CP even but contain a small admixture of a CP odd component.

A small CP odd contribution to the $hZZ$ coupling can affect the threshold behavior. The right-hand side of Fig. 2.9 shows the determination of this angle at a center of mass energy of 350 GeV from the value of the total cross section and from an appropriately defined optimal observable [79].

Tests of mixed CP property using the $hZZ$ coupling may not be the most effective ones, since the CP odd $hZZ$ coupling is of higher dimension and may be generated only through loops. It is more effective to use a coupling for which the CP even and CP odd components are on the same footing. An example is the $h$ coupling to $\tau^+\tau^-$, given by

$$\Delta\mathcal{L} = -\frac{m_\tau}{v} h\, \overline{\tau}(\cos\alpha + i\sin\alpha\gamma^5)\tau \qquad (2.33)$$

for a Higgs boson with a CP odd component. The polarizations of the final state $\tau$s can be determined from the kinematic distributions of their decay products; the CP even and odd components interfere in these distributions [80, 81]. In [82], it is estimated that the angle $\alpha$ can be determined at the ILC to an accuracy of 6°.

## 2.4.2 Inclusive cross section

Whereas all Higgs boson measurements at the LHC are measurements of $\sigma \cdot BR$, the ILC allows us to measure the absolute size of a Higgs inclusive cross section. This can be done by applying the recoil technique discussed above to the measurement of ($\sigma_{Zh}$) for the $e^+e^- \to Zh$ process. The measurement gives the cross section to a relative accuracy of $2.5\,\%$ at $250\,\text{fb}^{-1}$ without looking at the $h$ decay at all. This cross section is indispensable for extracting branching ratio ($BR$) from the event rate, which is proportional to $\sigma_{Zh} \cdot BR$, and limits its precision.

It is worth noting that the inclusive cross section is a direct measure of the $h$ to $ZZ$ coupling ($g_{hZZ}$). This single measurement at the ILC is capable of determining this coupling to $1.3\,\%$. If the $h$ particle is a scalar particle, this coupling must originate from a gauge-kinetic term of the form given by Eq.(2.5) with one $\Phi$ leg replaced by the vacuum expectation value associated with the $h$ particle. The observation of this coupling is, therefore, a strong evidence of the existence of a vacuum condensate associated with the $h$ particle. Moreover, the vacuum expectation value here has no solid reason to saturate the standard model value, $v = 246$ GeV. The $g_{hZZ}$ coupling hence measures to what extent the vacuum expectation value associated with the multiplet to which the $h$ particle belongs explains the mass of the $Z$ boson. This measurement, even considered alone, has extraordinary power to address the most basic issues in the breaking of electroweak symmetry.

As noted above, the ILC will not be capable of directly observing the width of the Higgs boson if it is as small as the Standard Model prediction of 4 MeV. However, because the ILC experiments can make this inclusive cross section measurement, they can also determine the width of the Higgs boson in a completely model-independent way. As a first step, note that the recoil technique gives Higgs boson branching ratios directly. We identify a $Z$ boson at the correct lab energy to be in recoil against the Higgs and count events on the opposite side in every final state. Then the total width of the Higgs is given by the formula

$$\Gamma_{\rm tot} = \frac{\Gamma(h \to ZZ)}{BR(h \to ZZ)} , \tag{2.34}$$

The quantity $\Gamma(h \to ZZ)$ is directly proportional to the inclusive cross section. The Higgs branching ratio to $ZZ$ is unfortunately quite small, so the direct measurement of this quantity at 250 GeV is statistics limited. In Section 2.5, we will explain how this quantity can be determined more accurately from data at higher energy. We will demonstrate there that, with 500 GeV data, the ILC should achieve an unambiguous measurement of the Higgs boson width to 6% accuracy.

### 2.4.3 Branching ratios and couplings

As we have just explained, the measurement of the inclusive cross section of the $e^+e^- \to Zh$ process allows us to directly extract the $h$ particle's branching fractions. A precise measurement of the absolute branching ratios of the Higgs bosons is an important test of the mass generation mechanism and provides a window into effects beyond the SM. For the branching ratio measurements we again use the $e^+e^- \to Zh$ process, but this time exploiting all the decay modes of the $Z$ boson including the $Z \to q\bar{q}$ and $Z \to \nu\bar{\nu}$ decays. The use of fully hadronic final states is possible only in a very clean environment of an $e^+e^-$ collider. In the clean environment of the ILC we can also use a high performance micro-vertex detector, placed very close to the interaction point, which makes it possible to identify not only $h \to b\bar{b}$ but also $h \to c\bar{c}$ decays. Figure 2.10 shows a lego plot of the multivariate estimate of $b$-likeness vs. $c$-likeness for the template samples of the signal and the SM background events. We can see the clear differences between the different decay modes of the Higgs boson. Thanks to these clear differences, a fit using these templates hence provides separate measurements of the cross section times branching fraction for the Higgs decays to $b\bar{b}$, $c\bar{c}$, and $gg$ with negligible mutual correlation. Together with the measurement of the $h \to \tau^+\tau^-$ decays, we can access the Yukawa couplings of both up-type and down-type fermions and test the coupling-mass proportionality. The loop-induced $h \to gg$ decay is indirectly sensitive to the top Yukawa coupling and possibly other new strongly interacting particles that couple to the Higgs particle but are too heavy to produce directly. By the same token, the $h \to \gamma\gamma$ and the $h \to Z\gamma$ decays are also important as tools to probe heavy particles with electroweak charges. The expected accuracies on the branching ratios are summarized in Table 2.2. It is worth noting that these full simulation results are consistent with the past fast simulation results [87–91].

The $h$ decay to invisible final states, if any, can be measured by looking at the recoil mass under the condition that nothing observable is recoiling against the $Z$ boson. Higgs portal models predict such decays and provide a unique opportunity to access dark matter particles [92]. The main background is $e^+e^- \to ZZ$ followed by one $Z$ decaying into a lepton pair and the other into a neutrino pair. With an integrated luminosity of 250 fb$^{-1}$ at $\sqrt{s} = 250$ GeV, the ILC can set a 95% CL limit on the invisible branching ratio to 4.8% using the golden $Z \to \mu^+\mu^-$ mode alone [93]. Using other modes including $Z \to q\bar{q}$, we could improve this significantly to 0.8% [94].

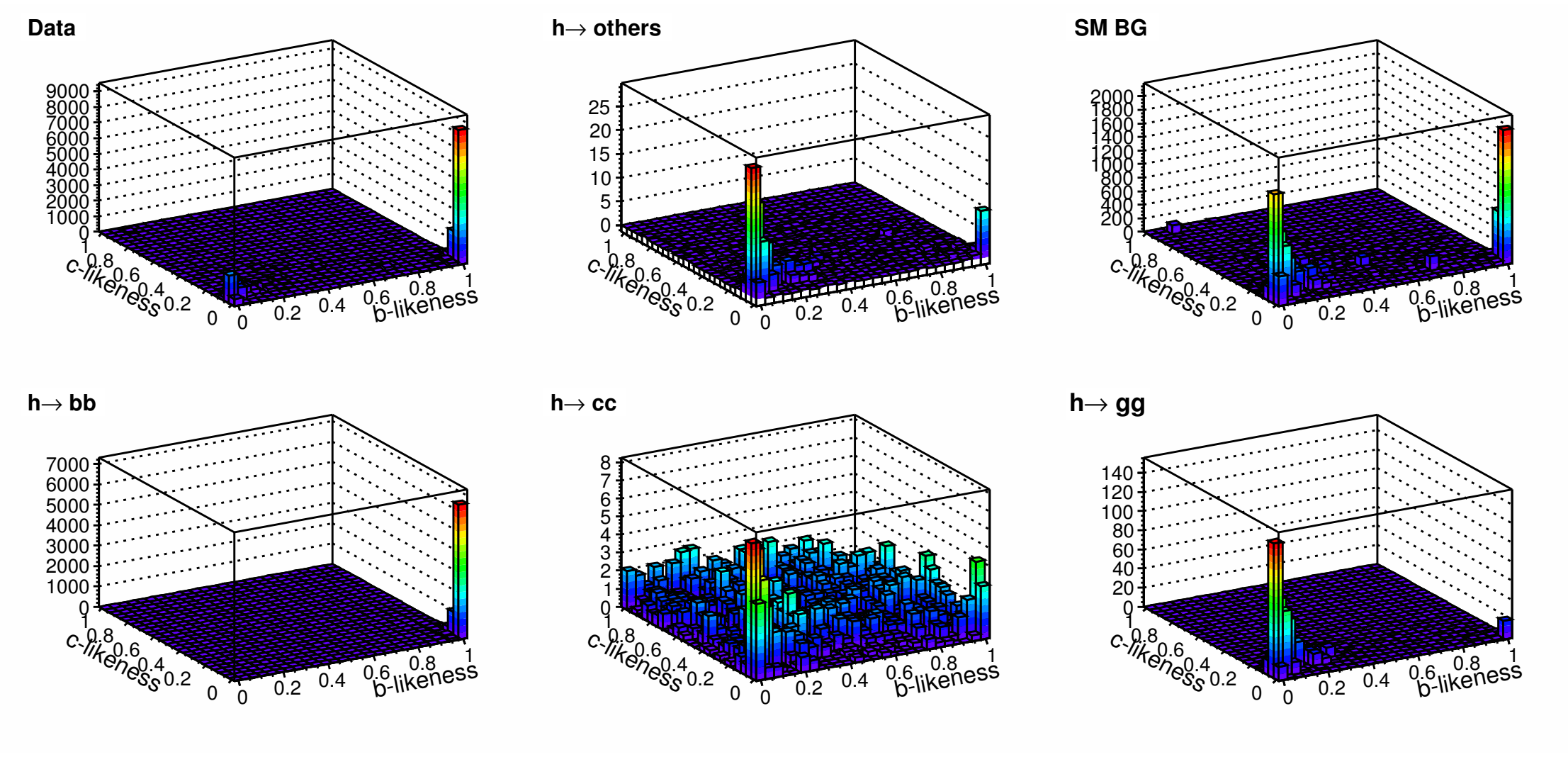

**Figure 2.10.** Two-dimensional images of the three-dimensional template samples as a function of $b$-likeness v.s. $c$-likeness. The bottom row shows Higgs decays, left to right, to $b\bar{b}$, $c\bar{c}$, and $gg$. The top row shows, left to right, the full Monte Carlo Higgs sample after event selection, the Higgs decays to non-2-jet modes, and the Standard Model background. From [83].

**Table 2.2.** Expected accuracies for the $h$ boson branching ratios for $m_h = 120$ GeV, obtained with full detector simulations at the $\sqrt{s} = 250$ GeV assuming $\mathcal{L} = 250$ fb$^{-1}$ and $(e^-, e^+) = (-0.8, +0.3)$ beam polarization [83–86]. The errors on $BR$ include the error on $\sigma$ of 2.5% from the recoil mass measurement.

| mode | $BR$ | $\sigma \cdot BR\,(\mathrm{fb})$ | $N_{evt}/250\,\mathrm{fb}^{-1}$ | $\Delta(\sigma\,BR)/(\sigma\,BR)$ | $\Delta BR/BR$ |
|---|---|---|---|---|---|
| $h \to b\bar{b}$ | 65.7% | 232.8 | 58199 | 1.0% | 2.7% |
| $h \to c\bar{c}$ | 3.6% | 12.7 | 3187 | 6.9% | 7.3% |
| $h \to gg$ | 5.5% | 19.5 | 4864 | 8.5% | 8.9% |
| $h \to WW^*$ | 15.0% | 53.1 | 13281 | 8.1% | 8.5% |
| $h \to \tau^+\tau^-$ | 8.0% | 28.2 | 7050 | 3.6% | 4.4% |
| $h \to ZZ^*$ | 1.7% | 6.1 | 1523 | 26% | 26% |
| $h \to \gamma\gamma$ | 0.29% | 1.02 | 255 | 23-30% | 23-30% |

The branching fraction measurements discussed so far are still statistics limited. If we are to improve the measurement precisions by increasing the integrated luminosity by doubling the number of bunches or by running longer, etc., we will need to estimate the systematic errors that may limit the measurement in particular for $h \to b\bar{b}$. The systematic error from the uncertainty in luminosity measurement should be less than $0.1\%$ and thus negligible. The dominant source of systematic errors is probably that from flavor identification and the separation of $Z$ plus jet signal from Standard Model backgrounds using the multivariate analysis described above. We are still in the process of optimizing this analysis, but we expect the systematic error due to flavor-tagging can be controlled by using the calibration processes $ZZ$, $Z\gamma$, and $WW$, all of which have large cross sections. These calibration samples will also allow us to calibrate and normalize the background estimate.

To determine the absolute normalization of Higgs boson partial widths from the measurements of branching ratios, we need to combine these with an accurate value of one partial width or cross section. As described above, the $250$ GeV running of the ILC for 250 fb$^{-1}$ will determine the cross section for $e^+e^- \to Zh$ very accurately, to 2.5%. The value can be directly converted to $g_{hZZ}$ or to the absolute partial width $\Gamma(ZZ)$. However, to use this value to normalize the other Higgs partial widths in a completely model-independent analysis, we would need to use the formula similar to (2.34)

$$\Gamma(A) = \Gamma(ZZ) \cdot \frac{BR(A)}{BR(ZZ)}, \tag{2.35}$$

and so we again need to measure the branching ratio for $h \to ZZ^*$. This is not easy to do at the ILC because it is a rare mode giving low statistics for a Higgs boson with $m_h \simeq 120$ GeV. No full simulation study of the $h \to ZZ^*$ branching ratio in $e^+e^- \to Zh$ is currently available. We will therefore use the result of the $h \to WW^*$ study [85] and scale accordingly. The error for the $h \to WW^*$ decay implies a 26% relative error for the $h \to ZZ^*$ branching ratio. The use of the formula (2.35) then implies that the uncertainties in absolute partial widths or Higgs couplings are those listed convolved with $2.5 \oplus 26\%$. This significantly degrades the precision information obtained at the ILC.

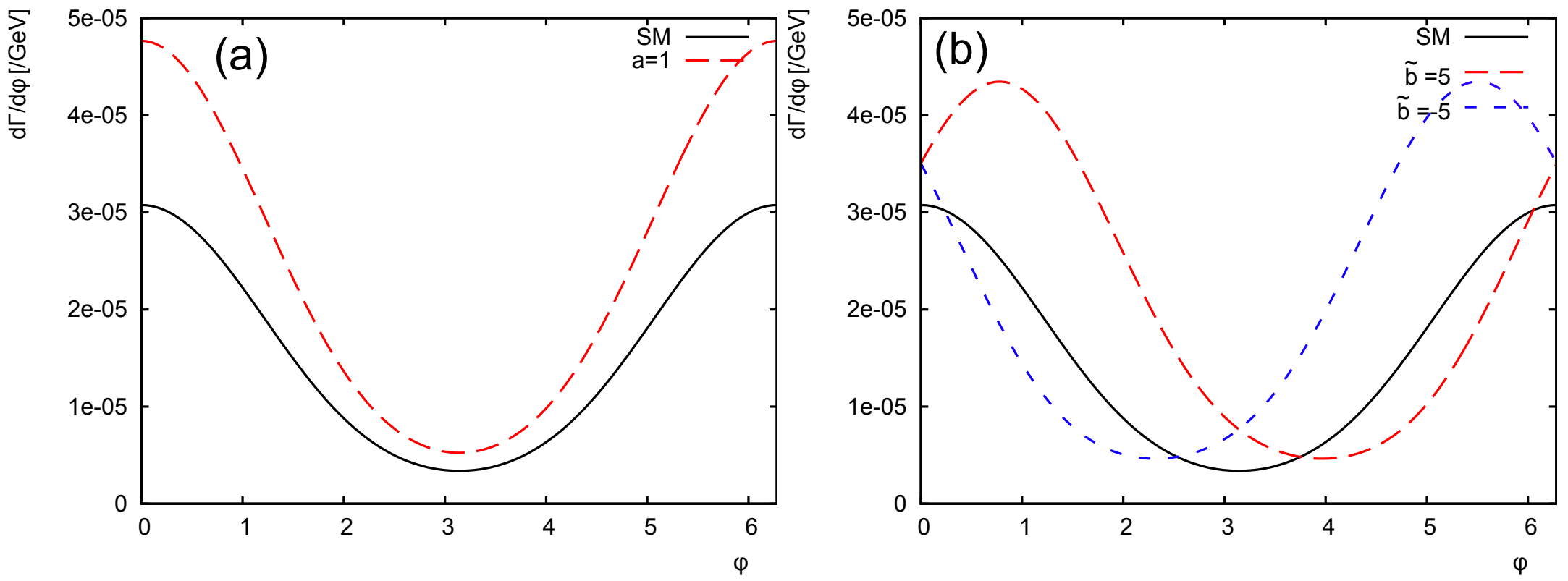

**Figure 2.11.** Distribution of the angle $\phi$ between two decay planes of $W$ and $W^*$ from the decay $H \to WW^* \to 4j$ with the inclusion of anomalous couplings [97]. (a) The SM curve along with that for $a = 1$, $b = \tilde{b} = 0$, $\Lambda = 1$ TeV; the position of the minimum is the same for both distributions. (b) The SM result with the cases $\tilde{b} = \pm 5$, $a = b = 0$, $\Lambda = 1$ TeV; the position of the minimum is now shifted as discussed in the text. From [97].

An alternative is to use the theoretical assumption

$$g(hWW)/g(hZZ) = \cos^2\theta_W \tag{2.36}$$

to tie together the $hZZ$ and $hWW$ couplings. Now $BR(WW^*)$ can be used in the denominator of Eq.(2.35). The error added in converting from branching ratios to partial widths is $2.5 \oplus 8.6\% = 9.0\%$.

A better way is to use the $WW$ fusion process, $e^+e^- \to \nu\bar{\nu}h$. The cross section for this process is proportional to $g^2(hWW)$ and thus to the $h \to WW^*$ partial width [95]. Although the $WW$ fusion cross section is small at $\sqrt{s} = 250$ GeV, $18\,\mathrm{fb}$ for $m_h = 120$ GeV and the standard left-hand beam combination, $(P_{e^-}, P_{e^+}) = (-0.8, +0.3)$, the expected yield exceeds 4k events and allows the measurement of the $WW$ fusion cross section to $\Delta\sigma(WW)/\sigma(WW) = 7.2\%$ for the $250\,\mathrm{fb}^{-1}$. Combining the $BR(WW^*)$ measurement, this implies that the total width can be determined to $11\%$ in a completely model-independent way from 250 GeV data alone [96]. As we will see below, the determination of the absolute strength of the Higgs coupling to $WW$ is expected to be further improved by a measurement of the $WW$ fusion cross section at $\sqrt{s} = 500$ GeV. The $500$ GeV data can also be used to improve the accuracy on the $BR(WW^*)$.

So far we have been dealing with the branching ratios and partial widths after phase space integration. The $h \to WW^*$ decay provides an interesting opportunity to study its differential width and probe the Lorentz structure of the $hWW$ coupling through angular analyses of the decay products. The relevant part of the general interaction Lagrangian, which couples the Higgs boson to $W$ bosons

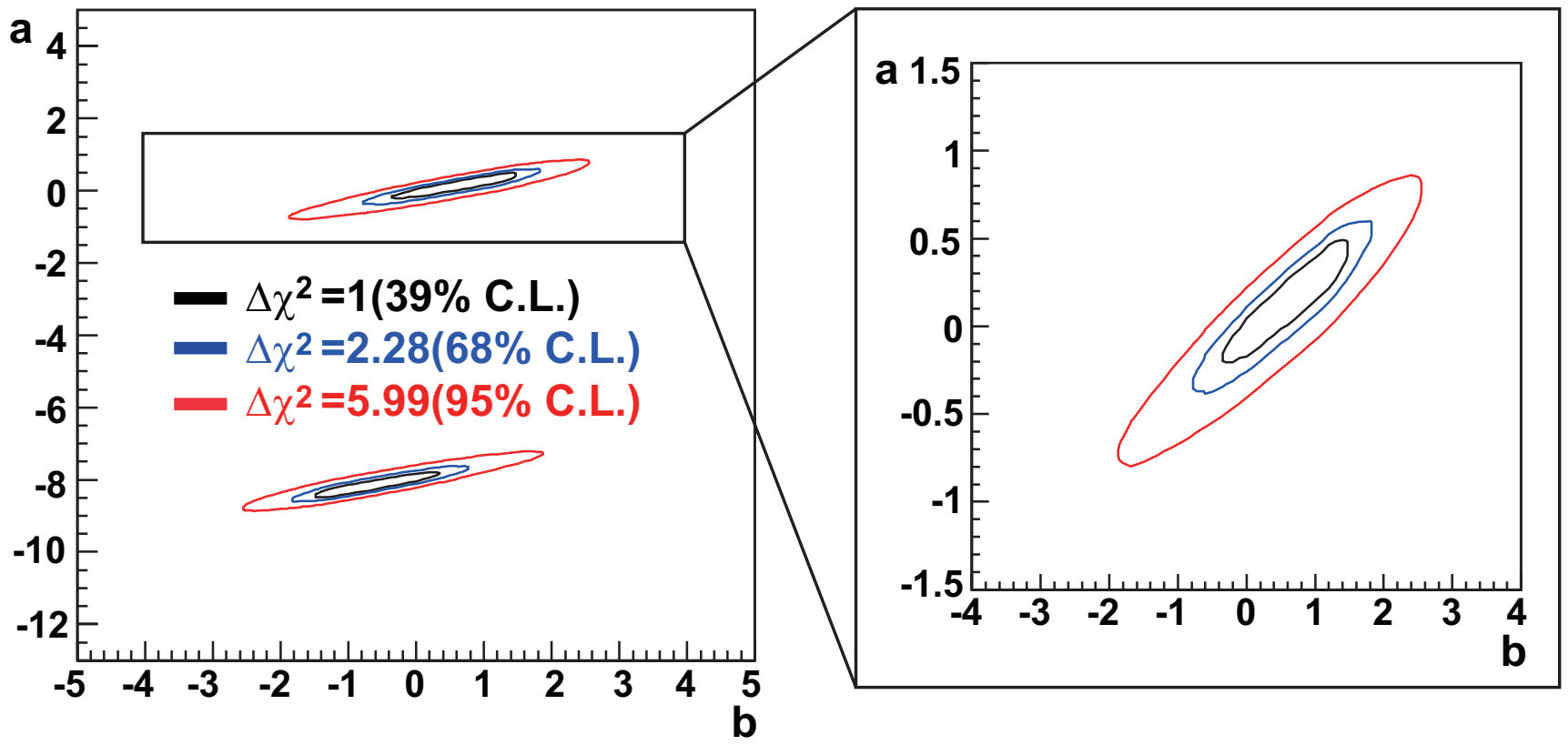

**Figure 2.12** Probability contours for $\Delta\chi^2 = 1$, 2.28, and 5.99 in the $a$-$b$ plane, which correspond to 39%, 68%, and 95% C.L., respectively.

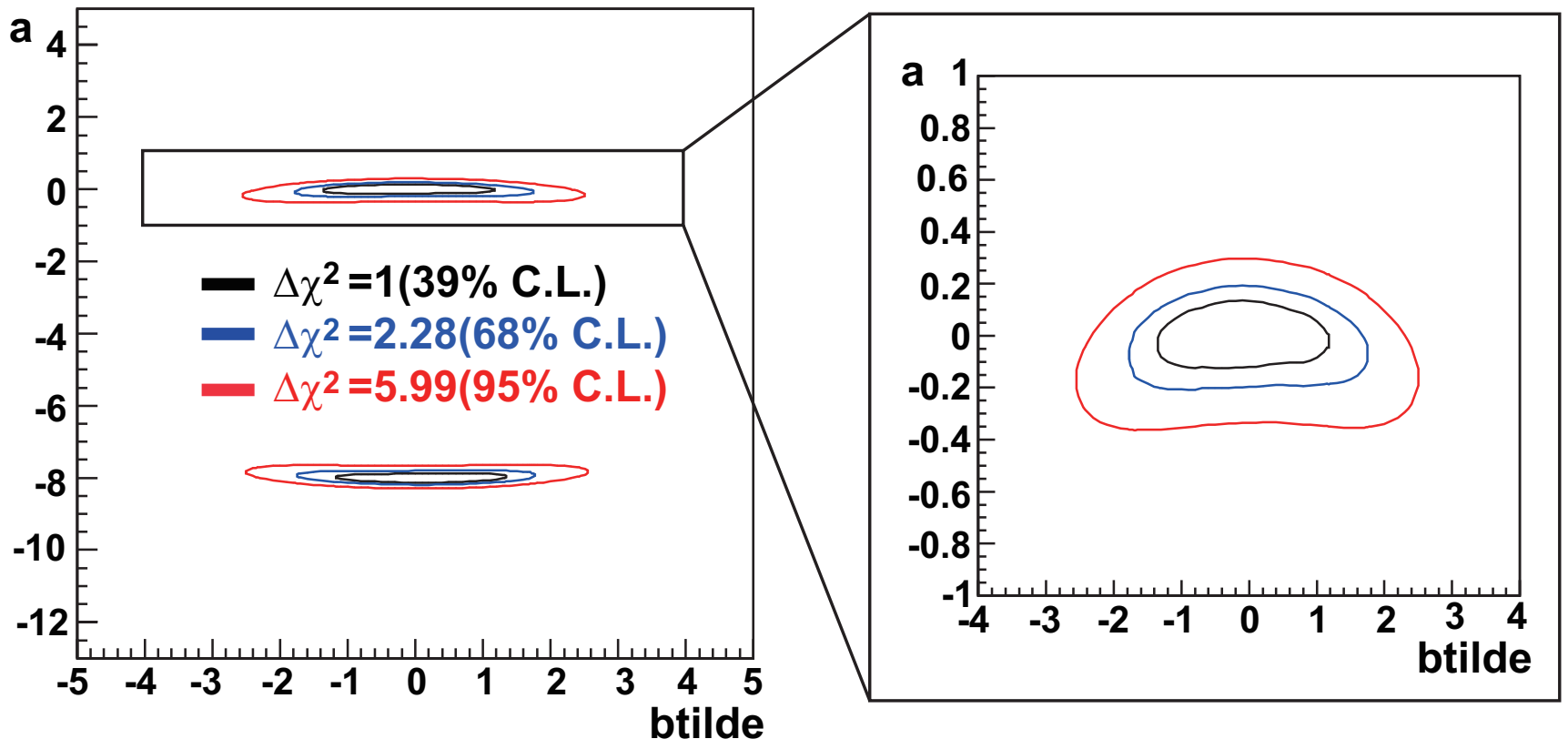

**Figure 2.13** Contours similar to Fig. 2.12 plotted in the $a$-$\tilde{b}$ plane.

in a both Lorentz- and gauge-symmetric fashion, can be parameterized as

$$\mathcal{L}_{\rm hWW} = 2m_W^2 \left(\frac{1}{v} + \frac{a}{\Lambda}\right) h\ W_\mu^+ W^{-\mu} + \frac{b}{\Lambda} h\ W_{\mu\nu}^+ W^{-\mu\nu} + \frac{\tilde{b}}{\Lambda} h\ \epsilon^{\mu\nu\sigma\tau} W_{\mu\nu}^+ W_{\sigma\tau}^- \ , \qquad (2.37)$$

where $W_{\mu\nu}^\pm$ is the usual gauge field strength tensor, $\epsilon^{\mu\nu\sigma\tau}$ is the Levi-Civita tensor, $v$ is the VEV of the Higgs field, and $\Lambda$ is a cutoff scale[1]. The real dimensionless coefficients, $a$, $b$, and $\tilde{b}$, are all zero in the Standard Model and measure the anomaly in the $hWW$ coupling, which arise from some new physics at the scale $\Lambda$. The coefficient $a$ stands for the correction to the Standard Model coupling. The coefficients $b$ and $\tilde{b}$ parametrize the leading dimension-five non-renormalizable interactions and corresponding to $(\boldsymbol{E}\cdot\boldsymbol{E} - \boldsymbol{B}\cdot\boldsymbol{B})$-type $CP$-even and $(\boldsymbol{E}\cdot\boldsymbol{B})$-type $CP$-odd contributions. The $a$ coefficient, if nonzero, would modify just the normalization of the Standard Model coupling, while the $b$ and $\tilde{b}$ coefficients would change the angular correlations of the decay planes. This effect is shown in Fig. 2.11 [97]. Nonzero $b$ and $\tilde{b}$ would also modify the momentum distribution of the $W$ boson in the Higgs rest frame. Simultaneous fits to $p_W$ and $\phi_{\rm plane}$ result in the contour plots in Figs.2.12 and 2.13.

[1] The Lagrangian (2.37) is not by itself gauge invariant; to restore explicit gauge invariance we must also include the corresponding anomalous couplings of the Higgs boson to $Z$ bosons and photons.

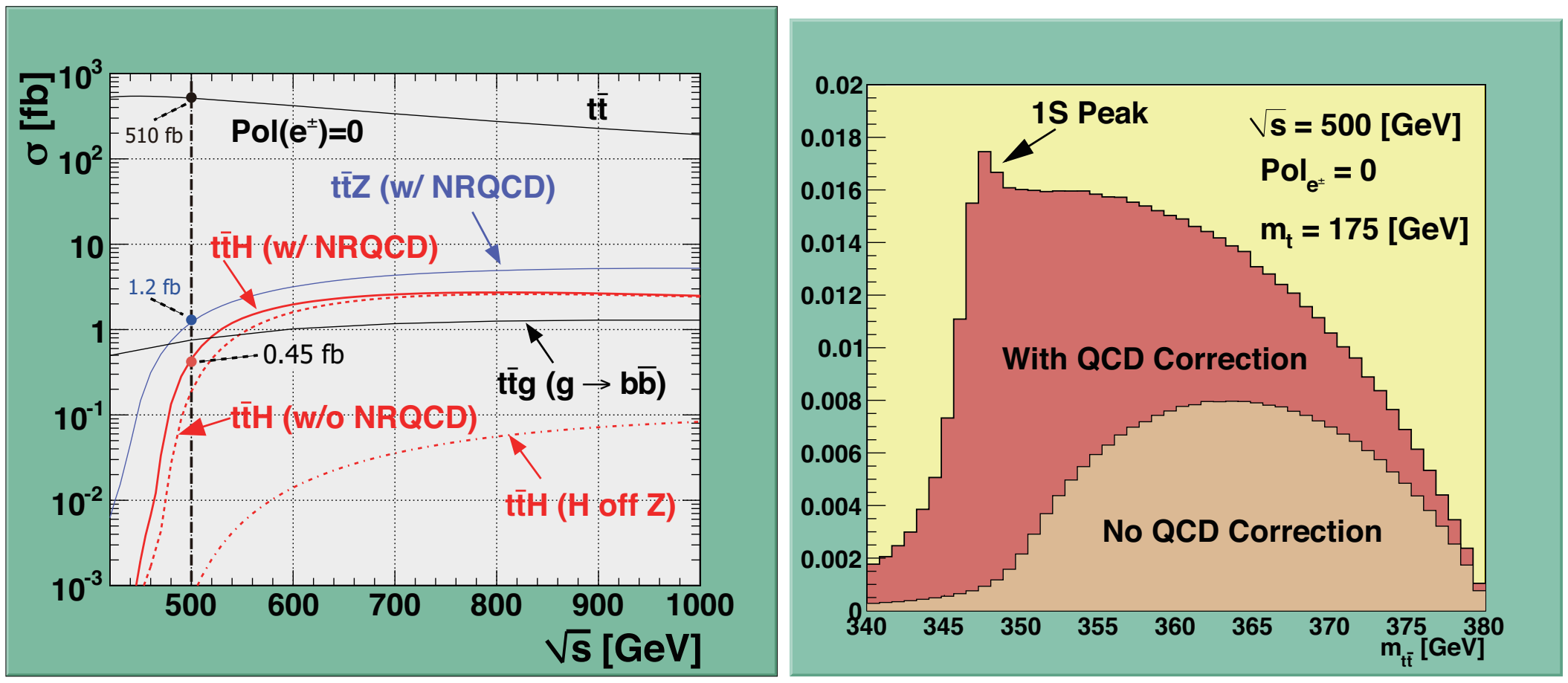

**Figure 2.14.** Left: Cross section for the $e^+e^- \to t\bar{t}h$ process as a function of $\sqrt{s}$, together with those of background processes, $e^+e^- \to t\bar{t}Z$, $\to t\bar{t}g^*$, and $\to t\bar{t}$. Right: The invariant mass distribution of the $t\bar{t}$ system from the $e^+e^- \to t\bar{t}h$ process with and without the non-relativistic QCD correction.

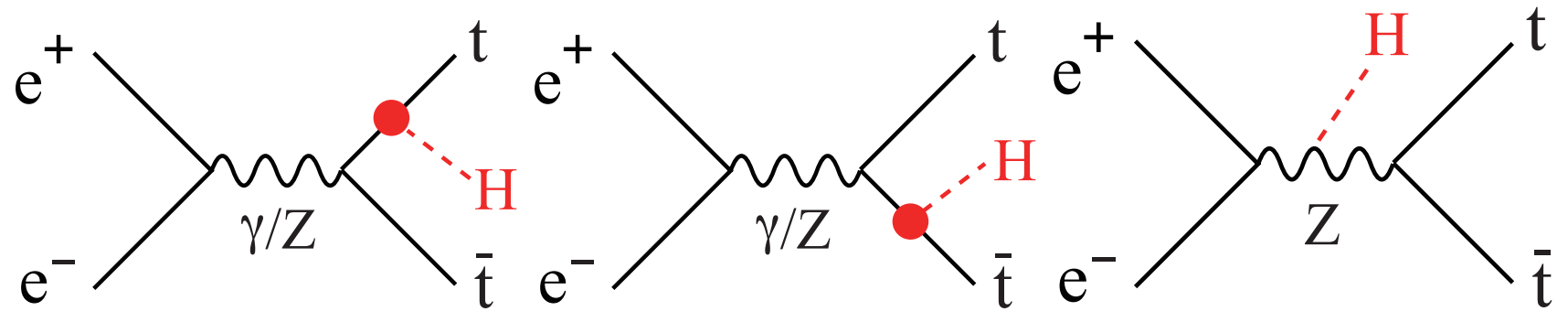

**Figure 2.15.** Three diagrams contributing to the $e^+e^- \to t\bar{t}h$ process. The $h$-off-$t$ or $\bar{t}$ diagrams, (a) and (b), contain the top Yukawa coupling while the $h$-off-$Z$ diagram (c) does not.

## 2.5 Higgs measurements at ILC at 500 GeV

The two very important processes will become accessible for the first time at $\sqrt{s} = 500$ GeV. The first is the $e^+e^- \to t\bar{t}h$ process [98, 99], in which the top Yukawa coupling will appear in the tree level for the first time at the ILC. The top quark, being the heaviest matter fermion in the Standard Model, would be crucial to understand the fermion mass generation mechanism. The second is the $e^+e^- \to Zhh$ process, to which the triple Higgs coupling contributes in the tree level. The self-coupling is the key ingredient of the Higgs potential and its measurement is indispensable for understanding the electroweak symmetry breaking.

### 2.5.1 Top quark Yukawa coupling

Past simulation studies for the $e^+e^- \to t\bar{t}h$ process were mostly made at around $\sqrt{s} = 800$ GeV, since the cross section attains its maximum there for $m_h \simeq 120$ GeV [100–102]. It was pointed out, however, that the cross section would be significantly enhanced near the threshold due to the bound-state effects between $t$ and $\bar{t}$ [103–109]. The effect is made obvious in the right-hand plot of Fig. 2.14. This enhancement implies that the measurement of the top Yukawa coupling might be possible already at $\sqrt{s} = 500$ GeV [110]. A serious simulation study at $\sqrt{s} = 500$ GeV was performed for the first time, with the QCD bound-state effects consistently taken into account for both signal and background cross sections, in [111].

The $e^+e^- \to t\bar{t}h$ reaction takes place through the three diagrams shown in Fig. 2.15 As shown in Fig. 2.14 (left), the contribution from the irrelevant $h$-off-$Z$ diagram is negligible at $\sqrt{s} = 500$ GeV, thereby allowing us to extract the top Yukawa coupling $g_t$ by just counting the number of signal events. By combining the 8-jet and 6-jet-plus-lepton modes of $e^+e^- \to t\bar{t}h$ followed by $h \to b\bar{b}$, the analysis of [111] showed that a measurement of the top Yukawa coupling to $\Delta g_t/g_t = 10\%$ is possible for $m_h = 120$ GeV with polarized electron and positron beams of $(P_{e^-}, P_{e^+}) = (-0, 8, +0.3)$ and an integrated luminosity of $1$ ab$^{-1}$. This result obtained with a fast Monte Carlo simulation has just recently been corroborated by a full simulation [112, 113].

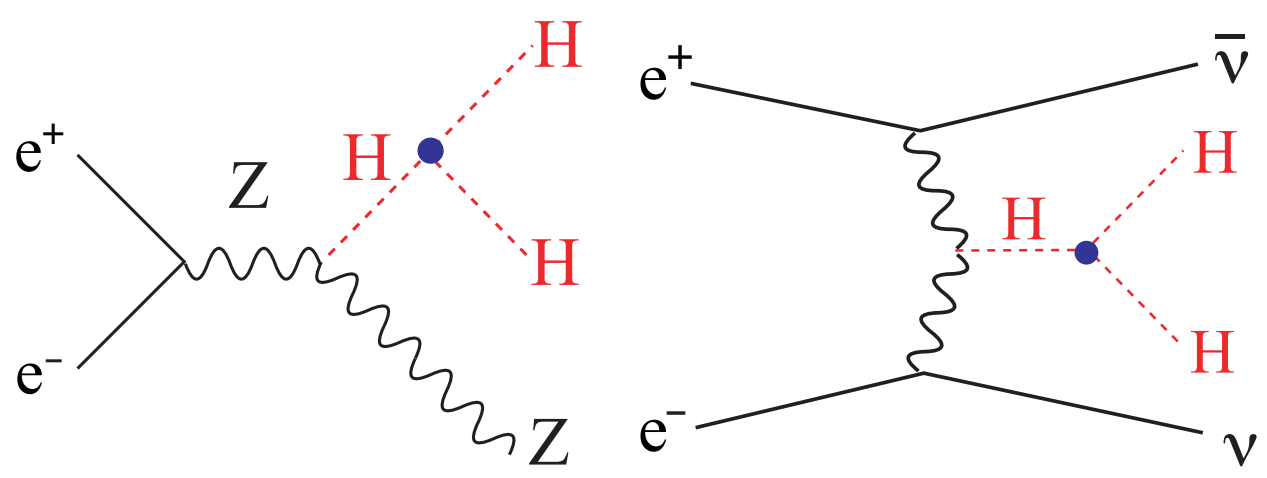

**Figure 2.16** Relevant diagrams containing the triple Higgs coupling for the two processes: $e^+e^- \to Zhh$ (left) and $e^+e^- \to \nu_e\overline{\nu}_e hh$.

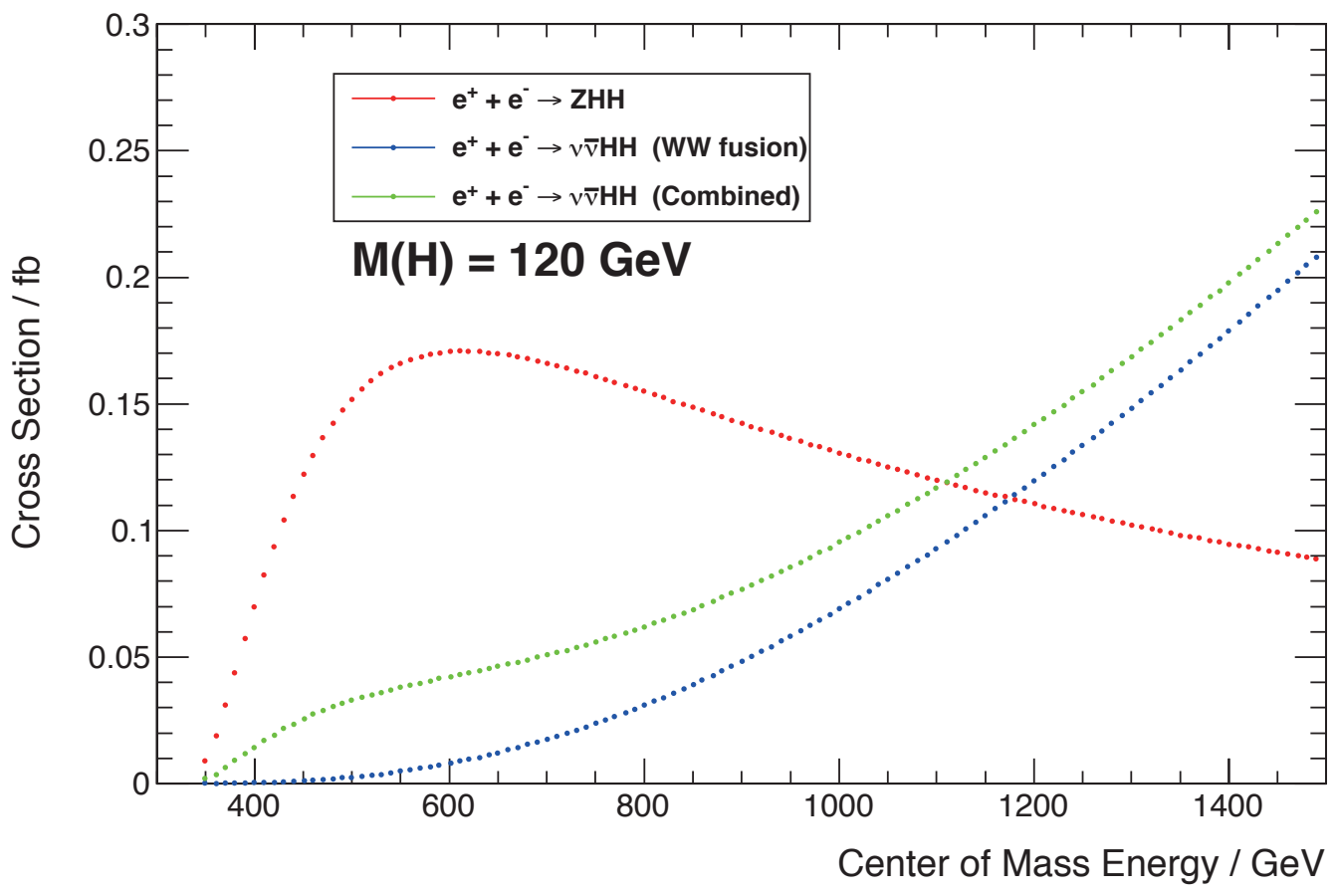

**Figure 2.17** Cross sections for the two processes $e^+e^- \to Zhh$ (left) and $e^+e^- \to \nu_e\overline{\nu}_e hh$ as a function of $\sqrt{s}$ for $m_h = 120$ GeV.

## 2.5.2 Higgs self-coupling

The triple Higgs boson coupling can be studied at the ILC through the processes $e^+e^- \to Zhh$ and $e^+e^- \to \nu_e\overline{\nu}_e hh$. The relevant Feynman diagrams are shown in Fig. 2.16 [114]. The cross sections for the two processes are plotted as a function of $\sqrt{s}$ for $m_h = 120\,\text{GeV}$ in Fig. 2.17. The cross section reaches its maximum of about $0.18\,\text{fb}$ at around $\sqrt{s} = 500\,\text{GeV}$, which is dominated by the former process. A full simulation study of the process $e^+e^- \to Zhh$ followed by $h \to b\bar{b}$ has recently been carried out in [115], making use of a new flavor tagging package (LCFIplus) [116] together with the conventional Durham jet clustering algorithm.

From the combined result of the three channels corresponding to different $Z$ decay modes, $Z \to l^+l^-$, $\nu\overline{\nu}$, and $q\overline{q}$, it was found that the process can be detected with an excess significance of $5$-$\sigma$ and the cross section can be measured to $\Delta\sigma/\sigma = 0.27$ for an integrated luminosity of $2\,\text{ab}^{-1}$ with beam polarization $(P_{e^-}, P_{e^+}) = (-0, 8, +0.3)$. Unlike the $e^+e^- \to t\bar{t}h$ case, however, the contribution from the background diagrams without the self-coupling is significant and the relative error on the self-coupling $\lambda$ is $\Delta\lambda/\lambda = 0.44$ with a proper event weighting to enhance the contribution from the self-coupling diagram. The result is not yet very satisfactory compared to the results from earlier fast simulation studies [117–121]. The major problem in the analysis is mis-clustering of color-singlet groups. Figure 2.18 compares the reconstructed invariant masses for the two Higgs candidates with Durham jet clustering (a) and with perfect jet clustering using Monte Carlo truth (b). We can see that the separation between the signal and the background is significantly improved if there is no mis-jet-clustering. A new jet clustering algorithm is now being developed.

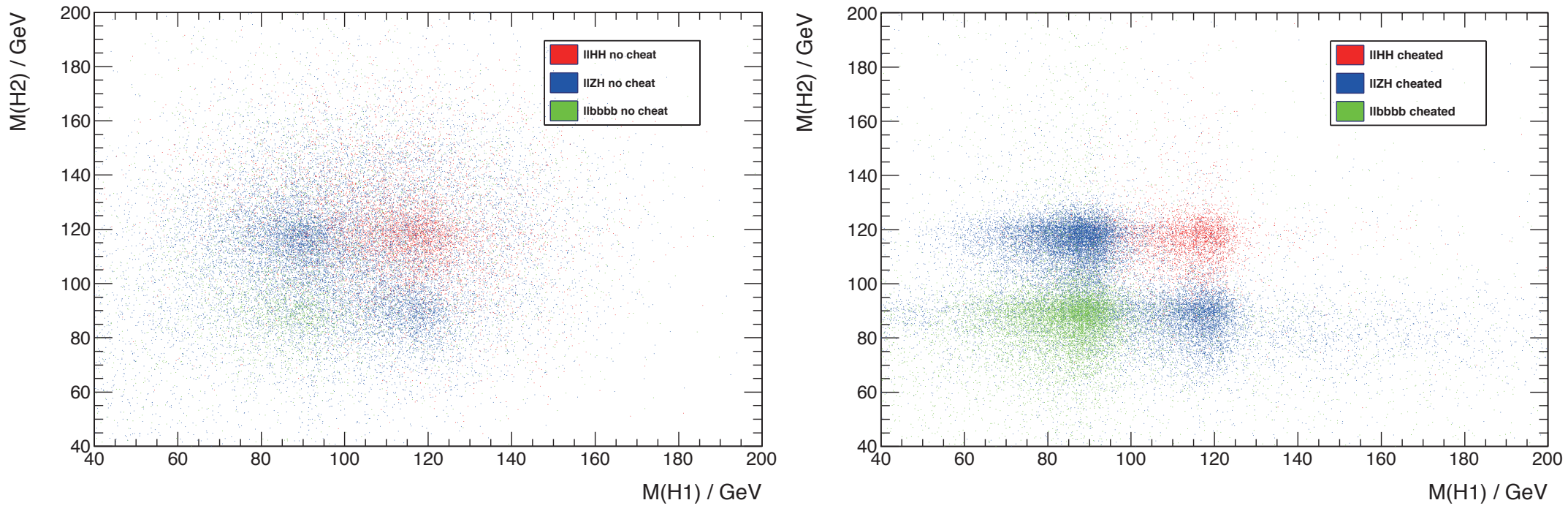

**Figure 2.18.** Scatter plot of the invariant masses of the two Higgs candidates. Left: with Durham jet clustering. Right: with perfect jet clustering using Monte Carlo truth on the color flow.

### 2.5.3 $WW$ fusion and the $hWW$ coupling

As shown in Fig.2.7, the $WW$ fusion process takes over the Higgs-strahlung process at around $\sqrt{s} = 450\,\text{GeV}$. The cross section for the fusion process is about $160$ fb at $\sqrt{s} = 500$ GeV for $m_h = 120\,\text{GeV}$. Thanks to this large cross section and the larger luminosity expected at this energy, the fusion process provides a unique opportunity to directly measure the $hWW$ coupling with high precision. With an integrated luminosity of $500\,\text{fb}^{-1}$, we can measure this cross section times the branching fraction to $b\bar{b}$ to a statistical accuracy of 0.60%. In terms of Higgs cross sections and branching ratios, the quantity measured is

$$\sigma(\nu\bar{\nu}h) \cdot BR(h \to b\bar{b}) \sim \Gamma(h \to WW^*) \cdot BR(h \to b\bar{b}) \ . \tag{2.38}$$

By combining this with the direct branching ration measurements at $\sqrt{s} = 250\,\text{GeV}$, we will be able to determine the cross section $\sigma(\nu\bar{\nu}h)$ to an accuracy of 2.7%, which translates to an expected error on the $hWW$ coupling of $\Delta g_{hWW}/g_{hWW} = 1.4\%$. The large data sample of the fusion process is also useful to improve the precision of the $h \to WW^*$ branching ratio. It is noteworthy that the background separation is easier at $\sqrt{s} = 500$ GeV than at $\sqrt{s} = 250$ GeV, enabling us to determine the cross section times branching ratio for $\sigma(\nu\bar{\nu}h) \cdot BR(WW^*)$ to 3.0% acccuracy. Applying Eq.(2.34) with $ZZ$ replaced by $WW$, we can determine the Higgs total width to $\Delta\Gamma_{\text{tot}}/\Gamma_{\text{tot}} \simeq 6\%$. The clean sample of $WW^*$ decays can be also used to investigate the Lorentz structure of the $hWW$ coupling as we discussed in the angular analysis of the $h \to WW^*$ decays in the $e^+e^- \to Zh$ process at $\sqrt{s} = 250$ GeV.

The measurement of the Higgs boson width can be further improved by using the full set of Higgs rate measurements and insisting that the observed branching ratios should sum to 1. Since *all* Higgs boson decay modes are observed in recoil against the $Z$, this assumption is justified at an $e^+e^-$ collider. At the end of this chapter, we will report on a global fit to the full set of Higgs couplings of a 125 GeV Higgs bosons. In that fit, the Higgs boson width is determined to an accuracy of 1.6%.

### 2.5.4 Expected improvements of branching ratio measurements

The Higgs sample from the $WW$ fusion and the Higgs-strahlung processes at $\sqrt{s} = 500\,\text{GeV}$ will enable us to significantly improve the branching ratio measurements described above for the $\sqrt{s} = 250$ GeV run. In particular we can do a template fitting similar to that employed for the $e^+e^- \to Zh$ sample at $\sqrt{s} = 250\,\text{GeV}$. The flavor-tagging performance at $\sqrt{s} = 500\,\text{GeV}$ will be similar, too. The expected relative errors on the cross section times branching ratios are summarized in Table 2.3. The table shows that the $WW$ fusion process contributes significantly, while the relative error on $\Delta BR(b\bar{b})/BR(b\bar{b})$ is limited by the error on the $Zh$ production cross section at $\sqrt{s} = 250$ GeV from the recoil mass measurement. If we need higher accuracy for $\Delta BR(b\bar{b})/BR(b\bar{b})$, we will need to run longer at $\sqrt{s} = 250$ GeV, though slight improvement is also expected from the recoil mass measurement at $\sqrt{s} = 500$ GeV.

**Table 2.3.** Expected accuracies for the $h$ boson branching ratios for $m_h = 120$ GeV when the 250 GeV measurements assuming $\mathcal{L} = 250\,\text{fb}^{-1}$ in Table 2.2 are combined with those at $\sqrt{s} = 500$ GeV assuming $\mathcal{L} = 500\,\text{fb}^{-1}$ and $(e^-, e^+) = (-0.8, +0.3)$ beam polarization. The errors on $BR$s include the error on $\sigma$ of 2.5% from the recoil mass measurement at $\sqrt{s} = 250\,\text{GeV}$.

| | $\Delta(\sigma \cdot BR)/(\sigma \cdot BR)$ | | | $\Delta BR/BR$ |
|---|---|---|---|---|
| mode | $Zh$ @ 250 GeV | $Zh$ @ 500 GeV | $\nu\overline{\nu}h$ @ 500 GeV | combined |
| $h \to b\bar{b}$ | 1.0% | 1.6% | 0.60% | 2.6% |
| $h \to c\bar{c}$ | 6.9% | 11% | 5.2% | 4.6% |
| $h \to gg$ | 8.5% | 13% | 5.0% | 4.8% |
| $h \to WW^*$ | 8.1% | 12.5% | 3.0% | 3.8% |
| $h \to \tau^+\tau^-$ | 3.6% | 4.6% | 11% | 3.6% |
| $h \to ZZ^*$ | 26% | 34% | 10% | 9.3% |
| $h \to \gamma\gamma$ | 23-30% | 29-38% | 19-25% | 13-17% |

## 2.6 Higgs measurements at ILC at 1000 GeV

Two out of the three processes selected as the DBD benchmark reactions at $\sqrt{s} = 1000$ GeV involve Higgs boson production: $e^+e^- \to t\bar{t}h$ and $e^+e^- \to \nu\overline{\nu}h$. We showed above that we would be able determine the top Yukawa coupling to an accuracy of about $10\,\%$ at $\sqrt{s} = 500$ GeV for $m_h = 120$ GeV, using the former process. The signal cross section grows to its maximum at around $\sqrt{s} = 700$ and only slowly decreases toward $\sqrt{s} = 1000$ GeV, while the $e^+e^- \to t\bar{t}$ background decreases much more rapidly. Thus, a more precise measurement of the top Yukawa coupling will be possible at this higher energy.

At the same time, the $WW$ fusion process $e^+e^- \to \nu\overline{\nu}h$ dominates the $s$-channel Higgs-sthrahlung process. Taking advantage of electron and positron beam polarization, the cross section for the $WW$ fusion process at 1000 GeV will be as large as $400$ fb for $(P_{e^-}, P_{e^+}) = (-0.8, +0.2)$ and $m_h = 125$ GeV, as shown in Fig. 2.19. Taking into account the higher luminosity expected at $\sqrt{s} = 1000$ GeV, this process will give us a high statistics Higgs boson sample: $4 \times 10^5$ events for $1$ ab$^{-1}$. This will allow us to improve the branching ratios to the various modes discussed above as well as to access the rare mode $h \to \mu^+\mu^-$. It is also noteworthy that one more process, $e^+e^- \to \nu\overline{\nu}hh$ process, will become sizable at $\sqrt{s} = 1000$ GeV. This reaction can be used together with the $e^+e^- \to Zhh$ process to improve the measurement of the Higgs self-coupling.

The $WW$ fusion processes occur only from the initial state $e^-_L e^+_R$, and the top quark production cross section is also much larger from this initial state. Thus, it is advantageous to spend most of the running time at 1000 GeV using the beam polarizations that favor $e^-_L e^+_R$. The accuracies estimated for ILC in this section will thus be based on 1000 fb$^{-1}$ taken *entirely* with the beam polarizations $(P_{e^-}, P_{e^+}) = (-0.8, +0.2)$.

### 2.6.1 Measurement of $h \to \mu^+\mu^-$ decay using $e^+e^- \to \nu\overline{\nu}h$

The branching fraction of the $h \to \mu^+\mu^-$ decay is as small as $0.03\,\%$ for the $120$ GeV Standard Model Higgs boson. Its measurement thus requires a very good invariant mass resolution for the $\mu^+\mu^-$ pair. The measurement of this rare mode is a challenge to the tracking detectors and hence was chosen as one of the benchmark processes. The SiD group performed a full simulation study of the $h \to \mu^+\mu^-$ decay at $\sqrt{s} = 250$ GeV with $250\,\text{fb}^{-1}$ for $m_h = 120$ GeV as one of its LOI studies [76]. The expected number of signal events was only 26 before any cuts. After a simple cut-and-count analysis, the expected number of signal events became 8 with 39 background events in the final sample of $e^+e^- \to Zh$ followed by $Z- > q\overline{q}$ and $h \to \mu^+\mu^-$. This corresponds to a statistical significance of $1.1\ \sigma$. The $WW$ fusion process at $\sqrt{s} = 1000$ GeV will provide a higher statistics sample of Higgs bosons, as discussed above. We thus expect about 100 events for the $h \to \mu^+\mu^-$ mode. Since the cross sections for the $e^+e^- \to W^+W^- \to \mu^+\nu_\mu\mu^-\overline{\nu}_\mu$ and $e^+e^- \to ZZ \to \mu^+\mu^- f\overline{f}$ backgrounds will decrease, while the signal cross section will increase at higher energies, we would expect a meaningful measurement of the muon Yukawa coupling. An earlier fast simulation result showed that a $5\ \sigma$ signal peak would be observed with a $1\,\text{ab}^{-1}$ sample for $m_h = 120$ GeV [122, 123]. More recent full simulations by SiD and ILD showed that indeed we would be able to measure $\sigma \times BR(h \to \mu^+\mu^-)$ to 32% for $m_h = 125$ GeV even with the full beam-induced backgrounds. Together with the tau Yukawa coupling from the $h \to \tau^+\tau^-$ branching ratio, this measurement will provide an insight into the physics of lepton mass generation. With the charm Yukawa coupling from the $h \to c\overline{c}$ branching fraction, this also will allow us to probe the mass generation mechanism for the second generation matter fermions.

**Figure 2.19** Production cross sections for the Higgs-strahlung, $e^+e^- \to Zh$, the $WW$ fusion, $e^+e^- \to \nu\overline{\nu}H$, and $ZZ$ fusion processes as a function of the center of mass energy for $m_h = 125$ GeV and beam polarization $(P_{e^-}, P_{e^+}) = (-0.8, +0.2)$.

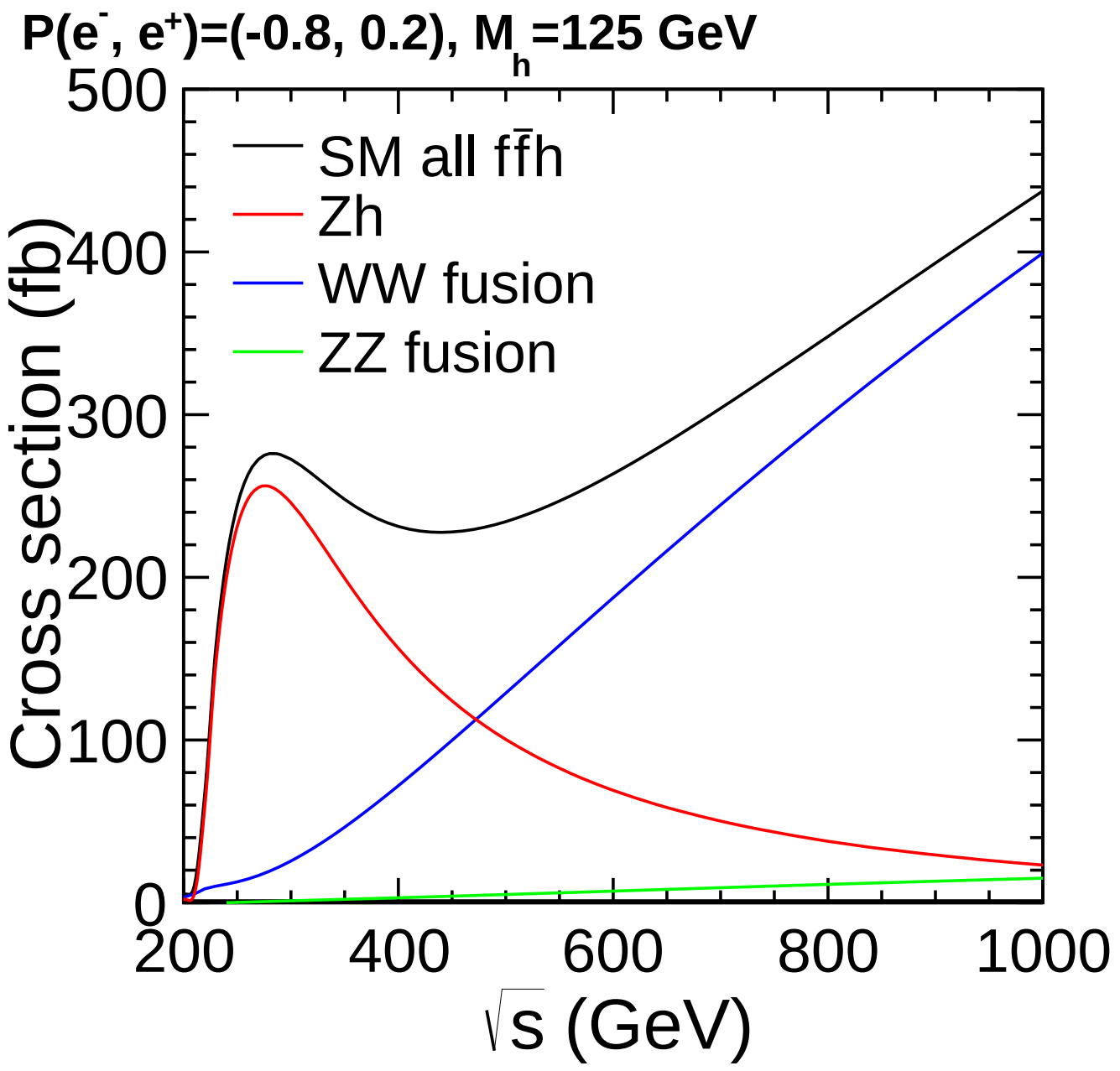

The new high-statistics sample of Higgs boson allows branching ratio measurements for the other decay modes to be improved. For example, we can achieve $\Delta BR(h \to \gamma\gamma)/BR((h \to \gamma\gamma) \simeq 5\,\%$ for $m_h = 120$ GeV with $1\,\text{ab}^{-1}$ taken at $(P_{e^-}, P_{e^+}) = (-0.8, +0.5)$ [124].

## 2.6.2 Top quark Yukawa coupling

The $10\%$ accuracy on the top quark Yukawa coupling expected at $\sqrt{s} = 500$ GeV can be significantly improved by the data taken at 1000 GeV, thanks to the larger cross section and the less background from $e^+e^- \to t\overline{t}$. Fast simulations at $\sqrt{s} = 800$ GeV showed that we would be able to determine the top Yukawa coupling to $6\%$ for $m_h = 120$ GeV, given an integrated luminosity of $1\,\text{ab}^{-1}$ and residual background uncertainty of $5\%$ [100, 101]. As described in the Detector Volume, Volume 4 of this report, full simulations just recently completed by SiD and ILD showed that the top Yukawa coupling could indeed be measured to a statistical precision of $4.0\%$ for $m_h = 125$ GeV with $1\,\text{ab}^{-1}$.

**Table 2.4.** Expected accuracies for cross section times branching ratio measurements for the $125$ GeV $h$ boson.

| | $\Delta(\sigma \cdot BR)/(\sigma \cdot BR)$ | | | | |
|---|---|---|---|---|---|
| $\sqrt{s}$ and $\mathcal{L}$ | 250 fb$^{-1}$ at 250 GeV | | 500 fb$^{-1}$ at 500 GeV | | 1 ab$^{-1}$ at 1 TeV |
| $(P_{e^-}, P_{e^+})$ | (-0.8,+0.3) | | (-0.8,+0.3) | | (-0.8,+0.2) |
| mode | $Zh$ | $\nu\overline{\nu}h$ | $Zh$ | $\nu\overline{\nu}h$ | $\nu\overline{\nu}h$ |
| $h \to b\overline{b}$ | 1.1% | 10.5% | 1.8% | 0.66% | 0.47% |
| $h \to c\overline{c}$ | 7.4% | - | 12% | 6.2% | 7.6% |
| $h \to gg$ | 9.1% | - | 14% | 4.1% | 3.1% |
| $h \to WW^*$ | 6.4% | - | 9.2% | 2.6% | 3.3% |
| $h \to \tau^+\tau^-$ | 4.2% | - | 5.4% | 14% | 3.5% |
| $h \to ZZ^*$ | 19% | - | 25% | 8.2% | 4.4% |
| $h \to \gamma\gamma$ | 29-38% | - | 29-38% | 20-26% | 7-10% |
| $h \to \mu^+\mu^-$ | 100% | - | - | - | 32% |

### 2.6.3 Higgs self-coupling in the $e^+e^- \to \nu_e\overline{\nu}_e hh$ process

At $\sqrt{s} = 1000$ GeV, the $e^+e^- \to \nu\overline{\nu}hh$ process will become significant and open up the possibility of measuring the triple Higgs coupling in the $WW$ channel [120]. The cross section for this process is only about $0.07$ fb$^{-1}$, but the sensitivity to the self-coupling is potentially higher since the contribution from the background diagrams is smaller, leading to the relation $\Delta\lambda/\lambda \simeq 0.85 \times (\Delta\sigma_{\nu\overline{\nu}hh}/\sigma_{\nu\overline{\nu}hh})$, as compared to $\Delta\lambda/\lambda \simeq 1.8 \times (\Delta\sigma_{Zhh}/\sigma_{Zhh})$ for the $e^+e^- \to Zhh$ process at 500 GeV. An early fast simulation study of $e^+e^- \to \nu\overline{\nu}hh$ showed that one could determine the triple Higgs coupling to an accuracy of $\Delta\lambda/\lambda \simeq 0.12$ [121], assuming $1$ ab$^{-1}$ luminosity and 80% left-handed electron polarization. A more recent fast simulation study indicated an accuracy $\Delta\lambda/\lambda \simeq 0.17$ for $2$ ab$^{-1}$ with $(P_{e^-}, P_{e^+}) = (-0.8, +0.2)$. The difference could be attributed to the more realistic analysis based on jet-clustering after parton showering and hadronization, as well as more background processes considered in the latter study. Finally, the measurement of the self-coupling has now been studied at 1 TeV will full simulation. That analysis is described in the Detector Volume, Volume 4 of this report. The result, for for $2$ ab$^{-1}$ with $(P_{e^-}, P_{e^+}) = (-0.8, +0.2)$, is $\Delta\lambda/\lambda \simeq 0.21$.

In addition to the fusion process, we also can use the $e^+e- \to Zhh$ process at $\sqrt{s} = 1000$ GeV. This process has somewhat less sensitivity, $\Delta\lambda/\lambda \simeq 2.8 \times (\Delta\sigma_{Zhh}/\sigma_{Zhh})$. The analysis gives $\Delta\lambda/\lambda \simeq 0.53$. Combining all of the measurements, assuming the nominal integrated luminosities of $500$ fb$^{-1}$ at $\sqrt{s} = 500$ GeV and $2000$ fb$^{-1}$ at $\sqrt{s} = 1000$ GeV with the left-handed beam combination: $(P_{e^-}, P_{e^+}) = (-0.8, +0.2)$, we expect that the Higgs self-coupling could be measured to $\Delta\lambda/\lambda \simeq 20.\%$.

**Table 2.5.** Expected accuracies for top Yukawa and self-coupling measurements of the $125$ GeV $h$ boson, with the specified energies and luminosity samples. The current analyses use the $h \to b\bar{b}$ mode only.

| process | $\sqrt{s}$ [GeV] | $\mathcal{L}$ [fb$^{-1}$] | $(P_{e^-}, P_{e^+})$ | $\Delta(\sigma \cdot BR)/(\sigma \cdot BR)$ | $\Delta g/g$ |
|---|---|---|---|---|---|
| $t\bar{t}h$ | 500 | 500 | (-0.8,+0.3) | 35% | 18% |
| $Zhh$ | 500 | 500 | (-0.8,+0.3) | 64% | 104% |
| $t\bar{t}h$ | 1000 | 1000 | (-0.8,+0.2) | 8.7% | 4.0% |
| $\nu\bar{\nu}hh$ | 1000 | 1000 | (-0.8,+0.2) | 38% | 28% |

**Table 2.6.** Expected accuracies for Higgs boson couplings derived from the accuracy estimates for measured rates given in Tables 2.4 and 2.5. For the invisible branching ratio, the numbers quoted are 95% confidence upper limits. The four columns refer to: LHC, 300 fb$^{-1}$, 1 detector; ILC at 250 GeV, with 250 fb$^{-1}$; ILC at 500 GeV, with 500 fb$^{-1}$; ILC at 1000 GeV, with 1000 fb$^{-1}$. Each column includes the stated data set and all previous ones [65].

| Mode | LHC | ILC(250) | ILC500 | ILC(1000) |
|---|---|---|---|---|
| $WW$ | 4.1 % | 1.9 % | 0.24 % | 0.17 % |
| $ZZ$ | 4.5 % | 0.44 % | 0.30 % | 0.27 % |
| $b\bar{b}$ | 13.6 % | 2.7 % | 0.94 % | 0.69 % |
| $gg$ | 8.9 % | 4.0 % | 2.0 % | 1.4 % |
| $\gamma\gamma$ | 7.8 % | 4.9 % | 4.3 % | 3.3 % |
| $\tau^+\tau^-$ | 11.4 % | 3.3 % | 1.9 % | 1.4 % |
| $c\bar{c}$ | – | 4.7 % | 2.5 % | 2.1 % |
| $t\bar{t}$ | 15.6 % | 14.2 % | 9.3 % | 3.7 % |
| $\mu^+\mu^-$ | – | – | – | 16 % |
| self | – | – | 104% | 26 % |
| BR(invis.) | $<$ 9% | $<$ 0.44 % | $<$ 0.30 % | $<$ 0.26 % |
| $\Gamma_T(h)$ | 20.3% | 4.8 % | 1.6 % | 1.2 % |

## 2.7 Summary of measurement precisions expected at ILC

For historical reasons, most of the full simulation studies we discussed above were done for $m_h = 120$ GeV. Given the likelihood that the new particle discovered at the LHC is a Higgs boson, we would like to know the ILC capabilities for a Higgs boson of mass 125 GeV. These can be obtained by extrapolation of the full-simulation results, taking into account the changes in the signal and background as well as the changes in the pattern of Higgs boson branching ratios as the assumed mass is changed. The extrapolated results for the $\sigma \cdot BR$ measurements at different energies are summarized in Table 2.4. In the extrapolation, we scaled the signal and the background with the effective cross sections calculated with the new TDR beam parameters and, for the signal, applied the LHC-recommended branching ratios for $m_h = 125$ GeV. For the 1 TeV results, there are some differences between ILD and SiD as seen in the benchmark results described in the corresponding DBD chapters. We listed the SiD values here to be conservative.

We performed a similar exercise for the top Yukawa coupling and the self-coupling measurements and tabulated the results of the extrapolation in Table 2.5, where we just scaled the signal with the background unchanged. Since the mass separation from $W$ and $Z$ bosons should be better for $m_h = 125$ GeV than for $m_h = 120$ GeV, these estimates should be conservative.

The measurements in Tables 2.4 and 2.5 imply a very high level of precision for the various Higgs boson couplings. To quantify this, we have carried out a global fit to these measurements, assuming the errors given in these tables with the Standard Model as the central value in all cases. The fit is done in parallel to the analysis reported above for the LHC in Fig. 2.4, with 9 parameters representing independent Higgs boson couplings to $WW$, $ZZ$, $b\bar{b}$, $gg$, $\gamma\gamma$, $\tau^+\tau^-$, $c\bar{c}$, $t\bar{t}$, and invisible final states. The results for the errors on Higgs couplings are shown in Table 2.6. The four columns represent the errors from LHC (300 fb$^{-1}$, 1 detector) only, and then, cumulatively, ILC at 250 GeV, ILC at 500 GeV, and ILC at 1000 GeV [65]. The result of this fit are shown graphically in Fig. 2.20.

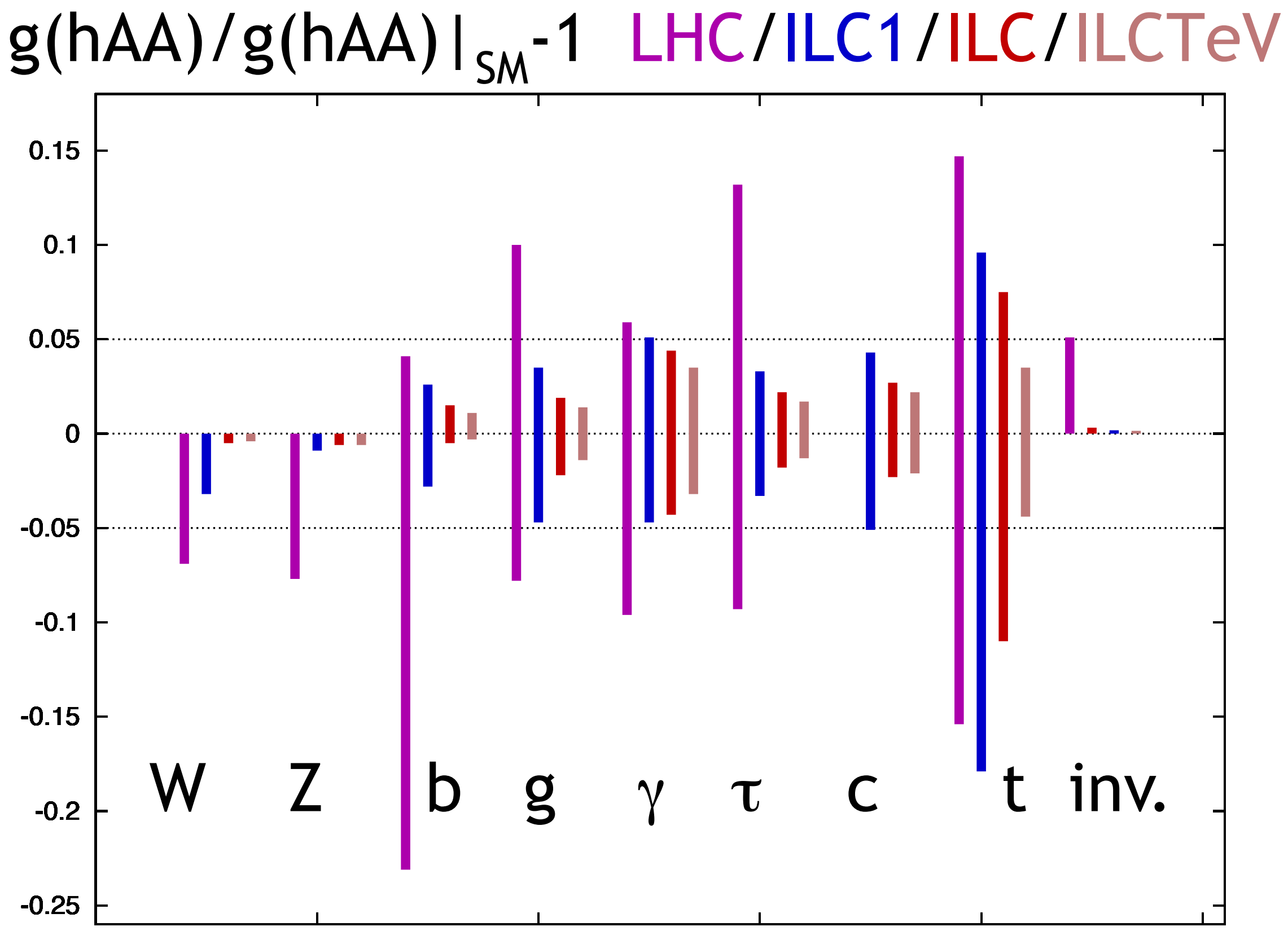

**Figure 2.20.** Estimate of the sensitivity of the ILC experiments to Higgs boson couplings in a model-independent analysis. The plot shows the 1 $\sigma$ confidence intervals as they emerge from the fit described in the text. Deviation of the central values from zero indicates a bias, which can be corrected for. The upper limit on the $WW$ and $ZZ$ couplings arises from the constraints (2.31). The bar for the invisible channel gives the 1 $\sigma$ upper limit on the *branching ratio*. The four sets of errors for each Higgs coupling represent the results for LHC (300 fb$^{-1}$, 1 detector), the threshold ILC Higgs program at 250 GeV, the full ILC program up to 500 GeV, and the extension of the ILC program to 1 TeV. The methodology leading to this figure is explained in [65].

## 2.8 Conclusion

The landscape of elementary particle physics has been altered by the discovery by the ATLAS and CMS experiments of a new boson that decays to $\gamma\gamma$, $ZZ$, and $WW$ final states [2, 3]. The question of the identity of this bosons and its connection to the Standard Model of particle physics has become the number one question for our field. In this section, we have presented the capabilities of the ILC to study this particle in detail. The ILC can access the new boson through the reactions $e^+e^- \to Zh$ and through the $WW$ fusion reaction $e^+e^- \to \nu\overline{\nu}h$. Though our current knowledge of this particle is still limited, we already know that these reactions are available at rates close to those predicted for the Higgs boson in the Standard Model. The ILC is ideally situated to give us a full understanding of this particle, whatever its nature.

The leading hypothesis for the identity of the new particle is that it is the Higgs boson of the Standard Model, or a similar particle responsible for electroweak symmetry breaking in a model that includes new physics at the TeV energy scale. We have argued that, if this identification proves correct, the requirements for experiments on the nature of this boson are extremely challenging. Though there are new physics models that predict large deviations of the boson couplings from the Standard Model predictions, the typical expectation in new physics models is that the largest deviations from the Standard Model are at the 5–10% level. Depending on the model, these deviations can occur in any of the boson's couplings. Thus, a comprehensive program of measurements is needed, one capable of being interpreted in a model-independent way. Our estimate of the eventual LHC capabilities, given

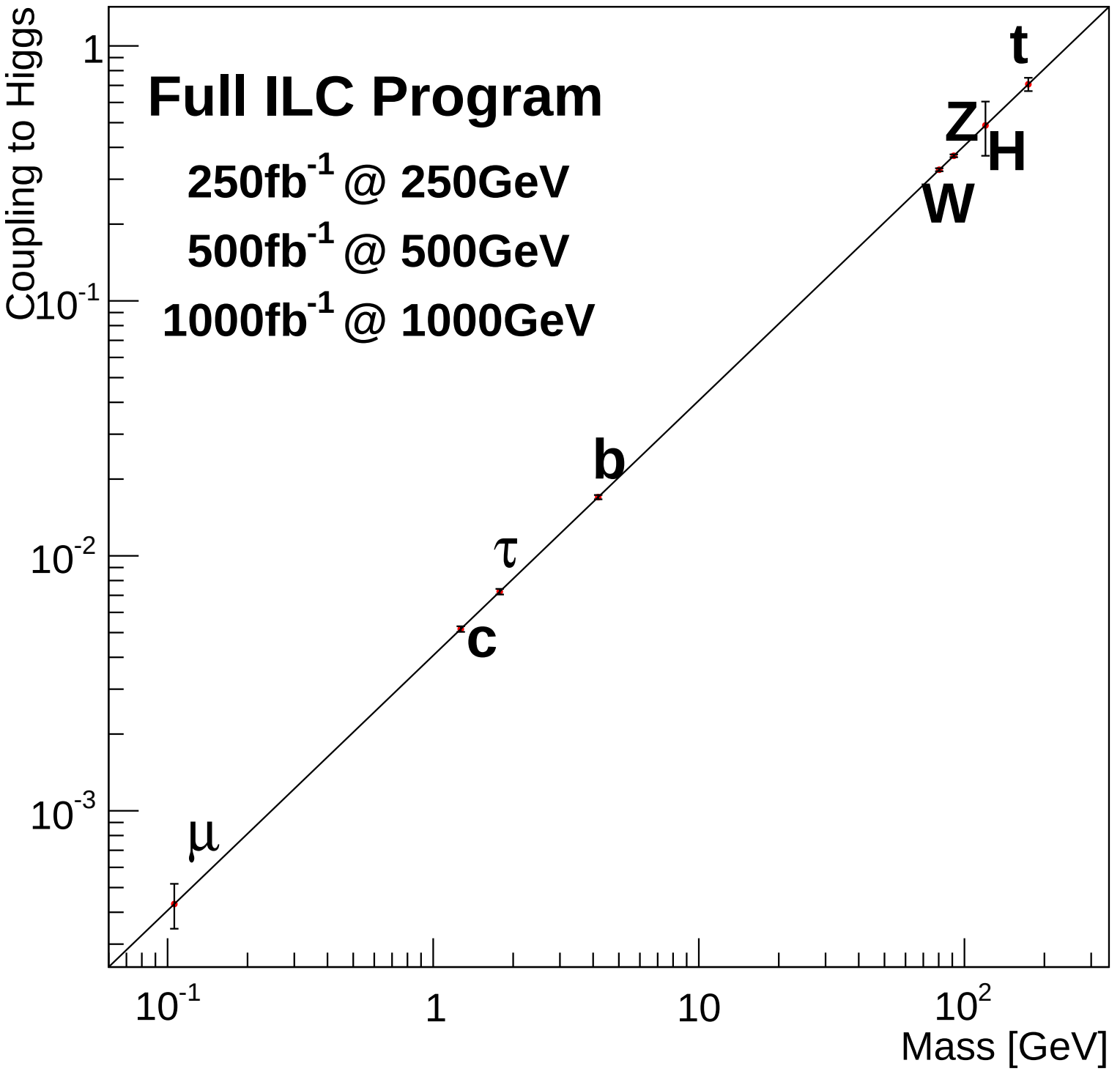

**Figure 2.21** Expected precision from the full ILC program of tests of the Standard Model prediction that the Higgs coupling to each particle is proportional to its mass.

in Fig. 2.4, falls short of that goal.

We then presented the capabilities of the ILC for precision measurements of the Higgs boson couplings. The ILC program for Higgs couplings can begin at a center of mass energy of 250 GeV, near the peak of the cross section for $e^+e^- \to Z^0h^0$. This program allows a direct measurement of the cross section, rather than measurement that includes branching ratios, already eliminating an important source of ambiguity from the LHC data. The program also allows the measurement of individual branching channels, observed in recoil against the $Z^0$ boson. The excellent flavor tagging capabilities of the ILC experiments allow access to the $c\bar{c}$ decay mode of the Higgs boson and sharpen the observation of many other modes. The ILC experiments are highly sensitive to possible invisible or other unexpected decay modes of the Higgs boson, with sensitivity at the percent level.

A later stage of ILC running at the full energy of 500 GeV will enhance these capabilities. At 500 GeV, the $W$ fusion reaction $e^+e^- \to \nu\bar{\nu}h$ turns on fully, giving a very precise constraint on the Higgs boson coupling to $WW$. The increased statistics sharpens the measurement of rare branching channels such as $\gamma\gamma$. Higher energy also gives improved $g/c/b$ separation in the hadronic decay models. Running at 500 GeV allows the first direct measurements of the Higgs coupling to $t\bar{t}$ and the Higgs self-coupling.

The technology of the ILC will eventually allow extended running at higher energies, up to 1 TeV in the center of mass. A 1 TeV program will add further statistics to the branching ratio measurements in all channels, using the increasing $e^+e^- \to \nu\bar{\nu}h$ cross section. It also very much increases the sensitivity of the determinations of the Higgs coupling to $t\bar{t}$ and the Higgs self-coupling.

The progression of this program is shown graphically in Fig. 2.20. For each Higgs boson coupling, four sets of error bars are shown, always assuming that the underlying value of the coupling is that of the Standard Model. The first is the estimate of the LHC capability, from Fig. 2.4. The second is the error that would be obtained by adding the data from a 250 fb$^{-1}$ run of the ILC at 250 GeV. The third is the error that would be obtained by adding to this the data from a 500 fb$^{-1}$ run of the ILC at 500 GeV. The final error bar would be the result of adding a 1 $ab^{-1}$ data set at 1 TeV. Not shown, but also relevant, are the capabilities of the ILC to measure the Higgs self-coupling to about 24% accuracy and the Higgs coupling to $\mu^+\mu^-$ to about 20% accuracy in the 1 TeV program.

The results of this program can also be represented as precision tests of the Standard Model relation that the Higgs coupling to each particle is exactly proportional to the mass of that particle. The expected uncertainties in those tests from the measurements described above are shown in Fig. 2.21.

This is the program that is needed to fully understand the nature of the newly discovered boson and its implications for the puzzle of electroweak symmetry breaking. The ILC can provide it.

# Chapter 3
# Two-Fermion Processes

The reactions $e^+e^- \to f\overline{f}$, where $f$ could be leptons or quarks, provide a powerful tool to search for and characterize physics beyond the Standard Model at the ILC. These processes are distinguished by clean, simple final states, and precise perturbative predictions of the SM contributions are available. As a result, ILC experiments will be sensitive to even small deviations from the SM predictions in these channels, enabling them to study new physics at energy scales far above the center-of-mass energy of the collider.

## 3.1 Systematics of $e^+e^- \to f\overline{f}$

Despite the simplicity of the two-fermion final state, the process $e^+e^- \to f\overline{f}$ offers a large number of methods with which to probe for deviations from the Standard Model. In this section, we will review the observables that the ILC will make available. In the following sections, we will review how these observables can be applied to discover and then to analyze many signals of new physics that can appear in these reactions.

For all channels except $e^+e^- \to e^+e^-$, helicity conservation implies that the process $e^+e^- \to f\overline{f}$ is dominated by $s$-channel spin 1 exchange. This assumption applies whenever fermion mass effects can be neglected, and this is an excellent approximation at 500 GeV for pair-production of all Standard Model fermions except the top quark. In this case, the angular distribution of $e^+e^- \to f\overline{f}$ is simply written as

$$\frac{d\sigma}{d\cos\theta} = \frac{\pi\alpha^2}{2s}[A_+(1+\cos\theta)^2 + A_-(1-\cos\theta)^2] \ . \tag{3.1}$$

The coefficients $A_+$, $A_-$ depend on the electron polarization. Models with gravitational effects at the TeV scale (for example, Randall-Sundrum models [1]) will add terms from $s$-channel spin 2 exchange that are higher polynomials in $\cos\theta$.

In (3.1), the term multiplying $A_+$ is generated by the polarized reactions $e^-_L e^+_R \to f_L\overline{f}_R$ and $e^-_R e^+_L \to f_R\overline{f}_L$, the term multiplying $A_-$ is generated by $e^-_L e^+_R \to f_R\overline{f}_L$ and $e^-_R e^+_L \to f_L\overline{f}_R$. All other polarized cross sections are zero in the absence of mass corrections. This means that by measuring the cross sections and forward backward asymmetries with highly polarized $e^-_L$ and $e^-_R$, we obtain 4 independent pieces of information on the $s$-channel amplitudes. In principle, only the electron beam needs to be polarized. However, even a small polarization of the positron beam improves the effective initial-state polarization according to

$$P_{eff} = \frac{P(e^-)+P(e^+)}{1+P(e^-)P(e^+)} \ . \tag{3.2}$$

Thus, a measurement with 80% polarization in the electron beam and 30% polarization in the positron beam yields an effective initial-state polarization of almost 90%. At the ILC, polarization is monitored externally, but in addition the actual polarization in collisions can be determined from the high-rate processes of Bhabha scattering and forward $W^-W^+$ production. Theoretical calculations of the 2-fermion cross sections are controlled to below the part-per-mil level.

The four observables described in the previous paragraph are available for any final state that can be distinguished at the ILC. That is, these quantities can be measured separately for light quarks, $c$ quarks, $b$ quarks, $e$, $\mu$, and $\tau$. The typical $c$, $\tau$ and $b$ identification efficiencies expected at the ILC are 35%, 60%, and over 96%, respectively [2]. In addition, the final state $\tau$ lepton polarization can be determined as a cross-check on the leptonic coupling measurements [3, 4].

The dominant contributions to $e^+e^- \to f\overline{f}$ at 500 GeV will probably come from Standard Model $s$-channel $\gamma$ and $Z^0$ exchange. However, additional effects may arise from new gauge bosons, from contact interactions associated with fermion compositeness, or from effects of extra dimensions. These terms can be seen at the ILC as corrections to the $e^+e^- \to f\overline{f}$ cross sections and asymmetries, arising from interference of new physics with the Standard Model amplitudes. In addition, for example in the case of extra dimensions, these effects can add additional dependence on $\cos\theta$ related to the spin-2 graviton exchange. We will now review the expected sensitivity of the ILC experiments to these effects.

## 3.2 $Z'$ physics

A canonical, well-motivated example of new physics that can be discovered and studied in $e^+e^- \to f\overline{f}$ is a new, heavy, electrically neutral gauge boson, commonly denoted by $Z'$. There are many extensions of the SM that predict one or more such particles (for reviews and references, see [5, 6]). For example, Grand Unified Theories (GUTs) based on groups such as $SO(10)$ or $E_6$ contain extra $U(1)$ factors in addition to the SM gauge group, and hence $Z'$ bosons. Similarly, superstring constructions often involve large gauge symmetries that contain extra $U(1)$ factors. Since the $Z'$ couplings conserve baryon and lepton numbers, the mass of the $Z'$ may be well below the GUT or string scale, as low as the TeV, without conflict with experiment. In fact, in many supersymmetric GUT and string models, the $Z'$ mass is tied to the soft supersymmetry breaking scale, expected to be at the TeV scale. The motivation for a TeV-scale $Z'$ is particularly strong in supersymmetric models with additional particles that are singlets of the SM $SU(2) \times U(1)$. One of these models, the next-to-minimal supersymmetric standard model (NMSSM), has recently attracted much interest, since it provides a simple way to reduce the fine-tuning associated with a 125 GeV Higgs [7]. The weak-scale mass of the SM singlet field can be naturally explained if this field is charged under a new $U(1)$ symmetry broken at TeV energies; in addition, the domain-wall problem of the NMSSM is avoided in this case. Among non-supersymmetric possibilities, a very interesting example of a model containing a $Z'$ is the Little Higgs model, where extra gauge bosons are introduced to cancel quadratic divergences in the Higgs mass renormalization by the SM gauge bosons (for reviews and references, see [8]). Naturalness of electroweak symmetry breaking requires that these new gauge bosons appear at the TeV scale.

Searches for $Z'$ have been conducted, most recently, at LEP and the Tevatron, and are currently in progress at the LHC. The negative results of these searches preclude the possibility of on-shell $Z'$ production at the ILC. Indeed, the LHC now excludes the appearance of large $Z'$ resonances over most of the range of proposed 3 TeV lepton colliders. This makes it likely that our most important tool for the characterization of any $Z'$ discovered at the LHC will be through indirect effects uncovered through the precision measurement of $e^+e^- \to f\overline{f}$ processes. The dominant effects of new physics in this case come from the interference between the diagrams involving the SM $\gamma/Z^0$ and those involving the $Z'$. Thanks to the high precision of the ILC, its capabilities to discover the $Z'$ and measure its couplings actually exceed those of the LHC in most cases.

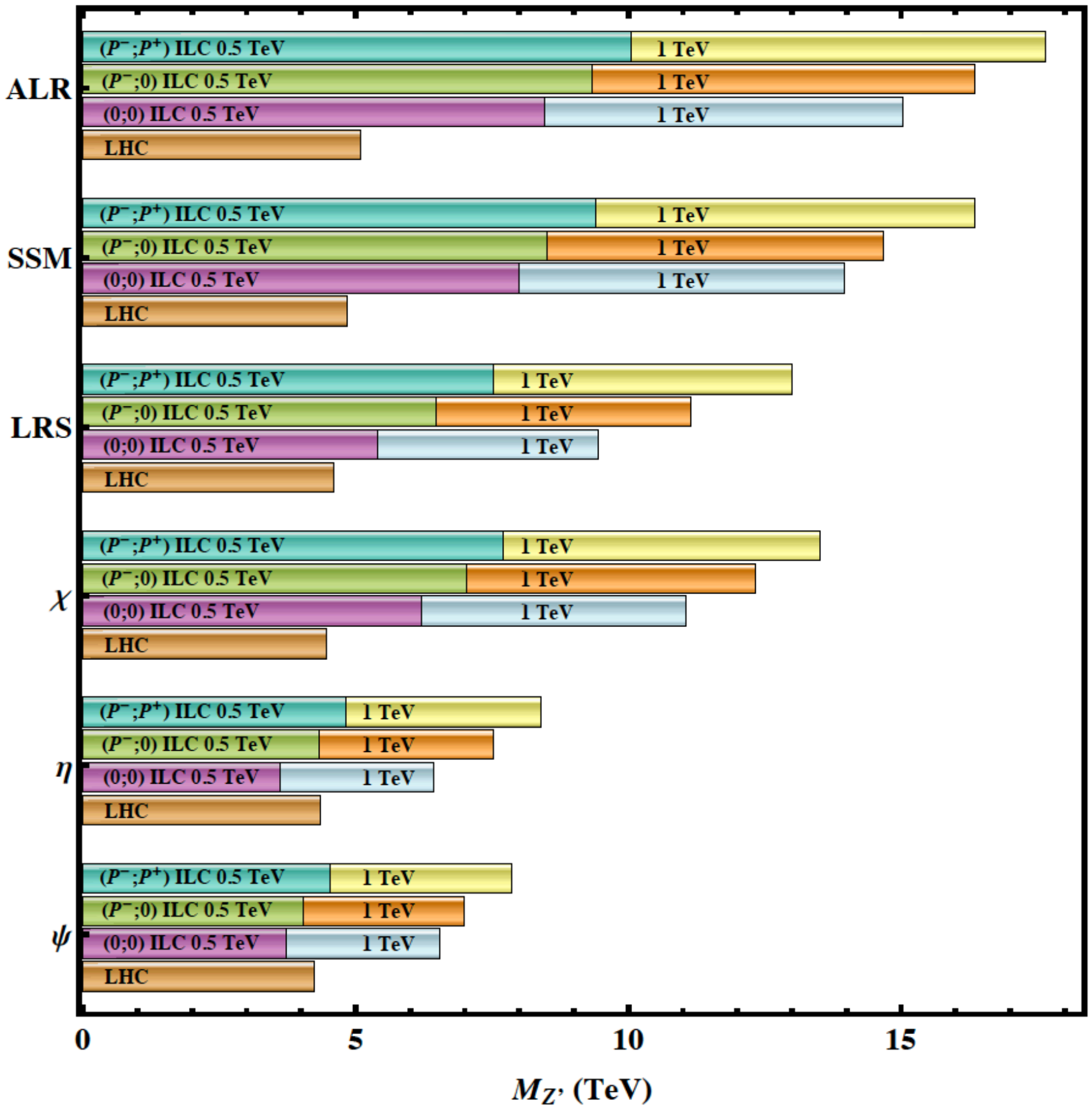

**Figure 3.1.** Sensitivity of the ILC to various candidate $Z'$ bosons (quoted as 95% confidence limits for exclusion), with $\sqrt{s} = 0.5$ (1.0) TeV and $\mathcal{L}_{\rm int} = 500$ (1000) fb$^{-1}$. The sensitivity of the LHC-14 via Drell-Yan process $pp \to \ell^+\ell^- + X$ with 100 fb$^{-1}$ of data are shown for comparison. For details, see [14].

## 3.2.1 Benchmark $Z'$ models

Predictions for the contribution of a $Z'$ to any observable depend on the boson's mass $M_{Z'}$ and its couplings to the SM fermions, which are model-dependent. While a very large variety of models have been proposed, a few canonical benchmark cases have been extensively studied and provide a set of reference points for comparisons between experiments. The Sequential Standard Model (SSM) assumes that all $Z'$ couplings are the same as for the SM $Z$. The left-right symmetric (LRS) model extends the SM electroweak gauge group to $SU(2)_L \times SU(2)_R \times U(1)_{B-L}$, with the $SU(2)_R \times U(1)_{B-L} \to U(1)_Y$ breaking at the TeV scale. The $Z'$ couples to the linear combination of $T_{3R}$ and $B-L$ currents orthogonal to the SM hypercharge. Another set of popular benchmark models is based on the $E_6$ GUT, where the TeV-scale $Z'$ is generally a linear combination of the two extra $U(1)$ gauge bosons $Z_\psi$ and $Z_\chi$: $Z' = Z_\chi \cos\beta + Z_\psi \sin\beta$. Some well-motivated possibilities are $\beta = 0$ (the "$\chi$-model"), $\beta = \pi/2$ (the "$\psi$-model"), and $\beta = \pi - \arctan\sqrt{5/3}$ (the "$\eta$-model", which occurs in Calabi-Yau compactification of the heterotic string if $E_6$ breaks directly to a rank-5 group). It is also possible to embed a left-right symmetric model in $E_6$, leading to the so-called "alternative" left-right (ALR) model. The $Z'$ couplings to the SM fermions in each of these models can be found, for example, in Table 1 of [9]. Well-studied Little Higgs models that contain $Z'$ candidates include the original "Littlest Higgs" (LH) [10], as well as the Simplest Little Higgs (SLH) [11].

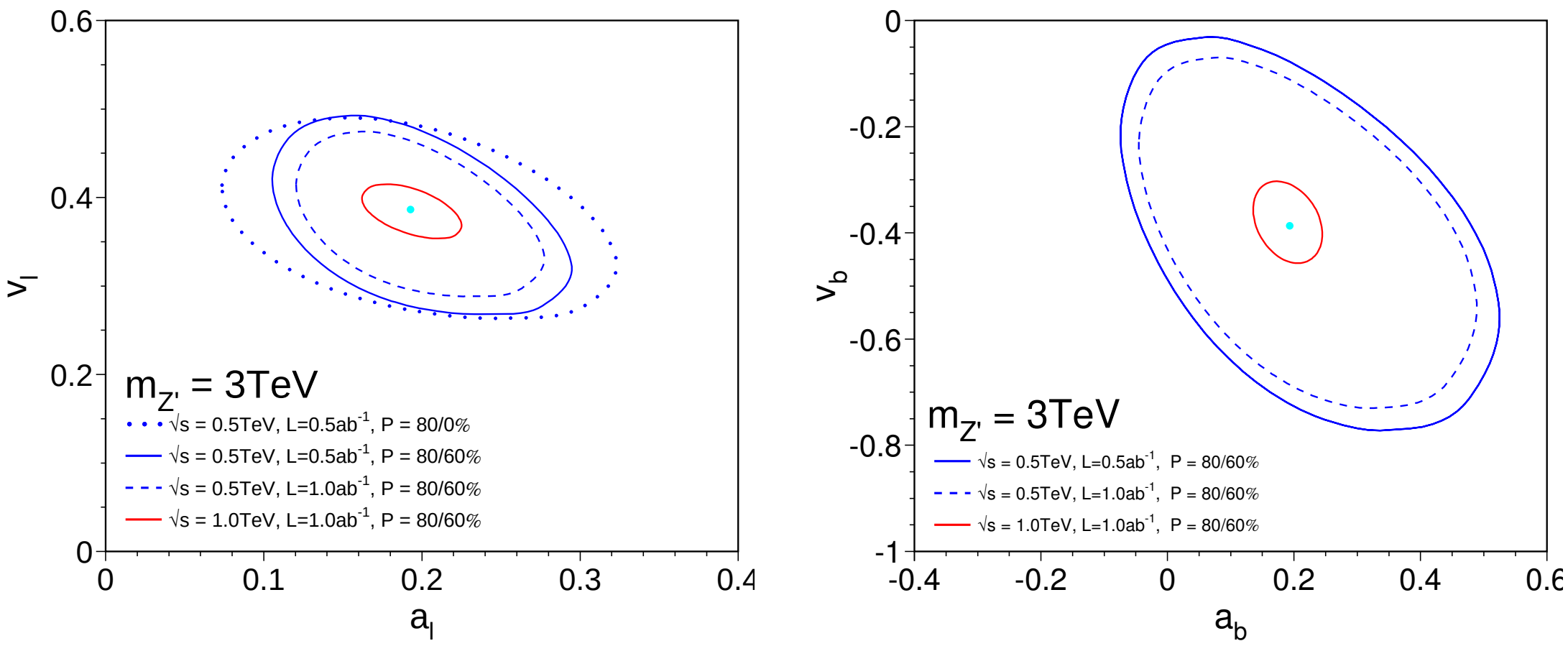

**Figure 3.2.** Derived coupling of a $SO(10)$ $Z'$ boson with $m_{Z'} = 3$ TeV to leptons (left) and b-quarks (right). Only the results for positive values of the vector leptonic coupling $v_l$ are shown; there is a reflection with all $Z'$ couplings reversed. The two solid curves correspond to ILC at 500 (1000) GeV with 500 (1000) fb$^{-1}$ and $P_{eff}$ = 95%.

### 3.2.2 Current limits on $Z'$ and the ILC reach

The most restrictive bounds on most $Z'$ models currently come from the LHC experiments. Both ATLAS and CMS published bounds using the 20 fb$^{-1}$ dataset at $\sqrt{s} = 8$ TeV, with dielectron and dimuon final states [12, 13]. For the SSM, the current bound on the $Z'$ mass is $2.9$ TeV, stronger than the indirect LEP-2 bound. A wide range of $E_6$ models have been excluded for $Z'$ masses below $2.4 - 2.6$ TeV, indicating that the model-dependence is rather weak.

The current LHC bounds rule out the possibility of on-shell production of a $Z'$ at the ILC. However, the ILC will be sensitive to $Z'$ even at $\sqrt{s} \ll M_{Z'}$, via contact-interaction corrections to 2-fermion processes. A recent estimate of the ILC reach in various $Z'$ models [14], compared to the LHC reach at 14 TeV [9], is shown in Fig. 3.1. The reach of a 500 GeV ILC exceeds the LHC reach in most models, while a 1 TeV ILC will significantly improve on the LHC performance in all cases, with sensitivity well above 10 TeV in many models.

### 3.2.3 Measurement of $Z'$ couplings

If a $Z'$ is discovered, the next task would be to measure its couplings to SM fermions. In this section, we present a case study illustrating the capabilities of the ILC to perform this measurement. We assume that the $Z'$ arises from $SO(10) \to SU(5)$ gauge symmetry breaking (the $\chi$-model), and has a mass of 3 TeV. Such a $Z'$ would be discovered at the LHC, and its mass and spin can be measured there. The ILC's role would be to complete the characterization of this particle by a precise measurement of its couplings.

The values of the $Z'$ vector and axial-vector couplings $v_f$ and $a_f$ are primarily determined by the measurement of the cross section of the process $e^+e^- \to f\overline{f}$. Measurements of the forward-backward charge asymmetry and of the left-right asymmetry shrink the range for axial-vector coupling and the left- and right-handed couplings, respectively. More details can be found in [5, 15]. Assuming lepton universality, the combination of all leptonic final states allows a precise measurement of the leptonic $Z'$ couplings. Here, the role of beam polarization is important. Without polarized beams, a 4-fold sign ambiguity for the couplings $a_l$ and $v_l$ is obtained. With a polarized electron beam, only a twofold ambiguity remains, and the $Z'$ couplings are determined with higher precision. Having both beams polarized improves the results further: the effective luminosity is increased and the error on the $A_{LR}$ measurement can be decreased due to the reduced uncertainty of the effective polarization. Once the leptonic couplings are measured, the $Z'$ couplings to quarks can be obtained from hadronic final states. Excellent flavor tagging at the ILC with high efficiency and purity is essential to achieve high

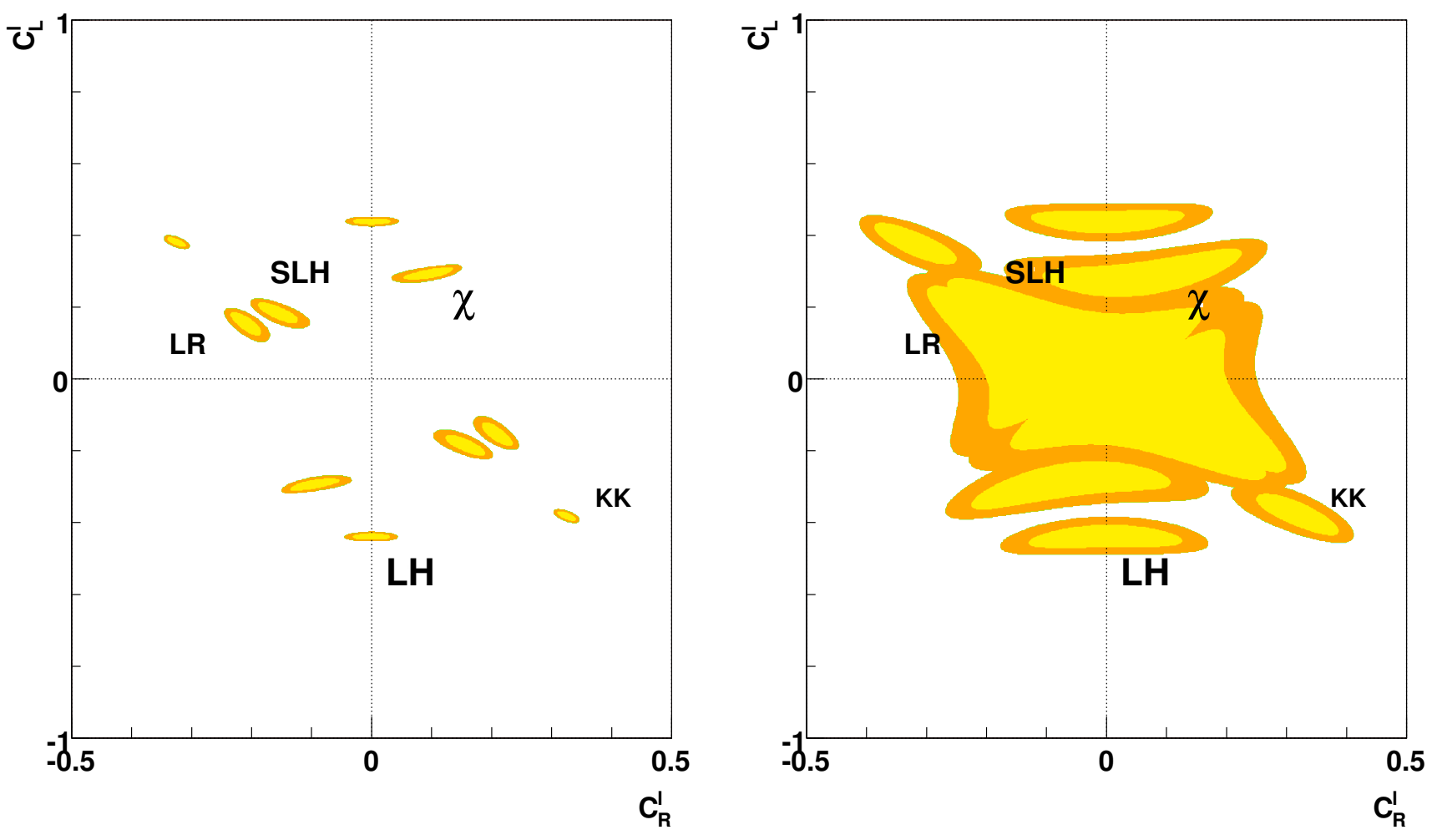

**Figure 3.3** 95% confidence regions in the plane of the couplings of left- and right-handed leptons to a $Z'$ boson, for the ILC with $\sqrt{s} = 500$ GeV and 1000 fb$^{-1}$ and 80%/60% electron and positron polarization, for $M_{Z'} = 2$ TeV (left panel) and 4 TeV (right panel). For further details, see [16].

precision measurements.

The results for the measurement of leptonic $Z'$ couplings are presented in Fig. 3.2. Systematic uncertainties of 0.2% for the leptonic observables and the luminosity are taken into account; a 0.25% error on beam polarization measurement is assumed. The $Z'$ coupling to b-quarks resulting from a simultaneous fit to lepton and $b\bar{b}$ final states is shown in Fig. 3.2, where a systematic uncertainty of 0.5% is assumed for b-quark observables.

It is evident that increasing the center-of-mass energy is more efficient than collecting more luminosity. At high luminosities systematic uncertainties limit the sensitivity, and even in case of negligible systematic errors doubling the luminosity would improve the range for the $Z'$ couplings only by a factor 0.84. A rough scaling for $Z'$ couplings and mass with energy and luminosity is given by the relation $g/m_{Z'} \propto (s \cdot L_{int})^{-1/4}$.

### 3.2.4 $Z'$ model discrimination

Since every model predicts a particular pattern of $Z'$ couplings to SM fermions, a measurement of these couplings makes it possible to distinguish between models. For example, expected accuracy of the measurement of the $Z'$ couplings to charged leptons, in a variety of popular $Z'$ models, is shown in Fig. 3.3 (from [16]). The predictions of the benchmark models are quite distinct. Most models can be readily distinguished even for a $Z'$ as heavy as 4 TeV, at a 500 GeV ILC. It should be emphasized that beam polarization plays a crucial role in this analysis.

## 3.3 Quark and lepton compositeness

In many extensions of the SM, quarks and leptons themselves are composite particles, resolved into more fundamental constituents at an energy scale $\Lambda$. The effect of such compositness in $2 \to 2$ fermion scattering processes at energies well below $\Lambda$ is to induce contact-interaction type corrections, similar to the corrections due to a heavy resonance discussed above. The effects can be parametrized by adding four-fermion operators to the Lagrangian with coefficients proportional to inverse powers of $\Lambda$ [17]. Currently, the strongest bounds on four-lepton and $eeqq$ operators are $\Lambda \gtrsim 10$ TeV [18, 19]. These bounds come from experiments at LEP. The LHC is unlikely to improve these limits, since at the LHC we have only limited polarization observables in 4-fermion reactions and we do not know the flavor of initial state quarks. The ILC can dramatically increase the reach, with sensitivity to scales as high as $50 - 100$ TeV depending on the helicity structure of the operators (see Fig. 3.4.)

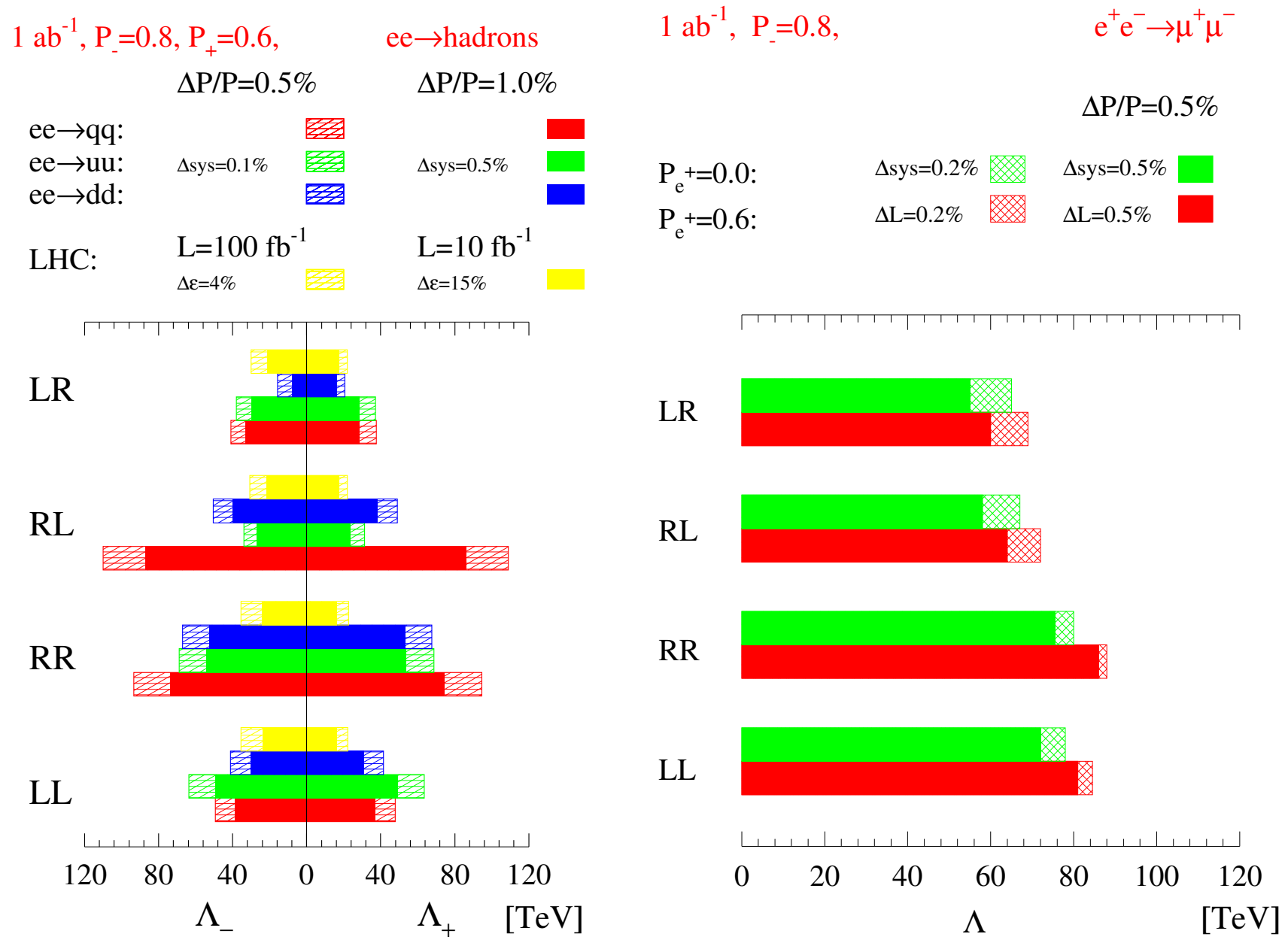

**Figure 3.4** Sensitivities (95% c.l.) of a 500 GeV ILC to contact interaction scales $\Lambda$ for different initial helicities, from [20]. Left: $e^+e^- \to$ hadrons. Right: $e^+e^- \to \mu^+\mu^-$.

## 3.4 Extra dimensions

Many interesting extensions of the SM postulate the existence of extra spatial dimensions, beyond the familiar three, which are usually assumed to be compact. Motivation for extra dimensions comes from two sides. From the top-down point of view, consistency of string theory requires that the full space-time be 10-dimensional, and additional dimensions must be compactified. From the bottom-up perspective, models with extra dimensions can address some of the theoretical shortcomings of the SM, such as the gauge hierarchy problem. While the extra dimensions of string theory can have any size, in all phenomenologically interesting models the extra dimensions become experimentally manifest at the TeV scale, within the range of the ILC experiments.

Phenomenologically, the most important feature of models with extra dimensions is the appearance of Kaluza-Klein (KK) resonances. Each SM particle (including the graviton) that is allowed to propagate beyond 4D is accompanied by a tower of KK excitations, particles of the same spin and progressively higher masses. In the simplest case of toroidal compactification of radius $R$, the $n$-th KK mode has mass $m_n = n/R$. The effect of the KK modes on $e^+e^- \to f\bar{f}$ are similar to that of a $Z'$. They produce contact interactions, or, if collision energy is sufficient, resonances.

### 3.4.1 Flat, TeV-sized extra dimensions

The simplest extension is to add $k$ extra dimensions compactified on a torus $T^k$, and allow all SM fields to propagate in the full space. The most popular model of this type is the "universal extra dimension" (UED) model [21], with $k = 1$ and radius $R \sim 1/$TeV. This model assumes a $\mathcal{Z}_2$ symmetry under which the $n$-th KK mode has KK-parity $(-1)^n$. As a result, production of a single first-level KK partner in SM collisions is not possible, and the phenomenology of the first-level KK states is similar to that of supersymmetric models with R-parity. The even-level KK states, on the other hand, may be singly produced via KK-number violating interactions, induced by loops [22]. This leads to resonances or contact-interaction corrections in $e^+e^- \to f\bar{f}$ [23, 24]. An estimated sensitivity of the ILC to the UED model is shown in Fig. 3.5; values of $1/R \sim 1$ TeV can be probed. The reach is significantly lower than for a conventional $Z'$, due to loop-suppressed couplings. However, it should be noted that the same suppression severely limits the ability of the LHC to search for the single KK-mode production. Any resonance for which the coupling to quarks is suppressed by a factor of 10 would contribute a fluctuation below 1% in the Drell-Yan mass spectrum, and this will be indistinguishable even for rather light KK masses. Small mass splittings among the KK states at the first level make the LHC searches for pair-production very difficult as well.

**Figure 3.5**
Discovery reach of the ILC, with $\mathcal{L}_{\text{int}} = 1000$ fb$^{-1}$ and energy indicated on the plot, for the UED model in the 2-fermion channel. Polarization of 80%/60% for electrons/positrons is assumed. Leptonic and hadronic final states are combined. The scale $\Lambda$ is the cutoff of the theory. From [23].

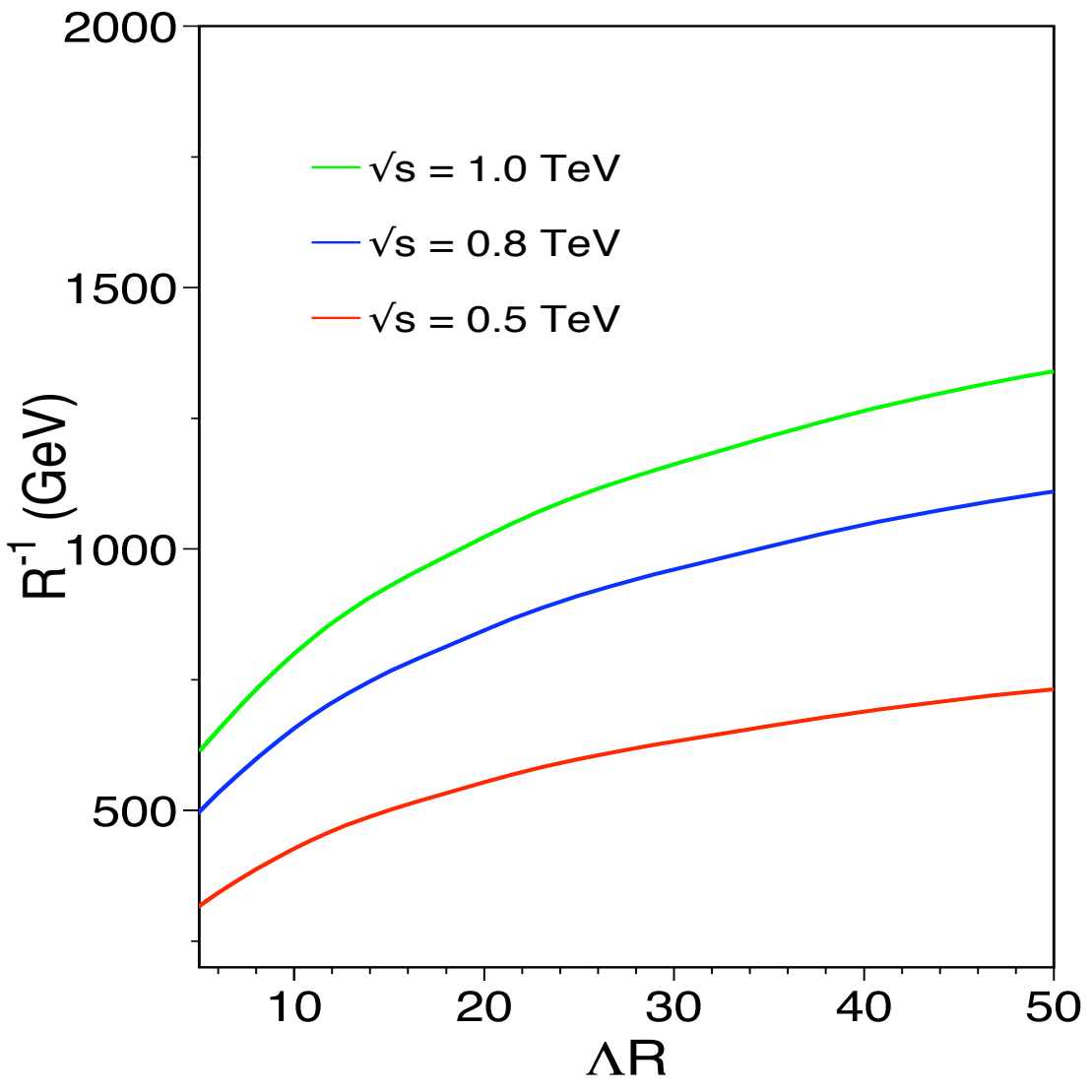

### 3.4.2 Large extra dimensions

The extra dimensions may have sizes much larger than TeV$^{-1}$, if *only* gravity can propagate in them, while the SM fields are confined on a 4D "brane" inside the full space. Arkani-Hamed, Dimopoulos and Dvali (ADD) [25] proposed that such models can provide an alternative solution to the gauge hierarchy problem: gravity is weaker than other forces due to the larger space in which it propagates. The ADD model is characterized by the fundamental Planck scale $M_D$ (required to be $\sim$TeV to solve the hierarchy problem); and the number of extra dimensions $k$. Constraints on macroscopic modifications of Newtonian gravity imply that only cases $k \geq 2$ are phenomenologically relevant.

The model predicts a tower of KK gravitons $G_{KK}$, with very small spacing in mass, of order $1/R$. While each of the $G_{KK}$ couples to the SM with gravitational strength, their large multiplicity may yield observable effects in $e^+e^- \to G_{KK} \to f\overline{f}$, although no individual resonances can be observed. Instead, the effect is a contact-interaction correction, parametrized as a dimension-8 operator [26]

$$\mathcal{L} = \frac{4\lambda}{\Lambda_H^4} T_{\mu\nu} T^{\mu\nu} \,, \tag{3.3}$$

where $T_{\mu\nu}$ is the SM fermion energy-momentum tensor, $\lambda = \pm 1$, and $\Lambda_H \sim M_D$ is the effective Planck scale.

The strongest bounds on the ADD model currently come from the LHC. A search for anomalous jet+$\not{E}_T$ events at CMS with 20 fb$^{-1}$ at 8 TeV [27] constrains $M_D > 3.0 - 5.0$ TeV for $k = 2 \ldots 6$ (with lower bounds for higher $k$). In addition, searches for operators of the form (3.3) in $\ell^+\ell^-$ final states [28] provide a bound $\Lambda_H \gtrsim 4.0$ GeV, approximately independent of $k$. The estimate of the discovery reach of the 500 GeV ILC is $\Lambda_H \approx 5.0 - 5.5$ TeV [29]. Since the KK graviton is a spin-2 object, the angular distribution of the final-state fermions in the ADD model is quite distinct from the case of a spin-1 $Z'$ or KK gauge boson. A unique identification of the spin-2 origin of the contact-interaction correction at a 500 GeV ILC is possible for $\Lambda_H$ up to about 3.0 TeV [30]; however, the LHC is likely to have an even higher reach using the dilepton final states [31]. Another crucial test of the gravitational nature of the contact interaction would be an independent determination of the size of the effect in a variety of four-fermion channels. Gravity couples to the total energy-momentum tensor, resulting in a set of four-fermion operators independent of the fermion type. Alternative models for spin-2 contact interactions, such as the exchange of string-Regge excitations of the SM gauge bosons [32], predict effects of different sizes for up-type and down-type quarks and leptons. The ILC will provide an ideal environment to perform this test.

### 3.4.3 Randall-Sundrum warped extra dimensions

While the ADD model eliminates the usual gauge hierarchy, it faces its own hierarchy problem: the large ratio of the size of the extra dimensions and their natural scale, TeV$^{-1}$, must be explained. This difficulty is avoided in the Randall-Sundrum (RS) model [1], which extends the space by a single extra dimension, compactified on an orbifold $S_1/\mathcal{Z}_2$, effectively an interval. The characteristic feature of this model is the non-flat "warped" metric, which can be used to generate the observed large hierarchy between the Planck and the weak scale without assuming any hierarchies among the input parameters. Interestingly, AdS/CFT duality has been used to argue that the RS model is simply a weakly-coupled description of a strongly-coupled four-dimensional model with a composite Higgs boson.

In the original RS model, only gravity was assumed to propagate in the full 5D space, while all SM fields were confined on the 4D boundary. As in ADD, potentially observable KK modes of the graviton are predicted; however, their masses are spaced by $\mathcal{O}$(TeV), and their couplings to the SM are suppressed by a scale of $\mathcal{O}$(TeV) and not the Planck scale. The LHC experiments search for RS KK graviton resonances in the $\ell^+\ell^-$ and $\gamma\gamma$ final states. The graviton couplings to the SM depend on the curvature of the extra dimension $k$. The dimensionless ratio $k/\overline{M}_{\rm Pl}$ is expected to be in a range between 0.01 and 0.1 on naturalness grounds. The current LHC bounds on the KK graviton mass vary from 2.1 TeV for $k/\overline{M}_{\rm Pl} = 0.1$ to 0.9 TeV for $k/\overline{M}_{\rm Pl} = 0.01$ [12, 13]. The LHC reach with $\sqrt{s} = 14$ TeV, $L_{\rm int} = 100$ fb$^{-1}$ is expected to be $2.5 - 4.5$ TeV, for the same range of $k/\overline{M}_{\rm Pl}$ [33]. At the ILC, these resonances appear as interference effects both in $e^+e^-$ and in $\gamma\gamma$ annihilation processes. As with $Z'$ resonances, the sensitivity is comparable to that for direct resonance searches at the LHC.

# Chapter 4
# $W$ and $Z$ Boson Physics

The ILC will yield a new level of precision in measurements of the $W$ and $Z$ boson masses, widths, and couplings. Several different ILC processes contribute to these measurements. These include the continuum production of two vector bosons, $e^+e^- \to W^+W^-$ and $e^+e^- \to ZZ$, production of weak bosons in 2-photon reactions, and triple boson production $e^+e^- \to VVV$, where the final state can be $WWZ$, $ZZZ$, or $WW\gamma$. In addition, the ILC can study vector boson scattering at high energy. Furthermore, the ILC offers the possibility of dedicated low-energy runs at the $Z$ and at the $WW$ threshold. In all cases, these measurements will supersede the precision of existing measurements from the previous colliders, including SLC, LEP and the Tevatron, and are expected also to surpass the accuracies that will be available from the LHC.

As we will explain in detail in this section, these measurements will allow us to go beyond simple tests of the description of the $W$ and $Z$ bosons in the Standard Model. Through the Higgs mechanism of mass generation, massive $W$ and $Z$ bosons contain states that belong to the Higgs boson sector and exhibit possible new couplings associated with Higgs boson compositeness or strong interactions. Precise measurements of the $W$ and $Z$ properties can reveal these effects.

Many models of new physics beyond the Standard Model predict new couplings of the $W$ and $Z$ bosons. These include models with additional heavy vector bosons such as technicolor and topcolor, Little Higgs models, extra-dimensional models with Kaluza-Klein recurrences of the $W$ and $Z$ boson, and Twin Higgs models. In many of these cases, the additional gauge bosons would be fermiophobic and would thus evade direct searches at the LHC. The new bosons must then be found through their mixing with the $W$ and $Z$ bosons. Such mixing effects could be detected by the precision measurements described in this section.

## 4.1 Beyond the SM $W/Z$ sector: the EW chiral Lagrangian

To interpret the results of precision measurements of the various $W$ and $Z$ processes that will be studied at the ILC, it is useful to have a common theoretical framework to which all of these measurements can be related. Frameworks of two different types are commonly used. The first is based on an effective field theory (EFT) that includes the most general modifications of the $W$ and $Z$ couplings that might be induced by adding higher-dimension operators to the Standard Model Lagrangian. Such effective field theories are presented in the literature in [1, 2]. A complementary approach is to postulate resonances with various quantum numbers and couple these to the $W$, $Z$, and Higgs bosons [3]. It is rather easy to switch between the two descriptions. Then limits on anomalous $W$ and $Z$ couplings parametrized in the EFT language can be expressed in terms of limits on the mass and width parameters of physical resonances.

In this section, we will summarize the description of the electroweak (EW) effective Lagrangian and its parameters from these two points of view and define the parameters of this Lagrangian that can be constrained by experiment. In the remainder of this chapter, we will quote constraints on this effective Lagrangian that can be obtained from the ILC experiments.

### 4.1.1 Formalism of the EW chiral Lagrangian

The EW effective chiral Lagrangian consists basically of the $SU(2)_L \times U(1)_Y$-invariant SM Lagrangian (without the Higgs field) and a non-linear sigma model describing the Goldstone bosons (which provide the longitudinal degrees of freedom of $W$ and $Z$). Though this formulation of the electroweak Lagrangian was originally constructed for models containing very heavy, composite, or no Higgs bosons, it can be easily extended to include a 125 GeV Higgs boson as indicated by the 2012 LHC discovery [4]. The lowest-order EW chiral Lagrangian contains the kinetic terms for the weak and hypercharge bosons as well as the kinetic term for their longitudinal degrees of freedom, which also yields the gauge boson mass terms. There is one additional possible dimension 2 operator, $\mathcal{L}'_0$. At the next order in mass dimension, there are ten possible dimension 4 operators (assuming $C$ and $CP$ conservation), $\mathcal{L}_{i=1,\ldots,10}$, with corresponding operator coefficients, $\alpha_{i=1,\ldots,10}$. The operators give all possible Lorentz- and gauge-invariant combinations of the transverse and longitudinal electroweak gauge fields. The detailed form of the Lagrangian can be found in [1–3, 5].

All of these operators modify the 2-, 3- and 4-point functions of the EW gauge bosons: $\mathcal{L}'_0, \mathcal{L}_1, \mathcal{L}_8$ give the oblique corrections which modify the gauge-boson propagators, while $\mathcal{L}_2, \mathcal{L}_3, \mathcal{L}_9$ induce anomalous triple gauge couplings (TGCs). The remaining five operators ($\mathcal{L}_4$–$\mathcal{L}_7$ and $\mathcal{L}_{10}$) only affect the quartic gauge couplings (QGCs). The coefficient of the extra dimension-2 operator, the parameter $\beta_1$, is directly related to the $\rho$ or $T$ parameter, and thus is rather special. Experimentally it is well-known that the deviation from $\rho = 1$ is quite small, such that the leading-order Lagrangian possesses a custodial isospin symmetry which is broken only at next-to-leading order by the non-vanishing EW mixing angle and the mass splittings inside the fermionic isospin doublets. Sometimes such custodial isospin conservation is assumed. This would then eliminate the operators $\mathcal{L}_6$–$\mathcal{L}_{10}$.

At the next order in mass dimension, there are five dimension-6 operators, $\mathcal{L}^{\lambda}_{1,\ldots,5}$, with coefficients $\alpha^{\lambda}_{i=1,\ldots,5}$ [1–3, 5]. These operators, which can be interpreted as contributions to anomalous magnetic moments of the EW gauge bosons, appear in the same order in the power counting of the perturbative expansion as the operators listed above. The first two, containing three field-strength tensors, induce also anomalous TGCs, while the last three, containing two field-strengths and two longitudinal bosons, only contribute to the QGCs. Including a light Higgs boson leads to more operators containing the Higgs field, some of which are however redundant and can be eliminated via equations of motion [6].

As we have discussed already, all of these ten plus five operators can be generated when integrating out one or more heavy particles beyond the SM. It is not unlikely that heavy particles that could contribute to the EW effective Lagrangian in this way would be discovered at the LHC in its run at 14 TeV.

We will see in subsequent sections that the ILC experiments can make precise statements about the values of the $\alpha_i$ parameters. Though the ILC measurements are done on the electroweak gauge bosons, the Equivalence Theorem [7] implies that the longitudinal polarization states of massive gauge bosons have couplings associated with the Higgs sector responsible for their mass generation. Thus, measurements of the $W$ and $Z$ couplings, codified by the $\alpha_i$ parameters, have a direct interpretation as Higgs sector interactions and can be used to constrain models of Higgs dynamics. For example, the values of the $\alpha_i$ constrain the existence of possible resonances, associated with composite Higgs sectors, strong weak interactions or similar models. We will describe this connection below.

First, however, it will be useful to explain how the formalism presented in the previous section is connected to the trilinear and quartic vector boson couplings. Within the SM, the trilinear and quartic couplings are specified by the constraints of gauge invariance. Beyond the SM, additional couplings may appear. Often, these are represented by effective Lagrangians with many parameters. The systematic effective Lagrangian approach of the previous section organizes these parameters in a useful way.

The EW chiral Lagrangian provides an off-shell formulation for a general electroweak sector combining all possible operators up to dimension 4. Complete (fermionic) matrix elements for $2 \to 6$ processes can be computed using the Feynman rules derived from this Lagrangian. These Feynman rules include EW boson interactions with anomalous couplings. In the next few paragraphs, we will give the relation between a general parametrization of the anomalous couplings and the effective Lagrangian parameters $\alpha_i$.

In unitarity gauge, the trilinear gauge interactions are conventionally written

$$\begin{aligned} L_{\mathrm{WWV}} = {} & g_{\mathrm{WWV}}[ \\ & \mathrm{i} g_1^{\mathrm{V}} V_\mu \left( W_\nu^- W_{\mu\nu}^+ - W_{\mu\nu}^- W_\nu^+ \right) + \mathrm{i}\kappa_{\mathrm{V}} W_\mu^- W_\nu^+ V_{\mu\nu} + \mathrm{i}\frac{\lambda^{\mathrm{V}}}{m_{\mathrm{W}}^2} W_{\lambda\mu}^- W_{\mu\nu}^+ V_{\nu\lambda} \\ & + g_4^{\mathrm{V}} W_\mu^- W_\nu^+ \left( \partial_\mu V_\nu + \partial_\nu V_\mu \right) + g_5^{\mathrm{V}} \epsilon_{\mu\nu\lambda\rho} \left( W_\mu^- \partial_\lambda W_\nu^+ - \partial_\lambda W_\mu^- W_\nu^+ \right) V_\rho \\ & + \mathrm{i}\tilde{\kappa}^{\mathrm{V}} W_\mu^- W_\nu^+ \tilde{V}_{\mu\nu} + \mathrm{i}\frac{\tilde{\lambda}^{\mathrm{V}}}{m_{\mathrm{W}}^2} W_{\lambda\mu}^- W_{\mu\nu}^+ \tilde{V}_{\nu\lambda}] \, , \end{aligned} \tag{4.1}$$

Similarly, the quartic gauge interactions are expressed as

$$\begin{aligned} \mathcal{L}_{QGC} = {} & e^2 \left[ g_1^{\gamma\gamma} A^\mu A^\nu W_\mu^- W_\nu^+ - g_2^{\gamma\gamma} A^\mu A_\mu W^{-\nu} W_\nu^+ \right] \\ & + e^2 \frac{c_w}{s_w} \left[ g_1^{\gamma Z} A^\mu Z^\nu \left( W_\mu^- W_\nu^+ + W_\mu^+ W_\nu^- \right) - 2 g_2^{\gamma Z} A^\mu Z_\mu W^{-\nu} W_\nu^+ \right] \\ & + e^2 \frac{c_w^2}{s_w^2} \left[ g_1^{ZZ} Z^\mu Z^\nu W_\mu^- W_\nu^+ - g_2^{ZZ} Z^\mu Z_\mu W^{-\nu} W_\nu^+ \right] \\ & + \frac{e^2}{2 s_w^2} \left[ g_1^{WW} W^{-\mu} W^{+\nu} W_\mu^- W_\nu^+ - g_2^{WW} \left( W^{-\mu} W_\mu^+ \right)^2 \right] + \frac{e^2}{4 s_w^2 c_w^4} h^{ZZ} (Z^\mu Z_\mu)^2 \quad . \end{aligned} \tag{4.2}$$

The overall prefactors are $g_{WW\gamma} = e$ and $g_{WWZ} = e \cos\theta_W / \sin\theta_W$. The symbols $V_{\mu\nu}$ and $\tilde{V}_{\mu\nu}$ are defined as:

$$V_{\mu\nu} = \partial_\mu V_\nu - \partial_\nu V_\mu \qquad \tilde{V}_{\mu\nu} = \epsilon_{\mu\nu\rho\sigma} V^{\rho\sigma}/2 \, . \tag{4.3}$$

The SM values of the trilinear couplings in (4.1) are given by

$$g_1^{\gamma,Z} = \kappa^{\gamma,Z} = 1, \quad g_4^{\gamma,Z} = g_5^{\gamma,Z} = \tilde{\kappa}^{\gamma,Z} = 0 \quad \text{and} \quad \lambda^{\gamma,Z} = \tilde{\lambda}^{\gamma,Z} = 0 \qquad , \tag{4.4}$$

The deviations of the couplings from the SM values are expressed in terms of the $\alpha_i$ parameters as

$$\Delta g_1^\gamma = 0 \qquad \Delta\kappa^\gamma = g^2(\alpha_2-\alpha_1)+g^2\alpha_3+g^2(\alpha_9-\alpha_8) \tag{4.5}$$

$$\Delta g_1^Z = \delta_Z + \frac{g^2}{c_w^2}\alpha_3 \qquad \Delta\kappa^Z = \delta_Z - g^2(\alpha_2-\alpha_1)+g^2\alpha_3+g^2(\alpha_9-\alpha_8) \tag{4.6}$$

and

$$\lambda^\gamma = -\frac{g^2}{2}\left(\alpha_1^\lambda+\alpha_2^\lambda\right) \qquad \lambda^Z = -\frac{g^2}{2}\left(\alpha_1^\lambda - \frac{s_w^2}{c_w^2}\alpha_2^\lambda\right) \tag{4.7}$$

where $\delta_Z$ is determined by the precision electroweak corrections. Note that in this setup only the C- and P-conserving parameters $g_1$, $\kappa$ and $\lambda$ can be generated. The parameters $g_5$, which violate C and P separately but leaves CP intact, and $g_4$, $\tilde{\kappa}$ and $\tilde{\lambda}$, which violate CP, are not shifted.

The SM values of the quartic couplings in (4.2) are given by

$$g_1^{VV'} = g_2^{VV'} = 1 \quad (VV' = \gamma\gamma, \gamma Z, ZZ, WW), \qquad h^{ZZ} = 0. \tag{4.8}$$

Deviations from these SM values in the quartic couplings are introduced through the corrections induced by the $\alpha_i$ to the couplings that preserve custodial $SU(2)$ symmetry,

$$\Delta g_1^{\gamma\gamma} = \Delta g_2^{\gamma\gamma} = 0 \qquad \Delta g_1^{\gamma Z} = \Delta g_2^{\gamma Z} = \frac{g^p p}{c_w^2 - s_w^2}\alpha_1 + \frac{g^2}{c_w^2}\alpha_3 \tag{4.9a}$$

$$\Delta g_1^{ZZ} = 2\Delta g_1^{\gamma Z} + \frac{g^2}{c_w^4}\alpha_4 \qquad \Delta g_2^{ZZ} = 2\Delta g_1^{\gamma Z} - \frac{g^2}{c_w^4}\alpha_5 \tag{4.9b}$$

$$\Delta g_1^{WW} = 2c_w^2\Delta g_1^{\gamma Z} + g^2\alpha_4 \qquad \Delta g_2^{WW} = 2c_w^2\Delta g_1^{\gamma Z} - g^2\left(\alpha_4+2\alpha_5\right) \tag{4.9c}$$

$$h^{ZZ} = g^2\left(\alpha_4+\alpha_5\right). \tag{4.9d}$$

Since we have consistently generated the trilinear and quartic couplings from a theory with exact but spontaneously broken $SU(2)\times U(1)$ symmetry, the vertices described in this section fit together into a unified formalism that can be used to compute the scattering amplitudes for complete electroweak processes. In particular, this formalism gives a consistent definition to off-shell propagators and vertices that appear in processes containing the quartic gauge boson vertices. The results of all experiments are expressed in terms of the parameters $\alpha_i$.

### 4.1.2 EW chiral Lagrangian and Higgs sector resonances

We now return to the question of the interpretation of the $\alpha_i$ parameters in terms of possible resonances in the electroweak sector. A formalism complementary to the chiral Lagrangian approach summarized above, based on adding resonances to the SM Lagrangian has been described in [3]. Note that this formalism can easily include a light Higgs boson as one of the resonances included. We review this formalism briefly here.

There are three different combinations of spin and isospin for which resonances can couple to the EW gauge boson system. The spin of these resonances can be 0, 1, or 2 (scalar, vector, or tensor), and, similarly, the value of the isospin, under the custodial isospin symmetry, can be 0, 1, or 2 (in this context, labeled singlet, triplet, and quintet). To couple invariantly to a pair of weak bosons, the parity in spin and isospin must be equal; hence we consider resonances with the quantum numbers:

- scalar singlet $\sigma$, scalar quintet $\phi$,
- vector triplet $\rho$,
- tensor singlet $f$, tensor quintet $a$.

**Table 4.1.** Coefficients $\xi$ appearing in the formula (4.10) for the partial widths for resonances with various quantum numbers to decay into longitudinally polarized vector bosons.

| Resonance | $\sigma$ | $\phi$ | $\rho$ | $f$ | $a$ |
|---|---|---|---|---|---|
| $\Gamma$ | 6 | 1 | $4v^2/3M^2$ | 1/5 | 1/30 |

In the model, these resonances are allowed to have arbitrary masses and widths, including both extreme cases of extremely heavy ($M \to \infty$) or extremely broad ($\Gamma \sim M \to \infty$) resonances. We might also list $\pi$ (scalar triplet) and $\omega$ (vector singlet), but their couplings to pairs of weak bosons violate custodial isospin. Then either their couplings are small, so that we can ignore them, or they require unnatural cancellations to preserve the SM value of the $\rho$ parameter.

An example of such a resonance of the type $\sigma$ is the SM Higgs boson itself. The techni-rho resonance of technicolor models is an example of the vector triplet $\rho$. This set of quantum numbers also appears in an extra-dimensional context as a Kaluza-Klein $W'$ or $Z'$ [8]. An example of the tensor $f$ is the graviton resonance in Randall-Sundrum models [9].

For the purposes of this section, we will assume that resonances in the EW sector have fermionic couplings very suppressed compared to the couplings to the EW sector. The opposite case has been discussed already in Chapter 3. For resonances that do not couple strongly to fermions, the dominant decays are to longitudinal EW gauge bosons. The widths are given by formulae

$$\Gamma_i = \frac{g_i^2}{64\pi}\frac{M^3}{v^2} \cdot \xi \; , \tag{4.10}$$

where the coefficients $\xi$ are displayed in Table 4.1. The couplings $g_i$ are the elementary couplings appearing in the resonance Lagrangian. With increasing number of spin and isospin components, the resonance width decreases. Note that, with our normalization convention for the dimensionless couplings $g_i$, the width of a vector resonance has a scaling behavior different from that of the other cases. If we want to work in a purely phenomenological approach, it is useful to eliminate the couplings $g_i$ in terms of the resonance widths using (4.10).

At the ILC, we are mainly concerned with (precision) measurements of electroweak processes at energies below the first resonance in an extended electroweak/Higgs sector. Any deviations observed from the Standard Model predictions can be interpreted in terms of the $\alpha_i$ parameters. To understand the relation of these parameters to the system of resonances, we can integrate out the resonances and expand the resulting effective Lagrangian in powers of $E/M$. The terms resulting from this integration out shift the parameters of the Standard Model Lagrangian, shift the parameters $\beta_1$ and $\alpha_2$, and shift the other $\alpha_i$ parameters. The shifts of the Standard Model couplings are absorbed into the renormalized electroweak parameters. The shifts of $\alpha_2$ and $\beta_1$ appear in the $S$ and $T$ parameters of electroweak interactions. The remaining shifts of the $\alpha_i$ provide new information. The most important effects appear as shifts of $\alpha_4$ and $\alpha_5$. The translation from the resonance masses to $\alpha_4$ and $\alpha_5$ is given by the relation

$$\Delta\alpha_i = \frac{16\pi\Gamma}{M}\frac{v^4}{M^4} \cdot \zeta \tag{4.11}$$

where the coefficients $\zeta$ are displayed for each type of resonance in Table 4.2.

Figure 4.1 shows the shifts in $\alpha_4$ and $\alpha_5$ induced by each particular type of Higgs sector resonance. There is an ambiguity in the values of the $\alpha_i$ associated with a change in the renormalization scale of

**Table 4.2.** Coefficients $\zeta$ in the relation (4.11) between the parameters of a Higgs sector resonance and the chiral Lagrangian coefficients $\alpha_4$ and $\alpha_5$ that result from integrating out that heavy resonance.

| Resonance | $\sigma$ | $\phi$ | $\rho$ | $f$ | $a$ |
|---|---|---|---|---|---|
| $\Delta\alpha_4$ | $0$ | $\frac{1}{4}$ | $\frac{3}{4}$ | $\frac{5}{2}$ | $-\frac{5}{8}$ |
| $\Delta\alpha_5$ | $\frac{1}{12}$ | $-\frac{1}{12}$ | $-\frac{3}{4}$ | $-\frac{5}{8}$ | $\frac{35}{8}$ |

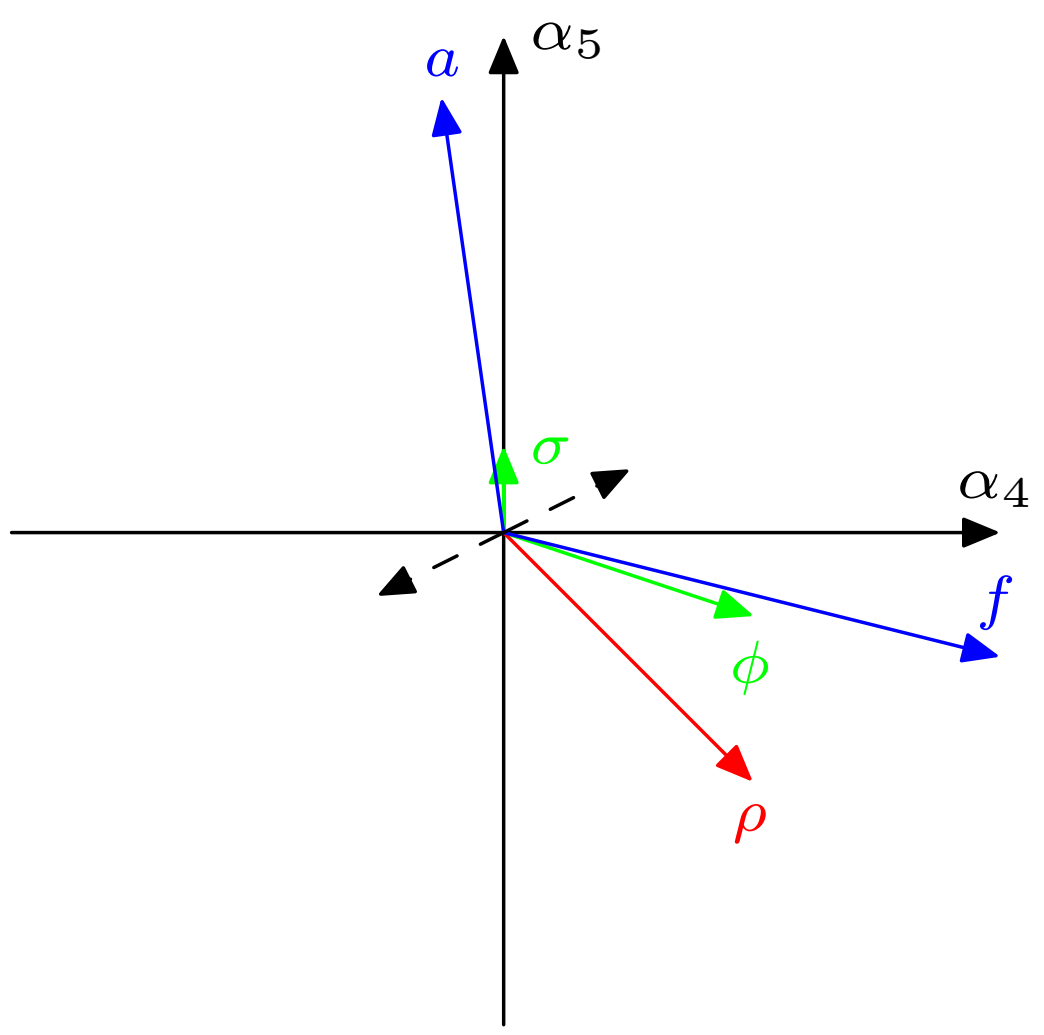

**Figure 4.1** Anomalous couplings $\alpha_{4/5}$ in the low-energy effective theory coming from the different resonances under the assumption of equal masses and widths, $M \sim \Gamma$ (Table 4.2). The dashed arrow indicates the shift due to renormalization scale variation.

the effective low-energy Lagrangian

$$
\begin{aligned}
\alpha_4(\mu) &= \alpha_4(\mu_0) - \frac{1}{12}\frac{1}{16\pi^2}\ln\frac{\mu^2}{\mu_0^2} \\
\alpha_5(\mu) &= \alpha_5(\mu_0) - \frac{1}{24}\frac{1}{16\pi^2}\ln\frac{\mu^2}{\mu_0^2} \, ,
\end{aligned}
\tag{4.12}
$$

where $\mu_0$ is a reference scale. This shift is plotted as a dashed arrow in Fig. 4.1. Fortunately, this small shift is almost orthogonal, in the $(\alpha_4, \alpha_5)$ plane, to the direction of the shift induced by a resonance. It should be interpreted as a theory uncertainty in the prediction for these shifts.

In the case that there is only one dominant resonance present, a combined fit to both $\alpha$ parameters allows us to disentangle isosinglet from isotriplet or isoquintet resonances. A worked example is given in [10]. The angular distributions of final vector bosons provide further information on the nature of a resonance. For example, a $\rho$ resonance multiplet would have the characteristic feature that the $ZZ$ decay channel is absent, by virtue of the Landau-Yang theorem

There is one more important issue to discuss in setting up the theory of strong interaction corrections to the electroweak sector. This is the question of high-energy behavior and unitarity. At the ILC, experiments on trilinear and quartic couplings in $e^+e^- \to VV$ and related processes can be analyzed by using the low energy effective Lagrangian directly. This is even correct in the study of vector boson scattering, $VV \to VV$, where corrections to the effective Lagrangian description come in only at the highest subprocess energies near 1 TeV. Measurements of these effects at hadron colliders probe a region of higher energies in which expressions derived from the effective Lagrangian must be greatly modified. The reason for this is that vertices due to higher-dimension operators grow dramatically at high energy and, if left unmodified, violate unitarity. In reality, unitarity can never be violated, but the restoration of unitarity requires additional higher-order effects or a proper UV completion of the theory. This introduces new parameters into the description. Even at the Tevatron, the analysis of measurements of the trilinear couplings must include form factors or other modifications so that the theory used to fit the data is internally consistent and avoids violation of unitarity. This is the flip side of the observation that, because it accesses higher energies, the LHC offers the opportunity to discover new states of an extended electroweak/Higgs sector as resonances. If resonances are not observed, or are not prominent, or if there are additional resonances beyond the reach of the LHC, there is no definite theoretical framework, and so results from the LHC will have ambiguity or model-dependence.

Thus, some heuristics are needed to define a complete formalism in which EFT descriptions like the EW chiral Lagrangian can be used as the basis of a formalism that can produce simulations to be compared to collider data and translate search limits between LHC and ILC experiments. In setting up this formalism, it would be advantageous to include possible first resonances explicitly, since these might be within the kinematical reach of the LHC. Such resonances would appear in strongly interacting Higgs sector models or in extra-dimensional models. On the other hand, the formalism must give amplitudes that preserve unitary. This second task can be achieved either by introducing momentum-dependent form factors for the low-energy scattering amplitudes and regularizing them, or using unitarization methods like the K matrix [11]. Details of a formalism that accomplishes this, and the translation between LHC and ILC results within this formalism, can be found in [3]. This method of unitarization can be combined with the generic off-shell parameterization of EW boson scattering given in (4.1) and (4.2) to give a complete description of Goldstone boson scattering amplitudes. For that purpose, the constant parameters $\alpha_{4/5}$ are replaced by energy-dependent (i.e., $s$-dependent) form factors. The technical details of that implementation can be found in [3]. This prescription does break crossing symmetry, but in fact that is broken already by the K-matrix prescription for unitarization. In principle, anomalous couplings for resonances might also be included. Such couplings are not considered here. We assume that they are subleading in the high-energy regime of a 1 TeV ILC or at LHC.

With this formalism in hand one can easily switch between the high-energy measurements on $VV$ scattering in the LHC environment and the much more precise measurements possible at the ILC and consider at the same time the parallel information from di- or triboson production. In the following sections we describe diboson production in the channels $WW$ and $ZZ$, the corresponding photon-induced processes, triboson production, EW boson scattering. We also discuss low-energy precision measurements at the $Z$ and at the $WW$ threshold.

## 4.2 Vector boson pair production

The major weak processes to be studied at an ILC are pair production of electroweak gauge bosons, $e^+e^- \to W^+W^-$ and $e^+e^- \to ZZ$. The ILC will be the first collider to provide $W$ pair production in lepton collisions with polarized beams. Due to the $V-A$ structure of the $W$ boson interactions, polarization of the beams radiating the electroweak boson can substantially enhance or suppress their production. Note, that there is also as a competing process, single $W$ production, originating mostly from photon-$W$ fusion (cf. Fig. 4.2). Since pair production is dominated by the $s$-channel pole, its cross section falls off linearly with energy. ILC will be the first lepton collider to enter that regime. On the other hand, single production is kinematically enhanced through the $t$-channel propagators and rises logarithmically with energy. 1 TeV is roughly the energy where the cross section for single production starts to exceed that of pair production.

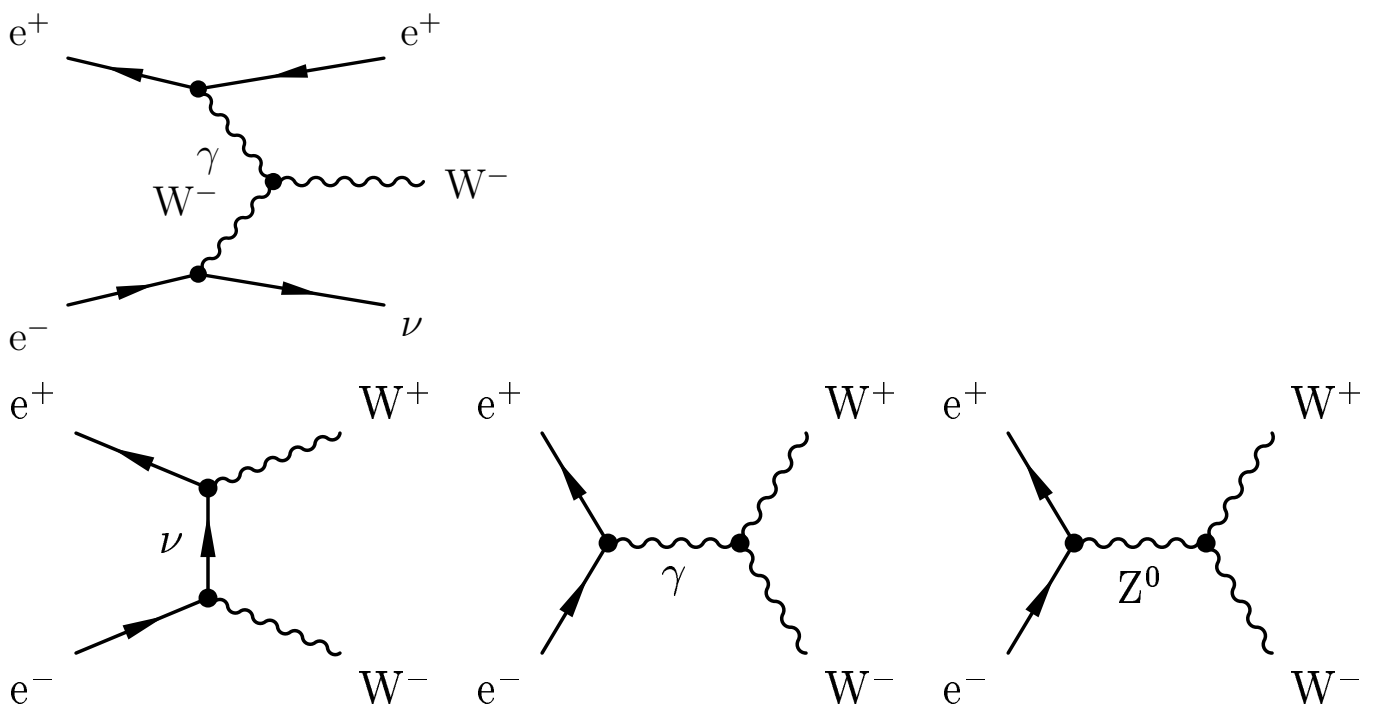

**Figure 4.2** Dominant Feynman diagrams for $W$ boson production at the ILC. Top: single $W$ production. Bottom: $W^+W^-$ pair production.

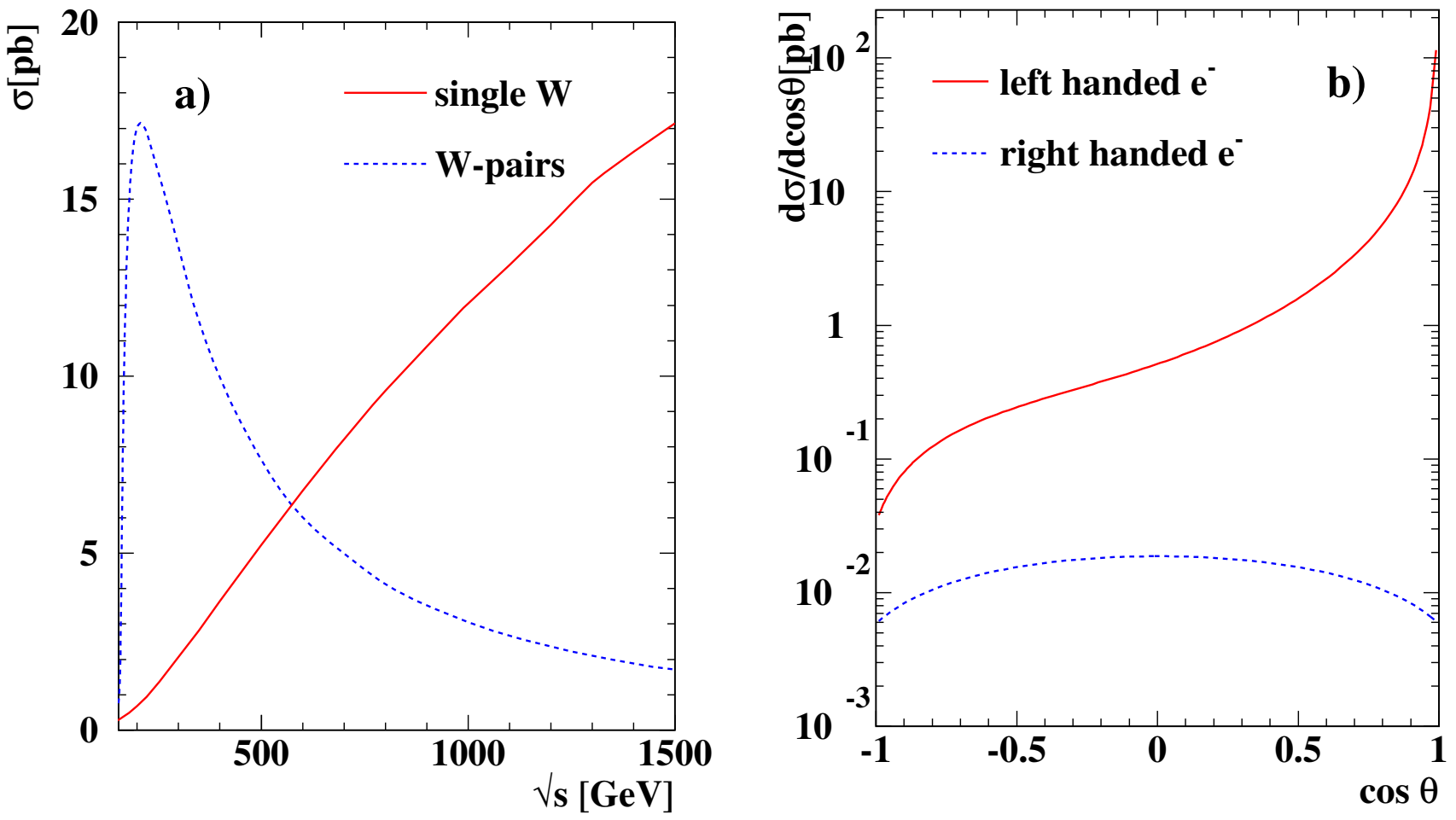

**Figure 4.3** Left: Total cross section for single $W$ [20–22] and $W$ pair production [12] as a function of the center of mass energy. Right: Differential cross section for $W$ pair production as a function of the $W$ polar angle for different beam polarizations.

## 4.2.1 $e^+e^- \to W^+W^-$

$WW$ production at a lepton collider is a theoretically well-studied process for which full next-to-leading (NLO) electroweak corrections are available, including the $W$ decays both in the double-pole approximation [12] and in a full $2 \to 4$ calculation [13]. These results have been cast into dedicated NLO Monte-Carlo programs, YFSWW3 [14] and RacoonWW [15]. The effects of finite fermion masses and different cuts on the cross section and distributions have also been studied in [16]. Furthermore, by means of effective field theory methods, the precise line-shape of $W$ pairs close to the thresholds have been investigated [17]. The leading NNLO corrections have recently been calculated in this framework [18]. The single $W$ production at a lepton collider is also available at NLO [19].

Fig. 4.3 shows, on the left, the cross sections for single $W$ and $W$ pair production at the ILC as a function of the center of mass energy. The right hand side of the figure shows the power of polarized beams at the ILC to enrich different helicity modes of the $W$s and hence their angular correlations.

The process of $WW$ production at the ILC allows for a sensitive measurement of triple gauge boson couplings, defined in (4.1). If one replaces the constant parameters by momentum-dependent form factors, (4.1) is in fact the most general parameterization. However, restricting again to the two lowest orders in the expansion of the EW chiral Lagrangian takes one back to constant coupling parameters. Note that there are some constraints to be fulfilled due to the unbroken electromagnetic gauge invariance, namely $g_1^\gamma(q^2=0)=1$ and $g_5^\gamma(q^2=0)=0$ at zero momentum transfer.

Measurements of $W$ pair production disentangle the various gauge structures contributing to the production amplitudes. The amplitudes depend on both sets of trilinear couplings, $WW\gamma$ and $WWZ$. The differential cross section with respect to the angle of the $W$ boson to the beam itself is sensitive to deviations from the SM values of the triple gauges couplings at the sub-percent level. The left-right asymmetry as function of the $W$ production angle adds to this information and enables one to discriminate between contributions to the anomalous $WW\gamma$ and $WWZ$ coupling. This can be seen from Fig. 4.4.

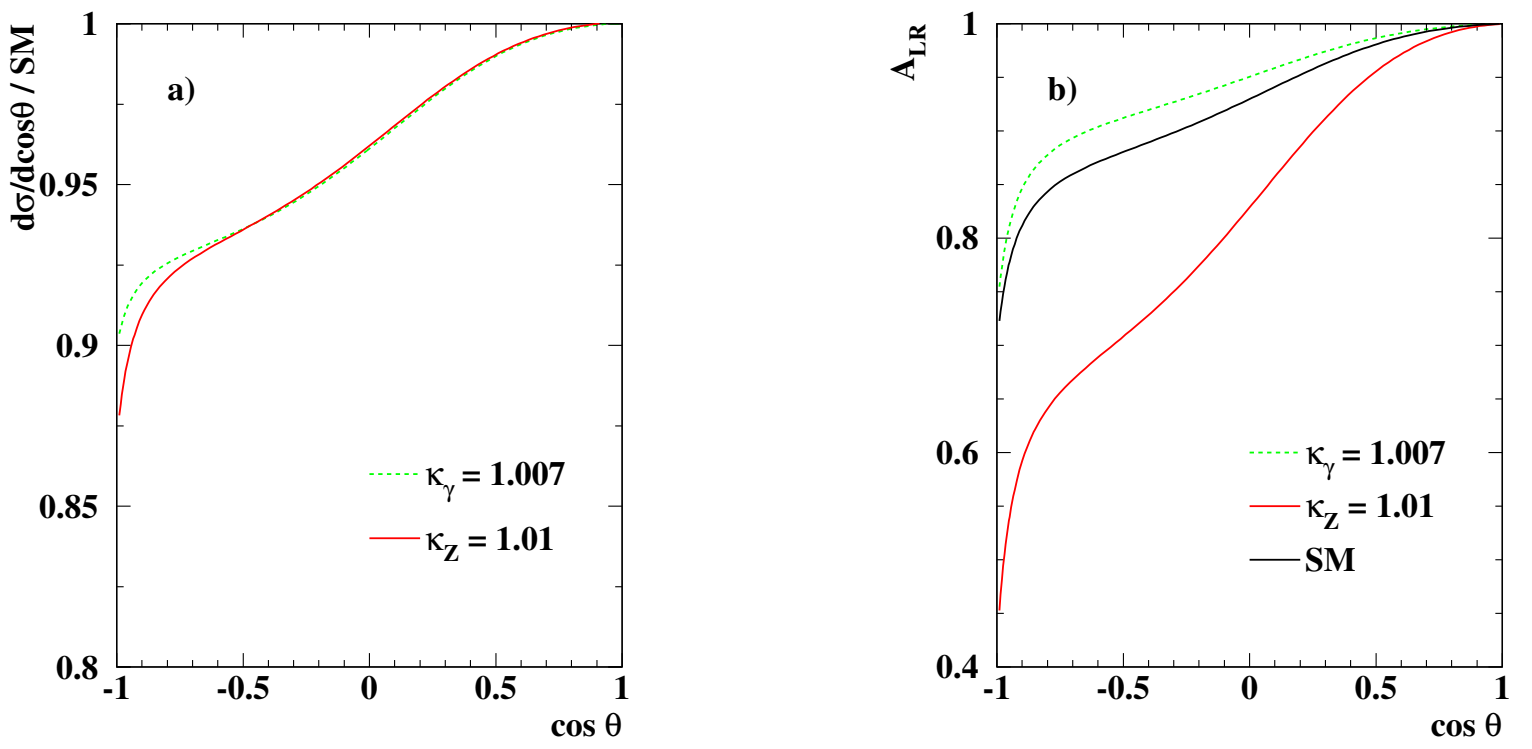

**Figure 4.4**
The effect of anomalous triple gauge couplings on the $W$ pair cross section, as a ratio to the SM prediction, as a function of the $W$ polar angle. Left: for the differential cross section. Right: for the left-right polarization asymmetry.

Analyses of $WW$ production to measure triple gauge couplings rely on a set of five different observables. We have already discussed the dependence on the polar angle $\theta_W$ of the outgoing $W^-$ with respect to the beam direction. This is the variable whose dependence is shown in Figs. 4.3 and 4.4. In addition, we can measure the polar angle $\theta^*$ of the decay fermion, with respect to the flight direction of the $W$ boson in the rest frame of the $W$ boson, for each of the two $W$ bosons. These variables are sensitive to the longitudinal polarization of the $W$ bosons. Finally, the transverse polarization of the $W$ bosons can be accessed via the azimuthal angles $\phi^*$ of the fermions in the plane constructed from the beam and the $W$ flight direction.

The most frequent decay mode of a $WW$ pair is the semileptonic one, which constitutes 44 % of all $WW$ decays. In semileptonic events, the polar angle of the negatively charged $W^-$ can be unambiguously reconstructed from the jet momenta and the lepton charge. Furthermore, the fermionic decay angles can be uniquely determined in case of the leptonically decaying $W$. For the hadronically decaying $W$ there is a twofold ambiguity, $(\cos\theta^*, \phi^*) \longleftrightarrow (-\cos\theta^*, \phi^* + \pi)$, arising from the fact that quarks and antiquarks cannot be distinguished, except possibly in $W \to c$ decays. While the semileptonic event sample is by far the most sensitive to triple gauge couplings, the largest sample is the fully hadronic one, which constitutes 46% of all decays. Here the sign ambiguity for the production angle of the $W$s cannot be resolved, since there is no means to determine the $W$ charges from the jet measurements. Even with sophisticated existing methods to get the correct pairing of jets, the sensitivity to the triple gauge couplings is smaller than from the semileptonic sample. The fully leptonic samples are smaller and more difficult to analyze. In roughly the half of these events, one lepton is a $\tau$, so a complete kinematic reconstruction is not completely possible. For the rest, there is a twofold ambiguity because of the missing information from the two neutrinos, and measurements from those samples are also limited by statistics.

Mixed leptonic and hadronic decays from $W$ pairs at the ILC can be selected very efficiently, and they also profit from a rather low background. As can be seen from Fig. 4.4, the cross section exhibits a large forward peak stemming from the $t$-channel neutrino exchange. This peak cross section events are not sensitive to triple gauge couplings at all, and are even partially lost in the beam pipe. Because the boost is much larger than at LEP, the $W$ production angle can be measured with much higher accuracy than in the LEP experiments. The detector resolution in those measurements is expected to be sufficiently good that there are almost no detector effects on the measurements; this is shown in [23]).

In most studies one marginalizes over some of the variables, since it is cumbersome to work with five independent variables. Many of the studies up to now have made use of the spin density matrix formalism [24]. It has been shown that this formalism leads to close to optimal results. In that formalism, it is possible to clearly separate signals from C, P, or CP-violating couplings from the corresponding C- or P-conserving ones. As one example to illustrate how this works, note that imaginary parts of off-diagonal elements of the spin-density matrix are only populated if there are nonzero CP-violating couplings. It has been shown that there are only negligible correlations between the different sets of couplings, hence, the fits can be done separately. These single parameter fits are quite useful to test models beyond the SM, though in principle a multi-variate analysis allows one to determine all five different C- and P-conserving couplings separately with the data from different beam polarization settings. Usually, one assumes full electroweak $SU(2) \times U(1)$ gauge invariance among the parameters, which leads to the following relations among the different parameters:

$$\begin{aligned} \Delta\kappa_\gamma &= -\cot^2\theta_W(\Delta\kappa_Z - g_1^Z) \\ \lambda_\gamma &= \lambda_Z \,. \end{aligned} \tag{4.13}$$

**Table 4.3**
Accuracies, quoted as 1 $\sigma$ errors, from single parameter fits for the different triple gauge couplings, $\sqrt{s} = 500$ GeV with $\mathcal{L} = 500\ \text{fb}^{-1}$ and for $\sqrt{s} = 800$GeV with $\mathcal{L} = 1000\ \text{fb}^{-1}$. For both energies, $\mathcal{P}_{e^-} = 80\%$ and $\mathcal{P}_{e^+} = 60\%$ has been used.

| coupling | error $\times 10^{-4}$ | |
|---|---|---|
| | $\sqrt{s} = 500$ GeV | $\sqrt{s} = 800$ GeV |
| C,P-conserving, $\mathrm{SU}(2) \times \mathrm{U}(1)$ relations: | | |
| $\Delta g_1^Z$ | 2.8 | 1.8 |
| $\Delta\kappa_\gamma$ | 3.1 | 1.9 |
| $\lambda_\gamma$ | 4.3 | 2.6 |
| C,P-conserving, no relations: | | |
| $\Delta g_1^Z$ | 15.5 | 12.6 |
| $\Delta\kappa_\gamma$ | 3.3 | 1.9 |
| $\lambda_\gamma$ | 5.9 | 3.3 |
| $\Delta\kappa_Z$ | 3.2 | 1.9 |
| $\lambda_Z$ | 6.7 | 3.0 |
| not C or P conserving: | | |
| $g_5^Z$ | 16.5 | 14.4 |
| $g_4^Z$ | 45.9 | 18.3 |
| $\tilde{\kappa}_Z$ | 39.0 | 14.3 |
| $\tilde{\lambda}_Z$ | 7.5 | 3.0 |

Table 4.3 shows the results from [5] for the sensitivity of the $WW$ measurement on the different anomalous triple gauge couplings, using integrated luminosities of 0.5 ab$^{-1}$ for 500 GeV CM energy and 1 ab$^{-1}$ for 800 GeV. This analysis assumed 80% polarization of the electron beam and 60% polarization for the positron beam. This corresponds to an effective polarization $P_{eff} =$ 95%, while $P_{eff} =$ 89% is more appropriate for the current ILC design; the change has only a minor effect on the final results. For the case of 800 GeV center of mass energy, the parameter fits which exhibit the largest correlations are illustrated in Fig. 4.5.

Note that these measurements are very precise and do not suffer from any significant systematic uncertainties, since detector effects, backgrounds, and smearing from beamstrahlung are almost negligible. The beam polarization can be determined in situ using the so-called Blondel scheme [25, 26]. Consequently, one can neglect additional systematic uncertainties from beam polarization. If there is no positron polarization at all, the statistical errors grow by about 50%. However, uncertainties are still completely under control. The forward peak is exclusively given by neutrino $t$-channel exchange, which only couples to left-handed electrons. Then the effective polarization $P_{eff}$ can still be determined from data alone [26].

To match the experimental precision, the theoretical errors need to be smaller than 0.5-1.0%. This is achieved in the predictions from the dedicated NLO programs RacoonWW and YFSWW3 [12]. In fact, the measurement at the ILC is so precise that it is smaller than the size of SM loop corrections. The errors are also smaller than some of the BSM loop corrections, for example, those from supersymmetry, as computed, for example, in [27]. With such precision it is possible to overconstrain the SM, and also to use the ILC measurements to search for deviations from the SM in virtual effects by new heavy particles. While the sensitivity to the dipole moment-like couplings $\Delta\lambda_\gamma$ are of the same order for the LHC and the ILC, estimates for the precision for different colliders for the trilinear coupling $\Delta\kappa_\gamma$ show that a 500 GeV ILC will supersede the LHC by roughly a factor of 10, increasing to a factor of 30 for 1 TeV running [5].

**Figure 4.5**
Two-dimensional sensitivity contours at $1\sigma$ and 95 % significance for several combinations of trilinear gauge couplings at a c.m. energy of 800 GeV for an integrated luminosity of 1 $ab^{-1}$, with 80 % electron and 60 % positron polarization. For all other variables the correlations are small.

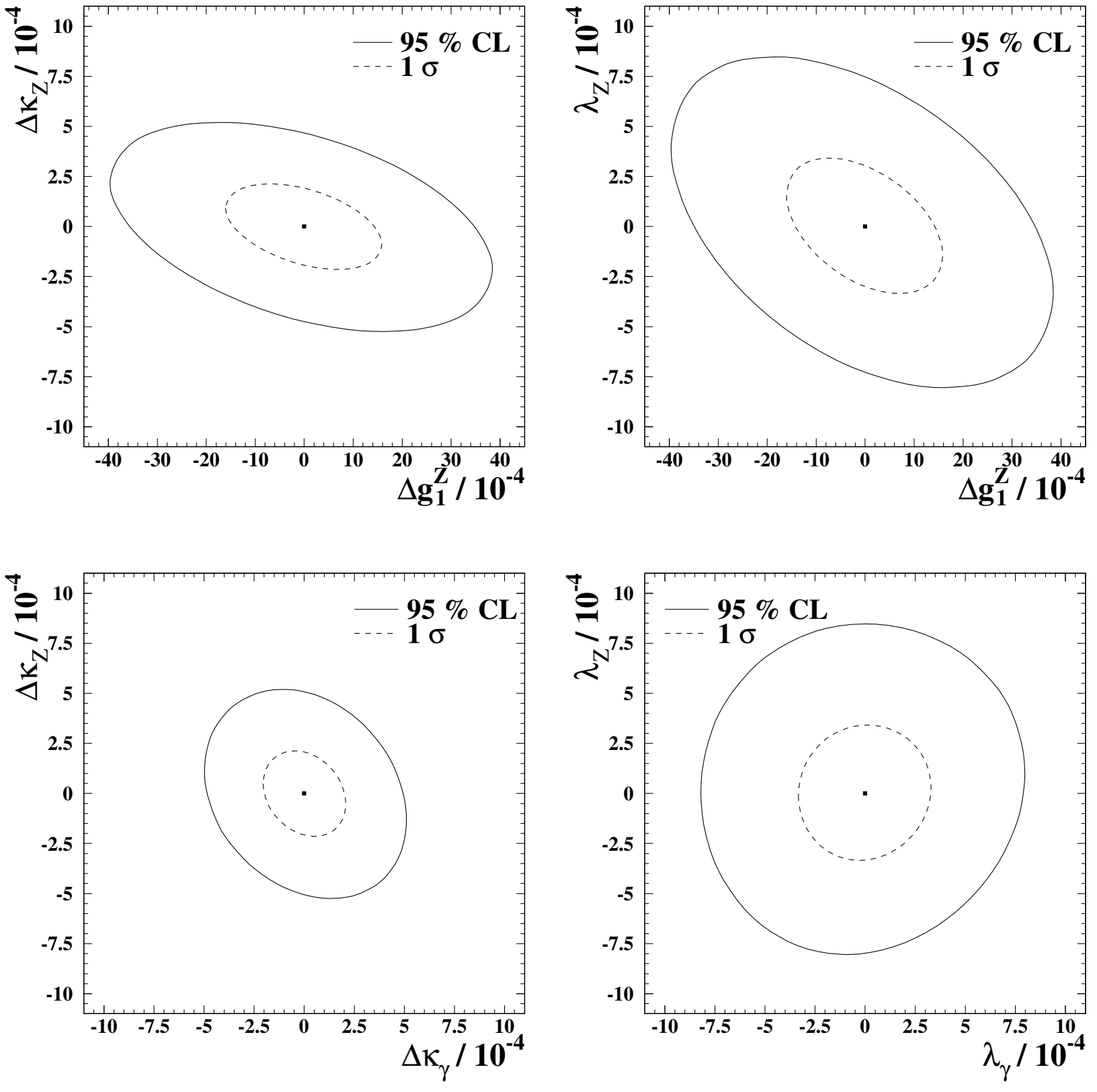

Some more details about measurements from the photon-induced channel as well as the precision measurement of the $W$ boson mass will be described in the following sections.

### 4.2.2 $e^+e^- \to ZZ$

This process is not used to do precision measurements at the ILC, since it is not sensitive the the leading EFT corrections. The measurement of this process mainly serves as a data-driven estimate of the background to the $WW$ production process. Many algorithms and details about how to separate the two processes can be found in [28].

### 4.2.3 $\gamma\gamma \to W^+W^-$

Though there is the specific option to construct a high-energy photon-photon collider by means of Compton backscattering, we do not discuss such measurements here. However, $\gamma$-induced processes also occur through photons from initial state radiation and beamstrahlung. These processes give a severe background for many new-physics searches, as discussed, for example, in [29]. But, on the other hand, they provide an opportunity to measure the $\gamma$-induced pair production of $W$ pairs, which has a large cross section of about 80 pb at 500 GeV. The physics of this process is similar to the single-$W$ production in $W\gamma$ fusion, whose cross section is roughly 30 pb at 500 GeV. The pair production process has been studied with the focus on the determination of possible anomalous gauge boson couplings, and its NLO corrections have been calculated in the double-pole approximation [30].

Using single $W$ and $W$ pair production from the photon substructure inside the electron beams adds an event sample of roughly the same order of magnitude to the sample from the $e^+e^-$ direct production mode. There are no studies on these modes using high luminosity from the point of view of anomalous coupling measurements. Low luminosity studies of the $W$ modes have focused on the total cross section measurement in the central part of the detector; this yields an order of magnitude less sensitivity than the direct production from the $e^+e^-$ mode. Adding angular correlations and the other observables mentioned in section 4.2 could possibly result in almost the same sensitivity as the $e^+e^-$ mode, thereby doubling the total statistics of the event samples.

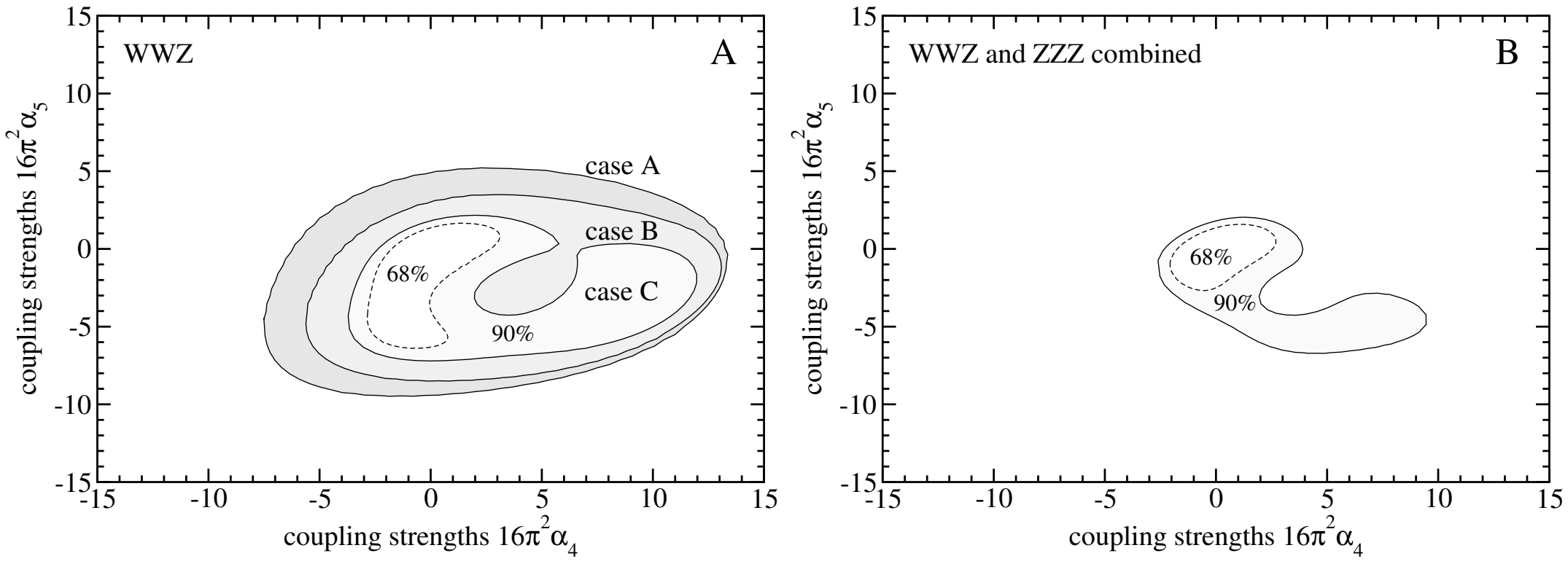

**Figure 4.6.** Expected sensitivity of a 1 TeV ILC for anomalous quartic gauge coupling parameters $\alpha_4/\alpha_5$, assuming an integrated luminosity of 1 ab$^{-1}$. Left: $WWZ$ alone. Right: $WWZ$ and $ZZZ$ combined. The solid lines show the 90% CLs. Cases A, B, and C refer, respectively, to the unpolarized case, the case with 80% electron polarization and the case with 80% electron plus 60% positron polarization. From [10].

## 4.3 Triple vector boson production

The production of three electroweak gauge bosons, mainly $e^+e^- \to W^+W^-Z$ and $e^+e^- \to ZZZ$, is an important precision test for the structure of the electroweak interactions. It has not been kinematically accessible at LEP. The measurement of these processes at the ILC allows a very clean and precise measurement of the triple and quartic gauge couplings and is complimentary to the corresponding observables in vector boson scattering processes. Though triboson production has already been measured at Tevatron and has and will be measured at the LHC, the process is much cleaner and offers a much higher precision at the ILC. For the ILC, the best dataset is that using the fully hadronic final state, which constitutes 32% of all $WWZ$ and $ZZZ$ events. Although, in principle, new-physics parameters that enter oblique corrections and triple gauge couplings can be determined in triple boson production, it is reasonable to assume that these have already been fixed by measurements of $WW$ production (or $VV$ scattering). Hence, they will be ignored in this section. In contrast to vector boson scattering, the different $\alpha$ parameters from the electroweak chiral Lagrangian cannot be completely disentangled in this measurement: the process $e^+e^- \to W^+W^-Z$ depends on the two linear combinations $\alpha_4 + \alpha_6$ and $\alpha_5 + \alpha_7$, while $e^+e^- \to ZZZ$ depends on the linear combination $\alpha_4 + \alpha_5 + 2(\alpha_6 + \alpha_7 + \alpha_{10})$.

The main SM background is rather large for the channel $W^+W^-Z$, coming from $t\bar{t}$ production with hadronically decaying $W$s. This background can be substantially reduced using right-handed electron polarization, which populates the longitudinal modes of the EW gauge bosons. For a 1 TeV ILC without polarization, the cross sections are 59 fb for $WWZ$ and 0.8 fb for $ZZZ$ production, respectively. Switching on electron polarization reduces the $WWZ$ cross section to 12 fb, for 80% right-handed electrons. For the neutral process, $ZZZ$, the SM background is negligible. Simulations of both processes are available at next-to-leading order [31–33]; in addition, most of the corrections are available in a dedicated Monte-Carlo program, LUSIFER [34].

The phenomenological analysis of these processes has been carried out in [10]. For the $WWZ$ process, three independent kinematical variables that are used, the invariant masses $M_{WZ}$ and $M_{WW}$ and the angle $\theta$ between the electron beam axis and the flight direction of the $Z$ boson. From the angular corrections as well as the diboson invariant masses, deviations from the SM can be determined (see Fig. 4.7), which then enable one to set limits on the anomalous couplings. Fig. 4.6 shows the expected sensitivity for the parameters $\alpha_4$ and $\alpha_5$ at the 90 and 68 per cent confidence level. The detailed values are give in Table 4.4.

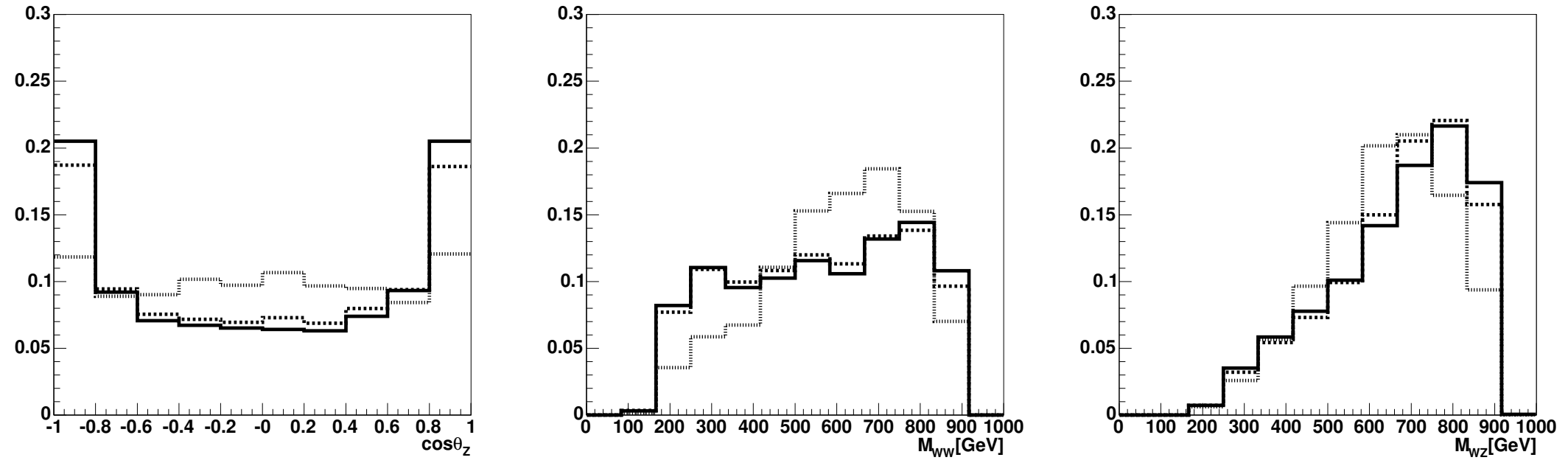

**Figure 4.7.** Reconstructed distributions of (left to right) $\cos\theta$, $M_{WW}$, and $M_{WZ}$ for $e^+e^- \to WWZ$, at the ILC at 1 TeV with 1 ab$^{-1}$, with 80% electron and 60% positron polarization. To show the shape dependence, the distributions are normalized to the respective total number of events for the SM. The solid, dashed, and dotted distributions are drawn for the SM, $\alpha_4 = 1.6\pi^2 \approx 15.8$ and $\alpha_5 \approx 15.8$, respectively [10].

**Table 4.4.** Sensitivity of $\alpha_4$ and $\alpha_5$, for the ILC at 1 TeV with 1 ab$^{-1}$, expressed as $1\sigma$ errors. The columns correspond to: $WWZ$: two-parameter fit; $ZZZ$: one-parameter fit; 'best': best combination of both. From [10].

| | | $WWZ$ | | | $ZZZ$ | best |
|---|---|---|---|---|---|---|
| | | no pol. | $e^-$ pol. | both pol. | no pol. | |
| $16\pi^2\Delta\alpha_4$ | $\sigma^+$ | 9.79 | 4.21 | 1.90 | 3.94 | 1.78 |
| | $\sigma^-$ | $-4.40$ | $-3.34$ | $-1.71$ | $-3.53$ | $-1.48$ |
| $16\pi^2\Delta\alpha_5$ | $\sigma^+$ | 3.05 | 2.69 | 1.17 | 3.94 | 1.14 |
| | $\sigma^-$ | $-7.10$ | $-6.40$ | $-2.19$ | $-3.53$ | $-1.64$ |

Further information comes from the process $e^+e^- \to W^+W^-\gamma$, which is complimentary to the $WWZ$ channel mentioned above. This channel is particularly interesting in the search for possible parity-violating operators. Because one does not have to pay the price for an additional weak boson, a considerable sensitivity could already be achieved at 500 GeV (or even 200 GeV) center-of-mass energy [35].

## 4.4 $WW$, $ZZ$ scattering at high energy

The process of $WW/ZZ$ scattering is at the heart of the study of the electroweak symmetry breaking mechanism because it reveals the self-interaction of both transversely and longitudinally polarized electroweak gauge bosons. The scattering of transversely polarized vector bosons is the equivalent of gluon-gluon scattering in QCD. The scattering of longitudinally polarized bosons is in fact the scattering of the Goldstone boson modes inside the electroweak gauge bosons, whose tree-level unitarity has been one of the most profound motivations for the existence of a Higgs boson [36]. In most studies, the scattering of weak gauge bosons has been seen specifically as a means to study the EW sector in the absence of a light Higgs boson, or, alternatively, to search for the presence of strong EW interactions. For an overview, see [37]. But even after the discovery of a light Higgs-like boson around 125 GeV [4], the scattering of EW gauge bosons remains one of the most important physical observables in the EW sector. Together with the precise measurements of the properties of the Higgs boson at the LHC and the ILC, $VV$ scattering allows us to overconstrain the EW sector and search for deviations from the EW structure of the Standard Model. Further, it offers by itself the possibility of searching for new physics in the EW sector beyond the Standard Model in a rather model-independent way. Any type of new physics that has considerable couplings to the SM fermions is very likely to show up earlier in Drell-Yan like processes at LHC or directly in electroproduction at the ILC. However, for new particles that couple only to the electroweak gauge sector (or have highly suppressed fermionic couplings), $VV$ scattering will be the primary production process. Furthermore, there are models, such as the strongly interacting light Higgs (SILH) [38], that give rise to a more or less SM-like Higgs boson, but nevertheless feature different physics at higher energies. For all of these reasons, the precision study of vector boson scattering has special importance.

The LHC will measure $VV$ scattering in the upcoming years; there are possibly even events in the final 2012 data set. On the other hand, the ILC offers the opportunity to use all vector boson final states, including the hadronic ones which cannot be used at the LHC because of trigger and background considerations. Furthermore, at the ILC, beam polarization allows the experiments to enrich longitudinal polarizations of the SM gauge bosons and to improve the ratio of longitudinal boson signal over transversely polarized boson background.

In order not to deal with a plethora of models, we will discuss the physics of $VV$ scattering in an approach as model-independent as possible. Most of our discussion is based on the approach of the EW chiral Lagrangian [1, 2]. In the original approach, this is understood formally as taking the limit of an infinitely heavy Higgs boson and removing it from the SM. The interactions left over give a nonlinear sigma model containing higher-dimensional operators coupling the transversal and longitudinal EW gauge bosons to each other. Such an approach was invented as a low-energy effective theory (LET) for the case of a heavy SM Higgs boson, for technicolor models featuring several strongly interacting resonances in the EW sector, or for Higgsless models (which are in some sense dual to the former class of models). In the light of the discovery of a light scalar boson at LHC, these specific models are now disfavored. However, such an electroweak chiral Lagrangian can be enlarged by the presence of possible resonances in the EW sector that could possibly couple to the EW sector. Such resonances can be classified to their spin and isospin quantum numbers. This classification has been performed in [3] including isoscalar, -vector, or -tensor resonances of spin 0, 1 and 2 that couple to a system of two weak gauge bosons. A light SM Higgs boson is just the isoscalar spin 0 case with particular couplings and is hence easily incorporated in that approach. The details have been summarized in the introductory Section 4.1.

The performance of a 1 TeV ILC for determining deviations from the triple and quartic gauge couplings of the SM has been studied in [10], extending an earlier analysis in [39]. These studies have been performed with full six-fermion matrix elements; hence, no simplifications such as the effective $W$ approximation (EWA), the Goldstone-boson equivalence theorem or the narrow-width approximation have been made. Note that a clear distinction of signal and backgrounds is rather intricate, since many EW processes (for example, triboson production) are intermingled with the pure $VV$ scattering process.

**Table 4.5**
Processes generated for the study of vector boson scattering in [10], giving the cross sections for signal and background for $\sqrt{s} = 1$ TeV, 80% left-handed polarization 80% for the electron beam 40% right-handed polarization for the positron beam. For each process, those final-state flavor combinations are included that correspond to the indicated signal or background subprocess.

| Process | Subprocess | $\sigma$ [fb] |
|---|---|---|
| $e^+e^- \to \nu_e\overline{\nu}_e q\bar{q}q\bar{q}$ | $W^+W^- \to W^+W^-$ | 23.19 |
| $e^+e^- \to \nu_e\overline{\nu}_e q\bar{q}q\bar{q}$ | $W^+W^- \to ZZ$ | 7.624 |
| $e^+e^- \to \nu\bar{\nu}q\bar{q}q\bar{q}$ | $V \to VVV$ | 9.344 |
| $e^+e^- \to \nu e q\bar{q}q\bar{q}$ | $WZ \to WZ$ | 132.3 |
| $e^+e^- \to e^+e^- q\bar{q}q\bar{q}$ | $ZZ \to ZZ$ | 2.09 |
| $e^+e^- \to e^+e^- q\bar{q}q\bar{q}$ | $ZZ \to W^+W^-$ | 414. |
| $e^+e^- \to b\bar{b}X$ | $e^+e^- \to t\bar{t}$ | 331.768 |
| $e^+e^- \to q\bar{q}q\bar{q}$ | $e^+e^- \to W^+W^-$ | 3560.108 |
| $e^+e^- \to q\bar{q}q\bar{q}$ | $e^+e^- \to ZZ$ | 173.221 |
| $e^+e^- \to e\nu q\bar{q}$ | $e^+e^- \to e\nu W$ | 279.588 |
| $e^+e^- \to e^+e^- q\bar{q}$ | $e^+e^- \to e^+e^- Z$ | 134.935 |
| $e^+e^- \to X$ | $e^+e^- \to q\bar{q}$ | 1637.405 |

**Table 4.6**
Sensitivity to quartic anomalous couplings in the various quasi-elastic weak-boson scattering processes accessible at the ILC.

| $e^+e^- \to$ | $\alpha_4$ | $\alpha_5$ | $\alpha_6$ | $\alpha_7$ | $\alpha_{10}$ |
|---|---|---|---|---|---|
| $W^+W^- \to W^+W^-$ | + | + | - | - | - |
| $W^+W^- \to ZZ$ | + | + | + | + | - |
| $W^\pm Z \to W^\pm Z$ | + | + | + | + | - |
| $ZZ \to ZZ$ | + | + | + | + | + |

For the simulation we assume a center of mass energy of 1 TeV and a total luminosity of 1000 fb$^{-1}$. Beam polarization of 80% for electrons and 40% for positrons is also assumed. Since the six-fermion processes under consideration contain contributions from the triple weak-boson production processes considered in the previous section ($ZZ$ or $W^+W^-$ with neutrinos of second and third generation as well as a part of $\nu_e\overline{\nu}_e WW(ZZ)$, $e\nu_e WZ$ and $e^+e^-W^+W^-$ final states), there is no distinct separation of signal and background. Signal processes are thus affected by all other vector boson processes as well as by pure background. The studies have been performed with event samples generated with `WHIZARD` [22], the shower and hadronization with `Pythia` [40] and the ILC detector response with `SimDet` [41]. Initial-state radiation (ISR) from the lepton beams is explicitly included. The processes studied and their cross sections are given in Table 4.5.

Possible observables sensitive to modifications in the (triple and quartic) couplings of longitudinal EW bosons are the total cross section as well as cross sections differential in the EW boson production and decay angles. In measuring properties of longitudinal gauge bosons, it is highly non-trivial if not impossible to measure observables like transverse momentum, since a cut has to be used to suppress the background from transverse gauge bosons, which drops off less fast than the contribution from longitudinal bosons. The general steps of this cut-based analysis use electron/positron tagging to identify background, with cuts on transverse momentum, missing mass and missing energy, as well as cuts around the EW boson masses to veto against events that are not tightly reconstructed. For the extraction of parameters like the triple and quartic gauge coupling, a binned likelihood fit has been used, in which events are described by a total of four kinematical variables.

We summarize the combined results for the measurements of anomalous EW couplings in Table 4.7 and Table 4.8. Both $SU(2)_c$ conserving and $SU(2)_c$ violating couplings are taken into account. The results are shown in Fig. 4.8 in graphical form, where projections of the multi-dimensional exclusion region in all $\alpha$s around the reference point $\alpha_i \equiv 0$ onto the two-dimensional subspaces $(\alpha_4, \alpha_5)$ and $(\alpha_6, \alpha_7)$ have been made. In order to transform these bounds on $\alpha_i$ parameters into more physical terms, and also in order to compare the capabilities of the ILC with direct resonance searches at the LHC, one can use the formalism described in the introductory section of this chapter to trade the anomalous couplings for parameters of physical resonances. These results for quartic gauge couplings in vector boson scattering can be combined with the ILC measurement results for triple gauge couplings and oblique corrections. Taking a single one of the resonances into account at each time, one could from the measured value of the $\alpha$ parameters reconstruct the properties and parameters of the resonance producing that particular value. From this, the sensitivity to new physics showing up as resonances in the high-energy region of EW boson scattering can be determined.

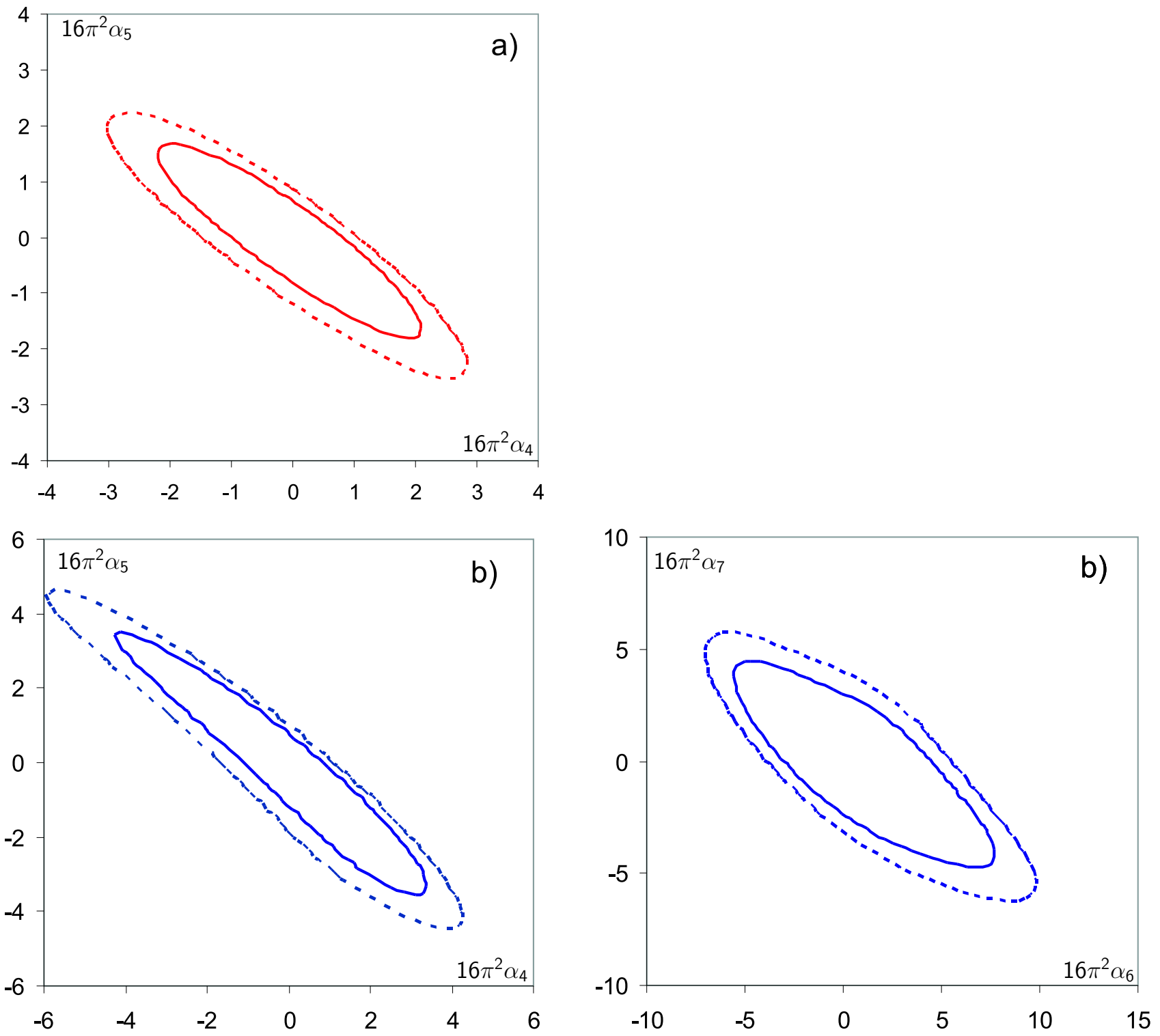

**Figure 4.8** The expected sensitivity to quartic anomalous couplings of a 1 TeV ILC with 1 ab$^{-1}$, in a combined fit for all sensitive processes. The dotted and solid lines show the 60% and 90% confidence regions. Top: $(\alpha_4, \alpha_5)$ in the case with $SU(2)_c$ conservation. Bottom: $(\alpha_4, \alpha_5)$ and $(\alpha_6, \alpha_7)$ in the case with broken $SU(2)_c$. From [10].

**Table 4.7.** The expected sensitivity to quartic anomalous couplings from the ILC with 1 TeV and 1 ab$^{-1}$ under the assumption of custodial $SU(2)_c$ conservation. The positive and negative 1 sigma errors are given separately. From [10].

| **coupling** | $\sigma-$ | $\sigma+$ |
|---|---|---|
| $\alpha_4$ | -1.41 | 1.38 |
| $\alpha_5$ | -1.16 | 1.09 |

The dependence of the different resonances on the $\alpha$ parameters as well as the correlation of the parameters and the technical points of the fit can be found in [10]. Here, we just give the scalar singlet as an example: in that case, $\alpha_4$ and $\alpha_6$ are zero, and, for the isospin-conserving case, in addition $\alpha_7$ and $\alpha_{10}$ are zero. If one uses the relation from integrating out the resonance, $\alpha_5 = g_\sigma^2 v^2/8M_\sigma^2$ and introduces the ratio between the width and the mass of the resonance, $f_\sigma = \Gamma_\sigma/M_\sigma$ one can solve for the mass of the resonance: $M_\sigma = v\,[4\pi f_\sigma/(3\alpha_5)]^{\frac{1}{4}}$. From the fit one can deduce the mass reach for scalar resonances at the ILC depending on scenarios with different widths. The results for the different masses for all cases are shown in Table 4.9. They can be summarized in the following numbers which hold for the $SU(2)_c$-conserving case: for spin-0 particles, the accessible reach is 1.39, 1.55, and 1.95 TeV for the isospin channels $I = 0$, $I = 1$, and $I = 2$, respectively, assuming a single resonance with optimal width to mass ratio that exclusively couples to the EW boson sector. For a vector resonance, the reach is 1.74 TeV for isosinglet and 2.67 TeV for isotriplets, respectively. Tensors provide the best reach because of the higher number of degrees of freedom participating. Here the ILC is sensitive to resonances of mass 3.00, 3.01, and 5.84 TeV for the isospin channels $I = 0$, $I = 1$, and $I = 2$, respectively. In the case of $SU(2)_c$ violation the effects on EW boson scattering are larger or more significant. In this sense, the $SU(2)_c$-conserving limit is a conservative estimate, though it is also favored by the EW measurements from SLC, LEP, Tevatron, and LHC.

**Table 4.8.** The expected sensitivity to quartic anomalous couplings from the ILC with 1 TeV and 1 ab$^{-1}$ for the case of broken $SU(2)_c$. The positive and negative 1 sigma errors are given separately. From [10].

| coupling | $\sigma-$ | $\sigma+$ |
|---|---|---|
| $\alpha_4$ | -2.72 | 2.37 |
| $\alpha_5$ | -2.46 | 2.35 |
| $\alpha_6$ | -3.93 | 5.53 |
| $\alpha_7$ | -3.22 | 3.31 |
| $\alpha_{10}$ | -5.55 | 4.55 |

**Table 4.9** Mass reach at a 1 TeV ILC in $VV$ scattering, assuming a data set of 1 ab$^{-1}$, for four different values of the ratio of width over mass for the resonances.

| $f_{\text{Res.}} = \Gamma_{\text{Res.}}/M_{\text{Res.}}$ | 1.0 | 0.8 | 0.6 | 0.3 |
|---|---|---|---|---|
| scalar singlet, $M_\sigma$ [TeV], $SU(2)_c$ cons. | 1.55 | 1.46 | 1.36 | 1.15 |
| scalar singlet, $M_\sigma$ [TeV], $SU(2)_c$ broken | 1.39 | 1.32 | 1.23 | — |
| scalar triplet, $M_{\pi^0}$ [TeV] | 1.39 | 1.32 | 1.23 | — |
| scalar triplet, $M_{\pi^\pm}$ [TeV] | 1.55 | 1.47 | 1.37 | 1.15 |
| scalar quintet, $M_\phi$ [TeV], $SU(2)_c$ cons. | 1.95 | 1.85 | 1.72 | 1.45 |
| scalar quintet, $M_{\phi^{\pm\pm}}$ [TeV], $SU(2)_c$ broken | 1.95 | 1.85 | 1.72 | 1.45 |
| scalar quintet, $M_{\phi^\pm}$ [TeV], $SU(2)_c$ broken | 1.64 | 1.55 | 1.44 | 1.21 |
| scalar quintet, $M_{\phi^0}$ [TeV], $SU(2)_c$ broken | 1.55 | 1.46 | 1.35 | 1.14 |
| vector singlet, $M_\omega$ [TeV], gen. case | 2.22 | 2.10 | 1.95 | 1.63 |
| vector triplet, $M_\rho$ [TeV], $SU(2)_c$ cons. | 2.49 | 2.36 | 2.19 | 1.84 |
| vector triplet, $M_{\rho^\pm}$ [TeV], no $SU(2)_c$, no mag. mom. | 2.67 | 2.53 | 2.35 | 1.98 |
| vector triplet, $M_{\rho^0}$ [TeV], no $SU(2)_c$, no mag. mom. | 1.74 | 1.65 | 1.53 | 1.29 |
| vector triplet, $M_{\rho^\pm}$ [TeV], special $SU(2)_c$ viol. | 3.09 | 2.92 | 2.72 | 2.29 |
| vector triplet, $M_{\rho^0}$ [TeV], special $SU(2)_c$ viol. | 1.78 | 1.69 | 1.57 | 1.32 |
| vector triplet, $M_{\rho^\pm}$ [TeV], gen. case | 2.54 | 2.41 | 2.34 | 1.88 |
| vector triplet, $M_{\rho^0}$ [TeV], gen. case | 1.71 | 1.62 | 1.51 | 1.27 |
| tensor singlet, $M_f$ [TeV], $SU(2)_c$ cons. | 3.29 | 3.11 | 2.89 | 2.43 |
| tensor singlet, $M_f$ [TeV], $SU(2)_c$ viol. | 3.00 | 2.84 | 2.64 | 2.22 |
| tensor triplet, $M_{a^0}$ [TeV] | 3.01 | 2.85 | 2.65 | 2.23 |
| tensor triplet, $M_{a^\pm}$ [TeV] | 2.81 | 2.66 | 2.47 | 2.08 |
| tensor quintet, $M_t$ [TeV], $SU(2)_c$ cons. | 4.30 | 4.06 | 3.78 | 3.18 |
| tensor quintet, $M_{t^c}$ [TeV], special $SU(2)_c$ viol. | 6.76 | 6.39 | 5.95 | 5.00 |
| tensor quintet, $M_{t^0}$ [TeV], special $SU(2)_c$ viol. | 4.53 | 4.28 | 3.98 | 3.35 |
| tensor quintet, $M_{t^{\pm\pm}}$ [TeV], gen. case | 5.17 | 4.89 | 4.55 | 3.83 |
| tensor quintet, $M_{t^\pm}$ [TeV], gen. case | 3.64 | 3.44 | 3.20 | 2.69 |
| tensor quintet, $M_{t^0}$ [TeV], gen. case | 5.84 | 5.52 | 5.14 | 4.32 |

## 4.5 Giga-$Z$

One of the main advantages of the ILC is its staged operation at almost arbitrary CM energies. This offers the opportunity to run the collider at rather low energies on the $Z$ resonance or at the $WW$ threshold to gather large amounts of data and perform precision measurements of the electroweak sector of the SM.

### 4.5.1 Precision measurement program at the $Z$

Running at a luminosity value of $\mathcal{L} = 10^{34}$ cm$^{-2}$ s$^{-1}$ allows to revisit the physics at LEP1 and SLC within a couple of days and study several billion $Z$ bosons within 1-2 months [42]. This is a especially important because two measurements at LEP1 and SLC constitute (together with the muon $g-2$) presently give the largest deviation from SM predictions. With the Giga-$Z$ option the tension between the two data sets from LEP1 and SLC could be resolved. Since these measurements are at the heart of the electroweak sector of the SM we describe them in detail.

The first of these measurements to be studied at the $Z$ pole is the left-right asymmetry

$$A_{LR} = \frac{1}{\mathcal{P}} \frac{\sigma_L - \sigma_R}{\sigma_L + \sigma_R} , \tag{4.14}$$

where $\sigma_{L/R}$ are the total cross sections for left- and right-handed polarized electrons and $\mathcal{P}$ is the longitudinal electron polarization. The measurement of the left-right asymmetry directly accesses the effective weak mixing angle, $\sin^2\theta^\ell_{eff}$, which in the case of a pure $Z$ exchange is given by

$$A_{LR} = \mathcal{A}_e = \frac{2 v_e a_e}{v_e^2 + a_e^2} , \tag{4.15}$$

where $v_e$ and $a_e$ are the vector and axial vector coupling of the $Z$ boson to electrons. Their ratio is given by $v_e/a_e = 1 - 4\sin^2\theta^\ell_{eff}$. At the ILC in the Giga-$Z$ option the left-right asymmetry $A_{LR}$ can be measured using the hadronic $Z$ decays on the $Z$ poles; this has a very high efficiency and almost no background. The technical details about this measurement and the other $Z$ pole observables can be found in [43]. Using a few billion events on the $Z$ pole translates into a statistical error of the order $\Delta A_{LR} = 10^{-5}$, with systematic uncertainties also under control at this level. The relative uncertainty on the polarization needs to be smaller than the corresponding uncertainty of the left-right asymmetry, $\Delta\mathcal{P}/\mathcal{P} < \Delta A_{LR}/A_{LR} = 10^{-4}$, which is only possible if both polarized electrons and positrons are available. In that case an in-situ polarization measurement is possible by means of the Blondel scheme [25]. Using the cross section with unpolarized beams, $\sigma_0$, and the polarization $\mathcal{P}_{e^-}$ and $\mathcal{P}_{e^+}$ for electrons and positrons, respectively, the polarized beam cross section can be expressed via the formula

$$\sigma = \sigma_0 \left\{1 - \mathcal{P}_{e^-}\mathcal{P}_{e^+} + A_{LR} \cdot (\mathcal{P}_{e^+} - \mathcal{P}_{e^-})\right\} . \tag{4.16}$$

If a method is used to externally determine all the four different combinations of beam polarizations, then the left-right asymmetry can be directly determined via

$$A_{LR} = \sqrt{\frac{(\sigma_{++} + \sigma_{-+} - \sigma_{+-} - \sigma_{--})(-\sigma_{++} + \sigma_{-+} - \sigma_{+-} + \sigma_{--})}{(\sigma_{++} + \sigma_{-+} + \sigma_{+-} + \sigma_{--})(-\sigma_{++} + \sigma_{-+} + \sigma_{+-} - \sigma_{--})}} . \tag{4.17}$$

Here $\sigma_{ij}$ is the cross section where the electron beam has the polarization $i$ and the positron beam the polarization $j$. In deriving this formula one has to assume that the absolute polarization values of the bunches with opposing helicity states are equal. Either to assure that this assumption is correct or in order to determine the corresponding corrections one needs polarimeters. Most systematics cancel out of this measurement since within each beam only relative measurements are necessary. Hence, this scheme allows one to achieve the desired accuracy in the polarization measurement. Note that because of helicity selection rules the cross sections for the combinations $(++)$ and $(--)$ are tiny, so that the collider needs to run only for one tenth of its luminosity on these helicity configurations in order to already reach optimal statistical precision. The in-situ polarization measurement with the Blondel scheme, on the other hand, yields a statistical error that is only slightly bigger than the one with the external polarimeter, if the degree of positron polarization exceeds $\mathcal{P}_{e^+} > 0.5$. If that value goes down to 20% the statistical error for $10^9$ $Z$ bosons on the peak reaches $\Delta A_{LR} = 8 \cdot 10^{-5}$. Another crucial ingredient for the precision of this measurement is the simultaneous knowledge of both the c.m. energy and the mass of the $Z$ boson, $M_Z$. This is because the $\gamma - Z$ interference generates a slope in the peak cross section of roughly $dA_{LR}/d\sqrt{s} = 2 \cdot 10^{-2}$/GeV. To suppress the dominance of the parametric uncertainty on the beam energy within the systematics one needs to calibrate the beam energy with the help of a spectrometer relative to the $Z$ mass with a precision of 1 MeV and allow for a scan around the vicinity of the $Z$ resonance peak. The second biggest systematics effect comes from the influence of the beamstrahlung which induces a shift in the value of $A_{LR}$ by $\Delta A_{LR} = 9 \cdot 10^{-4}$. For that scope the beamstrahlung spectrum needs to be known at a precision at the order of one per cent or even below, and studies show that this achievable [44–46]. All other systematic errors are very small, such that a quite conservative error estimate results in a final uncertainty of $\Delta A_{LR} = 10^{-4}$. That systematic uncertainty translates into an error of the weak mixing angle of $\Delta \sin^2 \theta^\ell_{eff} = 1.3 \cdot 10^{-5}$. However, in principle, the beamstrahlung spectrum should be the same both in the $A_{LR}$ measurement in the scan for the calibration. In that case the whole effect of beamstrahlung results in an obvious shift of the center of mass energy that cancels out in the uncertainties. Then, a precision in the measurement of the effective weak angle well below $10^{-5}$ could be achieved.

**Table 4.10**
Precision of several SM observables that can be achieved at the ILC from a high-luminosity low-energy run (GigaZ option). The left column gives the present status together with possible expectations from the LHC experiments; the right column gives the Giga-Z expectation. The values given for the $\Delta\rho$ parameter and for the determination of the strong coupling constant assume $N_\nu = 3$.

| | LEP/SLC/Tev/world av. [49] | ILC |
|---|---|---|
| $\sin^2\theta^\ell_{\rm eff}$ | $0.23146 \pm 0.00017$ | $\leq \pm 0.00001$ |
| $M_Z$ | $91.1876 \pm 0.0021$ GeV | $\pm 0.0016$ GeV |
| $\Gamma_Z$ | $2.4952 \pm 0.0023$ GeV | $\pm 0.0008$ GeV |
| $\alpha_s(m^2_Z)$ | $0.1184 \pm 0.0007$ | $\pm 0.0005$ |
| $\Delta\rho$ | $(0.55 \pm 0.10) \cdot 10^{-2}$ | $\pm 0.05 \cdot 10^{-2}$ |
| $N_\nu$ | $2.984 \pm 0.008$ | $\pm 0.004$ |
| $\mathcal{A}_b$ | $0.923 \pm 0.020$ | $\pm 0.001$ |
| $R^0_b$ | $0.21653 \pm 0.00069$ | $\pm 0.00014$ |
| $M_W$ | $80.385 \pm 0.015$ GeV | $\pm 0.006$ GeV |

As mentioned above, there is a discrepancy between the measured value of $A_{LR}$ and the value measured for a related quantity, the forward-backward asymmetry for bottom quarks $\mathcal{A}_b$. Since the ILC detectors will have $b$-tagging capabilities of an unprecedented excellence, the ILC can improve the precision of the $\mathcal{A}_b$ measurement by a factor of almost a factor of 20 [47]. A resolution of this discrepancy itself might allow improvement of the whole consistency and quality of the electroweak fit and open a door to precision searches for deviations from the SM.

The other observables that can be determined from the measurement of the $Z$ lineshape are the partial and the total width of the $Z$ boson, the $Z$ mass $M_Z$, the strong coupling constant at the scale of $M_Z$ ($\alpha_s(M_Z)$), the $\rho$ parameter, which is a measure of the modification of the strength of the fermionic $Z$ couplings due to radiative corrections, and the number of light, weakly-interacting neutrino species, $N_\nu$.

Concerning the partial widths and the total width of the $Z$ boson, there is still a considerable effect to the improved ILC measurements; however, it is less spectacular than the improvement for the weak mixing angle. The measurement of the total $Z$ width by the lineshape determination depends on the precision of the beam spectrometer and the calibration measurement of the beamstrahlung. This means that a total precision of the order $\Delta\Gamma_Z \approx 1$ MeV or better is possible. At the ILC, there will be a factor of up to three improvement for the selection efficiencies for hadrons, muons, and tau leptons compared to the LEP experiments [23, 48]. There will also be considerable improvement in the experimental systematics of the luminosity compared to LEP, such that the errors can be further reduced. Note that this has been accompanied by matching improvements in the theoretical predictions for the electroweak precision observables, which are now mostly at the two- or even three-loop level [50].

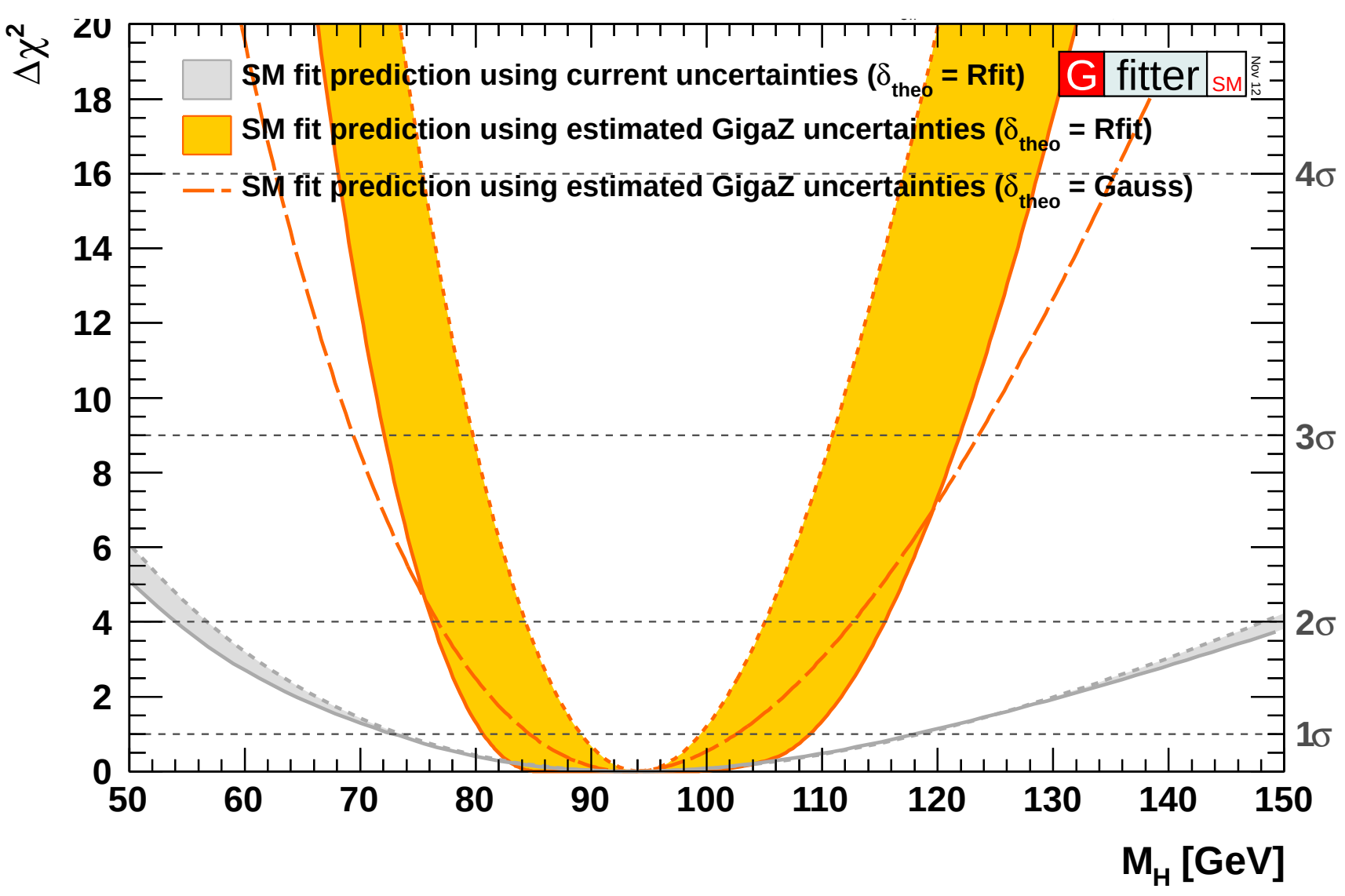

**Figure 4.9** Contour of $\chi^2$ in the precision electroweak fit as a function of the Higgs boson mass for the current values (grey band) and for improved uncertainties expected from the Giga-$Z$ program (orange band) [51]. The figure assumes that the current central value remains unchanged.

For almost all of these variables the ILC can considerably improve on the present-day precision. These improvements are summarized in Table 4.10. Taking into account the excellent $b$-tagging performance at the ILC detectors, even the ratio $R_b$ of the partial $Z$ width to bottom quarks to the full hadronic width can be improved at least by a factor of five.

A measure of the increased analytical power available from the Giga-$Z$ program is shown in Fig. 4.9. The figure shows the $\chi^2$ of the current electroweak fit as a function of the Higgs boson mass, and the $\chi^2$ curve that would result for the same central value and the measurement uncertainties that would result from Giga-$Z$. With these assumptions, the fit would give a Higgs boson mass of $m_h = 92.3^{+16.6}_{-11.6}$ GeV, with current theory errors, or $^{+5.3}_{-5.0}$ GeV, with negligible theory errors. Even in the former case, a mass of 126 GeV for the Higgs boson would be excluded at almost the 4 $\sigma$ level in a pure Standard Model fit, requiring additional contributions from new particles at the TeV mass scale [51, 52].

## 4.5.2 Precision measurement of the $W$ boson mass

The final physics point to be discussed here is the measurement of the $W$ boson mass from a threshold scan at the $W$ pair production threshold. The overall $WW$ cross section is proportional near threshold to the the (non-relativistic) velocity $\beta$ of the $W$ boson. Thus, the cross section around the threshold is highly sensitive to the exact value of the $W$ mass. It is important that the $s$-channel and $t$-channel diagrams contribute differently in this region. The $s$-channel contribution is suppressed by $\beta^3$, while the $t$-channel only by one power of $\beta$. The $t$-channel contribution depends only on the $We\nu_e$ coupling, which is well-known. Hence, the predictions for the threshold cross section are free from any possible contamination from unknown physics. Any new physics effects from the triple gauge couplings enter in the $s$-channel diagrams and are thus suppressed relative to the leading cross section by $\beta^2$. This guarantees a clean measurement of the $W$ mass from the threshold scan. This experimental setup is underlined by theoretical calculations in the last decade which provide full $2 \to 4$ calculations at next-to-leading order [13] and leading NNLO corrections to the total cross section [18], which allows to reduce the theory uncertainties to the same level as the experimental error estimates. Note that by using different polarization states an enhancement or suppression of the signal is possible such that the background can be directly estimated from the run by switching polarizations.

In an early study on such a scan [53], a scan in five steps between 160.4 and 162 GeV and an additional data point at $\sqrt{s} = 170$ GeV has been investigated. The analysis assumed an integrated luminosity of 100 fb$^{-1}$ and the same efficiency and purity values as obtained at LEP. With a total error of 0.25% on the luminosity and the selection efficiencies in that setup, $M_W$ can be determined with an error of $6-7$ MeV. The upper value comes from a fit where the efficiencies are not fixed but left free to float which shows the experimental stability of that method. If the detector performs much better than in the original study as would be expected for the ILC detectors [23], a precision of a few MeV can be achieved.

# Chapter 5
# Top Quark

The top quark is by far the heaviest particle of the Standard Model. Its large mass implies that the top quark is the Standard Model particle most strongly coupled to the mechanism of electroweak symmetry breaking. For this and other reasons, the top quark is expected to be a window to any new physics at the TeV energy scale. In this section, we will review the ways that new physics might appear in the precision study of the top quark and the capabilities of the ILC to discover these effects.

The top quark was discovered at the Tevatron proton-antiproton collider by the D0 and CDF experiments [1, 2]. Up to now, the top quark has only been studied at hadron colliders, at the Tevatron and, only in past three years, at the LHC. The Tevatron experiments accumulated a data sample of about $12\,\text{fb}^{-1}$ in Run I and Run II, at center of mass energies of $1.8$ TeV and $1.96$ TeV, respectively. About half of this data is fully analyzed. At the LHC, a data sample of about $5\,\text{fb}^{-1}$ has been recorded at a center-of-mass energy of $7$ TeV up to the end of 2011. In 2012, the machine has operated at a center of mass energy of $8$ TeV. In the following section, we will review the properties of the top quark determined so far at hadron colliders, based on the currently analyzed data sets. We will also discuss the eventual accuracies that will be reached in this program over the long term.

The ILC would be the first machine at which the top quark is studied using a precisely defined leptonic initial state. This brings the top quark into an evironment in which individual events can be analyzed in more detail, as we have explained in the Introduction. It also changes the production mechanism for top quark pairs from the strong to the electroweak interactions, which are a step closer to the phenomena of electroweak symmetry breaking that we aim to explore. Finally, this change brings into play new experimental observables—weak interaction polarization and parity asymmetries—that are very sensitive to the coupling of the top quark to possible new interactions. It is very possible that, while the top quark might respect Standard Model expectations at the LHC, it will break those expectations when studied at the ILC.

## 5.1 Top quark properties from hadron colliders

In this section, we will review the present and future capabilities of hadron colliders to study the top quark. This section is based largely on the review published in [3]. Where applicable, the information has been updated.

### 5.1.1 Top quark hadronic cross section

A central measurement for the top quark at hadron colliders is the $t\bar{t}$ production cross-section. At hadron colliders the following channels are typically measured: (1) lepton+jets channels, (2) dilepton channels, (3) fully hadronic channels, (4) channels with jets and missing transverse momentum (MET).For these channels the Tevatron experiments have published values between $7.2\,\mathrm{pb}$ and $7.99\,\mathrm{pb}$ [3]. The error on these values is typically 6–7%. The LHC experiments report values at 7 TeV [4, 5]

$$\begin{aligned} \sigma_{t\bar{t}} &= 177 \pm 3\,(\mathrm{stat.})^{+8}_{-7}\,(\mathrm{syst.}) \pm 7\,(\mathrm{lumi.})\,\text{pb} \quad \text{ATLAS} \\ \sigma_{t\bar{t}} &= 166 \pm 2\,(\mathrm{stat.}) \pm 11\,(\mathrm{syst.}) \pm 8\,(\mathrm{lumi.})\,\text{pb} \quad \text{CMS} \end{aligned} \tag{5.1}$$

This is to be compared with theoretical estimates from 'approximate NNLO' QCD predictions, for example, [6, 7]

$$\sigma_{t\bar{t}} = 163^{+7}_{-5}\,(\mathrm{scale}) \pm 9\,(\mathrm{PDF})\,\text{pb}. \tag{5.2}$$

A full NNLO QCD calculation should decrease the first error significantly. The agreement between theory and experiment is excellent at the present stage, both for the LHC and for the Tevatron results. Already at this early stage of data taking the LHC experiments are limited by the systematic uncertainty. For ATLAS, the dominant sources of the systematic error are those from predictions of different event generators together with the uncertainties of the parton distribution function of the proton. On the experimental side, the jet energy resolution constitutes an important source of systematic error. However, there are other sources of comparable influence, from the electron and muon identification. The quoted sources contribute roughly equally to the systematic error.

### 5.1.2 Top quark mass and width

The mass of the top quark is a fundamental parameter of the electroweak theory. In discussions of physics beyond the Standard Model, the top quark appears ubiquitously. To interpret particle physics measurements in terms of new physics effects, the top quark mass must be known very accurately. Two well known examples are the precision electroweak corrections, where the top quark contributions must be fixed to allow Higgs and other new particle corrections to be determined, and the theory of the Higgs boson mass in supersymmetry, in which the loop corrections are proportional to $(m_t/m_W)^4$.

Care must be taken in relating the measured top quark mass to the value of the top quark mass that is used as input in these calculations. Loop effects typically take as input a short-distance definition of the top quark mass such as the $\overline{MS}$ mass parameter. We will explain below that the determination of the top quark mass from the threshold cross section in $e^+e^-$ annhilation uses a precise short-distance definition of the top quark mass, though a different one from the $\overline{MS}$ mass.

Another frequently used definition of the top quark mass is provided by the position of the pole in the top quark propagator computed in perturbation theory. This top quark mass is greater than the $\overline{MS}$ mass by about 10 GeV. This difference contains a parametric ambiguity of order the QCD hadronization scale due to the asymptotic character of the perturbative series caused by the infrared sensitivity of the pole mass.

Current determinations of the top quark mass from kinematic distributions do not use either of these, in principle, well defined top quark mass definitions. Instead, they define the top quark mass as the input mass parameter of a Monte Carlo event generator, which is then constrained by measurements of the kinematics of the $t\bar{t}$ final state. At this time, there is no concrete analysis that relates this mass to either the short distance or the pole value of the top quark mass. For the case of $e^+e^-$ production of top quark pairs, it was shown in [8] how to relate event-shape variables that depend strongly on the top quark mass to an underlying short-distance mass parameter. The analysis requires center of mass energies much larger than $2m_t$. For hadron colliders, the corresponding analysis is much more difficult and has not yet been done.

Within the framework that is available now, the Tevatron and LHC experiments have achieved quite a precise determination of the top quark mass from kinematic observables. The value of the top quark mass $m_t$ as published by the Tevatron Electroweak Working Group is given to be $m_t = 173.2 \pm 0.9$ GeV [9]. This value has been obtained from the combined measurements of the Tevatron experiments. The LHC experiments report values of $m_t = 174.5 \pm 0.6 \pm 2.3$ GeV for the ATLAS collaboration [10] and $m_t = 172.6 \pm 0.4 \pm 1.2$ GeV for the CMS collaboration [11], where, in each case, the first error is statistical and the second is systematic. The dominant systematic errors come from jet energy resolution. In both cases, the mass definition used is that of the Monte Carlo event generator. Reduction of the error well below 1 GeV will require a more careful theoretical analysis giving the relation of the mass parameter used in these measurements to a more precise top quark mass definition.

Within the Standard Model the total decay width $\Gamma_t$ of the top quark is dominated by the partial decay width $\Gamma(t \to Wb)$. The top quark width is predicted to be approximately $1.5$ GeV, which is substantially larger than the hadronization scale $\Lambda_{\mathrm{QCD}}$. On the other hand, this value is small enough that it is not expected to be directly measured at the LHC.

At hadron colliders, the decay width can be determined via

$$\Gamma_t = \Gamma(t \to Wb)/BR(t \to Wb) \ . \tag{5.3}$$

The partial width $\Gamma(t \to Wb)$ is determined from the cross section for single top events while the branching ratio $BR(t \to Wb)$ is derived from top pair events. D0 gives a value of $\Gamma_t = 1.99^{+0.69}_{-0.55}$ [12]. CDF uses only the top quark mass spectrum and reports the 68% confidence interval to be $0.3 < \Gamma_t < 4.4$ GeV [13]. It is interesting to note here that D0 has published for the ratio of branching ratios $BR(t \to Wb)/BR(t \to Wq)$ a value of $0.9 \pm 0.04$ [14], which is about $2.5\sigma$ away from the Standard Model expectation.

### 5.1.3 Helicity of the $W$ boson

The top quark has a very short lifetime of about $10^{-25}\,\mathrm{s}$. Since this is about $10$ times shorter than typical scales for long range QCD processes, the top quark decays long before hadronization can affect it. Therefore, the structure of the top quark decay is very close to that of a bare quark. Within the Standard Model, the top quark decays almost exclusively via $t \to W^+b$. The V-A nature of the weak decay dictates that the resulting $b$ quark is almost completely left handed polarized. It also dictates the polarization of the $W$ boson, which in turn can be measured by observing the $W$ decay. The prediction is that the $W$ is produced only in the left-handed and longitudinal polarization states, with the fraction of longitudinal $W$ bosons predicted to be

$$f_0 = \frac{m_t^2}{2m_W^2 + m_t^2} \ . \tag{5.4}$$

The Standard Model predicts a value of $f_0 = 0.703$. The CDF experiment measures this value to be $f_0 = 0.78^{+0.19}_{-0.20}(\mathrm{stat.}) \pm 0.06(\mathrm{syst.})$ [15], in agreement with the Standard Model. The most precise measurements of this value have been achieved with events in which both the $W$ boson from the $t$ and the one from the $\bar{t}$ decay into leptons.

### 5.1.4 Top coupling to $Z^0$ and $\gamma$

It is particularly interesting to study the coupling of the top quark to the photon and the $Z^0$ boson to search for effects of new physics. Both of these couplings are subdominant effects at hadron colliders. The electroweak production of $t\bar{t}$ is suppressed with respect to QCD production, and this is especially true at the LHC where most of the $t\bar{t}$ production comes from gluon-gluon fusion. Radiation of photons from $t\bar{t}$ has been observed at the Tevatron. So far no precision measurements on the coupling of top quarks to the $Z^0$ boson have been reported.

Constraints on the top quark couplings to $\gamma$ and $Z^0$ have been reported using the expression for the couplings [16]

$$\Gamma_\mu^{ttX}(k^2, q, \bar{q}) = ie\left\{\gamma_\mu\left(\widetilde{F}_{1V}^X(k^2) + \gamma_5\widetilde{F}_{1A}^X(k^2)\right) + \frac{(q-\bar{q})_\mu}{2m_t}\left(\widetilde{F}_{2V}^X(k^2) + \gamma_5\widetilde{F}_{2A}^X(k^2)\right)\right\}. \quad (5.5)$$

where $X = \gamma, Z$ and the $\widetilde{F}$ are related to the usual form factors $F_1$, $F_2$ by

$$\widetilde{F}_{1V}^X = -\left(F_{1V}^X + F_{2V}^X\right), \qquad \widetilde{F}_{2V}^X = F_{2V}^X, \qquad \widetilde{F}_{1A}^X = -F_{1A}^X, \qquad \widetilde{F}_{2A}^X = -iF_{2A}^X. \quad (5.6)$$

In the Standard Model the only form factors which are different from zero are $F_{1V}^\gamma(k^2)$, $F_{1V}^Z(k^2)$ and $F_{1A}^Z(k^2)$. The quantities $F_{2V}^{\gamma,Z}(k^2)$ are the electric and weak magnetic dipole moment (EDM and MDM) form factors.

$F_{2A}^\gamma(k^2)$ is the CP-violating electric dipole moment form factor of the top quark, and $F_{2A}^Z(k^2)$ is the weak electric dipole moment (WDM). These two form factors violate CP. In the Standard Model they receive contributions only from the three loop level and beyond.

In the case of the $t\bar{t}Z^0$ final state, relatively clean measurements are expected at the LHC when the $Z^0$ decays leptonically. However, the cross section is quite small, so that meaningful results with precision of about 10% for $F_{1A}^{Z^0}$ and 40% for $F_{2V,A}^{Z^0}$ can only be expected after a few 100 fb$^{-1}$. At the HL-LHC, with an integrated luminosity of about 3000 fb$^{-1}$, the precision of this measurement is expected to improve by factors between 1.6 for $F_{2V,A}^{Z^0}$ and 3 for $F_{1A}^{Z^0}$. The situation is considerably better for measurements of the $t\bar{t}\gamma$ vertex. Already for 30 fb$^{-1}$ at the LHC, measurements with a precision of about 20% to 35% can be expected. These measurements may improve at the HL-LHC to values between 2% and 10%

For the related question of the coupling of the top quark to the Higgs boson, both the LHC expectations and the projections for the ILC are discussed in Chapter 2 of this report.

### 5.1.5 Asymmetries at hadron colliders

The last few years were marked by a number of publications from the Tevatron experiments which reported on tensions with Standard Model predictions in the measurement of forward backward asymmetries $A_{FB}$. This observable counts the difference in the number of events in the two hemispheres of the detector. In hadronic collisions, the polar angle is typically reported in terms of the rapidity $y$, which is invariant under longitudinal boosts and more descriptive at very forward and backward angles. For the analyses here and at the LHC, at least one member of the $t\bar{t}$ pair is required to decay leptonically to assure the particle identification. The average asymmetry reported by CDF is $0.201 \pm 0.065\,(\mathrm{stat.}) \pm 0.018\,(\mathrm{syst.})$ [17] which agrees with $0.196 \pm 0.060\,(\mathrm{stat.})^{+0.018}_{-0.026}\,(\mathrm{syst.})$ as reported by DO [18]. These values can be compared with an asymmetry of about 0.07 predicted by the to Standard Model from NLO QCD and electroweak effects. This result is difficult to verify at the LHC. The LHC is a proton-proton collider, so the two hemispheres are intrinsically symmetric. Further, at the LHC at 7 TeV, only 15% of the interactions arise from $q\bar{q}$ collisions; the 85% from $gg$ collisions can have no intrinsic asymmetry. Still, in $q\bar{q}$ collisions at the LHC, it is likely that the $q$ is a valence quark while the $\bar{q}$ is pulled from the sea. This implies that $t\bar{t}$ pairs produced from $q\bar{q}$ are typically boosted in the direction of the $q$. This offers methods to observe a forward backward asymmetry in $q\bar{q} \to t\bar{t}$. For example, a forward-backward asymmetry in the $q\bar{q}$ reaction translates into a smaller asymmetry $A_C$ in the variable $\Delta|y| = |y_t| - |y_{\bar{t}}|$. For this observable, CMS measures $A_C = 0.004 \pm 0.010\,(\text{stat.}) \pm 0.012\,(\text{syst.})$ [19], which agrees with the Standard Model predictions within the relatively large uncertainties. So far, the LHC experiments have not provided any independent evidence for asymmetries outside the Standard Model predictions [3, 20]. The theoretical interpretation of these asymmetries is also very uncertain. Many plausible models of the $t\bar{t}$ asymmetry predict effects in top quark physics at high energy that are excluded at the LHC. For a review of the current situation, see [21, 22]. It is possible that the tension between theory and experiment can be resolved by more accurate QCD calculation. For example, a lower choice of the QCD renormalization scale, argued for in [23], would increase the Standard Model prediction.

## 5.2 $e^+e^- \to t\bar{t}$ at threshold

One of the unique capabilities of an $e^+e^-$ linear collider is the ability to carry out cross section measurements at particle production thresholds. The accurately known and readily variable beam energy of the ILC makes it possible to measure the shape of the cross section at any pair-production threshold within its range. Because of the leptonic initial state, it is also possible to tune the initial spin state, giving additional options for precision threshold measurements. The $t\bar{t}$ pair production threshold, located at a center of mass energy energy $\sqrt{s} \approx 2m_t$, allows for precise measurements of the top quark mass $m_t$ as well as the top quark total width $\Gamma_t$ and the QCD coupling $\alpha_s$. Because the top is a spin-$\frac{1}{2}$ fermion, the $t\bar{t}$ pair is produced in an angular $S$-wave state. This leads to a clearly visible rise of the cross section even when folded with the ILC luminosity spectrum. Moreover, because the top quark pair is produced in a color singlet state and because the finite top lifetime provides an effective infrared cutoff for QCD corrections, the experimental measurements can be compared with very accurate and unambiguous analytic theoretical predictions of the cross section with negligible hadronization effects. The dependence of the top quark cross section shape on the top quark mass and interactions is computable to high precision with full control over the renormalization scheme dependence of the top mass parameter. In this section, we will review the expectations for the theory and ILC measurements of the top quark threshold cross section shape. The case of the top quark threshold is not only important in its own right but also serves as a prototype case for other particle thresholds that might be accessible at the ILC.

### 5.2.1 Status of QCD theory

The calculation of the total top pair production cross section makes use of the method of non-relativistic effective theories. The top quark mass parameter used in this calculation is defined at the scale of about 10 GeV corresponding to the typical physical separation of the $t$ and $\bar{t}$. This mass parameter can be converted to the $\overline{MS}$ mass in a controlled way. The summation of QCD Coulomb singularities, treated by a non-relativistic fixed-order expansion, is well known up to NNLO [24] and has recently been extended accounting also for NNNLO corrections [25]. Large velocity QCD logarithms have been determined using renormalization-group-improved non-relativistic perturbation theory up to NLL order, with a partial treatment of NNLL effects [26, 27]. Recently the dominant ultrasoft NNLL corrections have been completed [28]. The accuracy in this calculation is illustrated in Fig. 5.1.

Since the top quark kinetic energy is of the order of the top quark width, electroweak effects, which also include finite-lifetime and interference contributions, are crucial as well. This makes the cross section dependent on the experimental prescription concerning the reconstructed final state. Recently a number of partial results have been obtained [29, 30] which put approximate NNLL order predictions within reach. Theoretical predictions for differential cross sections such as the top momentum distribution and forward-backward asymmetries are only known at the NNLO level and thus are much less developed.

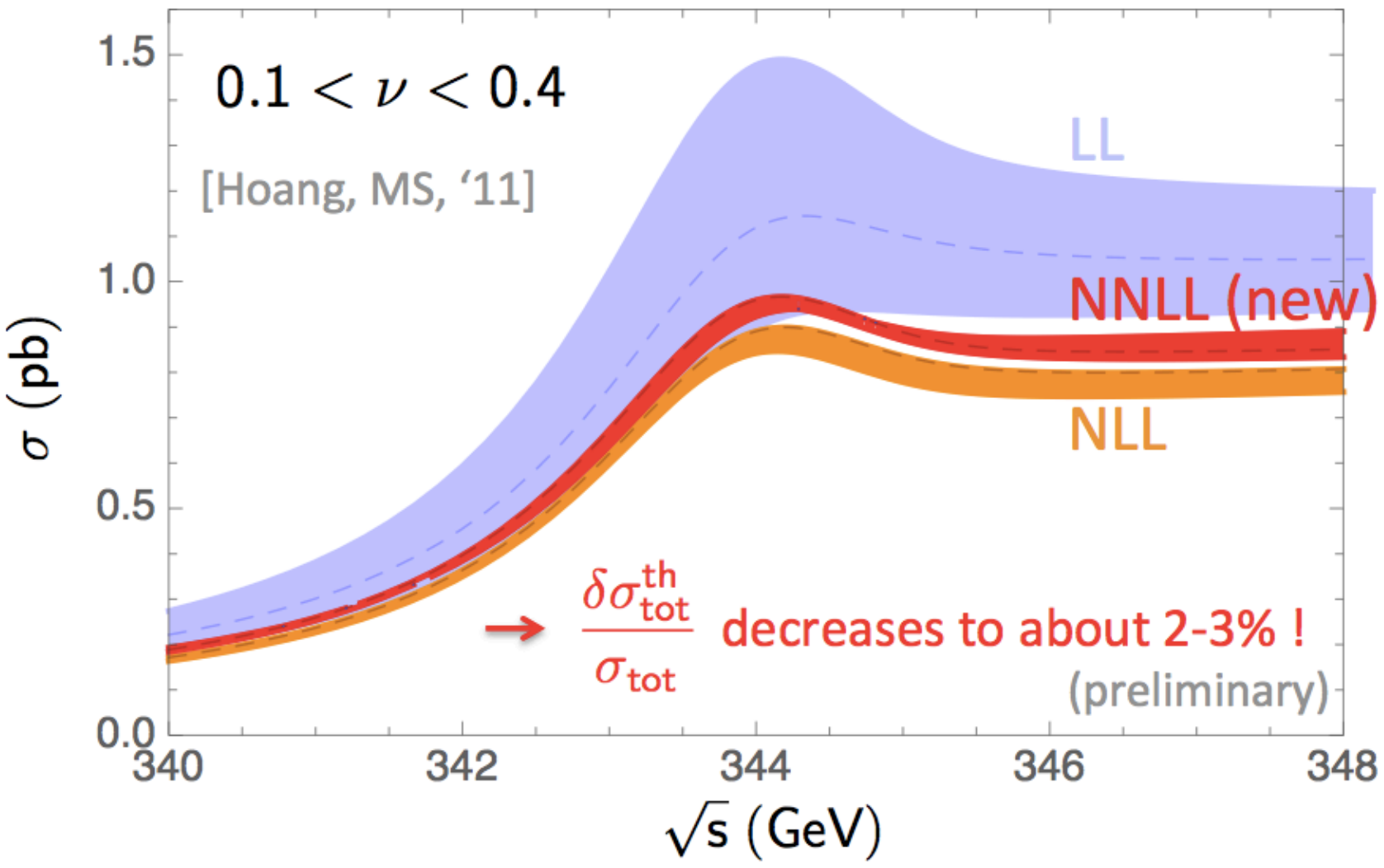

**Figure 5.1** Accuracy of the prediction of the top pair production cross section at the $t\bar{t}$ threshold at the ILC, in the 1S mass scheme, as achieved by QCD calculations with resummation of logarithmic corrections to leading (LL), next-to-leading (NLL), and next-to-next-to-leading (NNLL) order. From [31].

### 5.2.2 Simulations and measurements

The most thorough experimental study of the top quark threshold has been carried out by Martinez and Miquel in [32]. These authors assumed a total integrated luminosity of $300$ fb$^{-1}$, distributed over 10 equidistant energy points in a $10$ GeV range around the threshold, using the TELSA beam parameters. To treat the strong correlation of the input theory parameters, simultaneous fits were carried out for the top quark mass, the QCD coupling and the top quark width from measurments of the total cross section, the top momentum distributions and the forward-backward asymmetry. These were simulated based on the code TOPPIK with NNLO corrections [33]. The study obtained the uncertainties $\Delta m_t = 19$ MeV, $\Delta\alpha_s(m_Z) = 0.0012$ and $\Delta\Gamma_t = 32$ MeV, when all observables were accounted for Using just the total cross section measurements, the results were $\Delta m_t = 34$ MeV, $\Delta\alpha_s(m_Z) = 0.0023$ and $\Delta\Gamma_t = 42$ MeV. The difference shows the discriminating power of additional observables of the threshold region. The analysis included a theory uncertainty in the cross section codes of $3\%$, which at this time is only approached for total cross section computations. Although the analysis was only based on fixed order NNLO predictions, the quoted uncertainties should be realistic.

The analysis in [32] did not yet include a complete study of experimental systematic uncertainties, including, in particular, uncertainties in the knowledge of the luminosity spectrum. This last point is addressed in a more recent study by Seidel, Simon, and Tesar, for which the results are shown in Fig. 5.2 [34]. That study was carried with a full detector simulation using the ILD detector. It takes the initial state radiation and beamstrahlung of the colliding beams into account. The figure underlines the high sensitivity of the threshold region to the actual value of the top quark mass. The statistical precision obtained on the top quark mass in this study is of the order of 30 MeV. Due to the QCD corrections relevant for a precise calculation of the top quark mass, the threshold scan is sensitive to the value of $\alpha_s$. The error ellipse as obtained in a combined determination of $\alpha_s$ and $m_t$ is shown in the right-hand panel of Fig. 5.2.

The threshold top quark mass determined in this study must still be converted to the standard top quark $\overline{MS}$ mass. The conversion formula, to three-loop order, is given in [33]. The conversion adds an error of about 100 MeV from truncation of the QCD perturbation series and an error of

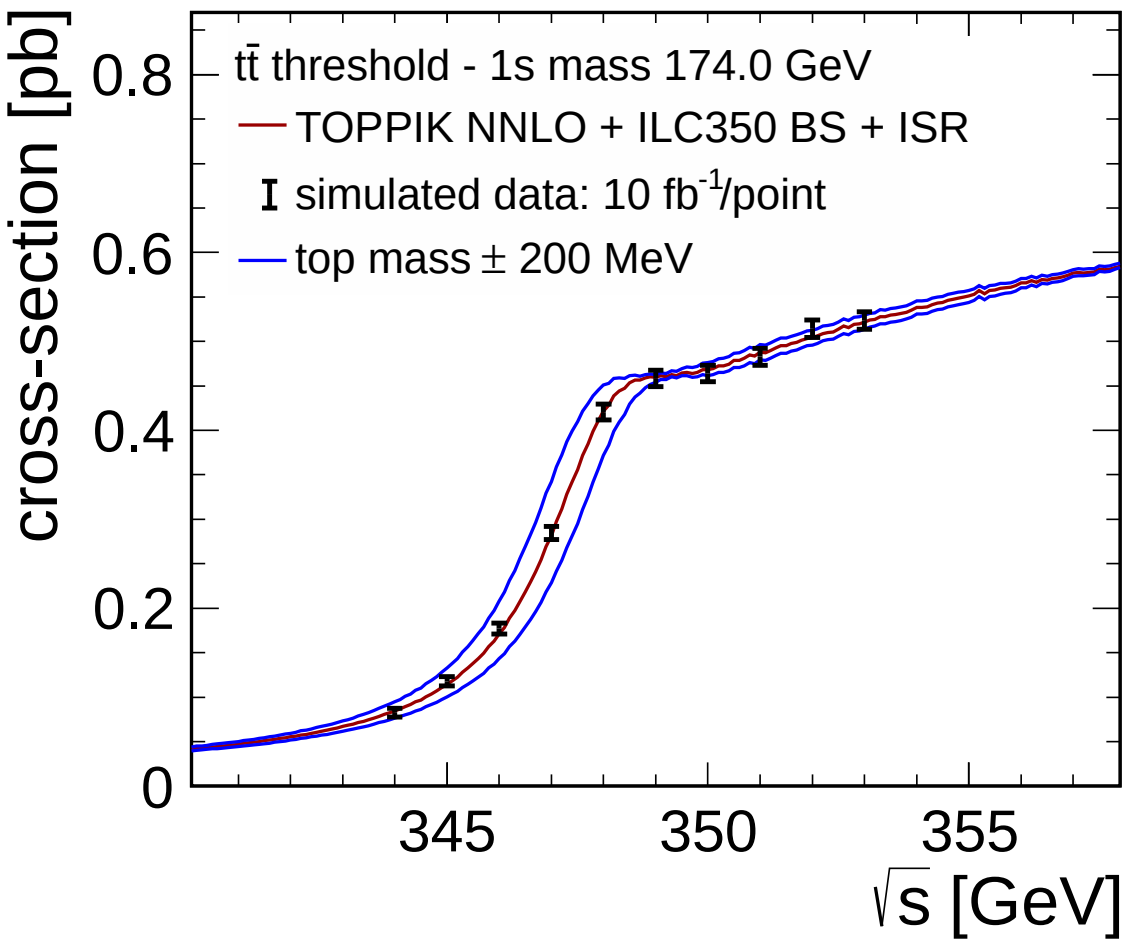

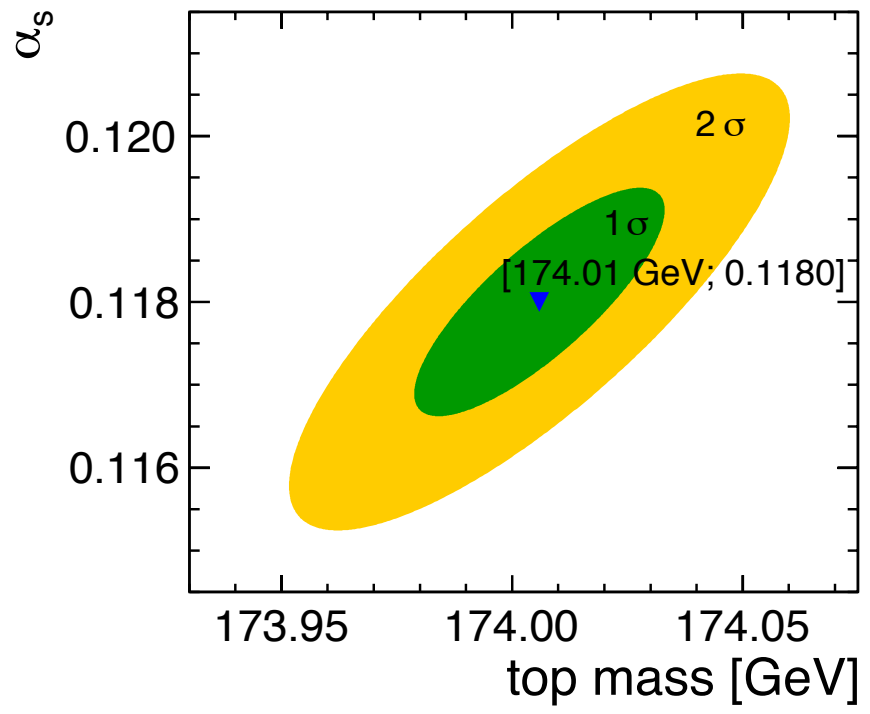

**Figure 5.2.** Illustration of a top quark threshold meausurement at the ILC. In the simulation, the top quark mass has been chosen to be 174. GeV. The blue lines show the effect of varying this mass by 200 MeV. The study is based on full detector simulation and takes initial state radiation (ISR) and beamstrahlung (BS) and other relevant machine effects into account: (left) the simulated threshold scan. (right) error ellipse for the determination of $m_t$ and $\alpha_s$. From [34].

70 MeV for each uncertainty of $0.001$ in the value of $\alpha_s$. Both sources of uncertainty should be reduced by the time of the ILC running. In particular, the study of event shapes in $e^+e^- \to q\bar{q}$ at the high energies available at ILC should resolve current questions concerning tensions between precision determinations of $\alpha_s$. It is important to note that these estimates of the accuracy of mass values are derived from a precision theory of the relation between the threshold mass and the top quark $\overline{MS}$ mass. A comparable theory simply does not exist for the conversion of the top quark mass measured in hadronic collisions to the $\overline{MS}$ value.

The precise determination of the top quark mass is likely to have important implications for fundamental theory. We have given one example at the end of Section 2.1. In that case, the value of the top quark mass, accurate at the level that ILC will provide, literally decides the fate of the universe.

In principle, the contribution of the Higgs exchange potential to the $t\bar{t}$ threshold makes it possible to measure that Higgs coupling to $t\bar{t}$. However, the precision of this measurement is strongly limited by the fact that the Higgs corrections are suppressed by the inverse square of the Higgs mass. For a Higgs mass of $m_H = 120$ GeV the study in [32] found that uncertainties of at least several $10\%$ should be expected in a measurement of the top quark Higgs Yukawa coupling. This coupling can be measured more accurately from the cross section for $e^+e^- \to t\bar{t}h$, as is explained in Section 2.6 and 2.7 of this report.

## 5.3 Probing the top quark vertices at the ILC

At higher energy, the study of $t\bar{t}$ pair production at the ILC is the ideal setting in which to make precise measurements of the the coupling of the top quark to the $Z^0$ boson and the photon. In contrast to the situation at hadron colliders, the leading-order pair production process $e^+e^- \to t\bar{t}$ goes directly through the $t\bar{t}Z^0$ and $t\bar{t}\gamma$ vertices. There is no concurrent QCD production of top pairs, which increases greatly the potential for a clean measurement. In the following section, we will review the importance of measuring these couplings precisely. Then we will describe studies of the experimental capabilities of the ILC to perform these measurements.

### 5.3.1 Models with top and Higgs compositeness

There are several classes of models that seek to answer the question of where the Higgs boson comes from and why it acquires a symmetry-breaking vaccum expectation value. Among these is supersymmetry, which will have its own discussion in Chapter 7 of this report. An alternative point of view is that the Higgs boson is a composite state within a larger, strongly interacting theory at the TeV scale. Though the first models of this type contained no light Higgs bosons, there are now many models that naturally contain a light Higgs boson very similar to the Higgs boson of the Standard Model coupling to new heavy particles at the TeV mass scale. In Chapters 2–4, we have described tests of models of this type at the ILC in Higgs boson, two-fermion, and $W$ boson measurements.

The top quark is the heaviest known particle that derives its mass entirely from electroweak symmetry breaking. Due to its high mass the top quark couples to the Higgs with a Yukawa coupling of strength $\lambda_t \approx 1$. It is therefore likely that any composite structure of the Higgs boson must be reflected in composite structure or non-Standard interactions of the top quark. While such interactions may exist, they may not be easy to find. The coupling of the top quark to the gluon and the photon are constrained at $Q^2 = 0$ by requirements from exact QCD and QED gauge invariance. However, the low-energy $t\bar{t}Z$ vertex is much less constrained. It is then likely that this is the crucial place to look for deviations from the Standard Model induced by a strongly interacting Higgs sector.

Models of composite Higgs bosons can be constructed in three ways that seem at first sight to be distinctly different. The Higgs bosons may be Goldstone bosons associated with strong-interaction symmetry breaking at the 10 TeV energy scale, as in Little Higgs models. They may arise as partners of gauge bosons in theories with an extra space dimension, as in Gauge-Higgs Unification. Or, they may arise in extra-dimensional theories as states confined to a lower-dimensional subspace or 'brane'. Randall and Sundrum constructed a model of the last type [35] but also argued that all three classes of models are related by strong coupling-weak coupling duality [36]. That is, it is possible to view the extra-dimensional models as tools that allow weak coupling calculations of effects that are intrinsically manifestations of strong coupling and composite state dynamics.

The Randall-Sundrum approach also includes a model explanation of the hierarchy of Higgs-fermion Yukawa couplings. This is one of the most mysterious aspects of the Standard Model, reflected in the fact that the top quark and the up quark have exactly the same quantum numbers but differ in mass by a factor of $10^5$. The extra dimension offers the possibility that the different flavors of fermion have wavefunction of different shape in the full space, and therefore different overlap with the wavefunction of the Higgs boson. In general, also, the right and left chiral components of each quark and lepton may have wavefunctions with different dependence on the extra dimensions. It is a typical prediction of Randall-Sundrum theories that the chiral components of the top quark have wavefunctions in the fifth dimension significantly different from those of the other quarks and from one another. The wavefunction of the right-handed top quark is shifted toward the low-energy boundary of the space, called the 'TeV brane', where the Higgs field is located. These differences of the wavefunctions are reflected directly in couplings of the top quark to the $Z^0$. These couplings are shifted from the values predicted in the Standard Model, with larger shifts specifically for the right-handed top quark. Figure 5.3 collects a number of predictions of the fractional shifts in the $t_L$ and $t_R$ couplings to the $Z^0$ in a variety of models proposed in the literature.

Models with extra dimensions may also be suited to explain the tensions observed at the Tevatron discussed in Section 5.1.5. The top forward-backward asymmetry may, for example, be explained by a new color octet vector boson $G_\mu$, which couples weakly to light quarks but strongly to the top quark. This difference is required in order to suppress ordinary dijet production from the new colour-octet state. The difference in the coupling can be realized by the arrangement of the top quark wavefunction along the extra-dimension [22].

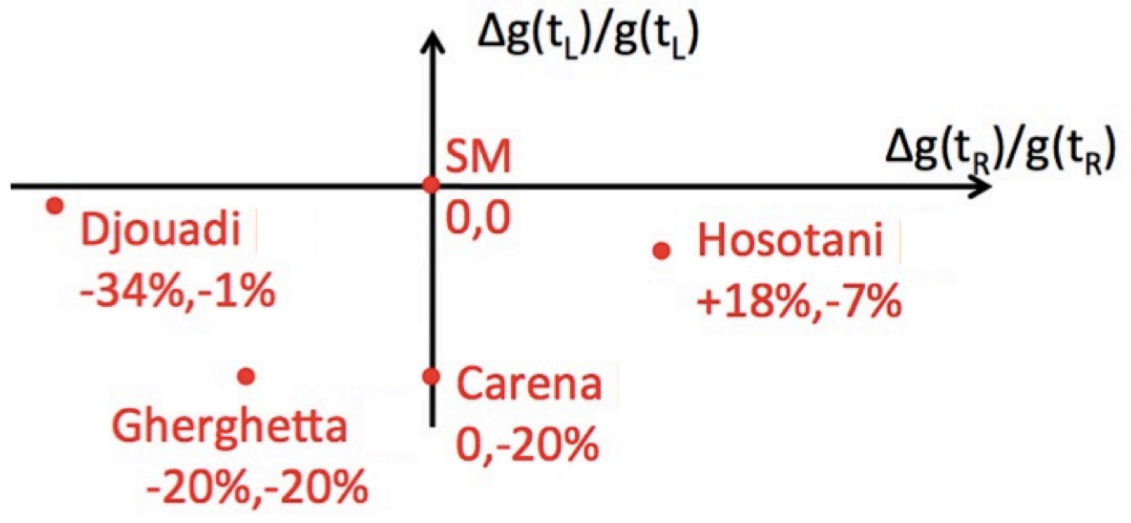

**Figure 5.3** Predictions of Randall-Sundrum extra dimensional models from various groups [37–40] for deviations from Standard Model couplings of the t quark to the $Z$ boson, from [44].

## 5.3.2 ILC measurements

In the previous section, we have described theories in which the top quark and Higgs boson are composite, with this compositeness being an essential element of the physics of electroweak symmetry breaking. A key test of this idea would come from the measurement of the $t\bar{t}Z$ couplings. Significant deviations from the predictions of the Standard Model would be expected. The ILC provides an ideal environment to measure these couplings. At the ILC, $t\bar{t}$ pairs would be copiously produced, with several 100,000 events for an integrated luminosity of $500\,\text{fb}^{-1}$. The production is by $s$-channel $\gamma$ and $Z$ exchange, so the $Z$ couplings enter the cross section in order 1. It is possible to almost entirely eliminate the background from other Standard Model processes. The ILC will allow for polarized electron and positron beams. This allows us to measure not only the total cross section for $t\bar{t}$ production but also the left-right asymmetry $A_{LR}$, the change in cross-section for different beam polarizations. For the $b$ quark, The most precise measurements for the $b$ quark of $A_{LR}$ at SLC and the forward-backward asymmetry at LEP result in a 3 $\sigma$ discrepancy of the effective electroweak mixing angle $\sin^2\theta_{eff}$ that has yet to be resolved [41]. If this effect is real, it is likely to be larger for the heavy top quark.

With the use of polarized beams, $t$ and $\bar{t}$ quarks oriented toward different angular regions in the detector are enriched in left-handed or right-handed top quark polarization [42]. This means that the experiments can independently access the couplings of left- and right-handed polarized quarks to the $Z$ boson. In principle, measurement of the cross sections and forward-backward asymmetries for two different polarization settings measures both the photon and $Z$ couplings of the top quark for each handedness. New probes of the top quark decay vertices are also available, although we expect that these will already be highly constrained by the LHC measurements of the $W$ polarization in top decay.

Recent studies based on full simulation of ILC detectors for a center of mass energy of $\sqrt{s} = 500\,\text{GeV}$ demonstrate that a precision on the determination of the couplings the left and the right chiral parts of the top quark wave function to the $Z^0$ to better than 1% can be achieved [44–46]. The most recent example of such a study, with full detector simulation, is shown in Figure 5.4. The figure demonstrates the clean reconstruction of the top quark direction, which allows for the precise determination of the forward-backward asymmetry. It has to be noted, however, that the final state gives rise to ambiguities in the correct association of the $b$ quarks to the $W$ bosons, see [46] for an explanation. These ambiguities can be nearly eliminated by requiring a high quality of the event reconstruction. The control of the ambiguities requires an excellent detector performance and event reconstruction. Another solution is the use of the vertex charge to separate the $t$ and $\bar{t}$ decays. It is shown in [45] that the high efficiency of vertex tagging in the ILC detectors will make this strategy available. The expected percent level independent measurements of the left- and right-handed top quark couplings will clearly discriminate the models shown in Fig. 5.3.

Even more incisive measurements than presented using optimised observables are investigated in [43]. These observables are the top pair production cross-section for left- and right-handed polarised beams and the fraction of right-handed ($t_R$) and left handed top quarks ($t_L$). Following a suggestion by [47] for the Tevatron, the fraction of $t_L$ and $t_R$ in a given sample can be determined with the helicity asymmetry. In the top quark rest frame the distribution of the polar angle $\theta_{hel}$ of a decay lepton is

$$\frac{1}{\Gamma}\frac{d\Gamma}{d\cos\theta_{hel}} = \frac{1 + a_t \cos\theta_{hel}}{2} \tag{5.7}$$

where $a_t$ varies between $+1$ and $-1$ depending on the fraction of right-handed $(t_R)$ and left handed top quarks $(t_L)$. The observable $\cos\theta_{hel}$ can easily be measured at the ILC. This observable is much less sensitive to ambiguities in the event reconstruction than the forward backward asymmetry. The slope of the resulting linear distribution provides a very robust measure of the net polarisation of a top quark sample. This net polarization is sensitive to new physics. The result of a full simulation study is shown in Fig. 5.5. It is demonstrated that over a range in $\cos\theta_{hel}$ the generated distribution is retained after event reconstruction. The reconstruction is nearly perfect for initial right handed electron beams. Remaining discrepancies in case of left handed electron beams can be explained by reconstruction inefficiencies for low final state lepton energies.

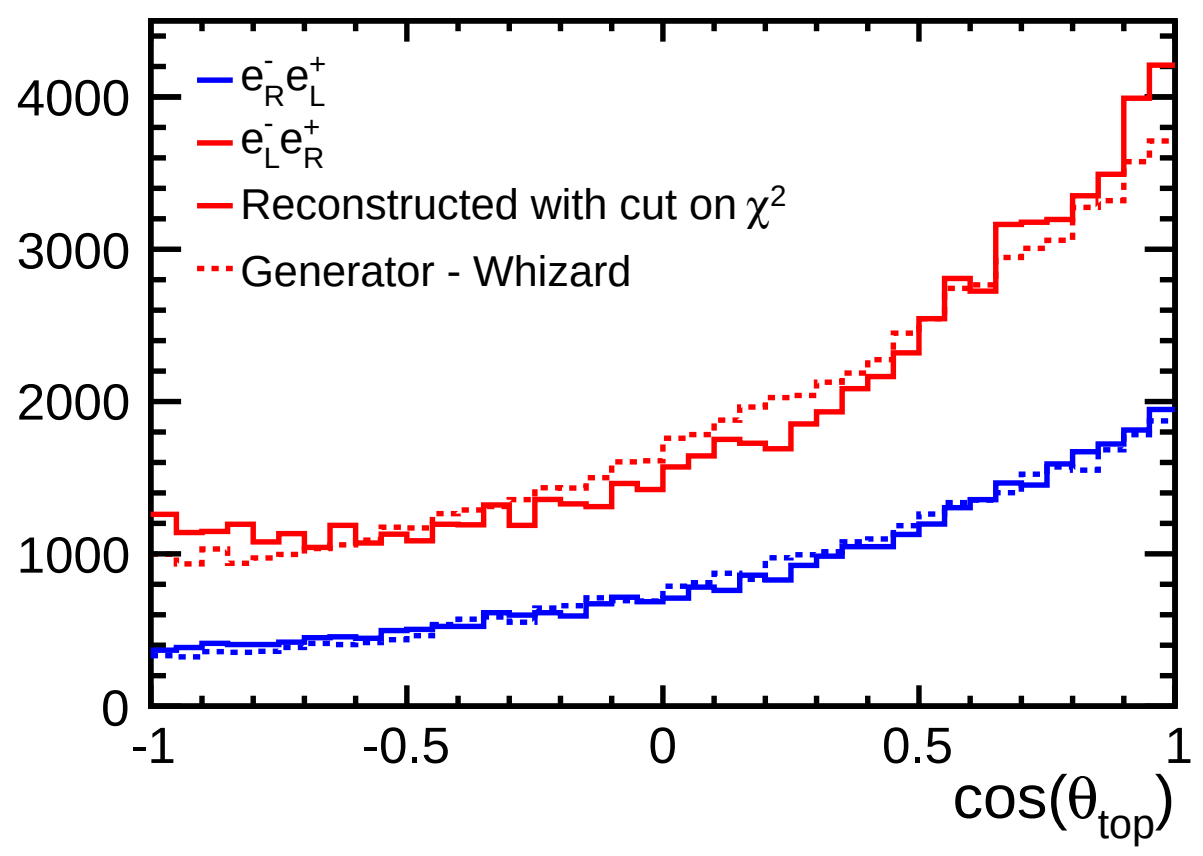

**Figure 5.4**
Reconstruction of the direction of the top quark in $t\bar{t}$ pair production for two different beam polarization [43]. It is known from [44] that the background is negligible.

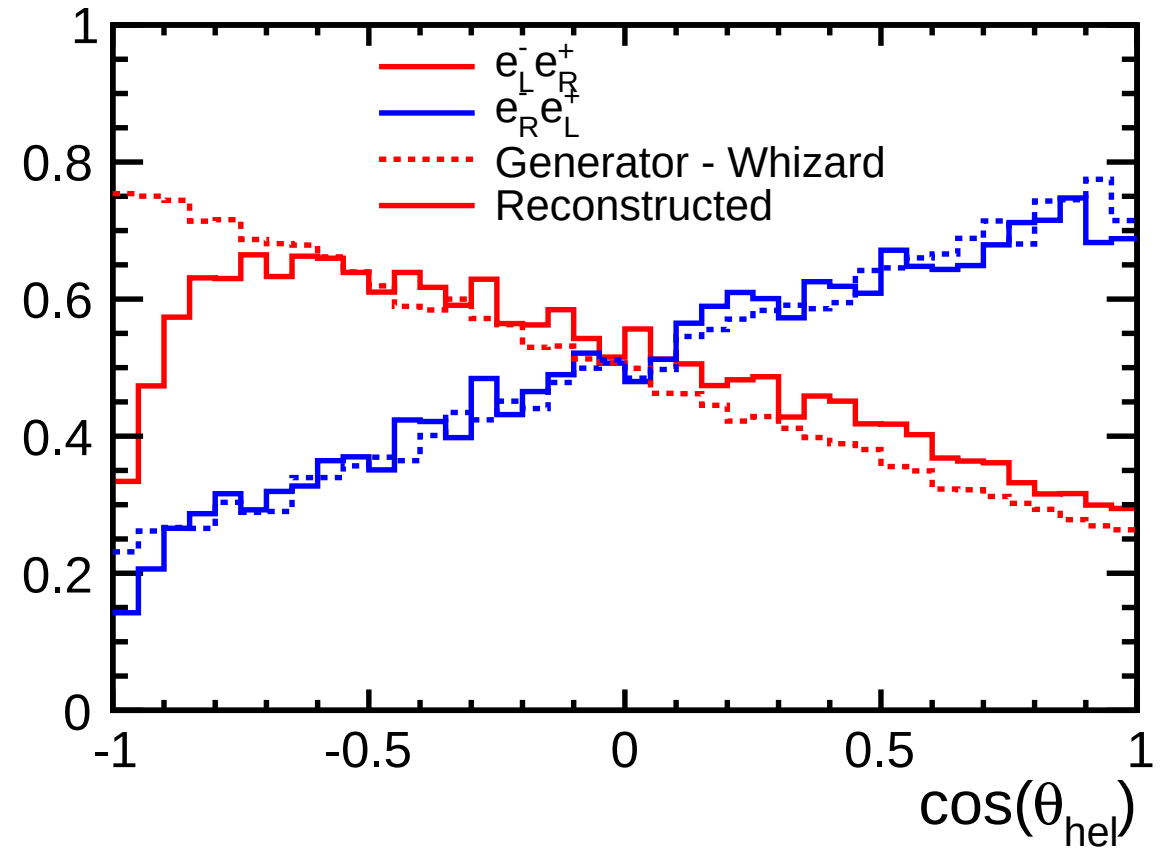

**Figure 5.5**
Top: Generated and reconstructed distributions of the top quark helicity angle $\cos\theta_{hel}$ in $e^+e^- \to t\bar{t}$ at the ILC at 500 GeV, from [43].

The observables $A_{FB}$, cross sections and helicity asymmetries are used to disentangle the coupling of the top quark to the photon and to the $Z$. Figure 5.6 compares the precision on the form factors expected from the LHC with that from the ILC.

Numerical values for the expected accuracies at linear $e^+e^-$ colliders, ILC and earlier on TESLA [49], on seven top quark form factors (due to QED gauge invariance the coupling $\widetilde{F}^{\gamma}_{1A}$ is fixed to 0), taken from the studies [43, 48, 49], are given in Tables 5.1 and 5.2, along with comparisons to the expectations from the LHC experiments.

**Figure 5.6**
Comparison of precisions for $CP$ conserving form factors of the top quark coupling to $\gamma$ and $Z$, $\widetilde{F}^{\gamma,Z}_{1V,A}$, expected at the LHC, taken from [16], and at the ILC. The LHC results assume an integrated luminosity of $\mathcal{L} = 300$ fb$^{-1}$. The results for ILC [43] assume an integrated luminosity of $\mathcal{L} = 500$ fb$^{-1}$ at $\sqrt{s} = 500$ GeV and 80% electron and 30% positron polarization.

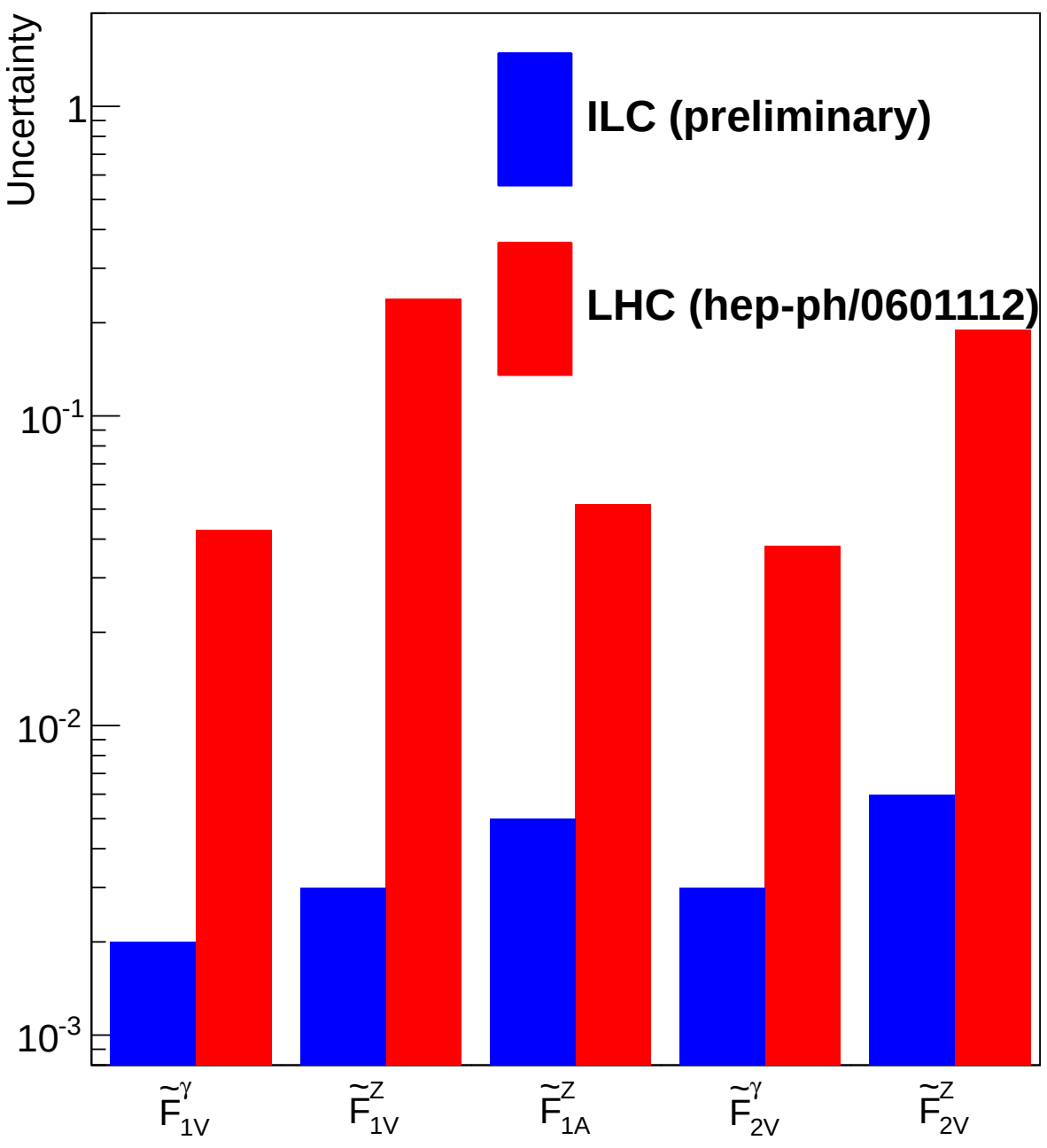

**Table 5.1.** Sensitivities achievable at 68% CL for the CP-conserving top quark form factors $\widetilde{F}^X_{1V,A}$ and $\widetilde{F}^X_{2V}$ defined in (5.5), at LHC and at the ILC. The assumed luminosity samples and, for ILC, beam polarization, are indicated. In the LHC studies and in the study [48], only one form factor at a time is allowed to deviate from its SM value. In the study [43], the form factors are allowed to vary independently.

| Coupling | LHC [16] $\mathcal{L} = 300$ fb$^{-1}$ | $e^+e^-$ [48] $P_{e^-} = \pm 0.8$ | $e^+e^-$ [43] $\mathcal{L} = 500$ fb$^{-1}$, $P_{e^-,+} = \pm 0.8, \mp 0.3$ |
|---|---|---|---|
| $\Delta\widetilde{F}^{\gamma}_{1V}$ | $^{+0.043}_{-0.041}$ | $^{+0.047}_{-0.047}$ , $\mathcal{L} = 200$ fb$^{-1}$ | $^{+0.002}_{-0.002}$ |
| $\Delta\widetilde{F}^{Z}_{1V}$ | $^{+0.24}_{-0.62}$ | $^{+0.012}_{-0.012}$ , $\mathcal{L} = 200$ fb$^{-1}$ | $^{+0.003}_{-0.003}$ |
| $\Delta\widetilde{F}^{Z}_{1A}$ | $^{+0.052}_{-0.060}$ | $^{+0.013}_{-0.013}$ , $\mathcal{L} = 100$ fb$^{-1}$ | $^{+0.005}_{-0.005}$ |
| $\Delta\widetilde{F}^{\gamma}_{2V}$ | $^{+0.038}_{-0.035}$ | $^{+0.038}_{-0.038}$ , $\mathcal{L} = 200$ fb$^{-1}$ | $^{+0.003}_{-0.003}$ |
| $\Delta\widetilde{F}^{Z}_{2V}$ | $^{+0.27}_{-0.19}$ | $^{+0.009}_{-0.009}$ , $\mathcal{L} = 200$ fb$^{-1}$ | $^{+0.006}_{-0.006}$ |

### 5.3.3 An example: the Randall-Sundrum scenario

The sensitivity of the top quark couplings to new physics can be paramerised by general dimension six operators contributing to the $t\bar{t}\gamma$ and $t\bar{t}Z$ vertex [50]. However, the potential of the ILC might be demonstrated more clearly by presenting a concrete example with one particular model. In the original model of Randall and Sundrum [35] there are additional massive gauge bosons in an assumed extra dimension. The model predicts increased couplings of the top quark, and perhaps also the $b$ quark, to these Kaluza Klein particles. Following the analysis in [37, 51], one can fix the parameters of the model so that these enhancements fit the two anomalies observed in the forward-backward asymmetry for $b$ quarks $A_{FB,b}$ at LEP1 and for top quarks $A_{FB,t}$ at the Tevatron. This gives a viable model of top quark interactions associated with top and Higgs compositeness. Figure 5.7 shows the expected modifications of the helicity angle distributions within this scenario.

Both the slopes and total cross sections are deeply modified in this scenario for the two polarizations. As explained previously, these observables are directly measured at the ILC, and the measurements allow one to fully disentangle the individual modifications of the $Z$ and photon couplings to top quarks. It can also be shown that by running at two energies, for instance 500 GeV and 1 TeV, one can fully extract the parameters of the model, for instance, the Kaluza Klein boson masses, which can be measured with about 1% precision.

**Table 5.2.** Sensitivities achievable at $68.3\%$ CL for the top quark CP-violating magnetic and electric dipole form factors $\widetilde{F}_{2A}^{X}$ defined in (5.5), at the LHC and at linear $e^+e^-$ colliders as published in the TESLA TDR. The assumed luminosity samples and, for TESLA, the beam polarization, are indicated. In the LHC studies and in the TESLA studies, only one form factor at a time is allowed to deviate from its SM value.

| Coupling | LHC [16] | $e^+e^-$ [49] |
|---|---|---|
| | $\mathcal{L} = 300$ fb$^{-1}$ | $\mathcal{L} = 300$ fb$^{-1}$, $P_{e^-,+} = -0.8$ |
| $\Delta\mathrm{Re}\, \widetilde{F}_{2A}^{\gamma}$ | $^{+0.17}_{-0.17}$ | $^{+0.007}_{-0.007}$ |
| $\Delta\mathrm{Re}\, \widetilde{F}_{2A}^{Z}$ | $^{+0.35}_{-0.35}$ | $^{+0.008}_{-0.008}$ |
| $\Delta\mathrm{Im}\, \widetilde{F}_{2A}^{\gamma}$ | $^{+0.17}_{-0.17}$ | $^{+0.008}_{-0.008}$ |
| $\Delta\mathrm{Im}\, \widetilde{F}_{2A}^{Z}$ | $^{+0.035}_{-0.035}$ | $^{+0.015}_{-0.015}$ |

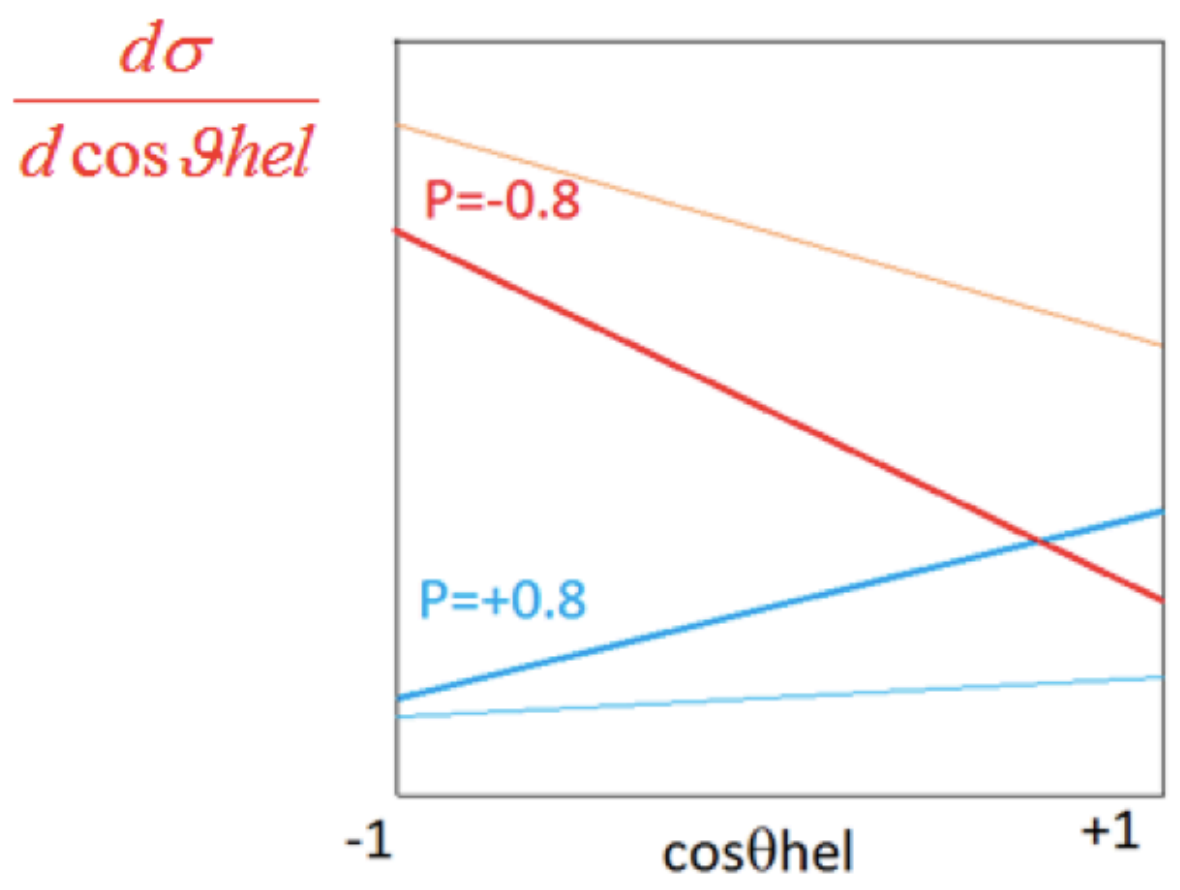

**Figure 5.7** Distributions of the helicity angle $\cos\theta_{hel}$ expected from the Standard Model (thick lines) and their modifications by the Randall-Sundrum model presented in [51]. The results are shown for electron and positron beam polarization equal to -80%/+30% and +80%/-30%.

When the Kaluza Klein particles become very heavy, ILC at 500 GeV can observe deviations in top couplings at greater than 3 $\sigma$ for masses which, depending on the details of the model, typically range between 4 and 48 TeV

## 5.3.4 Remarks on $(g-2)_t$

The determination of $\widetilde{F}_{2V}^{\gamma}$ gives access to the anomalous magnetic moment $(g-2)_t$ through the relation $\widetilde{F}_{2V}^{\gamma} = Q_t(g-2)_t/2$. The top quark $(g-2)$ receives Standard Model contributions from QED, QCD and EW [52]. One sees that this quantity will be measured to about 0.1% accuracy.

What is known about $(g-2)_t$ ? Limits from the reaction $b \to s\gamma$ giving a very crude constraint [53] :

$$-3.5 < g_t < 3.6 \tag{5.8}$$

In [54], it is argued that $(g-2)_t$ is a very sensitive measurement for compositeness. For leptons, this measure constrains compositeness at the 10000 TeV ( $10^{-18}\,\mathrm{cm}$) level. In other words, $e$ and $\mu$ are almost precisely elementary objects. But this need not be true for the top quark. The value of $(g-2)_t/2$ is proportional to $m_t/M$, where $M$ is the scale of top quark compositeness. It follows that, with the 0.1% accuracy expected at the ILC, the compositeness of the top quark can be tested up to about 100 TeV.

## 5.4 Concluding remarks

The top quark could be a window to new physics associated with light composite Higgs bosons and strong coupling in the Higgs sector. The key parameters here are the electroweak couplings of the top quark. We have demonstrated that the ILC offers unique capabilities to access these couplings and measure them to the required high level of precision. The mass of the top quark, which is a most important quantitiy in many theories can be measured in a model independent fashion to a precision of better than 100 MeV. It has however to be pointed out that all of these precision measurements require a superb detector performance and event reconstruction. The key requirements are the tagging of final state $b$ quarks with and efficiency and purity of better than 90% and jet energy reconstruction using particle flow of about 4% in the entire accessible energy range. These requirements are met for the ILC detectors described in the Volume 4 of this report.

# Chapter 6
# Extended Higgs Sectors

The Higgs sector in the Standard Model is of the simplest and most minimal form, containing one isospin doublet of scalar fields and one physical particle, the Higgs boson [1]. In Chapter 2, we have described the phenomenology of this minimal Higgs boson in some detail. However, it must always be kept in mind that the minimal model might not be the correct one. There is no principle that requires the Higgs sector to be of the minimal form. There are many possibilities for extension of the Higgs sector, corresponding to adding further multiplets of scalar fields, which might be singlets, doublets, or higher representations of $SU(2) \times U(1)$.

In fact, many new physics models, proposed to solve problems with the Standard Model or provide missing elements such as dark matter, naturally contain extended Higgs sectors. Among the models proposed to solve the gauge hierarchy problem and provide mechanism for electroweak symmetry breaking are supersymmetry, Little Higgs models, and models such as Gauge-Higgs unification that require new dimensions of space. Each of these models predicts a light Higgs boson similar to the Higgs boson of the Standard Model. In each case, however, this boson is a part of a larger Higgs sector with multiple scalar fields and, in the three cases, the details of the extension are different. Extended Higgs sectors are also introduced to build models for specific phenomena that cannot be explained in the SM, such as baryogenesis, dark matter, and neutrino masses.

Extended Higgs sectors can be searched for at hadron colliders, but often they are difficult to find. Higgs bosons have subdominant, electroweak-scale production cross sections. Their most prominent decay modes can be mimicked by background reactions from top and bottom quarks and other sources. At an $e^+e^-$ collider, on the other hand, extended Higgs bosons have pair-production cross sections that are as substantial as those for other particles with electroweak charges. The comprehensive search for extended Higgs bosons and the precision measurement of the properties of all accessible Higgs particles is thus an important goal for the ILC.

In Section 6.1 below, we give an orientation for models with extended Higgs sectors, defining the sometimes complex notation and clarifying the spectrum of physical Higgs states in various scenarios. In Section 6.2, we summarize the current constraints on these extended Higgs sectors, and the direct searches for extended Higgs bosons that can be carried out at the ILC. In Section 6.3, we discuss the ILC phenomenology of various exotic scenarios for neutrino mass, baryogenesis and dark matter which are relevant to extended Higgs sectors. Conclusions are given in Section. 6.4.

## 6.1 General description of extended Higgs sectors

The simplest examples of an extended Higgs sector are built by the addition of one $SU(2) \times U(1)$ singlet or one additional $SU(2) \times U(1)$ doublet scalar field. The case of an additional doublet is especially important. Supersymmetry requires distinct Higgs doublets to give mass to the $u$- and $d$-type quarks, and so the Minimal Supersymmetric Standard Model (MSSM) contains an extended Higgs sector [2]. In this section, we will describe the structure of these and more complicated Higgs sectors and define the parameters needed for a discussion of the phenomenology of these models.

### 6.1.1 The Two Higgs Doublet Model

The Two Higgs Doublet Model (THDM) includes two $SU(2) \times U(1)$ scalar doublets with $Y = 1$ [3]. The Higgs doublets can be parameterized as

$$\Phi_i = \left[ \begin{array}{c} w_i^+ \\ \frac{1}{\sqrt{2}}(v_i + h_i + i z_i) \end{array} \right], \quad (i = 1, 2). \tag{6.1}$$

The most general Higgs potential is parametrized by three mass parameters and 7 independent quartic coupling constants.

$$\begin{aligned} V = & m_1^2|\Phi_1|^2 + m_2^2|\Phi_2|^2 - (m_3^2\Phi_1^\dagger\Phi_2 + h.c.) + \tfrac{1}{2}\lambda_1|\Phi_1|^4 + \tfrac{1}{2}\lambda_2|\Phi_2|^4 + \lambda_3|\Phi_1|^2|\Phi_2|^2 \\ & + \lambda_4|\Phi_1^\dagger\Phi_2|^2 + \tfrac{1}{2}[\lambda_5(\Phi_1^\dagger\Phi_2)^2 + \lambda_6|\Phi_1|^2\Phi_1^\dagger\Phi_2 + \lambda_7|\Phi_2|^2\Phi_1^\dagger\Phi_2 + h.c.]. \end{aligned} \tag{6.2}$$

The Higgs potential in the MSSM is a special case of this potential in which the quartic couplings are related to the $SU(2)$ and $U(1)$ gauge couplings by supersymmetry. The model contains 3 degrees of freedom that are eaten by the $W^\pm$ and $Z^0$ when their masses are generated through the Higgs mechanism. This leaves over 5 physical Higgs bosons, two CP-even scalars $h$ and $H$, one CP-odd scalar $A$, and one pair of charged scalars $H^\pm$. The mass eigenstates are related to the fields in (6.1) by mixng angles $\alpha$ and $\beta$ according to

$$\begin{aligned} h &= -h_1 \sin\alpha + h_2 \cos\alpha, & H &= h_1 \cos\alpha + h_2 \sin\alpha \\ H^\pm &= w_1^\pm \cos\beta + w_2^\pm \cos\alpha, & A &= z_1 \cos\beta + z_2 \sin\beta, \end{aligned} \tag{6.3}$$

We define $h$ to be the lighter CP-even boson. The angle $\beta$ yields the parameter $\tan\beta = v_2/v_1$.

The two vacuum expectation values $v_1$ and $v_2$ satisfy

$$v_1^2 + v_2^2 = v^2 = (246 \text{ GeV})^2 \ . \tag{6.4}$$

The vector boson coupling constants for the lighter Higgs boson, $hZZ$ and $hWW$, are given by that of the SM Higgs boson times $\sin(\beta - \alpha)$, while those for $HZZ$ and $HWW$ are proportional to $\cos(\beta - \alpha)$. The scalars $h$ and $H$ thus share the Higgs field vacuum expectation value and share the strength of the coupling of $WW$ and $ZZ$ to scalar fields. The trilinear couplings $H^\pm W^\mp Z$, $H^\pm W^\mp \gamma$, $AW^+W^-$, $AZZ$ are zero at tree level.

Of the two mass parameters in (6.2), $m_1$ and $m_2$ are directly related to $v_1$ and $v_2$. The third parameter $m_3$ does not drive electroweak symmetry breaking and can potentially be much larger. When

$$M^2 \equiv m_3^2/\sin\beta\cos\beta \gg v^2 \ , \tag{6.5}$$

then we approach to the *decoupling limit* where the masses of the added scalar states $H$, $A$, and $H^\pm$

**Table 6.1.** Four possible $Z_2$ charge assignments that forbid dangerous flavor-changing neutral current effects in the THDM. [5].

| | $\Phi_1$ | $\Phi_2$ | $u_R$ | $d_R$ | $\ell_R$ | $Q_L$, $L_L$ |
|---|---|---|---|---|---|---|
| Type I | + | − | − | − | − | + |
| Type II (MSSM like) | + | − | − | + | + | + |
| Type X (lepton specific) | + | − | − | − | + | + |
| Type Y (flipped) | + | − | − | + | − | + |

**Table 6.2.** The mixing factors $\xi_X^f$ in the THDM Higgs interactions given in (6.7) [6].

| | $\xi_h^u$ | $\xi_h^d$ | $\xi_h^\ell$ | $\xi_H^u$ | $\xi_H^d$ | $\xi_H^\ell$ | $\xi_A^u$ | $\xi_A^d$ | $\xi_A^\ell$ |
|---|---|---|---|---|---|---|---|---|---|
| Type I | $\frac{\cos\alpha}{\sin\beta}$ | $\frac{\cos\alpha}{\sin\beta}$ | $\frac{\cos\alpha}{\sin\beta}$ | $\frac{\sin\alpha}{\sin\beta}$ | $\frac{\sin\alpha}{\sin\beta}$ | $\frac{\sin\alpha}{\sin\beta}$ | $-\cot\beta$ | $\cot\beta$ | $\cot\beta$ |
| Type II | $\frac{\cos\alpha}{\sin\beta}$ | $-\frac{\sin\alpha}{\cos\beta}$ | $-\frac{\sin\alpha}{\cos\beta}$ | $\frac{\sin\alpha}{\sin\beta}$ | $\frac{\cos\alpha}{\cos\beta}$ | $\frac{\cos\alpha}{\cos\beta}$ | $-\cot\beta$ | $-\tan\beta$ | $-\tan\beta$ |
| Type X | $\frac{\cos\alpha}{\sin\beta}$ | $\frac{\cos\alpha}{\sin\beta}$ | $-\frac{\sin\alpha}{\cos\beta}$ | $\frac{\sin\alpha}{\sin\beta}$ | $\frac{\sin\alpha}{\sin\beta}$ | $\frac{\cos\alpha}{\cos\beta}$ | $-\cot\beta$ | $\cot\beta$ | $-\tan\beta$ |
| Type Y | $\frac{\cos\alpha}{\sin\beta}$ | $-\frac{\sin\alpha}{\cos\beta}$ | $\frac{\cos\alpha}{\sin\beta}$ | $\frac{\sin\alpha}{\sin\beta}$ | $\frac{\cos\alpha}{\cos\beta}$ | $\frac{\sin\alpha}{\sin\beta}$ | $-\cot\beta$ | $-\tan\beta$ | $\cot\beta$ |

become much larger than the mass of $h$:

$$m_h^2 \simeq \lambda_i v^2 \text{ (SM like)}, \qquad m_\phi^2 \sim \lambda_i v^2 + M^2, \text{where } \phi = H, A, \text{and } H^\pm, \tag{6.6}$$

with $\sin(\beta-\alpha) \simeq 1$ [4] . In this case, the phenomenology of $h$ is similar to that of the SM Higgs boson except for small deviations in the Higgs boson couplings. However, it is not necessary that the additional bosons be heavy, and, in this case, there is room for substantial mixing between $h$ and $H$.

In the THDM, both of the doublets can in principle couple to fermions, and this can lead to dangerous flavor-changing neutral current couplings. A well-known way to suppress these couplings is to impose a softly broken $Z_2$ symmetry so that only one of the two Higgs doublets gives mass to the $u$-type quarks, to the $d$-type quarks, and to the leptons. The various possible assignments lead to four distinct models, displayed in Table 6.1 [5–7]. In the MSSM, supersymmetry requires the Type II assignment, with one doublet giving mass to the $u$ quarks and the other to the $d$ quarks and the charged leptons. In more general models, though, all four possibilities are open. The Yukawa interactions for these models are expressed as

$$\begin{aligned}\mathcal{L}_{THDM}^Y = \quad & -\textstyle\sum_{f=u,d,e}\left(\frac{m_f}{v}\xi_h^f \overline{f} f h + \frac{m_f}{v}\xi_H^f \overline{f} f H + i\frac{m_f}{v}\xi_A^f \overline{f}\gamma_5 f A\right) \\ & - \left[\sqrt{2}V_{ud}\overline{u}\left(\frac{m_u}{v}\xi_A^u P_L + \frac{m_d}{v}\xi_A^d P_R\right) d\, H^+ + \frac{\sqrt{2}m_\ell \xi_A^\ell}{v}\overline{\nu_L} e_R H^+ + h.c.\right],\end{aligned} \tag{6.7}$$

where $P_{L/R}$ are projection operators for left-/right-handed fermions, and the factors $\xi_\varphi^f$ are listed in Table 6.2.

The decays of the Higgs bosons in the THDM depend on the model chosen for the Yukawa interactions. When $\sin(\beta-\alpha) = 1$ [4], the decay pattern of $h$ is almost the same as that in the Standard Model. However, the decay patterns of $H$, $A$, and $H^\pm$ can vary over a large range. Figure 6.1 shows the decay branching ratios of $H$, $A$ and $H^\pm$ as a function of $\tan\beta$ for the four models, for boson masses of 150 GeV and $\sin(\beta-\alpha) = 1$. The decay pattern of $H$ is typically similar to that of $A$, but with some important exceptions. In the type I THDM, all fermionic decays, and the $gg$ decay mode, are suppressed at large $\tan\beta$. However, $H$, but not $A$, couples to $H^+H^-$, and this allows for $H$ a significant decay through a scalar loop to $\gamma\gamma$.

In general, the complexity of the $H$, $A$, $H^\pm$ decay schemes and in the four possible models make it difficult to determine the underlying model unless these bosons are created through a simple and well-characterized pair-production reaction. Thus, even if these bosons are discovered at the LHC, it will be important to study them in $e^+e^-$ pair-production at the ILC.

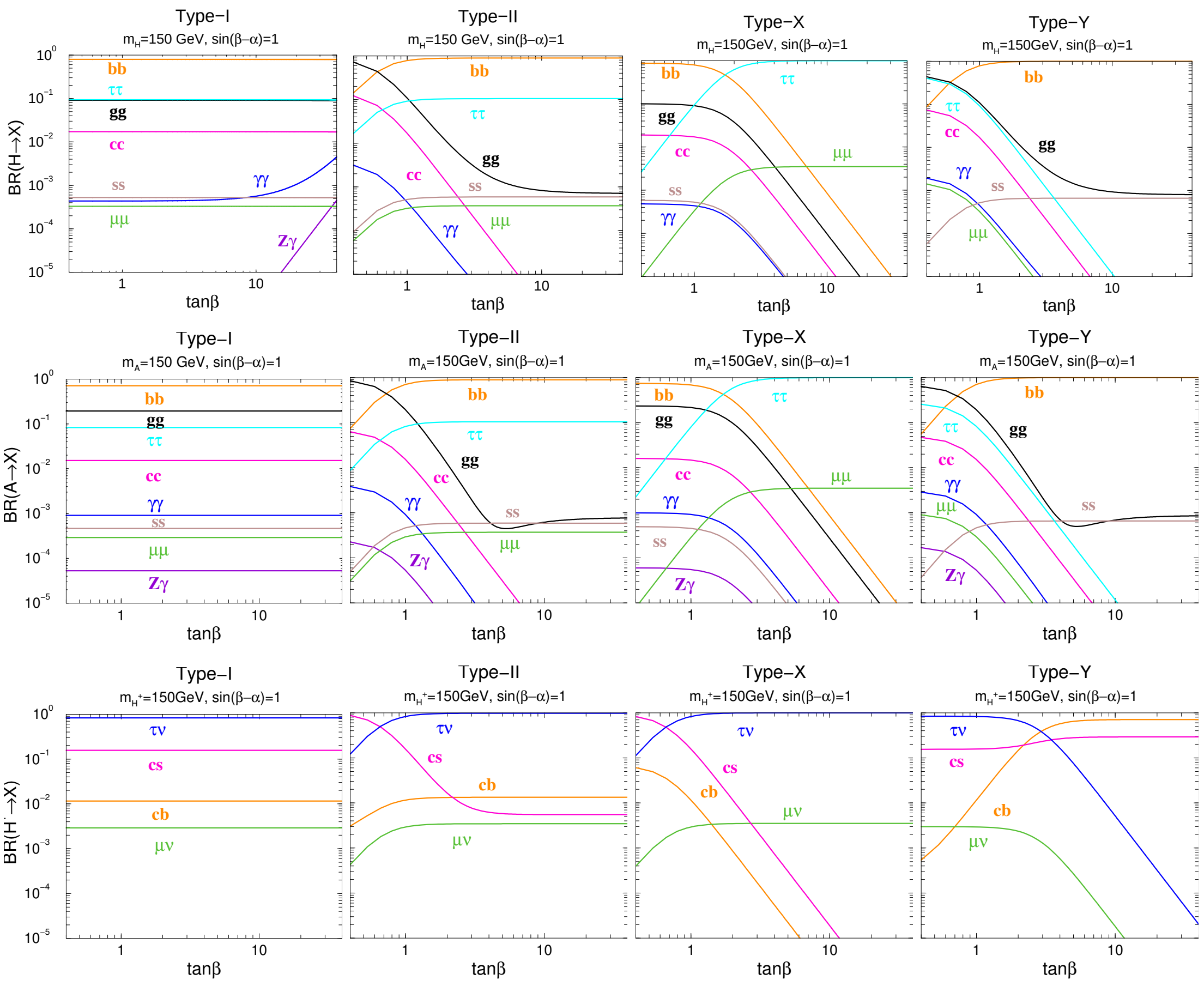

**Figure 6.1.** Decay branching ratios of $H$, $A$ and $H^{\pm}$ in the four different types of THDM as a function of $\tan\beta$ for $m_H = m_A = m_{H^{\pm}} = 150$ GeV. The SM-like limit $\sin(\beta-\alpha) = 1$ is taken.

## 6.1.2 Models with Higgs singlets

Another simple extension of the SM Higgs sector is the addition of a singlet scalar field $S$ with $Y = 0$. Such a singlet field is introduced in new physics models with an extra $U(1)$ gauge symmetry [8], for example, a theory with a $U(1)$ boson coupling to $B - L$ [9]. A neutral singlet scalar field is also introduced in the Next-to-Minimal SUSY Standard Model (NMSSM), along with the second Higgs doublet required in SUSY [10]. Singlet Higgs fields do not couple directly to quarks, leptons or gauge bosons of the SM.

In the model with only one additional neutral singlet scalar field to the SM, we parameterize the SM doublet $\Phi$ and $S$ as

$$\Phi = \begin{bmatrix} \varphi^+ \\ \frac{1}{\sqrt{2}}(v + \varphi + i\chi) \end{bmatrix}, \qquad S = \frac{1}{\sqrt{2}}(v_S + \varphi_S + i\chi_S), \tag{6.8}$$

where $v =$ 246 GeV, and $v_S$ is the vacuum expectation value of the singlet. The two CP-even mass eigenstates $h$ and $H$ are mixtures of $\varphi$ and $\varphi_S$,

$$h = \varphi\cos\theta - \varphi_S\sin\theta, \qquad H = \varphi\sin\theta + \varphi_S\cos\theta. \tag{6.9}$$

In models with an extra $U(1)$ gauge boson, this boson absorbs the CP-odd component field $\chi_S$. Then the difference from the SM is just one additional CP-even scalar boson $H$. Models with only added

Higgs singlets contain no physical charged Higgs bosons. All of the SM fields obtain mass from the VEV of the doublet $v$. Their coupling constants with $h$ and $H$ are obtained by the replacement $\phi_{\rm SM} \to h\cos\theta + H\sin\theta$.

In the decoupling regime $\theta \sim 0$. Then $h$ is SM-like with couplings reduced from their SM values by $\cos\theta \approx 1 - \theta^2/2$. On the other hand, when $\tan\theta \sim \mathcal{O}(1)$, both the $h$ and $H$ behave as SM-like Higgs bosons, sharing the SM couplings to gauge bosons and fermions. If $h$ and $H$ are almost degenerate in mass, the two bosons might appear as a single SM Higgs boson in the LHC experiments. At the ILC, the tagging of the Higgs mass by the $Z$ energy in $e^+e^- \to Z + (h, H)$ could allow the two Higgs bosons to be better separated.

The reduced couplings of $h$ and $H$ result in smaller production cross sections as compared to the SM predictions. Therefore, the mass bounds from the collider experiment can be milder. For example, the LEP experiments exclude the $h$ only to about 110 GeV for $\sin\theta = 1/\sqrt{2}$ while the exclusion in the SM is about 114 GeV [11]. Basso, Moretti and Pruna have surveyed the ILC phenomenology of the Higgs sector in the minimal $B-L$ model [12].

### 6.1.3 Models with Higgs triplets

We can go on to consider models that add scalar fields in higher representations of $SU(2)$, models with fields with $I = 1, \frac{3}{2}, \ldots$. There are many such models. However, these models are constrained by the requirement that they do not give sizable tree level corrections to the Standard Model relation

$$\rho = \frac{m_W^2}{m_Z^2\cos^2\theta} = 1\ . \tag{6.10}$$

When electroweak radiative corrections are included, (6.10) is in excellent agreement with the data, so it is dangerous to add to the model with fields that can modify it. In a general $SU(2)\times U(1)$ model with $n$ scalar multiplets $\phi_i$ with isospin $T_i$ and hypercharge $Y_i$, the $\rho$ parameter is given at the tree level by

$$\rho = \frac{\sum_{i=1}^n [T_i(T_i+1) - \frac{1}{4}Y_i^2]v_i^2}{\sum_{i=1}^n \frac{1}{2}Y_i^2 v_i^2}, \tag{6.11}$$

where $v_i$ are vacuum expectation values of $\phi_i$. So, singlets and doublets with $Y_i = \pm\frac{1}{2}$ preserve $\rho = 1$, while adding higher representation generally modifies this relation, unless those fields have very small vacuum expectation values [13].

As an example of a model that adds an isospin triplet, we review the case of a Higgs representation with $I = 1$ and $Y = 2$. A vacuum expectation value of this field can produce a Majorana neutrino mass [14].

A model with this triplet field will contain a Higgs doublet $\Phi$ in addition to the triplet $\Delta$. The component fields are

$$\Phi = \begin{bmatrix} \varphi^+ \\ \frac{1}{\sqrt{2}}(v_\varphi + \varphi + i\chi) \end{bmatrix}, \qquad \Delta = \begin{bmatrix} \Delta^+/\sqrt{2} & \Delta^{++} \\ \frac{1}{\sqrt{2}}(v_\Delta + \delta + i\eta) & -\Delta^+/\sqrt{2} \end{bmatrix}, \tag{6.12}$$

where $v_\varphi$ and $v_\Delta$ are the vacuum expectation values. The physical scalar states are two CP-even bosons ($h$ and $H$), a CP-odd boson ($A$), singly charged pair ($H^\pm$), and a doubly charged pair ($H^{\pm\pm}$). These are related to the original component fields by mixing angles $\alpha$, $\beta_0$ and $\beta_\pm$,

$$\begin{aligned} &h = \varphi\cos\alpha + \delta\sin\alpha, \quad H = -\varphi\sin\alpha + \delta\cos\alpha, \\ &A = -\chi\sin\beta_0 + \eta\cos\beta_0, \quad H^\pm = -\varphi^\pm\sin\beta_\pm + \Delta^\pm\cos\beta_\pm, \quad H^{\pm\pm} = \Delta^{\pm\pm}. \end{aligned} \tag{6.13}$$

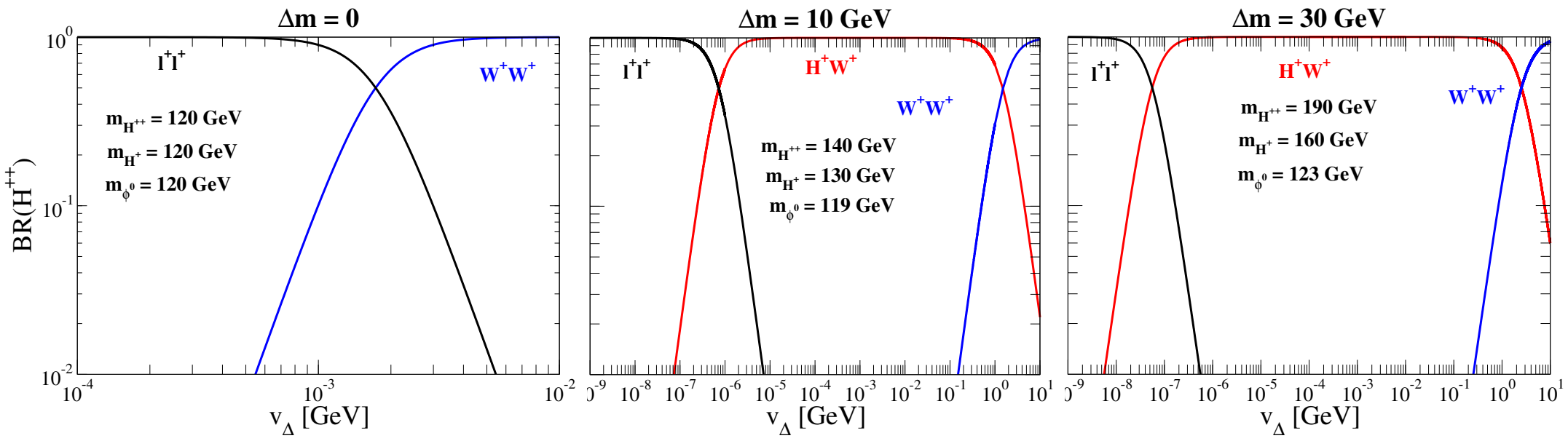

**Figure 6.2.** Decay branching ratio of $H^{++}$ as a function of $v_\Delta$. Left: $m_{H++}$ is set to be 120 GeV, $\Delta m = 0$. Center: $m_{H++}$ is 140 GeV, $\Delta m = 10$ GeV. Right: $m_{H++}$ is 190 GeV, $\Delta m = 30$ GeV.

We must arrange $v_\Delta \ll v_\varphi$ to preserve $\rho \simeq 1$. This constraint implies the mass relations

$$m_h^2 \simeq 2\lambda_1 v^2, \quad m_{H++}^2 - m_{H+}^2 \simeq m_{H+}^2 - m_A^2 \ , \quad \text{and} \quad m_H^2 \simeq m_A^2, \tag{6.14}$$

with $\alpha \ll 1$, $\beta_0 \ll 1$ and $\beta_\pm \ll 1$. Therefore, the model has a Standard Model-like Higgs boson $h$ and additional triplet-like scalar states whose masses become approximately equal in the decoupling limit.

The doubly charged Higgs bosons $H^{++}$ are the most characteristic feature of the model. The requirement that the vacuum expectation value of $\Delta$ gives a Majorana neutrino mass requires that this field must be assigned lepton number $L = 2$. Then, if the new Higgs bosons are degenerate, the dominant decays would be to lepton and neutrino pairs. In particular, $H^{++}$ would be expected to decay to $\ell^+\ell^+$. At the LHC, the search for $H^{\pm\pm}$ is underway using this decay mode. The exclusion of the signal implies a lower bound on the mass of $H^{++}$, $m_{H++} \gtrsim 400$ GeV [15], assuming a 100% branching ratio.

However, this analysis is correct only for a limited parameter region in which the vacuum expectation value of $\Delta$ is extremely small, $v_\Delta < 10^{-3}$ GeV. For larger, but still small, values of $v_\Delta$, a small mass splittings between $H^+$ and $H^{++}$ opens up that allows the decay to take advantage of the much larger coupling to $H^+W^+$ [16]. In Fig. 6.2, the decay branching ratios for $H^{\pm\pm}$ are shown as a function of $v_\Delta$ [17]. For $v_\Delta \sim 1$ GeV, corresponding to mass difference $\Delta m \sim 10$ GeV, the decay into $H^+W^+$ is dominant for a wide range of $v_\Delta$ when $m_{H++} > m_{H+} > m_{A,H}$. In this case, $H^{++}$ could be identified through its cascade decay. It is also possible to realize the opposite sign of the mass difference. In this case, the $H^{++}$ decays into $W^+W^+$.

This model gives another illustration that the properties of an extended Higgs boson can be highly sensitive to the parameter choices. In the most favorable cases, discovery is straightforward; other parameter choices, which might be equally or more likely, are more challenging. To work backwards from the data to the underlying parameters, we require a well-understood production mechanism and broad sensitivity to a wide range of final states.

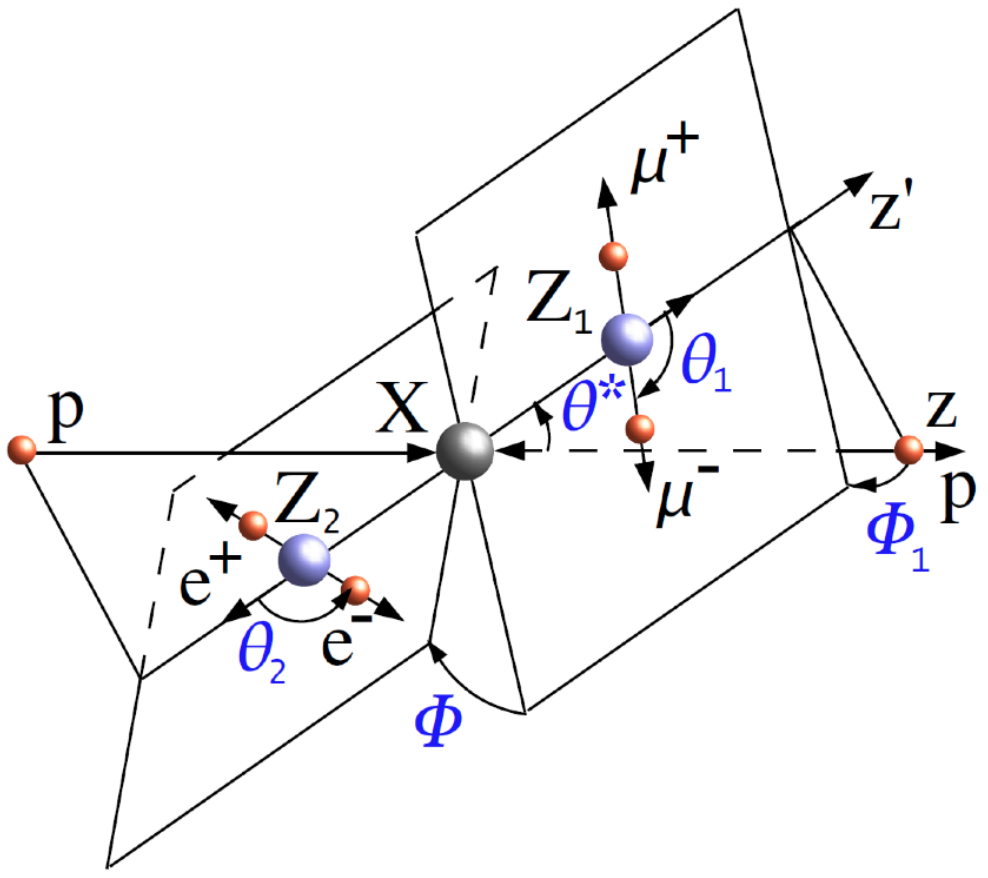

**Figure 6.3**
Schema of the angular analysis of the the Higgs decay into a pair of Z bosons that decay then into 4 leptons, as used by the CMS experiment [20].

## 6.2 Extended Higgs bosons searches at the ILC

The discovery of additional Higgs bosons such as $H$, $A$, $H^{\pm}$ and $H^{\pm\pm}$ would give direct evidence for extended Higgs sector. As already discussed, there are many possibilities for the decay branching ratios of these particles, illustrated by the various schemes presented in Section 6.2. The ongoing searches at LHC rely on specific production and decay mechanisms that occupy only a part of the complete model parameter space. At the ILC, the extended Higgs bosons are produced in electroweak pair production through cross sections that depend only on the $SU(2) \times U(1)$ quantum numbers and the mixing angles. Thus, the reach of the ILC is typically limited to masses less than $\sqrt{s}/2$, but it is otherwise almost uniform over the parameter space.

### 6.2.1 Constraints from the LHC experiments

The LHC is imposing several types of constraints in the exploration of the Higgs sector, but certainly the main constraint comes from the discovery of the resonance at 125 GeV by ATLAS [18] and CMS [19]. The resonance appears with particular significance in the decay channels into two $\gamma$'s and two $Z^0$ bosons. The exact nature of this new resonance has still to be confirmed. However there are some indications that it could well be the light Higgs neutral boson we have been so long looking for.

CMS has already performed an angular analysis of the channel $pp \to ZZ \to 4$ charged leptons (see Fig. 6.3). This analysis can potentially discriminate between a boson that decays mainly to longitudinally polarized $Z$ bosons, as expected if the boson is a scalar field with a vacuum expectation value, and a boson that decays only to transversely polarized $Z$ bosons, as expected for a $0^-$ boson and for other non-Higgs hypotheses. At present, the CMS analysis favors the $0^+$ SM hypothesis over the $0^-$ hypothesis by 2.5 $\sigma$ [20]. This gives hope that, with the full 2012 data set, we might have strong evidence that the resonance is a "Higgs boson".

In the context of extended Higgs models, this resonance might be interpreted as the $h$ or the $H$, or, if these bosons are within a few GeV of one another, both [21, 22]. The discrimination of these possibilities from the Standard Model will require much better measurements of the relative rates and, eventually, absolute branching ratios, into $\gamma\gamma$, $WW$, $ZZ$, $b\bar{b}$ and $\tau^+\tau^-$. The decay mode into $\tau$ lepton pairs, in particular, is quite important for many BSM cases [23, 24]. The current situation is consistent with the Standard Model, but the errors leave much room for other possibilities. The resolution of these questions will probably need to wait for the 14 TeV era at the LHC, or for measurements of even higher precision.

Beyond the search channels for a Standard Model Higgs boson, the LHC experiments are exploring additional channels that are specific to extended Higgs bosons. ATLAS and CMS have already performed a number of extended Higgs searches. The published results are only based on the 2011 data. Much more will become soon available by adding the first $5fb^{-1}$ data that are already recorded in 2012. The experiments have scanned a mass range up to 350-400 GeV/$c^2$ in a variety of interesting processes and BSM scenarios. There is presently no evidence for such new BSM heavy Higgs signals. The current results from the charged Higgs searches at hadron colliders are reported in section 6.2.3.

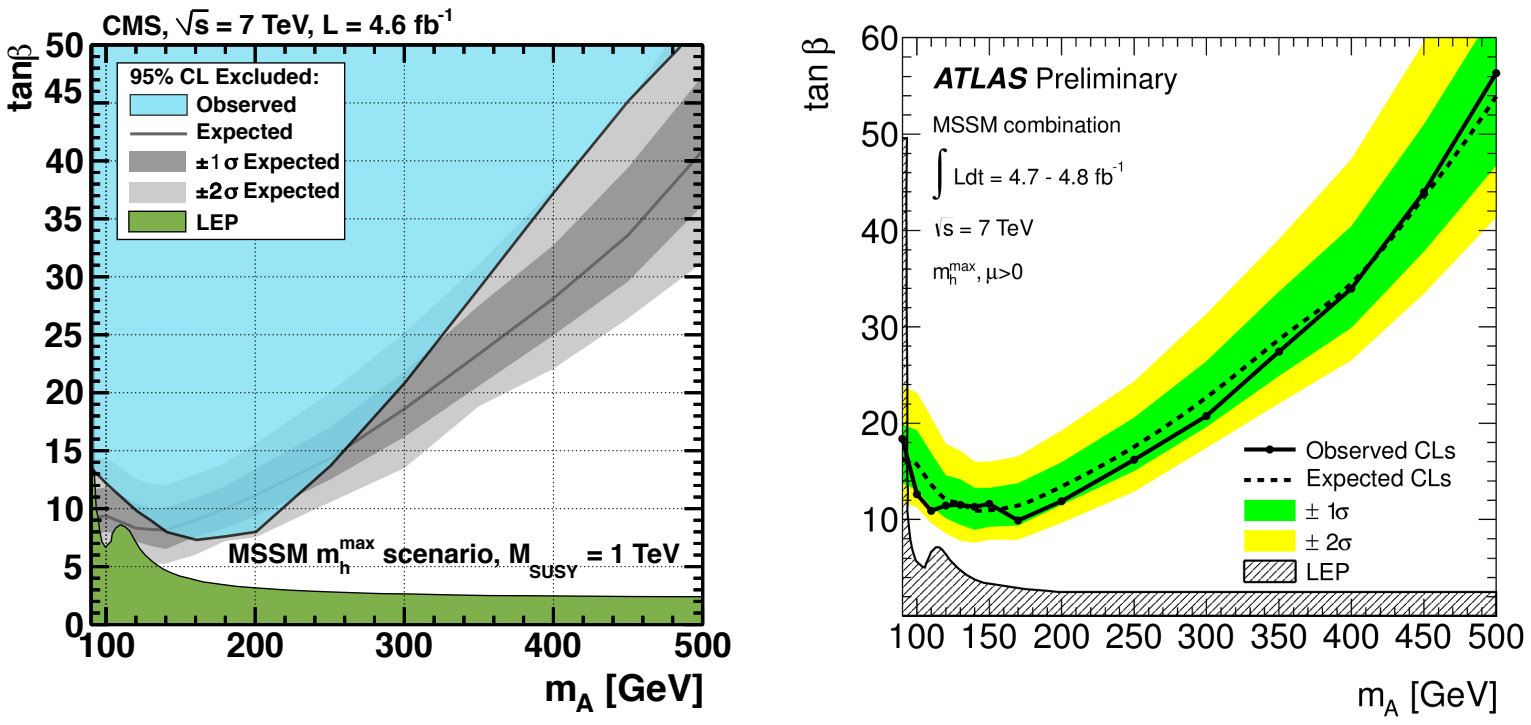

**Figure 6.4**
Limits on the signature with of extended Higgs particles decaying to two $\tau$ leptons, obtained by scanning $\tan\beta$ for each $M_A$ mass hypothesis and taking into account the dependence of $M_h$ and $M_H$ on $\tan\beta$, from Left: from CMS. Right: from ATLAS.

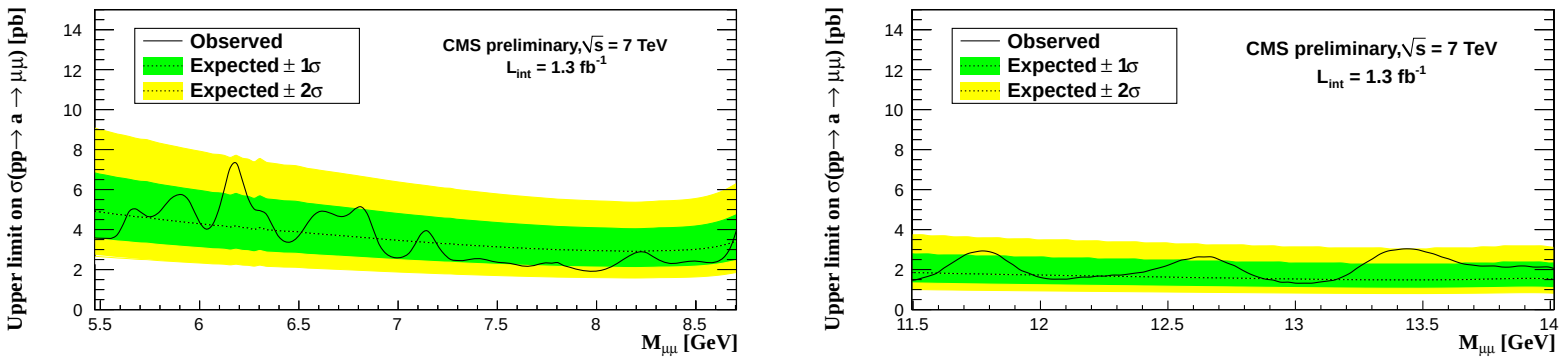

**Figure 6.5**
CMS search for a low mass Higgs decaying into two muons in a NMSSM scenario with the first 1.3 fb$^{-1}$ of data, from [25].

In the context of MSSM, the neutral Higgs bosons $h$, $H$ and $A$ are searched for in their decay into two $b$ quarks, two muons or two $\tau$ leptons. Doubly charged Higgs boson and Higgs boson in the SM reinterpreted with 4th generation of fermions are also investigated. The resonance at 126 GeV decaying into 2 photons is further reinterpreted in terms of a fermiophobic Higgs scenario. Some of the main present results at LHC on these searches are shown in Fig. 6.4. No significant excess is observed, and limits are set as low as $\tan\beta$ equal to 10. This is already a dramatic improvement compared to the Tevatron results.

The Next-to-Minimal Supersymmetric Standard Model (NMSSM) gives the possibility of a very light CP odd scalar boson that would decay to two muons. Both ATLAS and CMS have searched for a light extended Higgs boson of this type, but so far no signficant excess has been found. Figure 6.5 shows the results obtained by CMS based on 1.3 fb$^{-1}$ of data taken in 2011 [25]. This study demonstrates the potential of the LHC detectors to look for relatively low mass bosons produced as the result of high energy processes.

Other important constraints on extended Higgs bosons come from heavy flavor experiments, notably, meaurements of $b \to s\gamma$ and the process $B_s \to \mu^+\mu^-$ recently observed by LHCb [26]. Unfortunately, though deviations from the Standard Model predictions can clearly indicate a need for new physics, consistency of the Standard Model can result from cancellations among different contributions to loop-induced processes.

These examples, taken from the current early stages of the LHC program, demonstrate the great power that will eventually be available from the LHC in exploring for specific, even quite subtle, signatures of extended Higgs particles. We have argued, though, that this capability needs to be complemented by a broad program of searches based on a precisely understood production mechanism. We wil now describe how that such a program can be carried out at the ILC.

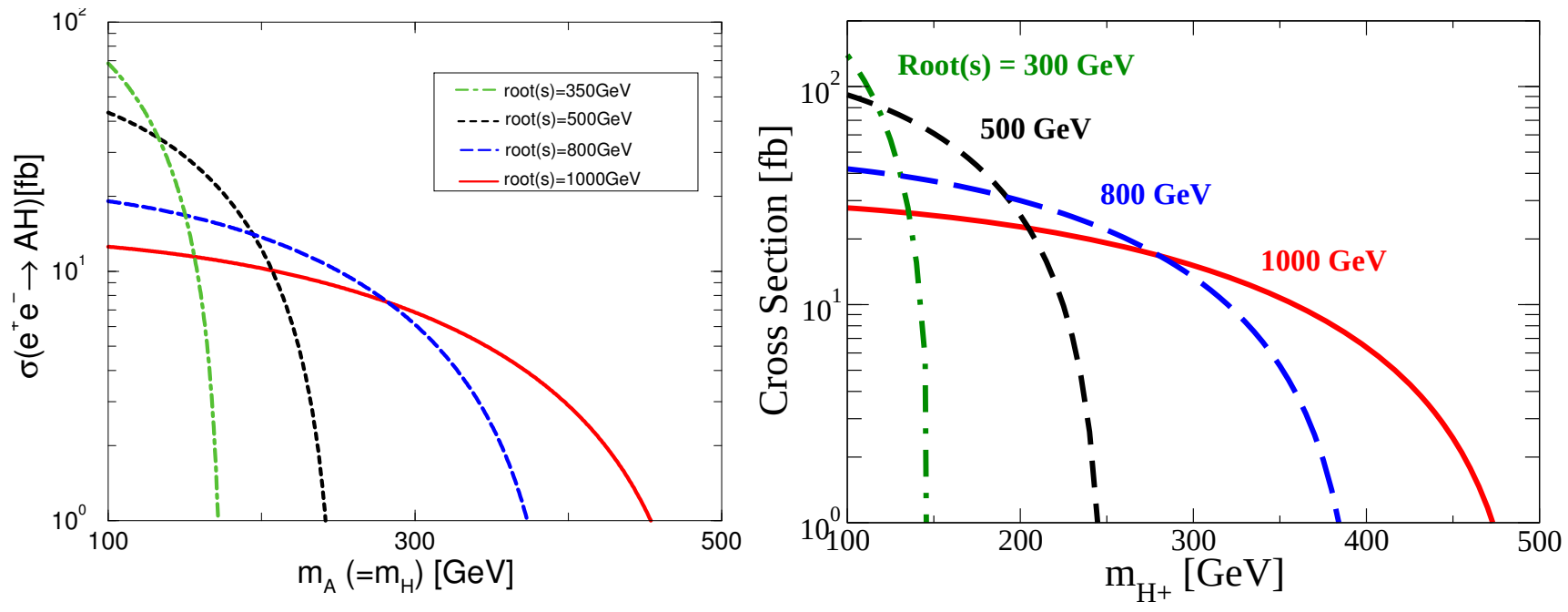

**Figure 6.6** Extended Higgs boson production cross sections as a function of the produced boson mass, at CM energies of 350, 500, 800, and 1000 GeV. Left: $e^+e^- \to AH$. Right: $e^+e^- \to H^+H^-$.

## 6.2.2 Neutral Higgs pair production at ILC

At the ILC, the pair production of extended Higgs bosons $e^+e^- \to AH$ in the THDM case, depends only on the boson masses in the decoupling limit. The production cross sections are shown in Fig. 6.6 for $\sqrt{s} = 350, 500, 800,$ and $1000$ GeV as a function of $m_A$ [27]. The decays of the extended Higgs state are mainly to fermion pairs. Thus, the observation of pair-produced Higgs bosons in various decay channels allows us to determine the type of Yukawa interaction, in the sense of Section 6.1.1, through the measurement of the corresponding branching ratios. For example, in the MSSM, which requires a Type II Higgs structure, the dominant final states for $HA$ production should be $bbbb$ and $bb\tau\tau$, while in the Type X (lepton specific) structure the dominant final state should be $\tau\tau\tau\tau$ for $\tan\beta > 2$. In a Type I Higgs model, the $bbjj$ final states signature is also important in addition to the $bbbb$ and $bb\tau\tau$ signatures, over a wide range of $\tan\beta$ values, while in Type Y (flipped) the $bbbb$ states dominate and the $bb\tau\tau$ and $bbjj$ states are suppressed for $\tan\beta > 2$.

The signals from $HA$ production in the $bbbb$ and $bb\tau\tau$ channels, in the context of the MSSM (Type-II THDM), was carried out in the studies of [28, 29]. A rather detailed detector simulation was performed in [29], including all the SM backgrounds at $\sqrt{s} = 500$, 800 and 1000 GeV. Using a kinematical fit which imposes energy momentum conservation and under the assumed experimental conditions, a statistical accuracy on the Higgs boson mass from 0.1 to 1 GeV is found to be achievable. The topological cross section of $e^+e^- \to HA \to bbbb$ ($e^+e^- \to HA \to \tau\tau bb$) could be determined with a relative precision of 1.5% to 7% (4% to 30%). The width of $H$ and $A$ could also be determined with an accuracy of 20% to 40%, depending on the mass of the Higgs bosons. Figure 6.7 shows, on the left, the $\tau^+\tau^-$ invariant mass obtained by a kinematic fit in $e^+e^- \to HA \to b\bar{b}\tau^+\tau^-$ for $m_A = 140$ GeV and $m_H = 150$ GeV, for $\sqrt{s} = 500$ GeV and 500 fb$^{-1}$ [29].

The $\tau^+\tau^-\tau^+\tau^-$ and $\mu^+\mu^-\tau^+\tau^-$ final states would be dominant for the type X (lepton specific) THDM. When $\sqrt{s} = 500$ GeV, assuming an integrated luminosity of $500$ fb$^{-1}$, one expects to collect 16,000 (18,000) $\tau^+\tau^-\tau^+\tau^-$ events in the type X (type II) THDM, and 110 (60) $\mu^+\mu^-\tau^+\tau^-$ events in the same models, assuming $m_H = m_A = m_{H^\pm} = 130$ GeV, $\sin(\beta - \alpha) = 1$ and $\tan\beta = 10$. These numbers do not change much for $\tan\beta \gtrsim 3$. It is important to recognize that the four-momenta of the $\tau$ leptons can be solved by a kinematic fit based on the known center of mass energy and momentum, by applying the collinear approximation to each set of $\tau$ lepton decay products [30, 31]. Figure 6.7 shows, on the right, the two dimensional invariant mass distribution of the $\tau$ lepton pairs from the neutral Higgs boson decays as obtained with a simulation at 500 GeV in which the masses of the neutral Higgs bosons are taken to be 130 GeV and 170 GeV [32].

Although the associated Higgs production process $e^+e^- \to HA$ is a promising one for testing the properties of the extended Higgs sectors, the kinematic reach is restricted by $m_H + m_A < \sqrt{s}$ and is not available beyond this limit. Above the threshold of the $HA$ production, the associated production processes $t\bar{t}\Phi$, $b\bar{b}\Phi$ and $\tau^+\tau^-\Phi$ ($\Phi = h, H, A$) could be used [33]. In particular, for $b\bar{b}\Phi$ and $\tau^+\tau^-\Phi$, the mass reach is extended almost up to the collision energy. The cross sections for these processes are proportional to the Yukawa interaction, so they directly depend on the type of Yukawa coupling in the THDM structure. In MSSM or the Type II THDM (Type I THDM), these processes are enhanced (suppressed) for large $\tan\beta$ values. In Type X THDM, only the $\tau^+\tau^-H/A$ channels could be significant while only $b\bar{b}H/A$ channels would be important in Type I and Type Y THDMs. These reactions can then be used to discriminate the type of the Yukawa interaction.

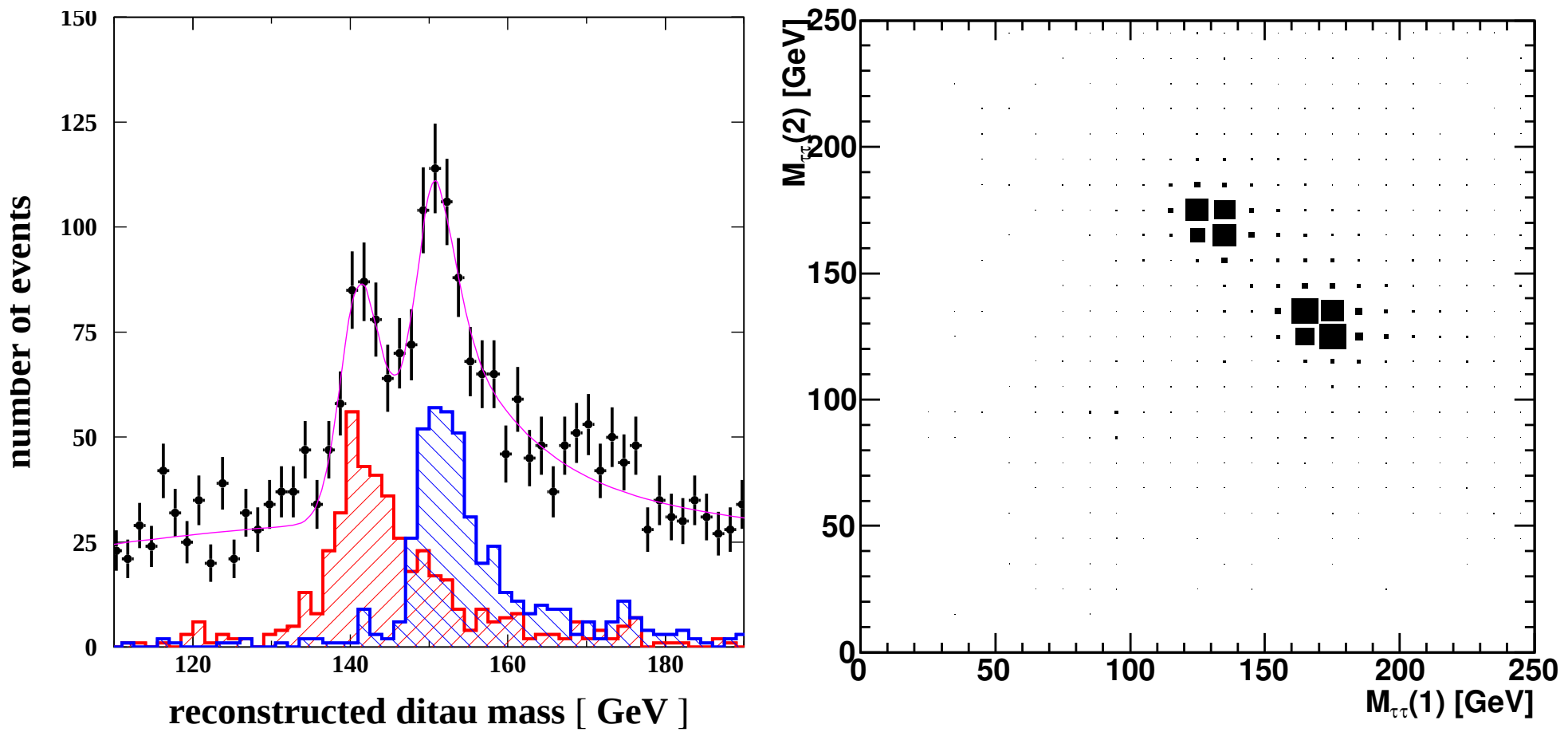

**Figure 6.7.** Left: Invariant mass reconstruction from the kinematical fit in the process $e^+e^- \to HA \to b\bar{b}\tau^+\tau^-$ in the Type-II (MSSM like) THDM for $m_A = 140$ GeV and $m_H = 150$ GeV at $\sqrt{s} = 500$ GeV and 500 fb$^{-1}$ [29] Right: Two dimensional distribution of ditau invariant mass in $e^+e^- \to HA \to \tau^+\tau^-\tau^+\tau^-$ in the Type X (lepton specific) THDM for $m_A = 170$ GeV and $m_H = 130$ GeV for $\sqrt{s} = 500$ GeV and 500 fb$^{-1}$ [32].

### 6.2.3 Charged Higgs boson production

The charged Higgs bosons $H^\pm$ are a clear signature for the extended Higgs sectors. They appear in most of the models except for those with only additional neutral singlets. Particular models imply constraints between the charged and neutral Higgs boson masses. In particular, in the MSSM, the mass $m_{H^\pm}$ is related to $m_A$ by $m_{H^\pm} = (m_A^2 + m_W^2)^{1/2}$ at the leading order. The precise measurement of the mass is very important in order to distinguish the MSSM from the other models, especially if the SUSY particles are rather heavy.

The direct lower bounds on $m_{H^\pm}$ come from LEP. The absolute lower bound is obtained as 79.3 GeV by ALEPH, and assuming the type II THDM, the bounds are 87.8 GeV for $\tan\beta \gg 1$ using the decay $\tau\nu$ mode, and 80.4 for relatively low $\tan\beta$ values. Using the characteristic relation in the MSSM, $m_{H^\pm} = (m_A^2 + m_W^2)^{1/2}$ with the absolute bounds $m_A > 92$ GeV, one obtains $m_{H^\pm} > 122$ GeV.

It is well known that $m_{H^\pm}$ in the Type II (and Type Y) THDM is stringently constrained by the precision measurements of the radiative decay of $B \to X_s\gamma$ by Belle, BABAR and CLEO. In these types of THDMs the loop contributions of $W^\pm$ and $H^\pm$ are always constructive, while this it not the case in the Type I and Type X. Consequently, a stringent lower bound on $m_{H^\pm}$ is obtained in the Type II (and Type Y); i.e., $295$ GeV $< m_{H^\pm}$ [34], while $m_{H^\pm} \sim 100$ GeV is not excluded unless $\tan\beta < 2$ in Type Y (Type X). The decay $B \to \tau\nu$ also can be used to constrain the charged Higgs parameters, being sensitive to $\tan\beta^2/m_{H^\pm}^2$ in the Type II THDM. The data already exclude $m_{H^\pm} < 300$ (1100) GeV for $\tan\beta > 40$ (100) at the 95% CL [35]. Similar but milder constraint on $m_{H^\pm}$ comes from tau leptonic decays in the Type II and Type X THDM: $m_{H^\pm} \sim 100$ GeV is excluded for $\tan\beta > 60$ in both models. These bounds can be relaxed in the MSSM through cancellation with loop diagrams involving supersymmetric partners.

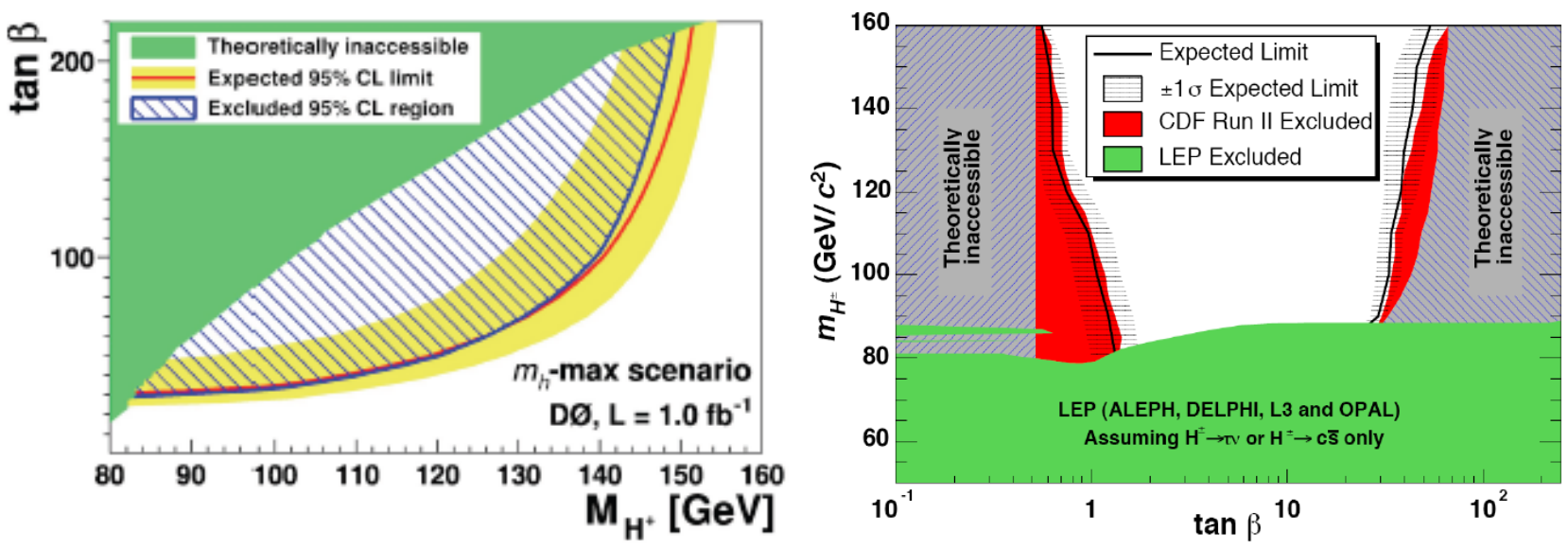

**Figure 6.8** Tevatron exclusions of charged Higgs bosons. Left: from the DO experiment [37], with 1 fb$^{-1}$. Right: from the CDF experiment [36], with 2.2 fb$^{-1}$.

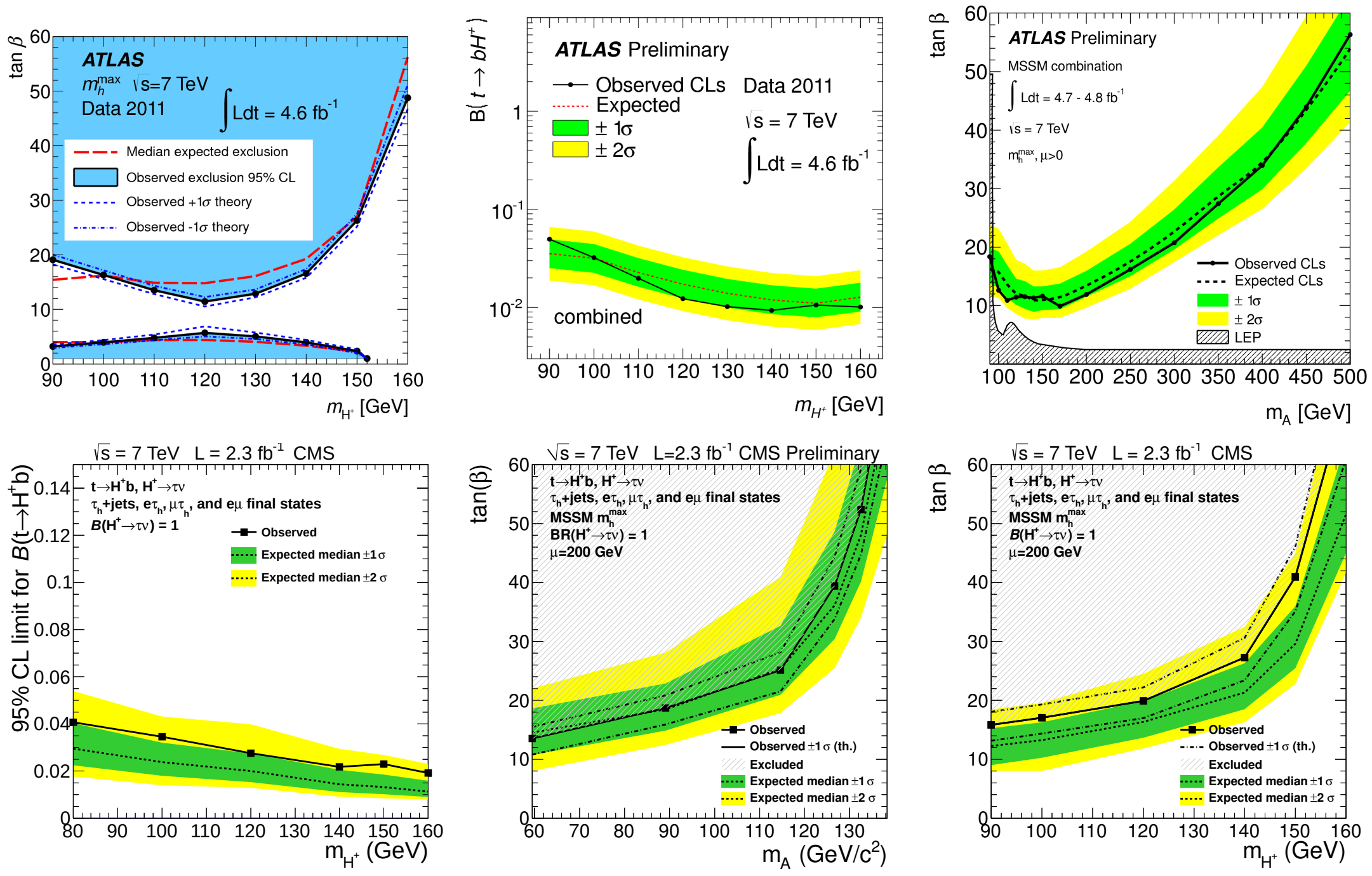

**Figure 6.9.** Top: Charged Higgs searches results from the ATLAS experiment at the LHC, based on only 4.6 fb$^{-1}$ of data [38]. Bottom: Charged Higgs searches results from the CMS experiment at the LHC, based on only 2.3 fb$^{-1}$ of data [39].

If a charged Higgs boson is lighter than the top quark, the decay $t \to H^+ b$ can compete with the SM decay $t \to W^+ b$. Both the Tevatron and the LHC experiments have searched for this process.

The Tevatron analyses look for top quark decays to $H^+ b$ in which the charged Higgs decays to $c\bar{s}$ or $\tau\nu$ [36, 37]. The results of these searches are shown in Fig. 6.8 as a function of $\tan\beta$ over a charged Higgs mass range between 90 and 160 GeV. In the case of the charged Higgs decay into a $\tau$ lepton, the search is carried out by measuring the branching ratio of the top into a $\tau$ lepton and by looking for a $\tau$ excess with respect to lepton universality. This measurement is effective for $\tan\beta > 1$. The search for the decay into $c\bar{s}$ is carried out by looking for a second bump in the two jet mass distribution of the events. This is effective for $\tan\beta < 1$.

The LHC experiments look for three possible final state signatures of a top pair production with a charged Higgs decay on one side and a standard $Wb$ decay on the other side. The three modes are lepton + jets, with the lepton coming from $\tau$ decay, $\tau$ + lepton, with the lepton coming from $W$ decay, and $\tau$ + jets, with the standard top decay purely hadronic. The results obtained by ATLAS, based only on the 2011 data [38], are shown in the top line of Fig. 6.9. No significant excess is observed, thus leaving very little room for a light charged Higgs with a mass below the top mass. Similarly, CMS, in an analysis with 2011 data corresponding to only to 2.3 fb$^{-1}$ of the recorded 2011 luminosity [39], obtains an upper limit on BR($t \to H^+ b$) that excludes a wide region of large $\tan\beta$ in the MSSM parameter space for $M_{H^+}/M_A > M_{\rm top}$. This is shown in the second line of Fig. 6.9.

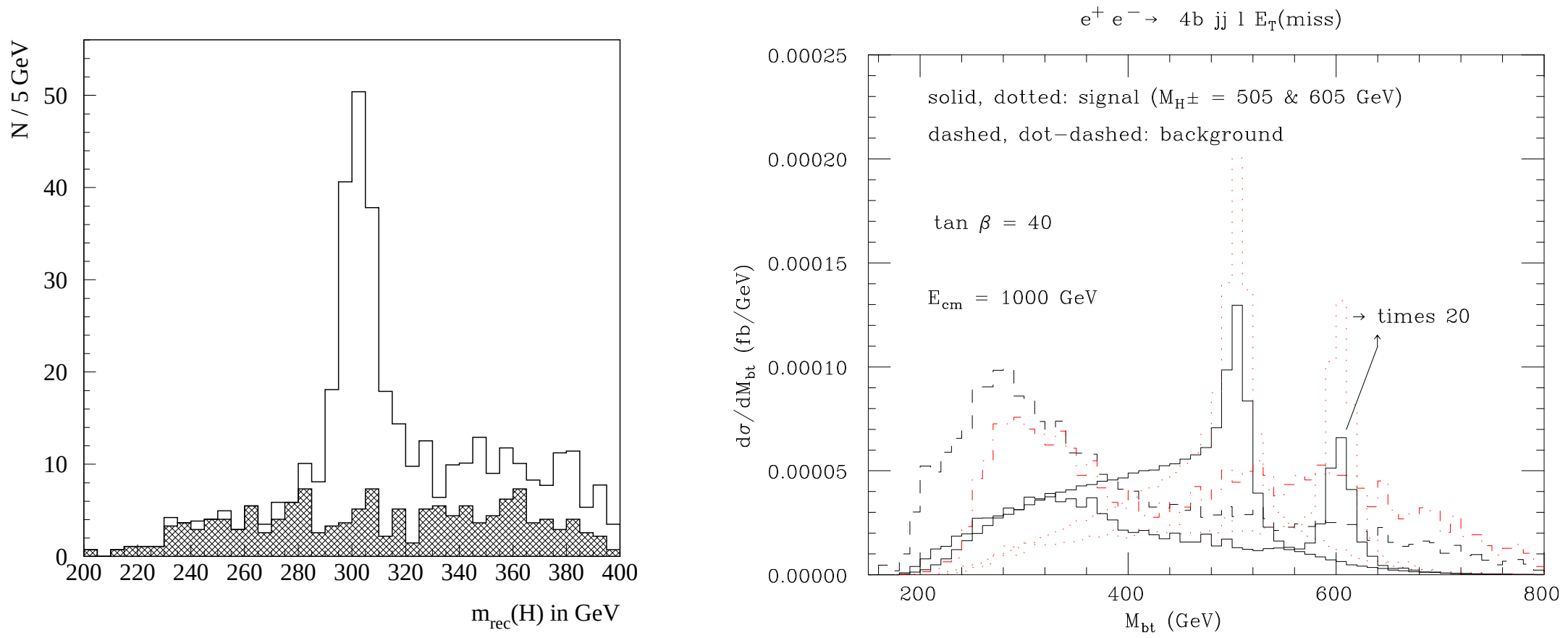

**Figure 6.10.** Left: Fitted charged Higgs boson mass in $H^+H^- \to (t\bar{b})(\bar{t}b)$ in the MSSM, with $m_{H^\pm} = 300$ GeV, measured at the ILC at CM energy 800 GeV with 1 ab$^{-1}$ of data. The background is shown by the dark histogram [41]. Right: Differential distribution of the reconstructed Higgs mass for the signal $e^+e^- \to b\bar{t}H^+ + t\bar{b}H^- \to t\bar{t}b\bar{b}$ and the background $e^+e^- \to t\bar{t}g^* \to t\bar{t}b\bar{b}$ in the MSSM or the Type II THDM [42].

At the ILC, charged Higgs bosons are produced in pairs in $e^+e^- \to H^+H^-$ [40]. The cross section is a function only of $m_{H^\pm}$ and is independent of the type of Yukawa interaction in the THDM. Therefore, as in the case of the $HA$ production, the study of the final state channels can be used to determine the type of Yukawa interaction. When $m_{H^\pm} > m_t + m_b$, the main decay mode is $tb$ in Type I, II and Y, while in Type X the main decay mode is $\tau\nu$ for $\tan\beta > 2$. When $H^\pm$ cannot decay into $tb$, the main decay mode is $\tau\nu$ except in Type Y for large $\tan\beta$ values. For $m_{H^\pm} < m_t - m_b$, the charged Higgs boson can also be studied via the decay of top quarks $t \to bH^\pm$ in THDMs except in Type X THDM case with $\tan\beta > 2$.

In the MSSM, a detailed simulation study of this reaction has been performed for the final state $e^+e^- \to H^+H^- \to t\bar{b}\bar{t}b$ for $m_{H^\pm} = 300$ GeV at $\sqrt{s} = 800$ GeV [41]. The final states is 4 $b$-jets with 4 non-$b$-tagged jets. Assuming an integrated luminosity of 1 ab$^{-1}$, a mass resolution of approximately 1.5% can be achieved (Figure 6.10 (left)). The decay mode $tbtb$ can also be used to determine $\tan\beta$, especially for relatively small values, $\tan\beta < 5$), where the production rate of the signal strongly depends on this parameter.

The pair production is kinematically limited to relatively light charged Higgs bosons with $m_{H^\pm} < \sqrt{s}/2$. When $m_{H^\pm} > \sqrt{s}/2$, one can make use of the single production processes $e^+e^- \to t\bar{b}H^+$, $e^+e^- \to \tau\bar{\nu}H^+$, $e^+e^- \to W^-H^+$, $e^+e^- \to H^+e^-\nu$ and their charge conjugates. The cross sections for the first two of these processes are directly proportional to the square of the Yukawa coupling constants. The others are one-loop induced. Apart from the pair production rate, these single production processes strongly depend on the type of Yukawa interaction in the THDM structure. In general, their rates are small and quickly suppressed for larger values of $m_{H^\pm}$. They can be used only for limited parameter regions where $m_H^\pm$ is just above the threshold for the pair production with very large or low $\tan\beta$ values.

In [42], a simulation study for the process $e^+e^- \to t\bar{b}H^- + b\bar{t}H^+ \to 4b + jj + \ell + p_T^{\rm miss}$ ($\ell = e$, $\mu$) has been done for $m_{H^\pm}$ just above the pair production threshold $m_{H^\pm} \simeq \sqrt{s}/2$. It is shown that this process provides a significant signal of $H^\pm$ in a relatively small region just above $\sqrt{s}/2$, for very large or very small values of $\tan\beta$, assuming a high $b$-tagging efficiency. The reconstructed $H^+$ mass distribution is shown in the right-hand side of Fig. 6.10.

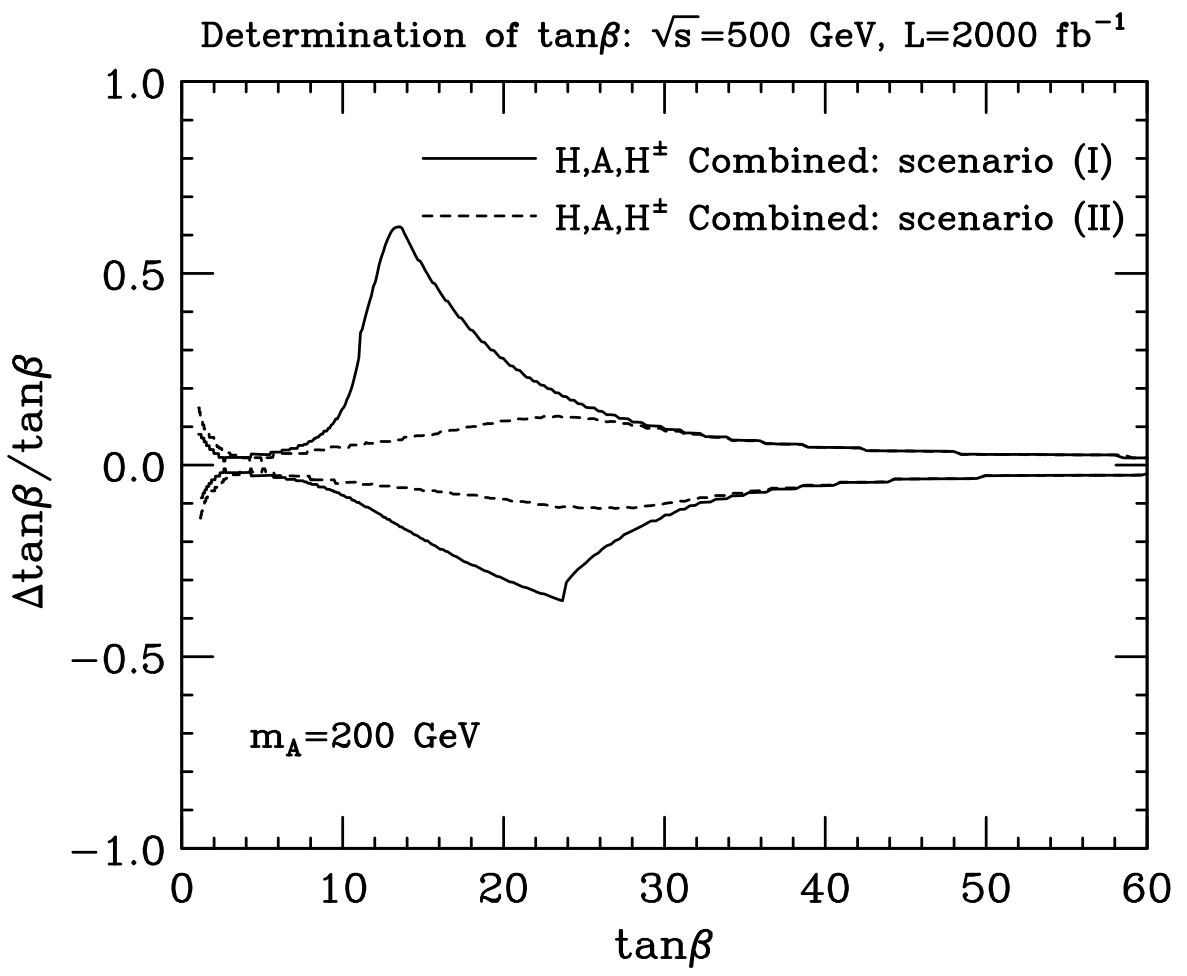

**Figure 6.11** Estimates of the 1 $\sigma$ statistical upper and lower bounds on $\tan\beta$ from ILC measurements, for an MSSM model with $m_{H^\pm} \sim m_A = 200$ GeV, assuming $\sqrt{s} = 500$ GeV and 2000 fb$^{-1}$ of data, from [43]. The quantity plotted is the relative error, $\Delta\tan\beta/\tan\beta$.

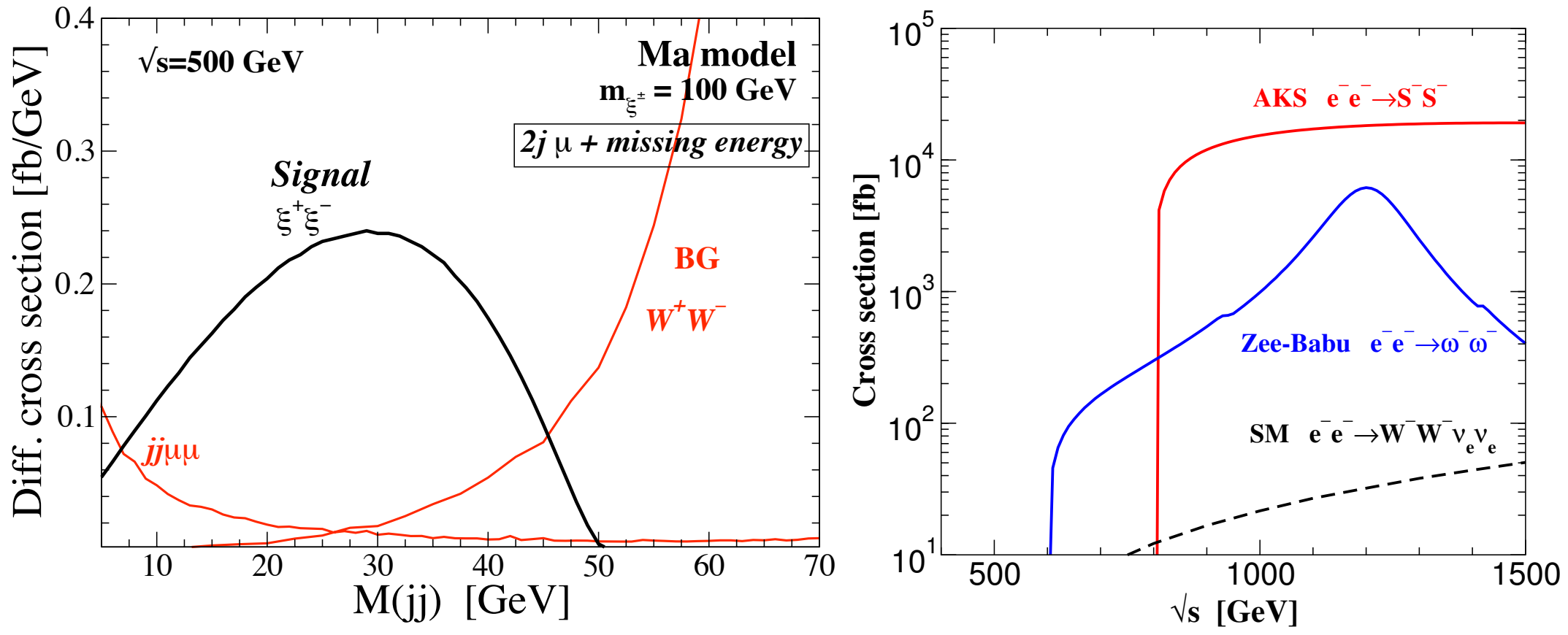

**Figure 6.12.** Left: The jet invariant mass distributions of the signal and background in the Ma model at $\sqrt{s} = 500$ GeV. The signal process is $e^+e^- \to \xi^+\xi^- \to jj\mu\nu\xi_r^0\xi_r^0$, with $m_{\xi^\pm} = 100$ GeV. Right: The cross sections of like-sign charged Higgs pair productions in the Zee-Babu model ($\omega^-\omega^-$) and in the AKS model ($S^-S^-$), shown as a function of the collision energy $\sqrt{s}$ [44].

## 6.2.4 Measurement of $\tan\beta$

The ILC measurements on charged and neutral extended Higgs bosons would be able to precisely determine $\tan\beta$, the most important parameter in the extended Higgs sector with two Higgs doublet fields. In Ref. [43], the sensitivity to $\tan\beta$ is studied by combining the measurements of production processes, branching ratios and decay widths of heavy Higgs bosons $H$, $A$ and $H^\pm$. The study is done in the context of the MSSM Type II scenario. In the case of $m_A = 200$ GeV with $\sqrt{s} = 500$ GeV and 2 ab$^{-1}$, the sensitivity is evaluated by using a large variety of complementary methods such as the production rates of $e^+e^- \to HA \to b\bar{b}b\bar{b}$ and $e^+e^- \to H^+H^- \to t\bar{b}\bar{t}b$, which provide a good sensitivity to $\tan\beta$ for relatively low $\tan\beta$, and the rate of $e^+e^- \to b\bar{b}A$, $b\bar{b}H \to b\bar{b}b\bar{b}$ and the measurement of the total widths of $H$, $A$ and $H^\pm$, which become important for large $\tan\beta$ values. For intermediate $\tan\beta$ values, the sensitivity is rather worse for the scenario (I) where heavy Higgs bosons only decay into the SM particles but it is much better for the scenario (II) where they can decay into superpartner particles via $H^\pm \to \tilde{\chi}^\pm\tilde{\chi}^0$ and similar processes. For $3 < \tan\beta < 5$, where the LHC does not have good sensitivity to extended Higgs bosons, the ILC can measure $\tan\beta$ quite accurately. The combined expected errors on $\tan\beta$ is shown in Figure 6.11. For low $\tan\beta$ regime, a good sensitivity (a few %) to $\Delta\tan\beta/\tan\beta$ can be achieved, while for $10 < \tan\beta < 30$ the accuracy would be 10–30%.

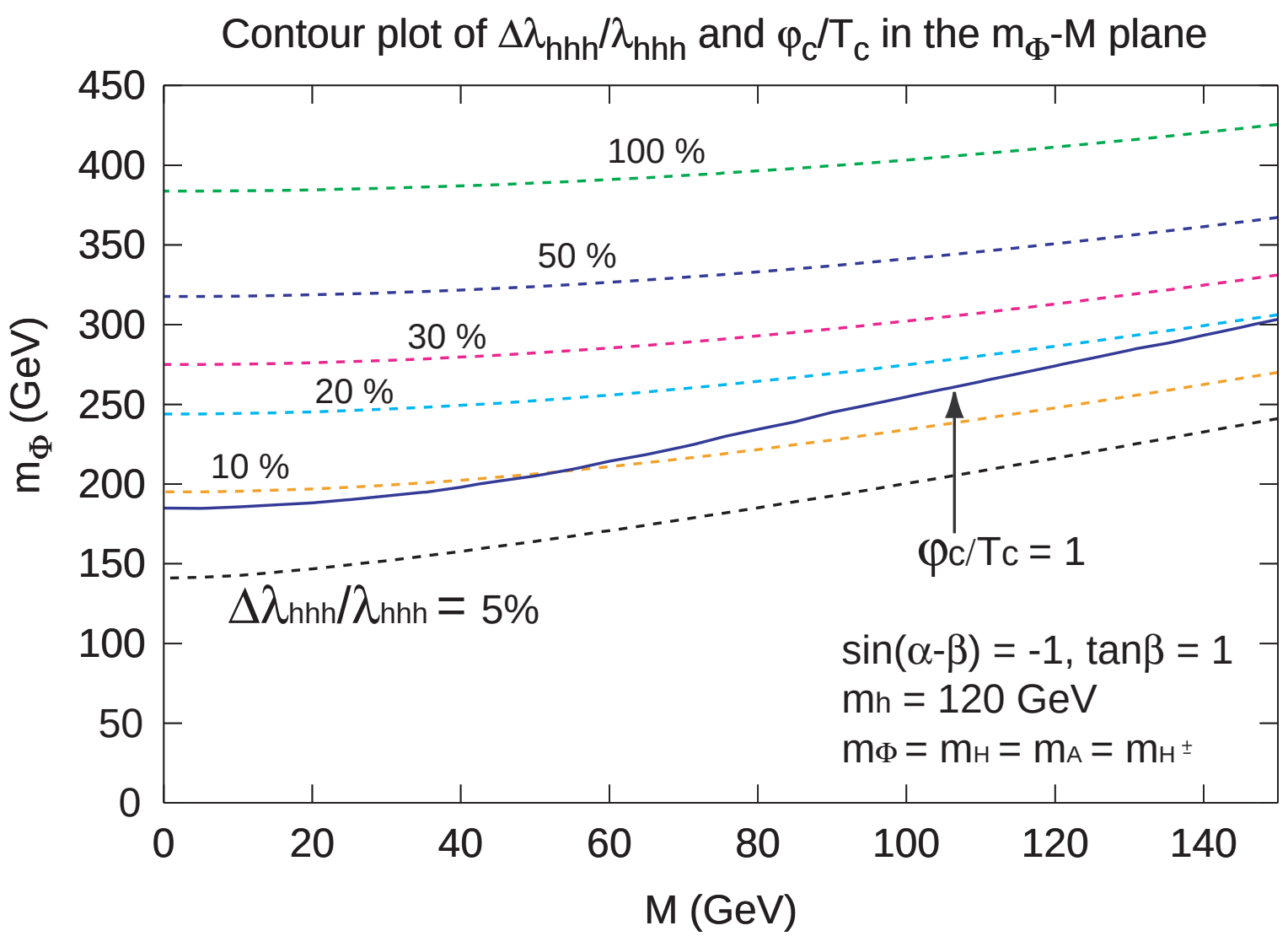

**Figure 6.13.** The region of strong first order phase transition ($\varphi_c/T_c > 1$) required for successful electroweak baryogenesis, shown as a contour plot of the deviation in the triple Higgs boson coupling from the SM prediction [51]. In this plot, $m_\Phi$ represents the degenerate mass of $H$, $A$ and $H^\pm$. The quantity $M$ is defined in (6.5).

## 6.3 Exotic Higgs bosons

Various exotic possibilities for the extended Higgs sector are motivated by other challenging problems of particle physics. We have little direct insight from experiment into the mechanisms that lead to neutrino masses, baryogenesis, and dark matter. The answers to each of these questions might arise in an extended Higgs boson sector. Models that address these questions have striking implications for extended Higgs processes that might be observed at the ILC.

We have already pointed out that neutrino masses might be associated with the addition to the Standard Model of a triplet Higgs boson multiplet. These models, described in Section 6.1.3, lead to novel reactions at the ILC, including $H^{++}$ pair production to modes that are very difficult to discover at the LHC. For example, for $m_{H^{++}} > m_{H^+} > m_{A,H}$ with the mass difference of $O(10)$ GeV and $v_\Delta \sim 10^{-5}$-$10^{-3}$ GeV, the main decay modes are $H^{\pm\pm} \to H^\pm W^\pm$, $H^\pm \to W^+H$ and $W^\pm A$, and $H, A \to \nu\,\overline{\nu}$ [16]. In this case, it is challenging to measure the signal at the LHC [17], but the ILC may be able to study it via $e^+e^- \to H^{++}H^{--} \to \ell^+\ell^+ jjjj\nu\nu\nu\nu$ if the background is reduced sufficiently. The cross section of $H^{++}H^{--}$ is about 100 fb for $m_{H^{\pm\pm}} = 200$ GeV, which implies that of the final state with a same sign dilepton signature with jets and missing energies can be around 10 fb, including the charge conjugated final state.

An alternative scenario for neutrino masses is based on radiative generation of neutrino masses by an extension of the Higgs sector [45–47]. The source of lepton number violation in these models is a coupling in the extended Higgs sector or the Majorana masses of $Z_2$-odd right-handed neutrinos. The ILC can test these models by measuring characteristic extra scalars. For example, in the Ma model [46], where neutrino masses are generated at the one-loop level by the $Z_2$ odd scalars and right handed neutrinos, the $Z_2$ odd scalar doublets $(\xi^+, \xi^0)^T$ would be observed at the ILC in a jets plus leptons final state, $e^+e^- \to \xi^+\xi^- \to jj\mu\nu\xi^0_r\xi^0_r$. The left side of Figure 6.12 shows the characteristic 2-jet mass distribution in this reaction. A striking test of these models would be the observation of double like-sign Higgs production in $e^-e^-$ collisions. The cross sections for this process in the Zee-Babu model [45] and the Aoki-Kanemura-Seto model [47] are shown in the right side Fig. 6.12.

Among the various scenarios for baryogenesis, electroweak baryogenesis [48] is attractive because of its testability in collider experiments. The parameter region for electroweak baryogenesis in the SM is already excluded. However, electroweak baryogenesis is possible within the THDM [49], which allows additional CP violating phases and a sufficiently strong 1st order electroweak phase transition. This scenario is compatible with a mass of 125 GeV for the $h$ boson by making use of loop effects of extra Higgs bosons. One of the interesting phenomenological predictions for such a scenario is a large quantum effect on the triple Higgs boson coupling [50, 51]. The requirement of sufficiently strong 1st order phase transition results in a large deviation in the triple Higgs boson coupling as seen in Fig. 6.13. The predicted effect would be clearly seen in the triple Higgs boson measurements described in Section 2.6.3. Electroweak baryogenesis will be discussed further in Section 8.1.

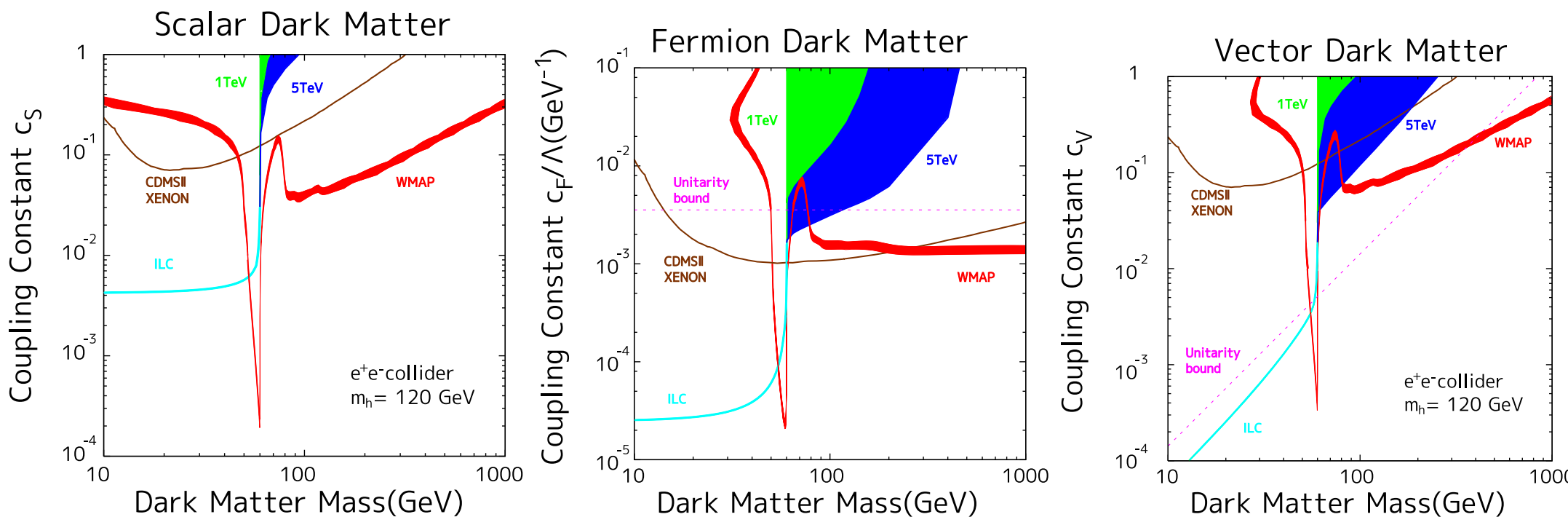

**Figure 6.14.** Sensitivities to detect the signal of scalar, fermionic, and vector dark matter signal at the ILC and CLIC, from [55]. The parameter regions with $N_S/\sqrt{N_S+N_B} > 5$ are shown in green for an $e^+e^-$ collider with CM energy 1 TeV and in blue for an $e^+e^-$ collider with CM energy and 5 TeV. In both cases, 1 ab$^{-1}$ of data is assumed, with $m_h = 120$ GeV. Constraints on direct detection experiments and the tree level unitarity limit for dark matter are also shown.

Dark matter requires a new stable particle with mass at the weak interaction scale. Though models involving supersymmetry and extra dimensions are more fashionable, there is no reason why this particle cannot come from an extended Higgs sector. The dark matter particle can be made stable by a $Z_2$ or higher discrete symmetry of this sector. Models realizing this scenario are given in [52–54].

An important phenomenological prediction of these scenarios is the invisible decay $h \to DD$ of the SM like Higgs boson in to a dark matter pair, if this decay is kinematically allowed. At the linear collider, these invisible decays can be well measured via $e^+e^- \to Zh$, as we have discussed in Section 2.4.3. The case $m_h < 2m_D$, where the above decay mode is not open, can be studied in the $ZZ$ fusion process. Nabeshima has analyzed the LHC and linear collider prospects for the study of this reaction as shown in Fig. 6.14. The dark matter consistent with the WMAP data would be tested at the ILC [55].

## 6.4 Summary

The Higgs sector is the window for new physics beyond the Standard Model. There is no reason to restrict this sector to the SM Higgs. There are several important theoretical frameworks that predict an enriched Higgs sector. These extended Higgs sector possibilities are very important to explore not only for clarifying the nature of the electroweak symmetry breaking but also for investigating more general schemes for physics beyond the Standard Model. The ILC brings important capabilities to this study.

First, the ILC offers increased potential for discovery. The LHC experiments have a strong potential for discovery if an extended Higgs sector; they will be able to cover a wide region in the parameter space including the possibility to reach relatively high masses. But the ILC covers all possibilities for pair-production of extended Higgs bosons uniformly up to the kinematic limit. This adds important capability for charged Higgs bosons and in the low $\tan\beta$ region that hadron colliders have difficulty in reaching.

Second, the ILC offers a program of comprehensive precision measurements. To understand the coupling scheme of extended Higgs bosons, we need measurements of rates to a variety of channels with a production cross section that is precisely known as a function of the boson masses. Electroweak pair production at the ILC provides just this setting.

Finally, the ILC has great power to discriminate between possible theoretical frameworks. We have emphasized that the phenomenology of extended Higgs models can be complex, with several new parameters and mixing angles, in each of many possible theoretical schemes. The experiments offered by the ILC provide a level of definiteness of interpretation that are not provided by hadron collider measurements of individual reaction rates.

The possibility of an extended Higgs sector is a key topic for models of physics beyond the Standard Model. In order to advance into this unknown field and to disentangle the many present proposed theoretical frameworks, it is essential to have complementary machines for comparing and combining their results. ILC is essential to LHC and vice and versa.

# Chapter 7
# Supersymmetry

In this Chapter, we discuss the opportunities the ILC provides for the detailed analysis of new sectors of particles by direct spectroscopy, taking Supersymmetry as perhaps the best studied example. The key assets of a Linear Collider, namely its clean environment, the flavour democracy of the electroweak interaction, the tunable center-of-mass energy and the adjustable beam polarisation offer unique potential which is in every sense complementary to the LHC. These unique features of a linear $e^+e^-$ collider allow for high precision studies of any new particles which might be discovered by the LHC. Perhaps more importantly, they allow for discovery of a variety of new matter states which can be produced at LHC, but which would lie hopelessly buried beneath formidable QCD backgrounds, and inaccessible to discovery.

Supersymmetry has long been considered a strong point of the case for the ILC. It has been known for a long time that the ILC has the ability to make precise measurements, not only of supersymmetry particle masses, but also of the underlying fundamental parameters of the model [1]. The precision measurement of masses is not degraded even in the presence of cascade decays of sparticles which are the norm in many models [2]. The ability of ILC measurements to complement and extend the information we will obtain on supersymmetric particles from the LHC has been studied explicitly in many examples [3, 4]. However, the first results from the LHC have shifted the ground under the theory of supersymmetry, ruling out many of the benchmark models and changing our perspective on what regions of the model space are the most relevant. In this Chapter, we present a new discussion of supersymmetry at the ILC relevant to the current LHC era. We will review the continued importance of supersymmetry as a principle for physics beyond the Standard Model. We will then discuss the classes of supersymmetry models that remain consistent with the LHC data and the particular role that the ILC will have in the exploration of new particles in these models.

Following an introduction, in 7.2 we will lay the basis for our discussion of the experimental capabilities of the ILC by summarizing the *recent change of paradigm* from very constrained models to considerations of naturalness and phenomenological approaches. In section 7.3 we continue with a brief dicussion of the state of direct and indirect constraints on SUSY (circa summer 2012) and it's implications for the ILC. As an illustration of these ideas and as guideline for the experimental discussion, we continue in section 7.4 by introducing two example scenarios compatible with current knowledge, but featuring very different phenomenology. The main part of this section, section 7.5 finally highlights possible key measurements for a variety of new particles, including remarks on model discrimination and parameter determination.

## 7.1 Introduction

While no direct evidence for the existence of non-Standard Model particles has emerged so far, there are many indications that the Standard Model (SM) is not valid up to the Planck scale. Among these, the most well-known is the gauge hierarchy problem, the instability of the weak scale against quantum corrections to fundamental scalar fields. Solutions to this problem require new particles to appear at or around the weak scale. Additional problems arise from cosmology. The SM does not contain any candidate particles to constitute the needed cold dark matter (CDM). It also lacks a sufficient source of $CP$ violation needed to explain baryogenesis. The SM is not sufficient as a part of a complete theory of nature at very small distance scales because the SM gauge couplings do not unify when extrapolated to high energies, and because the SM has no clear way to incorporate quantum gravity.

One approach which has the potential to address all these problems is Supersymmetry (SUSY), a quantum spacetime symmetry which predicts a correspondence between bosonic and fermionic fields [5–8]. SUSY removes the quadratic divergences of scalar field theory and thus offers a solution to the aforementioned gauge hierarchy problem. This allows for stable extrapolation of the Standard Model couplings into the far ultraviolet ($E \gg M_{weak}$) regime [9, 10], with the suggestion of gauge unification. SUSY provides an avenue for connecting the Standard Model to ideas of grand unification (GUTs) and/or string theory, and provides a route to unification with gravity via local SUSY, or supergravity theories [11–13]. SUSY theories offer several candidates [14] for dark matter, including the neutralino, the gravitino or a singlet sneutrino. In SUSY theories where the strong $CP$ problem is solved via the Peccei-Quinn mechanism, there is the added possibility of mixed axion-neutralino [15–17], axion-axino [18–20] or axion-gravitino cold dark matter. In order to explain the measured baryon to photon ratio $\eta \sim 10^{-10}$, SUSY offers at least three prominent possibilities including electroweak baryogenesis (now nearly excluded in the minimal theory by limits on $m_{\tilde{t}_1}$ and a light Higgs scalar with $m_h \sim 125$ GeV [21]), thermal and non-thermal leptogenesis [22], and Affleck-Dine baryo- or leptogenesis [23, 24].

There is good reason, then, to adopt SUSY as a well-motivated example of an extension of the Standard Model in order to discuss the potential of the ILC to solve the current puzzles of electroweak symmetry breaking, cosmology and grand unification. In this section, we will describe the capabilities offered by the ILC for the discovery of supersymmetric particles and the precision measurement of their properties. It should be stressed that the experimental capabilities of the ILC presented here apply to new particles with similar signatures whatever the nature of the high scale model.

## 7.2 Setting the scene

The simplest supersymmetric theory which contains the SM is known as the Minimal Supersymmetric Standard Model, or MSSM. To construct the MSSM, one adopts the gauge symmetry of the SM and promotes all SM fields to superfields. There is a unique generalization of the SM if one imposes the requirements of gauge symmetry, renormalizability, and $R$-parity conservation. This model requires two Higgs doublet superfields, and thus includes an extended Higgs sector as described in Section 6 as well as corresponding higgsino particles. To be phenomenologically viable, supersymmetry must be broken. SUSY breaking is implemented explicitly in the MSSM by adding all allowed *soft* SUSY breaking terms. The resulting model contains 178 parameters, many of which lead to flavor violation (FV) or CP violation (CPV). The pMSSM ignores the FV and CPV terms, and then contains just 19 or 24 weak scale parameters, depending on whether one does or does not assume universality between the masses of the first and second generation scalar superpartners [25, 26].

Because of the large number of parameters in the general MSSM, the phenomenology of SUSY has often been discussed in terms of a subspace of the more general theory with a reduced parameter set. For many years, the phenomenology of SUSY was described using the parameter space of a set of models called "minimal supergravity" [27], also known as mSUGRA or the cMSSM. These models assumed that the soft supersymmetry breaking parameters unified at the GUT scale, so that the model could be described by four parameters, a weak scale gravitino mass $m_{3/2}$ and universal scalar masses $m_0$, gaugino masses $m_{1/2}$ and trilinear terms $A_0$ at the GUT scale. Other similarly specific choices are given by the minimal gauge mediated SUSY breaking model [28] and the minimal anomaly-mediated SUSY breaking model [29, 30]. In all of these schemes, the unification assumption ties together the mass scales of the supersymmetric partners of quarks, gluons, gauge bosons, and Higgs bosons.

In fact, it was realized a long time ago that the constraints linking these scales are not necessary and might not yield the most attractive models. In 1996, Cohen, Kaplan, and Nelson discussed the "more minimal supersymmetric Standard Model" in which only the partners of the third generation particles are light [31]. Over the years, other authors have discussed models in which some or all of the squarks are very heavy with respect to the electroweak scale without disturbing the naturalness of electroweak symmetry breaking [32–34].

Now the first data from the LHC have weighed in on this issue. Searches at ATLAS and CMS have excluded minimal supergravity or the cMSSM for all models in which the squark and gluino masses are below 1 TeV [35, 36]. These powerful exclusions have, to our knowledge, not caused any theorists to abandon SUSY. However, they have led to a dramatic change in thinking about the parameter space of the MSSM.

Specifically, these exclusions have led theorists to rethink the expectations for the masses of supersymmetric particles that come from the idea that supersymmetry should naturally produce the scale of electroweak symmetry breaking. It is easy to arrange in a supersymmetric model that the Higgs bosons have a potential with a symmetry-breaking minimum. The condition for minimizing this potential can be written

$$\frac{1}{2}m_Z^2 = \frac{(m_{H_d}^2 + \Sigma_d) - (m_{H_u}^2 + \Sigma_u)\tan^2\beta}{(\tan^2\beta - 1)} - \mu^2 \ . \tag{7.1}$$

where, $\Sigma_u$ and $\Sigma_d$ arise from radiative corrections [37]. The largest contribution to $\Sigma_u$ comes from the mass of the top squarks $\tilde{t}_i$, $i = 1, 2$,

$$\Sigma_u(\tilde{t}_i) \sim -\frac{3y_t^2}{16\pi^2} \times m_{\tilde{t}_i}^2 \log(m_{\tilde{t}_i}^2/Q^2), \tag{7.2}$$

where $y_t$ is the top quark Yukawa coupling and $Q = \sqrt{m_{\tilde{t}_1} m_{\tilde{t}_2}}$. The negative sign of this radiative correction is typically the force that drives the Higgs mass term negative.

The MSSM is said to generate the electroweak scale "naturally" if the terms in (7.1) are all of roughly the same size, without large cancellations between the two terms on the right-hand side. By this criterion, the primary implication of the naturalness of the electroweak scale is that the parameter $\mu$, the higgsino mass parameter, should be of the order of 100 GeV [38, 39]. Other supersymmetric partners are required to be light only to the extent that they contribute to the parameters of (7.1) through radiative corrections. The particles primarily constrained by this criterion are the higgsinos themselves, the top squarks, which enter through (7.2), and the gluino, whose mass enters the radiative corrections to the top squark masses.

Imposing this criterion strictly leads to a very different spectrum from that of the cMSSM. In the cMSSM, $\mu$ is an output parameter and the values typically output are larger than the squark and gluino masses. Direct argumentation from (7.1), on the other hand, leads to a spectrum in which $|\mu| \sim 100 - 200$ GeV, so that the lightest neutralino is likely higgsino-like. The third generation squarks should have masses that are relatively small, though these masses might be as high as $\lesssim 1 - 1.5$

TeV [40]. The gluino could be heavier, up to a few TeV [41]. The superpartners of electroweak gauge bosons would be found at masses of 1-2 TeV, while the first and second generation scalar partners could be much heavier, possibly in the multi-TeV regime. This last condition is actually beneficial, giving at least a partial solution to the SUSY flavor, $CP$, proton decay, and gravitino problems. This region of the MSSM parameter space has been dubbed "natural SUSY" [42]. The extreme limit of this schema, in which only the higgsinos are light, has been studied in [43, 44]. A more general exploration of the parameter space of natural SUSY can be found in [45].

The push from the LHC results toward natural SUSY has motivated many theorists to find model-building explanations for this choice of SUSY parameters. Some interesting proposals can be found in [46–49]. Not only have the LHC results on SUSY not damped theorists' enthusiasm, but they have pushed theorists increasingly toward models with higgsino-like charginos and neutralinos with masses below 250 GeV that are ideal targets for the ILC experiments.

## 7.3 Direct and indirect experimental constraints

### 7.3.1 Particle sectors of a supersymmetric model

In this section, we present the current direct and indirect experimental constraints on SUSY models. We have emphasized in the previous section that a SUSY model consistent with the experimental constraints from the LHC probably does not belong to the subspace of artificially unified models such as the cMSSM. We find it most useful to analyze an MSSM model in terms of distinct particle sectors with different properties and influence. At generic points in the MSSM parameter space, these sectors can have masses very different from one another. It is important to keep track of which experimental constraints apply to which sector.

The new particle sectors of an MSSM model are:

1. The first and second generation squarks.
2. The first and second generation sleptons.
3. The third generation squarks and sleptons.
4. The gauginos.
5. The higgsinos.

We have already described the constraints on the masses of these particles from the theoretical consideration of naturalness. We now review the constraints from experiment.

### 7.3.2 Indirect constraints on SUSY models

The magnetic moment of the muon $a_\mu \equiv \frac{(g-2)_\mu}{2}$ was measured by the Muon $g-2$ Collaboration [50] and has been found to give a $3.6\sigma$ discrepancy with SM calculations based on $e^+e^-$ data [51]: $\Delta a_\mu = a_\mu^{meas} - a_\mu^{SM}[e^+e^-] = (28.7 \pm 8.0) \times 10^{-10}$. When $\tau$-decay data are used to estimate the hadronic vacuum polarization contribution rather than low energy $e^+e^-$ annihilation data, the discrepancy reduces to $2.4\sigma$, corresponding to $\Delta a_\mu = a_\mu^{meas} - a_\mu^{SM}[\tau] = (19.5 \pm 8.3) \times 10^{-10}$. The SUSY contribution to the muon magnetic moment is [52]

$$\Delta a_\mu^{SUSY} \sim \frac{m_\mu^2 \mu M_i \tan\beta}{m_{SUSY}^4} \ , \tag{7.3}$$

where $i = 1, 2$ labels the electroweak gaugino masses and $m_{SUSY}$ is the characteristic sparticle mass circulating in the muon-muon-photon vertex correction, one of: $m_{\tilde{\mu}_{L,R}}$, $m_{\tilde{\nu}_\mu}$, $m_{\tilde{\chi}_i^+}$ and $m_{\tilde{\chi}_j^0}$. Attempts to explain the muon $g-2$ anomaly using supersymmetry usually invoke sparticle mass spectra with relatively light smuons and/or large $\tan\beta$ (see *e.g.* Ref. [53]). Some SUSY models where $m_{\tilde{\mu}_{L,R}}$ is correlated with squark masses (such as mSUGRA) are now highly stressed to explain the $(g-2)_\mu$ anomaly. In addition, since naturalness favors a low value of $|\mu|$, tension again arises between a large contribution to $\Delta a_\mu^{SUSY}$ and naturalness conditions. These tensions motivate scenarios with non-universal scalar masses [54].

The combination of several measurements of the $b \to s\gamma$ branching fraction finds that $BF(b \to s\gamma) = (3.55 \pm 0.26) \times 10^{-4}$ [55]. This is somewhat higher than the SM prediction [56] of $BF^{SM}(b \to s\gamma) = (3.15 \pm 0.23) \times 10^{-4}$. SUSY contributions to the $b \to s\gamma$ decay rate come mainly from chargino-top squark loops and loops containing charged Higgs bosons. They are large when these particles are light and when $\tan\beta$ is large [57].

The decay $B_s \to \mu^+\mu^-$ occurs in the SM at a calculated branching ratio value of $(3.2 \pm 0.2) \times 10^{-9}$. The CMS experiment [58] has provided an upper limit on this branching fraction of $BF(B_s \to \mu^+\mu^-) < 1.9 \times 10^{-8}$ at 95% CL. The CDF experiment [59] claims a signal in this channel at $BF(B_s \to \mu^+\mu^-) = (1.8 \pm 1.0) \times 10^{-8}$ at 95% CL, which is in some discord with the CMS result. Finally, the LHCb experiment has reported a strong new bound of $BF(B_s \to \mu^+\mu^-) < 4.5 \times 10^{-9}$ [60]. In supersymmetric models, this flavor-changing decay occurs through exchange of the pseudoscalar Higgs $A$ [61, 62]. The contribution to the branching fraction from SUSY is proportional to $\tan^6\beta/m_A^4$.

The branching fraction for $B_u \to \tau^+\nu_\tau$ decay is calculated [63] in the SM to be $BF(B_u \to \tau^+\nu_\tau) = (1.10 \pm 0.29) \times 10^{-4}$. This is to be compared to the value from the Heavy Flavor Averaging group [64], which finds a measured value of $BF(B_u \to \tau^+\nu_\tau) = (1.41 \pm 0.43) \times 10^{-4}$, in agreement with the SM prediction, but leaving room for additional contributions. The main contribution from SUSY comes from tree-level charged Higgs exchange, and is large at large $\tan\beta$ and low $m_{H^+}$.

Finally, measurements of the cold dark matter (CDM) abundance in the universe find $\Omega_{CDM}h^2 \sim 0.11$, where $\Omega_{CDM}$ is the dark matter relic density scaled in terms of the critical density. Simple explanations for the CDM abundance in terms of thermally produced neutralino LSPs are now highly stressed by LHC SUSY searches, and are even further constrained if the light SUSY Higgs $h$ turns out to have mass $\sim 125$ GeV [65]. A higgsino LSP is not a good dark matter candidate, since it has too large an annihilation rate to vector boson pairs, leading to too small a thermal relic density. However, this deficit can be repaired in well-motivated extensions of the MSSM, including mixed axion-LSP dark matter and models with late decaying moduli fields. For purposes of considering ILC or LHC physics, it seems prudent not to take dark matter abundance constraints on SUSY theories too seriously at this point in time.

### 7.3.3 Impact of Higgs searches

The ATLAS and CMS experiments have reported the discovery of a narrow resonance with mass near 125 GeV [66, 67]. At the same time, they exclude a Standard Model-like Higgs boson in the mass ranges $110-123$ and $130-558$ GeV at 95% CL. The discovery is based on an excess of events mainly in the $\gamma\gamma$, $ZZ^* \to 4\ell$ and $WW^*$ decay channels. These excesses are also corroborated by recent reports from CDF and D0 at the Fermilab Tevatron of excess events in the $Wb\bar{b}$ and other channels over the mass range 115-130 GeV [68].

Searches by ATLAS and CMS for $H,\ A \to \tau^+\tau^-$ now exclude a large portion of the $m_A$ vs. $\tan\beta$ plane [69, 70]. In particular, the region around $\tan\beta \sim 50$, which is favored by Yukawa-unified SUSY GUT theories, now excludes $m_A < 500$ GeV. For $\tan\beta = 10$, only the range $120$ GeV $< m_A < 220$ GeV is excluded. ATLAS excludes charged Higgs bosons produced in association with a $t\bar{t}$ pair for $m_{H^\pm} < 150$ GeV for $\tan\beta \sim 20$ [71].

A Higgs mass of $m_h = 125 \pm 3$ GeV lies within the narrow mass range $m_h \sim 115 - 135$ GeV which is allowed between LEP searches for a SM-like Higgs boson and calculations of an upper limit to $m_h$ within the MSSM. However, such a large value of $m_h$ requires large radiative corrections and large mixing in the top squark sector. In models such as mSUGRA, trilinear soft parameters $A_0 \sim \pm 2m_0$ are thus preferred, and values of $A_0 \sim 0$ would be ruled out [72–74]. In other constrained models such as the minimal versions of GMSB or AMSB, Higgs masses of $125$ GeV require even the lightest of sparticles to be in the multi-TeV range [65], leading to enormous electroweak fine-tuning. In the mSUGRA/cMSSM model, requiring a Higgs mass of about $125$ GeV pushes the best fit point in $m_0$ and $m_{1/2}$ space into the multi-TeV range [72] and makes global fits of the model to data increasingly difficult [75]. This already motivates us to consider the prospects for precision measurements of new particles at the ILC in a more general context than the cMSSM.

### 7.3.4 Direct searches for supersymmetric particles

The most model-independent limits on SUSY particles, especially the uncoloured ones, have been set by the LEP experiments [76–80] on *sleptons, charginos and neutralinos*. The fact that these limits have not been superseded in the general case by LHC data illustrates the complementarity of $e^+e^-$ and $pp$ colliders as well as the fact that the interpretation of $e^+e^-$ data requires significantly fewer model assumptions.

The ATLAS and CMS collaborations have searched for multi-jet+$E_T^{\mathrm{miss}}$ events arising from gluino and squark pair production in 4.4 fb$^{-1}$ of 2011 data taken at $\sqrt{s} = 7$ TeV [81, 82] and in up to 5.8 fb$^{-1}$ of 2012 data taken at $\sqrt{s} = 8$ TeV [83]. In the limit of very heavy squark masses, they exclude $m_{\tilde{g}} \lesssim 1.1$ TeV, while for $m_{\tilde{q}} \simeq m_{\tilde{g}}$ then $m_{\tilde{g}} \lesssim 1.5$ TeV is excluded, assuming $m_{\tilde{\chi}_1^0} = 0$ GeV in both cases. $m_{\tilde{q}}$ refers to a generic first generation squark mass scale, since these are the ones whose production rates depend strongly on valence quark PDFs in the proton.

A recent ATLAS search for direct bottom squark pair production followed by $\tilde{b}_1 \to b\tilde{\chi}_1^0$ decay ($pp \to \tilde{b}_1\bar{\tilde{b}}_1 \to b\bar{b} + E_T^{\mathrm{miss}}$) based on 2 fb$^{-1}$ of data at $\sqrt{s} = 7$ TeV now excludes $m_{\tilde{b}_1} \lesssim 350$ GeV for $m_{\tilde{\chi}_1^0}$ as high as $120$ GeV. For larger values of $m_{\tilde{\chi}_1^0}$, there is no limit at present [84]. These constraints also apply to top squark pair production where $\tilde{t}_1 \to b\tilde{\chi}^+$ decay and the $\tilde{\chi}^+$ decays to soft, nearly invisible particles, as would be expected in natural SUSY.

In models with gaugino mass unification and heavy squarks (such as mSUGRA with large $m_0$), electroweak gaugino pair production $pp \to \tilde{\chi}_1^\pm\tilde{\chi}_2^0$ is the dominant SUSY particle production cross section at LHC7 for $m_{\tilde{g}} > 0.5$ TeV [85]. Two searches by ATLAS in the $3$ lepton final state using 2.1 fb$^{-1}$ of 7 TeV data [86] and in the $2$ lepton final state using 4.7 fb$^{-1}$ of $8$ TeV data [87] give results in the pMSSM and in a simplified model. Both cases assume that chargino and neutralino decay to intermediate sleptons, which enhances the leptonic branching fractions. The theoretically more interesting case of chargino and neutralino three-body decay through $W^*$ and $Z^*$ leading to a clean trilepton signature [88, 89] awaits further data and analysis.

The opposite-sign/same flavor dilepton final state [87] can also originate from direct production of slepton pairs. The resulting exclusion in the slepton-LSP mass plane is rather model-independent and extends the LEP2 limit to higher slepton masses of up to $200$ GeV for an LSP mass of $30$ GeV. For LSP masses larger than $80$ GeV, no slepton masses can be excluded beyond the LEP2 limit.

In addition, a wide variety of other searches for SUSY have been made – including searches for long-lived quasi-stable particles, electroweakinos with small mass difference, RPV SUSY, minimal gauge mediated SUSY etc. After 5 fb$^{-1}$ of data at LHC7 and a first glimpse into another 5 fb$^{-1}$ of data at LHC8, it is safe to say that no compelling signal for SUSY has yet emerged at LHC.

### 7.3.5 Impact of the constraints on the SUSY particle sectors

We can summarize the results of this section as constraints on the various sectors of an MSSM model set out in Section 7.3.1:

1. The first and second generation squarks: The particles in this sector are highly constrained by flavour and $CP$ violation limits and by LHC squark searches. Typically we expect $m_{\tilde{q}} \gtrsim 1.5$ TeV. This sector has little connection to the EW scale: indeed, in split SUSY models [90] the squark (and slepton) masses are sometimes pushed to the $10^{10}$ GeV level.

2. The first and second generation sleptons: The particles in this sector are favored by $(g-2)_\mu$ to have masses below 1 TeV. However, the absence of leptonic flavour violating processes (*e.g* $\mu \to e\gamma$ decay) push this sector to be much heavier.

3. The third generation squarks and sleptons: The particles in this sector are influenced by large Yukawa couplings. Naturalness favors their masses to be below a few TeV. $B$-meson decay data prefer top squarks with mass at or above the TeV scale.

4. The gauginos: The particles in this sector are in principle independent of the squark mass scale and might also be independent of one another. Simple SUSY GUT models favor gaugino mass unification $M_1 = M_2 = M_3 \equiv m_{1/2}$ at $M_{GUT}$, giving a $1:2:7$ ratio of masses at the weak scale. More general models allow for essentially independent gauginos masses. Electroweak fine-tuning prefers gaugino masses not too far above the TeV scale. As of today, $M_1$ and $M_2$ are not substantially constrained beyond the LEP limits, but $M_3$, the gluino mass, probably must be above 1 TeV.

5. The higgsinos: The masses of the particles in this sector are determined by the superpotential $\mu$ term, which is not a soft SUSY breaking term. In the context of the MSSM alone, it could be expected to occur at the $M_{GUT}$ or $M_{string}$ scale. This however would require immense fine-tuning in the corrections to the $Z$ mass: *c.f.* Eq'n 7.1. Naturalness arguments prefer a value of $|\mu|$ not far above $\sim 100$ GeV, close to but somewhat beyond the limits from LEP2 chargino searches.

Ironically, the LHC has its greatest capability—in terms of mass reach—to detect the first generation squarks and the gluinos. These are particles with indirect or no connection to the $Z$ mass scale. On the other hand, the ILC has an excellent capability to detect electroweakinos. In the case where the light electroweakinos are higgsinos, the ILC would be directly probing that sector which is most directly connected to the $Z$-mass scale via electroweak fine-tuning. The ILC also has excellent capabilities to study the sleptons, probing a sector that is very difficult to study at the LHC. It is possible that the third generation squarks and sleptons lie within the mass range of the ILC. In that case, the ILC would greatly enhance the knowledge of these sparticles gained from the LHC, since the ILC has the capability to precisely measure not only the masses but also the quantum numbers and mixing angles of these particles. We will present examples of these ILC capabilities in the next several sections.

## 7.4 Two benchmark points for the ILC

In Ref. [91], a variety of post LHC7 benchmark points for ILC physics were proposed. Here, we include two of these for reference in the discussion of supersymmetric particle discovery and measurement capabilities at the ILC. These models are completely viable in the face of the LHC supersymmetry searches and they address important questions in physics beyond the Standard Model. Many of the more specific scenarios discussed in Section 7.5 can be identified within their particle spectra. A very large number of additional viable supersymmetry models, illustrating models with both neutralino and gravitino LSPs, are presented in [92],

**Figure 7.1**
SUSY particle spectrum of the two benchmark scenarios discussed in Section 7.4: Top: Natural SUSY model; Bottom: $\delta M\tilde{\tau}$ model.

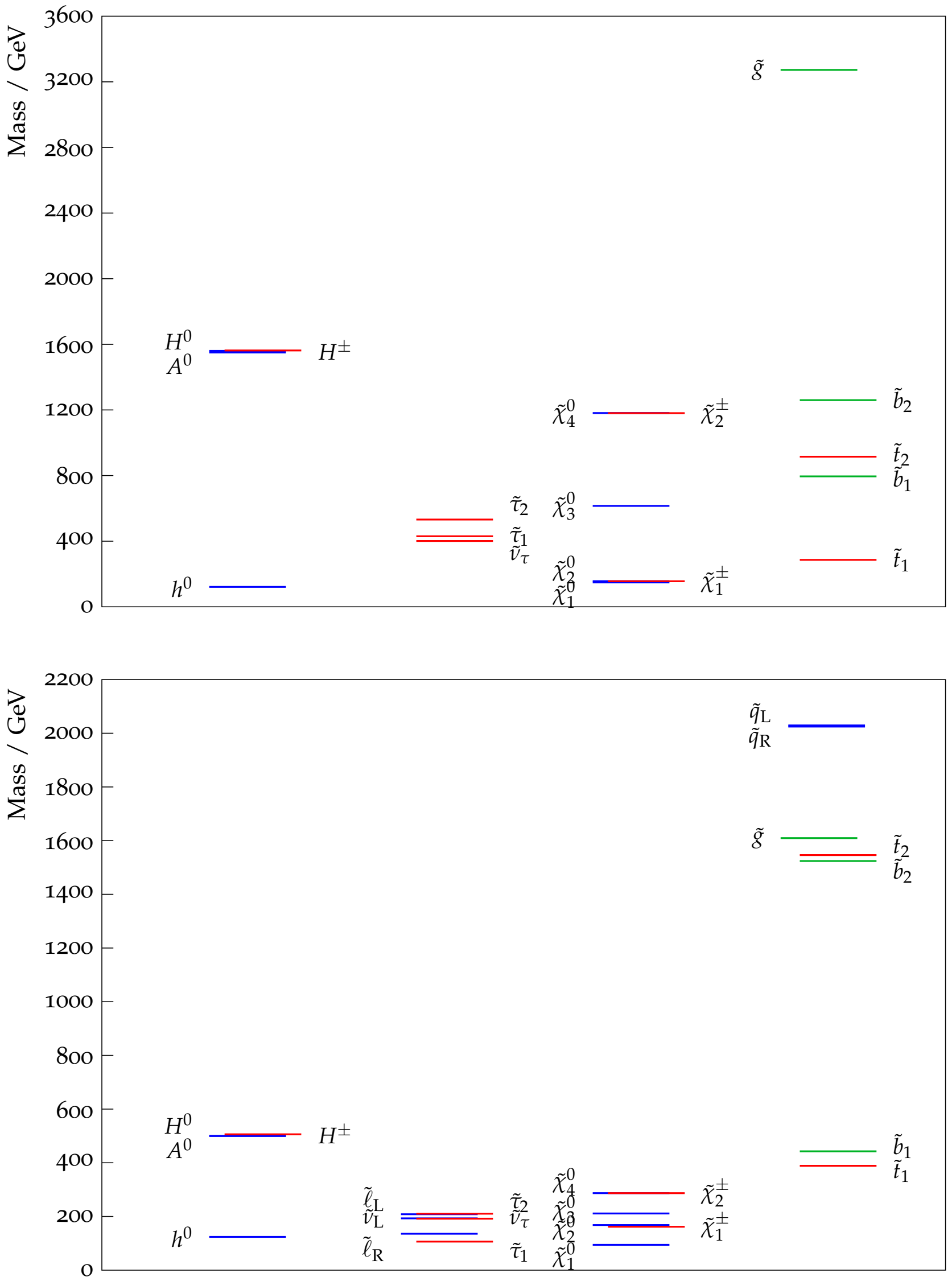

### 7.4.1 Natural SUSY and light higgsinos

For natural SUSY (NS), we adopt a benchmark point using input parameters $m_0(1,2) = 13500$ GeV, $m_0(3) = 760$ GeV, $m_{1/2} = 1380$ GeV, $A_0 = -167$ GeV, $\tan\beta = 23$ GeV, $\mu = 150$ GeV and $m_A = 1550$ TeV. The resulting mass spectrum is listed in Table 1 of Ref. [91] and shown in Figure 7.1.

The point contains higgsino-like $\tilde{\chi}^0_1$, $\tilde{\chi}^0_2$ and $\tilde{\chi}^\pm_1$ with masses $\sim \mu = 150$ GeV, where $m_{\tilde{\chi}_1} - m_{\tilde{\chi}^0_1} = 7.4$ GeV and $m_{\tilde{\chi}^0_2} - m_{\tilde{\chi}^0_1} = 7.8$ GeV. Due to the small energy release in their three body decays, the $\tilde{\chi}^\pm_1$ and $\tilde{\chi}^0_2$ will be difficult to detect at LHC [44]. Third generation squark masses are at $m_{\tilde{t}_1} = 286.1$ GeV, $m_{\tilde{t}_2} = 914.9$ GeV and $m_{\tilde{b}_1} = 795.1$ GeV. Since the mass difference $m_{\tilde{t}_1} - m_{\tilde{\chi}^0_1}$ is less than the top mass, the decay $\tilde{t}_1 \to b\tilde{\chi}_1^\pm$ dominates, thus yielding a signature for $\tilde{t}_1$ pair production of two acollinear $b$-jets plus missing transverse energy. It is likely that the LHC experiments will eventually find the $\tilde{t}_1$, though at the moment the searches are not sensitive. Resolving the $\tilde{\chi}_1^\pm$, $\tilde{\chi}_1^0$ (and $\tilde{\chi}_2^0$) as distinct states will be extremely difficult at the LHC. Most other sparticles lie well beyond LHC reach.

For ILC, the spectrum of higgsino-like $\tilde{\chi}_1^\pm$, $\tilde{\chi}_1^0$ and $\tilde{\chi}_2^0$ would be accessible for $\sqrt{s} \gtrsim 320$ GeV via $\tilde{\chi}^+\tilde{\chi}^-$ and $\tilde{\chi}_2^0\tilde{\chi}_2^0$ pair production and $\tilde{\chi}_1^0\tilde{\chi}_2^0$ mixed production. although the energy release from decays will be small at beam energies near the threshold. Top squark pair production would become accessible when $\sqrt{s}$ exceeds about $575$ GeV.

### 7.4.2 An MSSM model with light sleptons

Using the freedom in the MSSM to decouple the masses of squarks and sleptons, we generated a model in the 13-parameter pMSSM that gives a spectrum of color singlet particles close to that of the well-studied SPS1a′ point [93]. The SPA1a′ point is phenomenologically well-motivated in that it naturally reconciles the measured $(g-2)_\mu$ anomaly (which favors light smuons) with the measured $b \to s\gamma$ branching fraction (which favors rather heavy third generation squarks). It furthermore predicts a neutralino relic density compatible with cosmological observations, making use of stau coannihilation. The SPA1a′ point belongs to the cMSSM and so is now excluded by LHC searches for squarks and sleptons. But it is easy to find a more general MSSM point that shares its virtues and is not yet tested by LHC searches. We call this the $\delta M\tilde{\tau}$ model. The particle masses of this model are listed in Table 2 of Ref. [91] and displayed in Figure 7.1.

With gluino and first/second generation squark masses around $2$ TeV, the model lies beyond current LHC limits, especially since the gluino decays dominantly via $\tilde{t}_1 t$ or $\tilde{b}_1 b$. The tau sleptons $\tilde{\tau}_1$ have masses of $104$ GeV, so stau pair production would be accessible even at the first stage of ILC running. Right-selectrons and smuons with mass $135$ GeV would also be produced at the ILC during the early runs, while left-sleptons and sneutrinos, with mass about $200$ GeV, would be accessible when $\sqrt{s}$ exceeds 400 GeV. The $\tilde{\chi}_1^0\tilde{\chi}_2^0$ reaction opens up at $\sqrt{s} > 250$ GeV, and $\tilde{\chi}_1^+\tilde{\chi}_1^-$ pair production is accessible for $\sqrt{s} \gtrsim 310$ GeV. In addition, with $m_{A,H} \sim 400$ GeV, $hA$ production opens at $525$ GeV, stop pair production at $600$ GeV, sbottom pair production at $680$ GeV and finally charged Higgses and $HA$ appear at $800$ GeV.

## 7.5 Experimental capabilities and parameter determination

In this section, we will review the ILC's experimental capabilities for precision measurements of SUSY particle properties. These measurements allow to determine the parameters of the underlying theory and to test its consistency at the quantum loop level.

As discussed above, the highly constrained cMSSM/mSUGRA models of supersymmetry are under tension from several different types of LHC observations. Therefore, we will discuss SUSY measurements in the more general context of the $CP$ and $R$-parity conserving MSSM. At the ILC, we will study the lightest particles of the SUSY spectrum, so the measurements that we will discuss involve simple reactions without complex cascade decay chains [94]. Thus, these measurements involve only a few of the MSSM parameters and, typically, those parameters can be determined with high precision.

We start with the minimal case in which only the lighter neutralinos and charginos are kinematically accessible. We then proceed to discuss sleptons and squarks, especially those of the third generation. Finally, we discuss possible extensions of the theory, encompassing $R$-parity violation, $CP$ violation, the NMSSM and the MSSM with an additional gauge group. We close with comments on model discrimination and parameter determination.

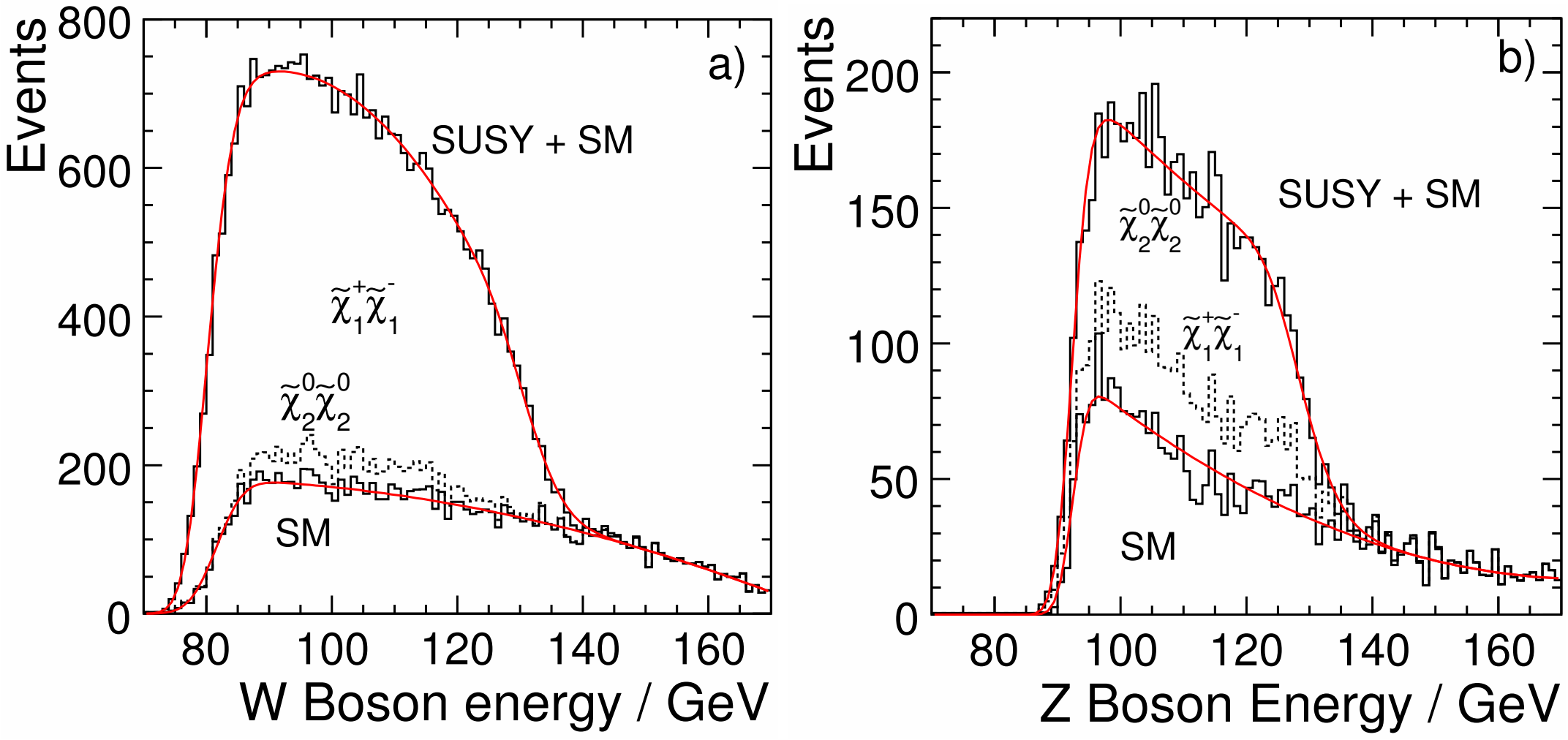

**Figure 7.2.** ILC measurements in chargino and neutralino pair production at 500 GeV. Top: Energy spectrum of the $W^\pm$ candidates reconstructed from events selected as $\tilde{\chi}^\pm_1$ pair production. Bottom: Energy spectrum of the $Z^0$ candidates reconstructed from events selected as $\tilde{\chi}^0_2$ pairs. From [103].

**Figure 7.3**
Measurement of the chargino mixing angles at the ILC at 500 GeV from the production cross-sections using $e^- Le^+ R$ and $e^-_R e^+_L$ polarized beams. From [104].

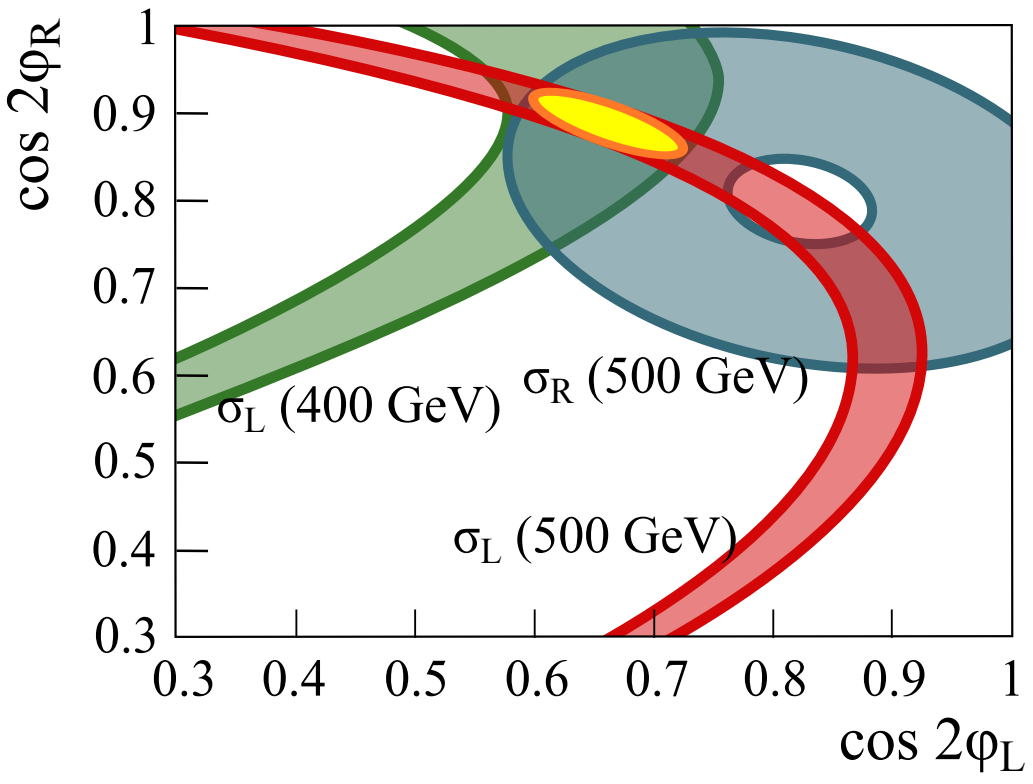

## 7.5.1 Neutralino and chargino sector

At the ILC, the electroweak gaugino sector can be probed in a model independent way up to masses of $\sqrt{s}/2$. Associated pair production can access masses above this value. The masses and couplings of the electroweak gauginos can be measured with high precision [95, 96]. Especially accurate values of the masses can be obtained through threshold scans, which have a precision below the per mil level [97, 98]. The relatively simple dynamics of pair production in $e^+e^-$ gives powerful methods for spin and quantum number determination [99].

Most of the SUSY models consistent with all experimental data feature light electroweakinos. These can either have dominant Bino/Wino components, or—as motivated by naturalness—dominant higgsino components. Examples of the latter case include the Natural SUSY benchmark introduced in section 7.4.1, as well as models with mixed gauge-gravity mediation [100], and the remaining points in the cMSSM parameter space. A more detailed overview of the light higgsino case is given in [91]. A characteristic pattern in all cases is a very small mass splitting between the $\chi^0_1$ and $\chi^\pm_1$ / $\chi^0_2$ of typically a few GeV or smaller. This small splitting is very difficult to resolve at the LHC. However, these states can be discovered and disentangled at the ILC by using ISR recoil techniques to overcome the background from 2-photon processes, and taking advantage of the capability of the detectors to observe the very soft visible decay products of the $\chi^\pm_1$ / $\chi^0_2$. These models can also lead to short displaced vertices that can be resolved thanks to the excellent vertex resolution at the ILC.

In the past, the case of small mass splitting between $\chi^\pm_1$ and $\chi^0_1$ has been studied in the context of AMSB models [101], where it has been shown that mass differences down to $50$ MeV can be resolved. For a $400$ MeV mass difference, it has been shown that the $\chi_1^\pm$ mass can be determined to $1.8$ GeV from the recoil against an ISR photon. Observing the energy of the single soft pion from the $\chi_1^\pm$ decay, the $\chi_1^\pm$–$\chi_1^0$ mass difference can be determined to $7$ MeV [102]. Although the minimal version of the AMSB is currently disfavoured due to its incompatibility with a Higgs mass of $125$ GeV, the fact that such small mass differences can be precisely measured at the ILC remains unchanged. In the Natural SUSY example discussed above, it is also true that the $\chi_2^0$ is nearly mass degenerate with the $\chi_1^\pm$. This creates an additional experimental complication, but on the other hand offers an additional handle for parameter determination. While a detailed experimental study is underway, the $\chi_2^0$ / $\chi_1^\pm$ separation should be possible when the various exclusive decay modes are exploited, which is feasible due to the clean environment and excellent detector resolutions available at the ILC. The measurement of the polarization and beam energy dependence of the cross-sections of these processes then allows us to establish the higgsino character of the particles and to precisely determine $\mu$.

If the mass difference between $\chi_1^\pm$ or $\chi_2^0$ and $\chi_1^0$ is larger than about $80$ GeV without sleptons in between, the decays of these particles will proceed via real $W^\pm$ or $Z$ bosons. In the challenging case where $\chi_1^\pm$ and $\chi_2^0$ are nearly mass degenerate, their decays can be disentangled even in the fully hadronic decay mode. This case has been studied both by SiD and ILD in full detector simulation. Figure 7.2 shows the energy spectra of the reconstructed gauge boson candidates from signal, SUSY and SM background for the chargino and neutralino event selection [103]. Assuming an integrated luminosity of $500$ fb$^{-1}$ at $\sqrt{s} = 500$ GeV and a beam polarization of $P(e^+, e^-) = (30\%, -80\%)$, the edge positions can be determined to a few hundred MeV. Due to sizable correlations, this translates into uncertainties of $2.9$, $1.7$ and $1.0$ GeV for the $\chi_2^0$, $\chi_1^\pm$ and the $\chi_1^0$ masses, respectively. The cross-sections can be measured to $0.8\%$ $(2.8\%)$ in the $\chi_1^\pm$ $(\chi_2^0)$ case from the hadronic channel alone.

Independently of the mass splitting, the polarized cross-section measurements at different center-of-mass energies can be employed to determine the mixing angles in the chargino sector, as illustrated in Figure 7.3 [104]. This example is based on simulations performed in the SPS1a scenario; the results also apply to the $\delta M\tilde{\tau}$ scenario introduced above. The bands include both statistical and systematical uncertainties, where the limiting contribution is the precision of the chargino mass.

More recently, it has been shown that the achievable experimental precision allows us also to determine the top squark masses and mixing angle via loop contributions to the polarized chargino cross-sections and the forward-backward asymmetries [105]. This allows us to predict and to constrain the heavier states of the SUSY model and to test its structure directly independently of the SUSY breaking scheme.

### 7.5.2 Gravitinos

If the gravitino is lighter than the lightest neutralino, the neutralino could decay into a photon plus a gravitino. In such a case, the lifetime of the neutralino is related to the mass of the gravitino: $\tau_\chi \sim m_{3/2}^2 M_{Pl}^2 / m_\chi^5$. Therefore the measurement of the neutralino lifetime gives access to $m_{3/2}$ and the SUSY breaking scale. A similar statement applies to models in which a different particle is the lightest Standard Model superpartner, decaying to the gravitino. A well-studied example is that of the $\tilde{\tau}$ NLSP. The experimental capabilities of a Linear Collider in scenarios with a gravitino LSP have been evaluated comprehensively many years ago [106], where it has been demonstrated that with the permille level mass determinations from threshold scans, the clean environment and the excellent detector capabilities, especially in tracking and highly granular calorimetry, fundamental SUSY parameters can be determined to $10\%$ or better.

Although this study was based on minimal GMSB models, which are currently disfavoured by their prediction of too low masses for the Higgs boson and the gluino, the signatures and experimental techniques remain perfectly valid. They could apply to other non-minimal scenarios including general gauge mediation. Aspects of the detector performance which were still speculative when the studies in [106] were performed have been established in the intervening time with testbeam data from prototype detectors. For instance, the performance of neutralino lifetime determination from non-pointing clusters in the electromagnetic calorimeter has recently been reevaluated based on full detector simulation gauged against Calice testbeam data. These confirm the estimates from [106] that lifetimes between $0.1$ and $10$ ns can be reconstructed with a few percent accuracy, although a calibration of the lifetime reconstruction is needed [107]. Similarly it has been shown in [108], that, in the case of a $\tilde{\tau}$ NLSP, the lifetime can be measured down to $10^{-5}$ ns, corresponding to gravitino masses of a few eV. Figure 7.4 shows the $1\sigma$ and $2\sigma$ uncertainty bands as a function of the lifetime of a $\tilde{\tau}$ with a mass of $120$ GeV.

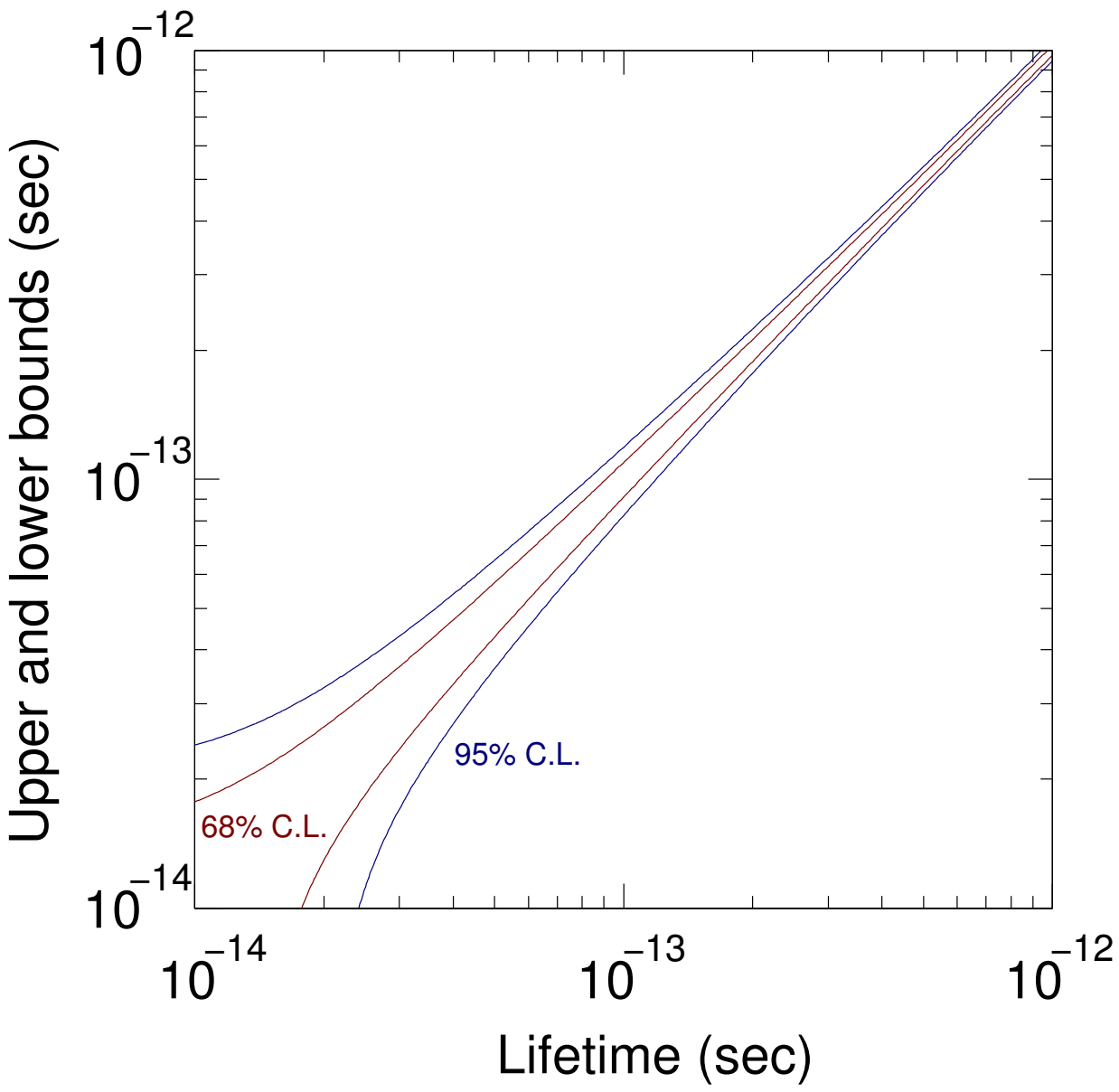

**Figure 7.4** ILC measurement of the lifetime of the gravitino in a GMSB scenario, from [108]. The $1\sigma$ and $2\sigma$ uncertainty bands are shown as a function of the lifetime of a $\tilde{\tau}$ with a mass of $120$ GeV.

Scenarios with very long-lived $\tilde{\tau}$ NLSPs which get trapped in the calorimeter and decay much later have been studied in [109]. It has been shown there that, with a suitable read-out of the ILC detectors, the gravitino mass and the SUSY breaking scale can also be determined in such cases. The first signs of these heavy, detector-stable charged particles would their large ionization losses in the tracking volume. This is a nearly background-free signature even at the LHC, so it is also possible there to discover electroweak production of very long-lived $\tilde{\tau}$ NLSPs or $\tilde{\chi}_1^{\pm}$ NLSPs. If this discovery were made, it would be important and fascinating to measure the polarized electroweak cross sections of these particles with high precision at the ILC.

### 7.5.3 Third generation squarks

At the ILC, the stop $\tilde{t}_1$ can be probed up to $m_{\tilde{t}_1} = \sqrt{s}/2$ regardless of its decay mode and the masses of other new particles. At $\sqrt{s} = 500$ GeV, the $\tilde{t}_1$ mass can be determined to $1$ GeV in the $\tilde{t}_1 \to c\tilde{\chi}_1^0$ decay mode, which dominates for small mass differences, and to $0.5$ GeV in the $\tilde{t}_1 \to b\tilde{\chi}_1$ mode [110]. At the same time, the stop mixing angle can be determined to $\Delta\cos\theta_t = 0.009$ and $0.004$ in the neutralino and chargino modes, respectively. A more recent study improved the mass resolution in the $\tilde{t}_1 \to c\tilde{\chi}_1^0$ decay to $0.42$ GeV, including systematic uncertainties estimated based on LEP experience by assuming data from two different center-of-mass energies [111]. In a top-squark co-annihilation scenario, the predicted dark matter relic density depends strongly on the stop-neutralino mass difference. The precise ILC mass measurenents give an uncertainty on the calculated dark matter relic density of $\Delta\Omega_{\mathrm{CDM}}h^2 = 0.015$, comparable to the current WMAP precision. Figure 7.5 shows the correlation between the stop mass and $\Omega_{\rm CDM}h^2$ and the respective precisions. This clearly shows that sub-GeV precision on the stop mass is mandatory to establish the $\tilde{\chi}_1^0$ as a cosmic relic. Although these studies were performed with slightly lower stop masses, one can expect similar precisions in the two scenarios introduced in section 7.4 if on the way to a $1$ TeV upgrade the ILC is operated at a center-of-mass energy of $600$ GeV or above. And, indeed, there is still much room for the $\tilde{t}_1$ to be found at the LHC at a mass below 250 GeV.

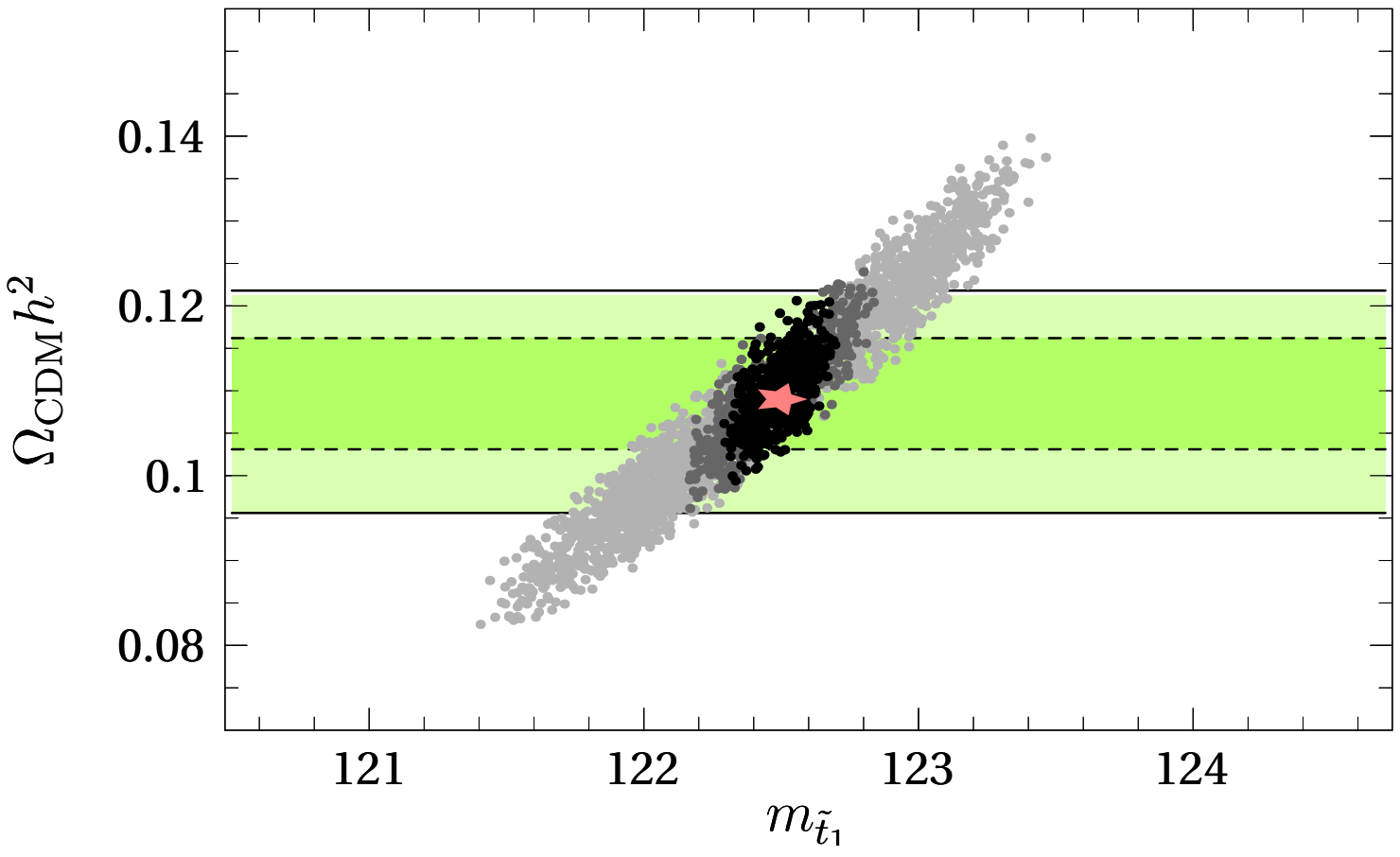

**Figure 7.5.** Predicted dark matter density $\Omega_{DM}$ in a stop coannihilation model. The scatter plot shows the values of $\Omega_{DM}$ and $m_{\tilde{t}_1}$ for models consistent with ILC observations within $1\sigma$ experimental precision, assuming a meaurement error on the mass splitting in the decay, $\delta\tilde{t}_1$, of 1.2 GeV (light gray), 0.42 GeV (dark gray) and 0.24 GeV (black). The bands show the current WMAP precision on $\Omega_{DM}$. The input value is marked with a star. From [111].

The polarized cross sections $\sigma(e^-_L e^+_R \to \tilde{t}_1\bar{\tilde{t}}_1)$ and $\sigma(e^-_R e^+_L \to \tilde{t}_1\bar{\tilde{t}}_1)$ allows a direct determination of the $(\tilde{t}_L, \tilde{t}_R)$ mixing angle with an accuracy of a few degrees. This is crucial information for the theory of electroweak symmetry breaking in SUSY and for the explanation for the Higgs boson mass at 125 GeV.

In sbottom-co-annihilation scenarios, which typically exhibit a sbottom-LSP mass difference of about $10\%$ of the LSP mass, the process $\tilde{b}_1 \to b\tilde{\chi}_1^0$ can be discovered for sbottom masses up to about $10$ GeV below the kinematic limit and for mass differences down to only $5$ GeV larger than the kinematic limit [112]. It will be extremely difficult to cover such small mass differences comprehensively at the LHC.

Additional interesting reactions arise if the stop and sbottom decay to Higgsinos. We have argued that, because of naturalness, this is the expected situation. Then the charginos are close in mass to the neutralinos, allowing the decay $\tilde{t} \to \tilde{\chi}^+ b$, with a subsequent decay of the $\tilde{\chi}^+$ with small missing energy release. The ability of the ILC to study decay chains with small energy differences will be important if this is a dominant mode.

### 7.5.4 Scalar charged leptons

For slepton masses below $\sqrt{s}/2$, sleptons could be produced copiously at the ILC without relying on cascades from heavier sparticles. The lighter sleptons typically decay directly into the corresponding lepton and the lightest neutralino, giving a very clear signature of two isolated same flavor opposite sign leptons and missing four-momentum. The lepton energy spectrum has a box-like shape, and its lower and upper edge give direct access to the slepton and neutralino mass. In practice, the box is slightly smeared by the beam energy spectrum, ISR, detector resolution and, in case of $\tau$ leptons, by the unmeasured neutrinos from the $\tau$ decay. Nevertheless, this technique works reliably down to very small mass differences of a few GeV. For mass differences below $\sim 10$ GeV, the lower edge is buried in background from 2-photon processes. Then an additional observable is needed to determine the lightest neutralino mass. The adjustable center-of-mass energy of the ILC allows us to achieve even higher precision by scanning the production thresholds.

In SUSY, the superpartners of the left- and right-handed leptons are distinct scalar particles with different electroweak quantum numbers. These particles can be distinguished at the ILC in a model-independent way by the measurement of their production cross sections from left- and right-polarized beams in $e^+e^-$ annihilation [113]. It is not expected that the left- and right-sleptons should be mass degenerate, but, even in this case, the two particles can be studied separately, since each has enhanced production in cases with electron beams of the same handedness. For the case of $\tau$ sleptons, the polarization of the $\tau$ leptons produced in the decay can be analyzed to provide another powerful probe of the slepton quantum numbers and couplings [114].

The heavier sleptons typically decay via intermediate charginos, neutralinos or sneutrinos, depending on the details of the spectrum [94]. By choosing an intermediate center-of-mass energy, the production of heavier superpartners and thus the background from their cascades can be switched off. This allows the ILC experiments to disentangle even rich spectra similar to the $\delta M\tilde{\tau}$ scenario discussed above.

The $\tilde{\tau}$ sector of a scenario very similar to $\delta M\tilde{\tau}$ has recently been studied in full simulation with the ILD detector [115], since the small $\tilde{\tau}$-$\tilde{\chi}_1^0$ mass difference provides an interesting challenge for the detector and the accelerator conditions. In this case, the beam energy spectrum was accounted for and also accelerator background from $e^+e^-$ pairs created from beamstrahlung was overlayed in order to verify the robustness of the reconstruction even of fragile final states such as soft $\tau$ leptons against spurious tracks and clusters from beam background.

With an integrated luminosity of $500$ fb$^{-1}$ at a center-of-mass energy of $\sqrt{s} = 500$ GeV and with $P(e^+, e^-) = (-30\%, +80\%)$, the following results were achieved for the $\tilde{\tau}$ masses using pair production cross-sections and the $\tau$ polarisation $\mathcal{P}_\tau$ from $\tilde{\tau}$ decays. Both of these quantities depend on the $\tilde{\tau}$ mixing angle, the higgsino component of the $\tilde{\chi}_1^0$ and $\tan\beta$ in a well-understood way.

$$\begin{aligned}
\delta M(\tilde{\tau}_1) &= {}^{+0.03}_{-0.05} \pm 1.1 \cdot \delta M(\tilde{\chi}_1^0) \text{ GeV (endpoint)} \\
\delta M(\tilde{\tau}_2) &= {}^{+11}_{-5} \pm 18 \cdot \delta M(\tilde{\chi}_1^0) \text{ GeV (endpoint)} \\
\frac{\delta\sigma}{\sigma}(\tilde{\tau}_1) &= 3.1\ \% \\
\frac{\delta\sigma}{\sigma}(\tilde{\tau}_2) &= 4.2\ \% \\
\mathcal{P}_\tau &= 91 \pm 6 \pm 5 \text{ (bkg)} \pm 3 \text{ (SUSY masses) } \% \text{ } (\pi \text{ channel}) \\
\mathcal{P}_\tau &= 86 \pm 5\ \% \text{ } (\rho \text{ channel}).
\end{aligned}$$

The measurement of the endpoint of the $\tau$ jet energy spectrum from $\tilde{\tau}_1$ decays is shown in Figure 7.6. The $\tilde{\tau}$ mixing angle can be determined independently of the $\tau$ polarisation from $\tilde{\tau}_1\tilde{\tau}_2$ associated production below the $\tilde{\tau}_2$ pair production threshold. With a dedicated threshold scan, the $\tilde{\tau}_2$ mass measurement can be improved to $\delta M(\tilde{\tau}_2) \approx 0.86$ GeV [116]. Even smaller mass differences have been studied in an earlier fast simulation analysis [117], which found $\delta M(\tilde{\tau}_1) \approx 0.15 - 0.3$ GeV depending on $\tilde{\tau}_1$ mass and the $\tilde{\tau}_1$-$\tilde{\chi}_1^0$ mass difference.

Since the measurement of isolated electrons and muons is straightforward for the ILC detectors, scalar electron and muon production have mainly been studied in fast detector simulations. In [117,118], a scenario similar to $\delta M\tilde{\tau}$ has been studied assuming an integrated luminosity of $200$ fb$^{-1}$ and beam polarisations of $P(e^+, e^-) = (-60\%, +80\%)$ at a center-of-mass energy of $\sqrt{s} = 400$ GeV. The study found precisions of $\delta M(\tilde{\mu}_R) \approx 170$ MeV and $\delta M(\tilde{e}_R) \approx 90$ MeV. Comparable values were found in [116], where in addition a precision of $20$ MeV was achieved for $M(\tilde{e}_R)$ from a threshold scan. This kind of precision below $100$ MeV can typically be obtained when no irreducible SUSY background from other cascades is present.

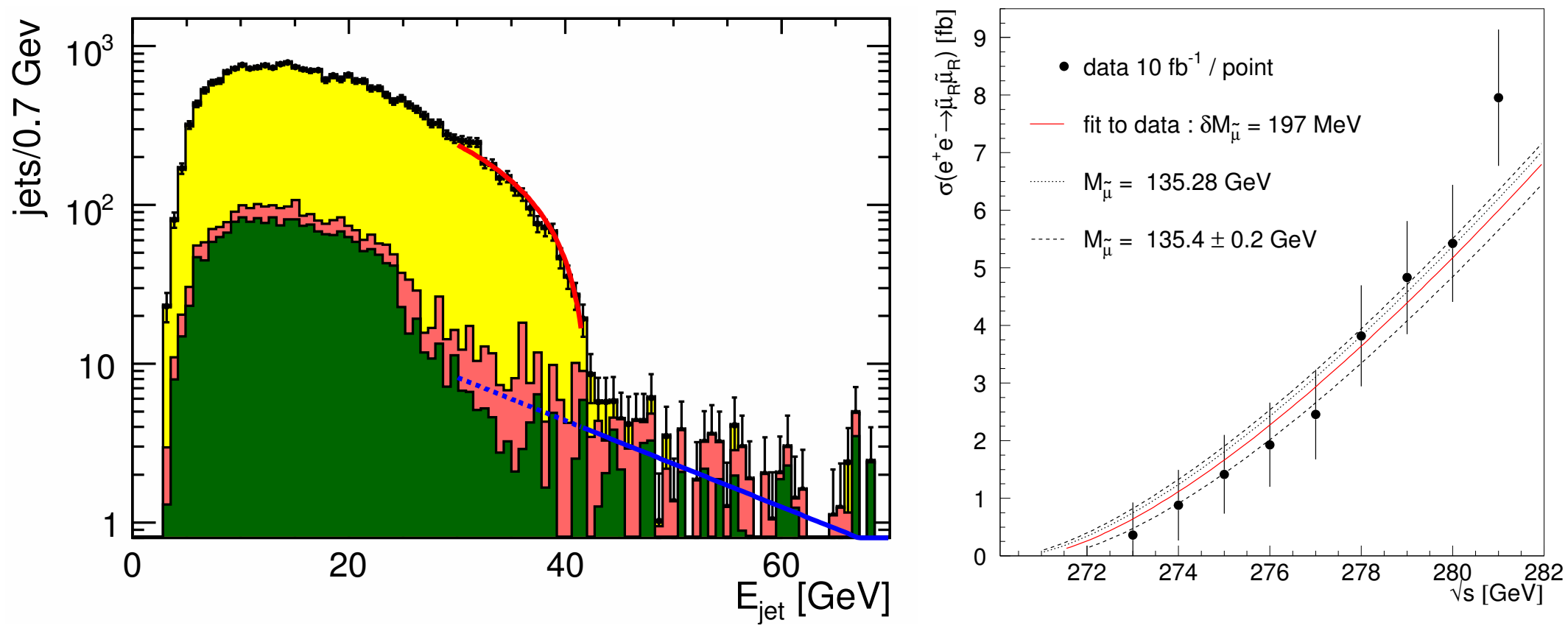

**Figure 7.6.** Left: Measurement of the $\tilde{\tau}_1$ mass from the endpoint of the $\tau$ jet energy spectrum in a scenario with small $\tilde{\tau}_1$-$\chi_1^0$ mass difference very similar to the $\delta M\tilde{\tau}$ scenario introduced in Section 7.4.2. The stacked histogram contains (from the bottom), SUSY background, SM background, signal. The background is fitted in the signal-free region to the right (solid portion of the line), and extrapolated into the signal region (dashed). From [115]. Right: Measurement of the $\tilde{\mu}_R$ mass from a threshold scan with a total integrated luminosity of $100$ fb$^{-1}$. The precision of about 200 MeV obtained in this study is limited by the assumed integrated luminosity [120].

The $\delta M\tilde{\tau}$ scenario is actually challenging in this respect, since substantial background from neutralino decays into muons is present at the $\tilde{\mu}_R$ pair production threshold. This case has recently been studied using the fast simulator SGV [119] tuned to the detector performance found in full simulation of the ILD detector concept. All relevant SM backgrounds, especially $W^+W^- \to l^+\nu l^-\overline{\nu}$, $ZZ \to$ 4 leptons, and $\mu$ and $\tau$ pairs, as well as all open SUSY channels were generated with Pythia 6.422 at 9 center of mass energies near the $\tilde{\mu}_R$ threshold. The simulations included beamstrahlung based on Circe 1 and the incoming beam energy spectrum according to the TDR design of the ILC. The measured cross-section as a function of the center of mass energy is shown in Figure 7.6 assuming $10$ fb$^{-1}$ per point with $P(e^+, e^-) = (-30\%, +80\%)$. A fit to the threshold yields a statistically limited uncertainty of about $200$ MeV on the $\tilde{\mu}_R$ mass [120].

In case of the heavier smuon $\tilde{\mu}_L$, a mass resolution of $100$ MeV has been achieved in full simulation for the ILD Letter of Intent assuming $500$ fb$^{-1}$ with $P(e^+, e^-) = (+30\%, -80\%)$ at $\sqrt{s} = 500$ GeV [121]. This is consistent with earlier fast simulation studies [98, 116].

All resolutions here are by far statistically limited. Masses or cross-sections critical for SUSY parameter determination in a certain scenario could therefore be measured with even better precision when more integrated luminosity is accumulated in the corresponding running configuration of the machine.

### 7.5.5 Sneutrinos

Depending on the properties of the sparticle spectrum, sneutrinos may decay visibly into modes such as $\tilde{\nu}_\ell \to \ell\tilde{\chi}_1^+$ [2], or they may decay invisibly via $\tilde{\nu}_\ell \to \nu_\ell\tilde{\chi}_1^0$. Even in this latter case, the sneutrino mass can be measured from cascade decays of other sparticles. For instance, in the $\delta M\tilde{\tau}$ scenario, the chargino has a 13% branching fraction into a sneutrino and the corresponding charged lepton. From these decays, the sneutrino mass can be reconstructed to $\delta M(\tilde{\nu}) \approx 0.5$ GeV [122, 123].

Sneutrinos which are too heavy to be produced directly still influence the cross section for chargino production and the forward-backward asymmetry of three-body chargino decays. The latter yields $\delta M(\tilde{\nu}) \approx 10$ GeV for sneutrino masses up to $1$ TeV at $\sqrt{s} = 500$ GeV [98]. The chargino pair production cross-section is sensitive to sneutrino masses of up to $12$ TeV at center-of-mass energies $\sqrt{s} \sim 1$ TeV [124].

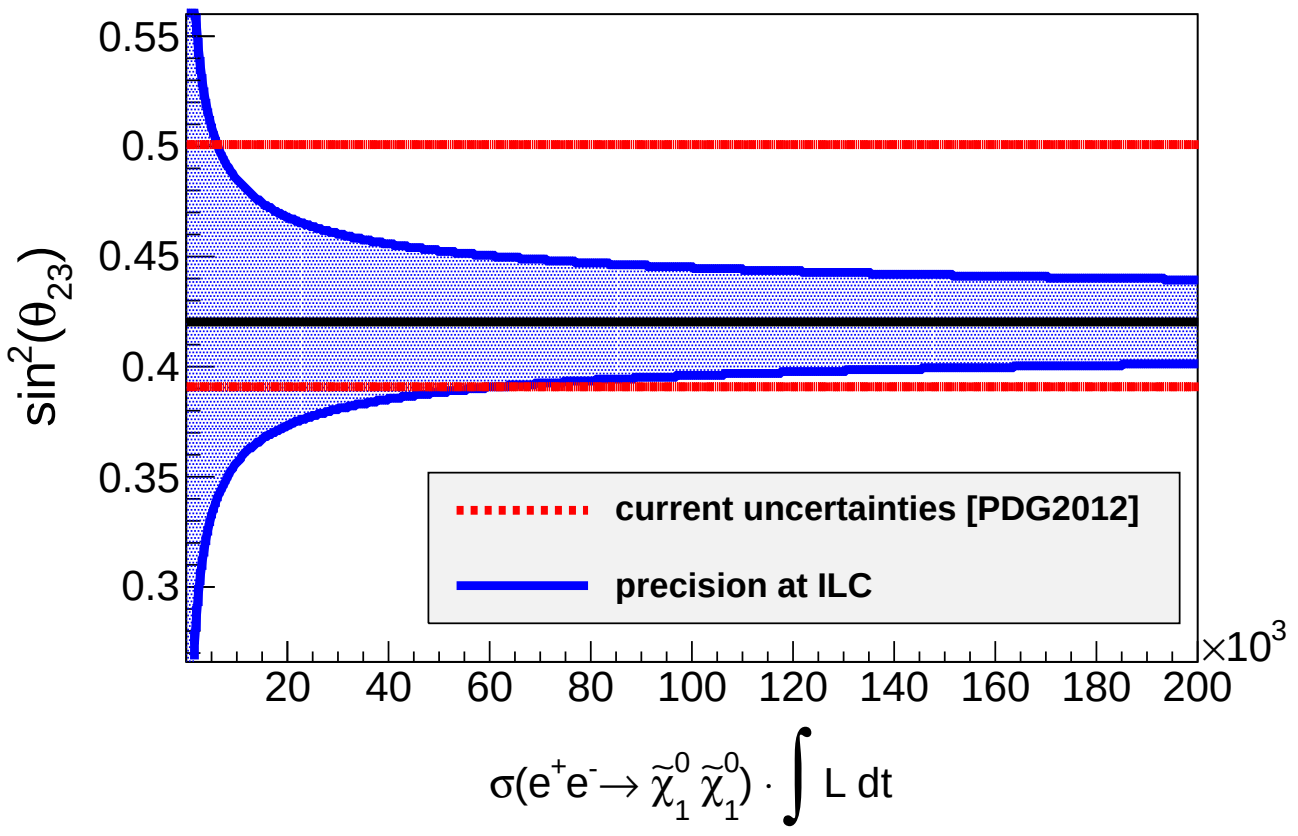

**Figure 7.7.** The precision expected from ILC measurements on the neutrino mixing angle $\sin^2\theta_{23}$ from RPV decays of the $\chi^0_1$, shown as a function of the number of neutralino pairs produced [129]. The dotted lines give the current precision from neutrino oscillation measurements. Over a large part of the $m_0$ vs. $m_{1/2}$ plane, the neutralino pair production cross-section is of order $100$ fb.

### 7.5.6 Beyond the CP and RP conserving MSSM

*$R$-Parity Violation:*

$R$-parity violation (RPV) has two important experimental consequences at colliders: it allows for single production of SUSY particles, and it allows the LSP to decay to purely SM particles. The latter aspect makes RPV SUSY much harder to detect at the LHC due to the absence of missing transverse energy, so that the currently explored region is significantly smaller than in the $R$-parity conserving case, even when assuming mass unification at the GUT scale as in the cMSSM [125].

Bilinear $R$-parity violation (bRPV) has phenomenological motivations in neutrino mixing [126] as well as in leptogenesis [127, 128]. In this case, the characteristic decay $\tilde\chi^0_1 \to W^\pm l^\mp$ will lead to background-free signatures at the ILC, possibly with a detectable lifetime of the $\tilde\chi^0_1$ depending on the strength of the RPV couplings. In the hadronic decay mode of the $W^\pm$, these events can be fully reconstructed and the $\tilde\chi^0_1$ mass can be measured to $\mathcal{O}(100)$ MeV depending on the assumed cross-section [129]. By measuring the ratio of the branching ratios for $\tilde\chi^0_1 \to W^\pm\mu^\mp$ and $\tilde\chi^0_1 \to W^\pm\tau^\mp$, the neutrino mixing angle $\sin^2\theta_{23}$ can be determined to percent-level precision, as illustrated in Figure 7.7. Agreement with measurements from neutrino oscillation experiments would then prove that RPV SUSY is the origin of the structure of mixing in the neutrino sector.

In the case of trilinear R-parity violation, $s$-channel sneutrino-exchange can interfere with SM Bhabha scattering. For $m_{\tilde\nu} < \sqrt{s}$, sharp resonances are expected. In addition, heavier sneutrinos could be detected via contact interactions, for example up to $m_{\tilde\nu} = 1.8$ TeV for $\lambda_{1j1} = 0.1$ at $\sqrt{s} = 800$ GeV [101].

*$CP$ violation:*

An attractive feature of supersymmetry is that it allows for new sources of $CP$ violation which are needed in order to explain the baryon-antibaryon asymmetry observed in the universe. The neutralino and chargino sector can accommodate two independent $CP$ phases, for instance on $M_1$ and $\mu$ when rotating away the phase of $M_2$ by a suitable redefinition of the fields. While the phase of $\mu$ is strongly constrained by EDM bounds, the phase of $M_1$ could lead to $CP$ sensitive triple product asymmetries up to $10\%$. These can be measured from neutralino two-body decays into slepton and lepton to $\pm1\%$. From a fit to the measured neutralino cross-sections, masses and $CP$-asymmetries, $|M_1|$ and $|\mu|$ can be determined to a few permille, $M_2$ to a few percent, $\Phi_1$ to $10\%$ as well as $\tan\beta$ and $\Phi_\mu$ to $16\%$ and $20\%$, respectively [130]. Other models of baryogenesis accessible to study at the ILC are discussed in Section 8.1.

*NMSSM:*

If indeed the higgsino is the LSP, as motivated by naturalness, then all by itself it is not a good dark matter candidate, since higgsino pairs annihilate rapidly into $WW$ and $ZZ$. However, if we invoke an extended Higgs sector (the NMSSM) to explain the value of the Higgs boson mass, this extension adds a new SUSY partner, the singlino, which might have mass below that of the higgsino. The decay width of the higgsino to the singlino is of order 100 MeV. The pattern of decay final states is rich, and the measurement of branching ratios will illuminate the Higgs sector [131]. These decay products are quite soft, however, and are invisible under the standard LHC trigger constraints. Whether or not these particles can be seen at the LHC, the ILC would again be needed for a complete study. The annihilation cross section of singlinos, which determines the singlino thermal dark matter density, depends on the singlino-higgsino mixing angle. This could be measured at the ILC by measurement of the higgsino width using a threshold scan or by precision measurments of the NMSSM mass eigenvalues.

The capabilities of the ILC to distinguish between the NMSSM and the MSSM when the observable particle spectrum and the corresponding decay chains are very similar has been studied for instance in [132] based on analytical calculations. The study showed that with data taken at three different center-of-mass energies the distinction is possible. When exploiting the available information even more efficiently by applying a global fit, even two center-of-mass energies can be sufficient [133]. If the full neutralino/chargino spectrum is accessible, sum rules for the production cross sections can be exploited that show a different energy behaviour in the two models.

In scenarios where the lightest SUSY particle is nearly a pure singlino, the higgino lifetimes are long, leading to a displaced vertex signature. The lifetimes can be precisely resolved thanks to the excellent vertex resolution of the ILC detectors.

### 7.5.7 Parameter determination and model discrimination

Beyond simply measuring the properties of new particles, a further goal of ILC is to fully uncover the underlying theory. This involves, among other issues, the measurement of the statistics of the new particles and the verification of symmetry predictions of the model. In this, we review some examples of such studies.

For example, if only the minimal particle content of a weakly interacting new particle $\chi^0$ and an electrically charged partner $\chi^\pm$ is observed, the behaviour of the production cross-section at threshold and the production angle distribution of $\chi^+\chi^-$ pair production can be employed to distinguish between SUSY, where the $\chi$'s are fermions, Littlest Higgs models, where they are vector bosons, and Inert Higgs models, where they are scalar bosons [134].

If the model is indeed SUSY, we would like to establish the basic symmetry relation of supersymmety experimentally. This can be done by examining whether the gauge couplings $g(Vff)$ and $\overline{g}(V\tilde{f}\tilde{f})$ of a vector boson $V$ and the Yukawa coupling $\tilde{g}(\tilde{V}f\tilde{f})$ for corresponding gauginos are equal [135]. From the various cross-section measurements in the slepton and gaugino sector, these couplings can be extracted and their equality checked with sub-percent precision [1, 2, 98].

In addition to the couplings, the mass measurements at ILC, at the per mille level, allow one to extract the weak scale MSSM parameters. Here the polarized beams play a crucial role since they allow us to determine the mixing character both in the gaugino and in the slepton sector, especially if left- and right-handed superpartners are close in mass and thus difficult to separate kinematically. These parameters can then be extrapolated to higher energy using the renormalization group equations [136].

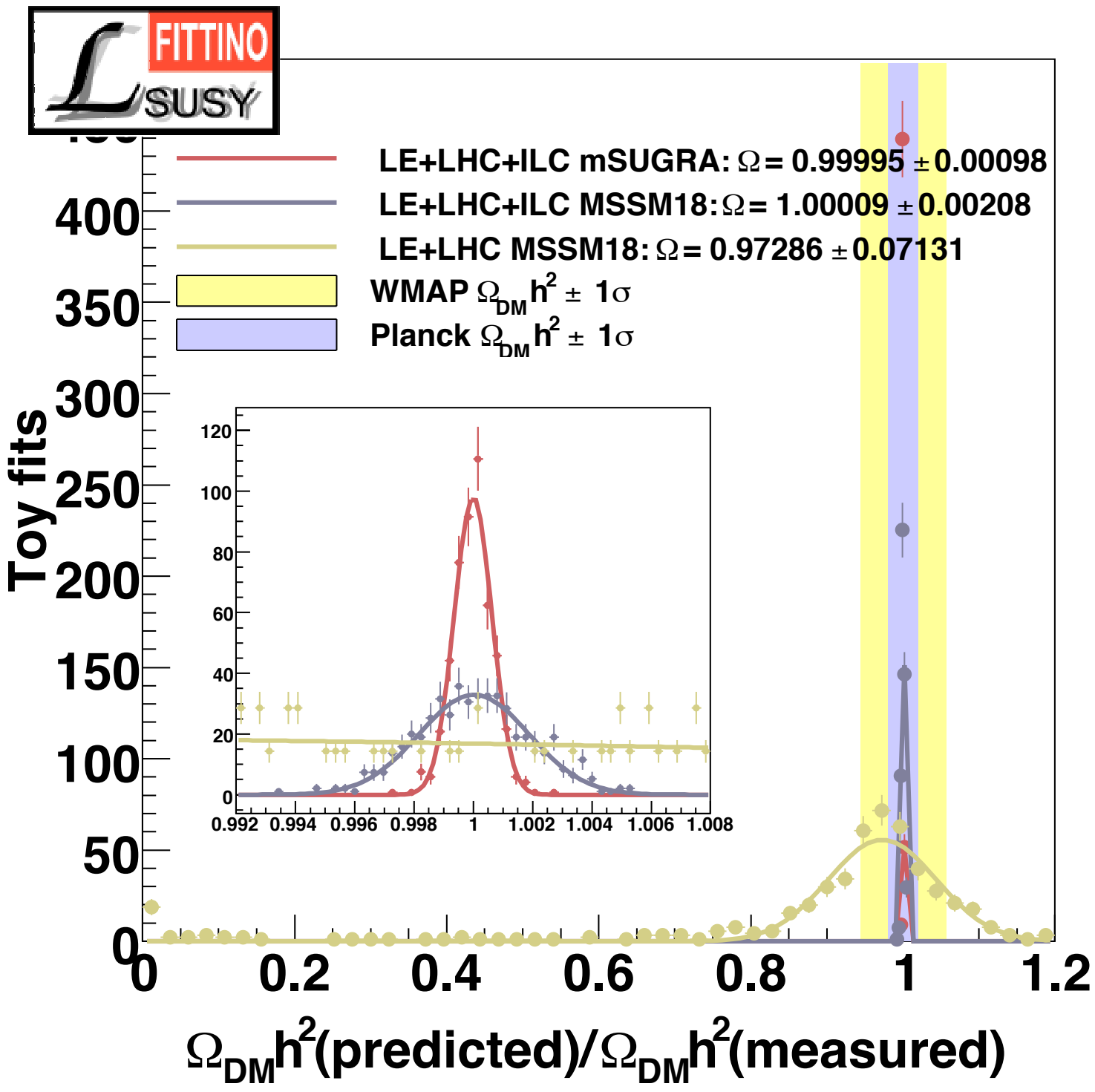

**Figure 7.8** Ratio of the predicted value of the dark matter density $\Omega_{\mathrm{pred}}h^2$ to the nominal value of $\Omega_{\mathrm{SPS1a}}h^2$ in the SPS1a scenario, for a variety of Fittino Toy Fits without using $\Omega_{\mathrm{CDM}}h^2$ as an observable, from [142]. The anticipated predictions for LHC and ILC measurements on this model are compared to current and projected cosmological observations.

This might reveal that groups of these parameters unify, for example, at the GUT scale. The impact of ILC precision on this procedure has been studied in detail in [137], based on a scenario in which the color singlet sector is nearly identical to that of the $\delta M\tilde{\tau}$ scenario. They found that the weak scale parameters can be determined to percent level precision, some even to the per mille level. They further showed that ILC precision, beyond that achievable at the LHC, is needed to establish whether the weak scale parameters are consistent with a certain SUSY breaking scheme (in this case mSUGRA) or not. MSSM parameter determinations, both analytically and employing global fits, have been studied also in various other scenarios in [93, 138–140].

Another crucial question to be answered is that of whether the lightest SUSY particle can account for some or all of the cosmological dark matter. Assuming that lightest SUSY particle was produced thermally in the early universe, its relic density can be computed from the Lagrangian parameters obtained from collider data and the result can be compared to the observed value of the dark matter density [141]. The Fittino collaboration has studied the prediction of the dark matter density from ILC data at the reference point $SPS1a'$, which, for this analysis, is very similar to the $\delta M\tilde{\tau}$ scenario [142]. Figure 7.8 shows the result of this comparison without assuming a specific SUSY breaking scenario, *i.e.* based on weak scale parameters. In this scenario, the ILC precision is needed to match the precision of the prediction to that expected from cosmological observations.

The $SPS1a'/\delta M\tilde{\tau}$ point is a rather special case in which $\Omega_{\mathrm{CDM}}h^2$ can be predicted with part per mille accuracy. More typically, the mechanisms that establish the dark matter relic density are more complex, and the accuracy of the prediction from collider data is less. We have seen an example already in Section 4.5.3 in our discussion of the stop coannihilation scenario. However, the more complex the physics of the dark matter density, the more important it is to make high precision measurements of the SUSY parameters. This important question will be discussed further in Section 8.2.

## 7.6 Conclusions

In this section, we have discussed the ILC capabilities for supersymmetry measurements in the light of the new information that we have gained from the LHC experiments. The discovery of a new boson at 125 GeV points to a mechanism of electroweak symmetry breaking that involves weakly coupled scalar fields. Supersymmetry is one of, if not the leading candidate, for such a model.

So far, the ATLAS and CMS experiments have found no evidence for supersymmetric particles. They have presented impressive limits on the masses of squarks and gluinos. However, these limits do not exclude the possibility of SUSY at the TeV scale. Rather, they push us to explore SUSY models in different parameter regions of the MSSM than those that have been given most attention in the past.

In particular, the LHC exclusions have focused much attention on models in which the first- and second-generation squarks are heavy while the naturalness of the electroweak symmetry breaking scale keeps color singlet particles light. Naturalness arguments, in particular, favor a low value of $\mu \sim M_Z$, with $\mu$ ranging perhaps as high as 200–300 GeV. This then leads to a spectrum including several light higgsino-like charginos and neutralinos. The lightest neutralino, which is a possible WIMP candidate, would be predominantly higgsino-like. The light higgsinos are automatically mass-degenerate with typical mass gaps of 10-20 GeV. The small energy release from higgsino decay would be very difficult to detect at LHC. In contrast, an ILC with $\sqrt{s} = 0.25 - 1$ TeV would be a *higgsino factory*, in addition to being a Higgs factory! These arguments, and also possibly the muon $g - 2$ anomaly, predict a rich array of new matter states likely accessible to the ILC.

In our review of the experiments at the ILC that would discover and measure the properties of these particles, we have emphasize the many tools that the ILC detectors will provide for exploring the nature of these new states of matter. These include the tunable beam energy, the use of beam polarization, precision tracking, vertex finding and calorimetry, which provide the ability to detect very low energetic particles as well as to observe and separate $W$ and $Z$ in hadronic modes. We have shown with many examples that all of these capabilities find new uses in the exploration of a new sector of particles.

The precision measurements available at the ILC will provide a window to physics at much higher energy scales, possibly those associated with grand unification and string theory. The ILC will also provide a key connection between particle physics and cosmology, especially in identifying the nature of dark matter and shedding light on possible mechanisms for baryogenesis.

Supersymmetry is challenged by the new results from the LHC, but this theory is still very attractive for the answers that it gives to the pressing theoretical problems of the Standard Model. The constraints from the LHC guide us to new regions of the large parameter space of supersymmetry. The ILC will explore these regions definitively and make precise measurements of new particles that may be found there. From this perspective, the construction of an ILC is more highly motivated now than ever before.

# Chapter 8
# Cosmological Connections

Two of the major puzzles of cosmology can be explained with new physics at the electroweak scale. These are the matter-antimatter asymmetry, which might be due to baryogenesis at the electroweak phase transition, and the dark matter of the universe, which might be composed of a stable weakly-interacting massive particle (WIMP) with a mass at the hundred GeV scale. We have seen references to both of these mechanisms that might act in the early universe in our discussions of the top quark, extended Higgs sectors, and supersymmetry. In this chapter, we review these topics in a more unified way.

Both electroweak baryogenesis mechanisms and WIMP candidates naturally arise within the two major paradigms for explaining electroweak symmetry breaking, supersymmetry and Higgs compositeness. To work correctly, these phenomena require quite specific details of the spectrum and parameter choices. These details must be verified if we are to understand whether the particles observed at the TeV scale indeed suffice to explain these major cosmological mysteries. The details that we must learn concern aspects of the TeV scale physics that are especially difficult to access at hadron colliders—knowledge of the Higgs spectrum and couplings and the properties of other color-singlet particles, and understanding of the more general new particle spectrum in situations with compressed spectra and small energy release in decays.

The capabilities of the ILC that we have described in earlier chapters are sufficient to meet these challenges. This might be the strongest motivation for the construction of the ILC, that it provides unique opportunities to understand the basic mechanism that form the universe we see around us.

## 8.1 Baryogenesis at the Electroweak Scale

Among the mechanism for creating the baryon number of the universe, an especially attractive one is the idea that this asymmetry was created as a result of the electroweak phase transition [1]. The high temperature phase of the Standard Model contains a mechanism for baryon number violation, the thermally activated sphaleron solution of the $SU(2)$ gauge theory, which has the ability to simultaneously violate baryon and lepton number. A net baryon asymmetry can be produced if the two other Sakharov conditions are satisfied, that is, if the theory has sufficient appropriate CP violation and if the electroweak phase transition is first-order [2]. The process is non-local, relying on the dynamics in the vicinity of expanding bubbles that grow the broken symmetry phase out of the supercooled high-temperature symmetric phase of the electroweak theory. The walles of these bubbles carry the CP violating interactions [3]. Because it involves electroweak scale physics only, this mechanism is particularly appealing and amenable to experimental test and verification.

EW baryogenesis has been investigated in detail in the Standard Model [4] and its supersymmetric extension [5–9]. Within the SM parametrization of the Higgs potential, the one loop effective potential at high temperature roughly reads

$$V(\phi,T) \approx \frac{1}{2}(\mu^2 + cT^2)\phi^2 + \frac{\lambda}{4}\phi^4 - ET\phi^3 \ , \tag{8.1}$$

where

$$-ET\phi^3 \subset -\frac{T}{12\pi}\sum_{i=W,Z,h} m_i^3(\phi) \tag{8.2}$$

The last term is responsible for a barrier separating the symmetric and broken EW vacua; this barrier gives the possibility of a first-order EW phase transition. The coefficient $E$ is due to *bosonic* degrees of freedom coupling to the Higgs. In the SM, $E$ is too small and the phase transition can be first-order only for a very light Higgs, a possibility that is excluded experimentally [10]. In the MSSM, new bosonic degrees of freedom with large couplings to the Higgs—in particular, the stop $\tilde{t}$—can enhance the value of $E$ and guarantee that $\phi/T$ can be large enough at the time of the transition to suppress sphaleron washout. This has led to the so-called light stop scenario for EW baryogenesis. Possible extensions of the Higgs sector, without or outside the MSSM, offer other possibilities to realize this mechanism.

### 8.1.1 Electroweak baryogenesis in supersymmetry

The correlation between the strength of the EW phase transition and the collider signatures of the Higgs boson were recently studied in [11] in the case of a simplified model including a new scalar field $X$ that couples to $H$ according to:

$$-\mathcal{L} = M_X^2|X|^2 + \frac{K}{6}|X|^4 + Q|X|^2|H|^2 = M_X^2|X|^2 + \frac{K}{6}|X|^4 + \frac{1}{2}Q(v^2 + 2vh + h^2)|X|^2 \tag{8.3}$$

These basic interactions describe a broad range of theories. In particular, they apply to the MSSM, where $X$ corresponds to a light, mostly right-handed scalar top quark responsible for one-loop thermally generated cubic Higgs interactions. However, it doe not apply to models where the strength of the EW phase transition is affected by other scalars. The quantity $Q$ parametrizes the strength of the $X$ coupling to the $H$ which will induce the potential barrier.

Analysis of the Higgs potential using this approach or more specific calculations indicates that there is a fine-tuned window of parameter space in the MSSM where EW baryogenesis is viable [12, 13]. It corresponds to a stop-split supersymmetric spectrum illustrated in Fig. 8.1. A light Higgs boson and a light $\tilde{t}_R$, with mass less than 115 GeV, are needed for the EW phase transition to be sufficiently first-order. At the same time, the $\tilde{t}_L$ should be heavy to produce a sufficiently heavy Higgs boson to agree with experimental observation. A generic difficulty of EW baryogenesis is that it requires large new sources of CP violation [14] which are typically at odds with experimental constraints from electric dipole moments. To evade these constraints the other sfermions should be also heavy. The mechanism does require a light higgsino and a light chargino to supply CP-violating scattering processes within the expanding bubble walls during the phase transition.

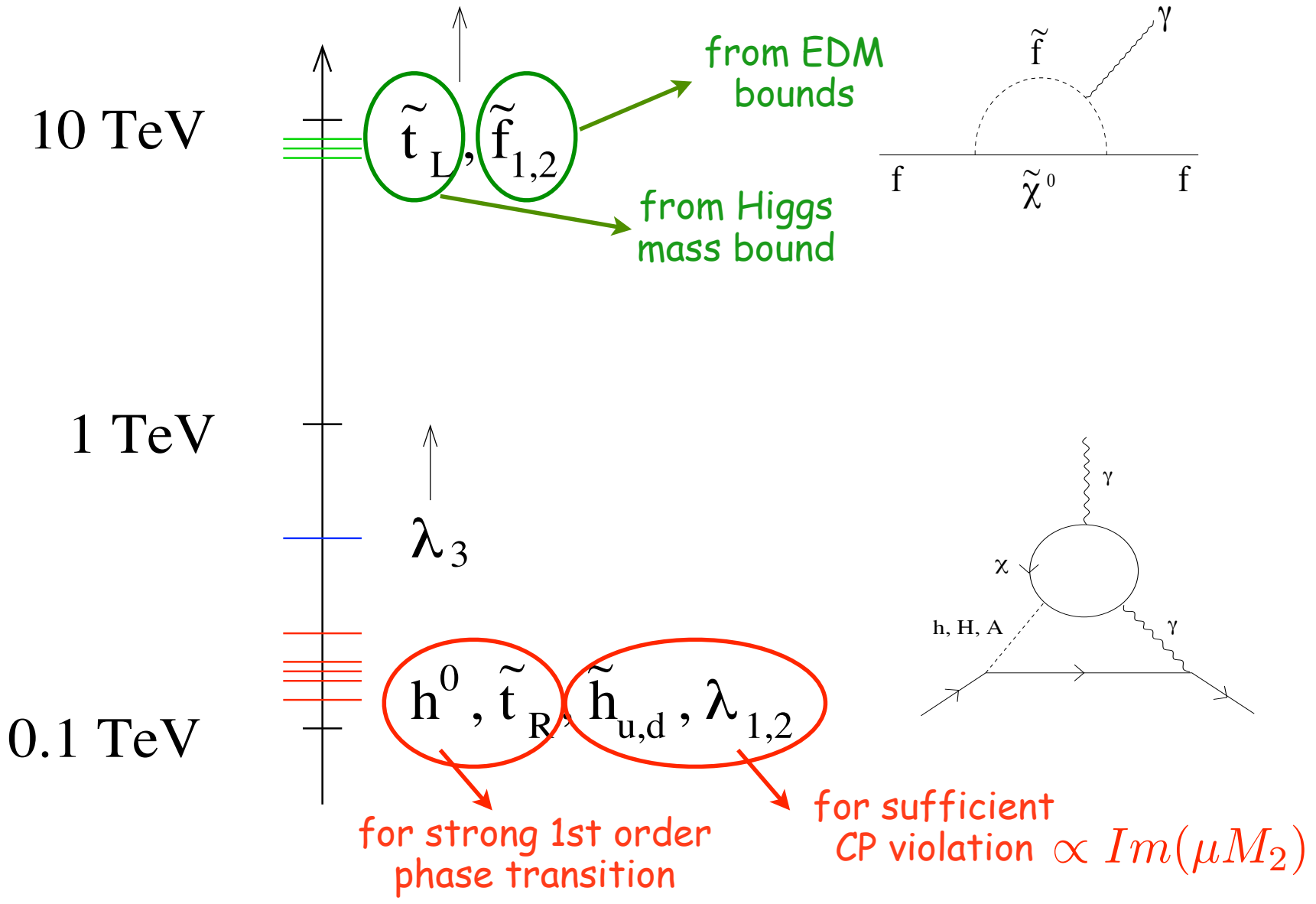

**Figure 8.1** The stop-split supersymmetric spectrum of MSSM EW baryogenesis.

The Higgs boson mass value of about 125 GeV is consistent with this scenario but narrows the parameter space. Additional constraints on the model will be derived once the Higgs branching ratios are measured with higher precision, since new fields that couple to the Higgs can lead to significant modifications of the rates for Higgs boson production and decay. The light stop or, more generally, the $X$ particle, appears in the loop diagrams that are responsible for the Higgs decays to $gg$ and $\gamma\gamma$ (discussed in Section 2.2.3) and can modify these rates by effects of order 1. A new scalar will typically interfere constructively with the top quark contribution to these loops, increasing the partial width to $gg$ but decreasing the width to $\gamma\gamma$. The effect on the rate for $gg \to h \to \gamma\gamma$ is plotted in Fig. 8.2 for the case of the scalar with the quantum numbers of the MSSM stop. From this plot, it is clear that, in the region where the phase transition is sufficiently strongly first-order $(\phi_c/T_c > 0.9)$, large deviations are expected with respect to the SM Higgs properties. Actually, it was concluded in [15] that EW baryogenesis in the MSSM can already be excluded using 2011 LHC data, see Fig. 8.3.

However, the MSSM, using only the stop and making no extension of the Higgs sector, may well be too restrictive a context. If the rate for Higgs production and decay to $\gamma\gamma$ remains high compared to the Standard Model, this scenario could remain in play due to new light Higgs particles discoverable at the ILC.

A difficulty with implementing electroweak baryogenesis within the MSSM is that the first-order phase transition appears only as a one-loop effect. It is much easier to obtain a strong first-order phase transition by modifying the Higgs potential at tree level. One straightforward example is to add a scalar singlet. There is an extensive literature on this possibility. A recent and complete study of this scenario was provided in [16]. Interestingly, such a scenario can be theoretically well-motivated in composite models where the Higgs arises as a pseudo-Nambu Goldstone boson of a new strongly interacting sector, as we discuss next.

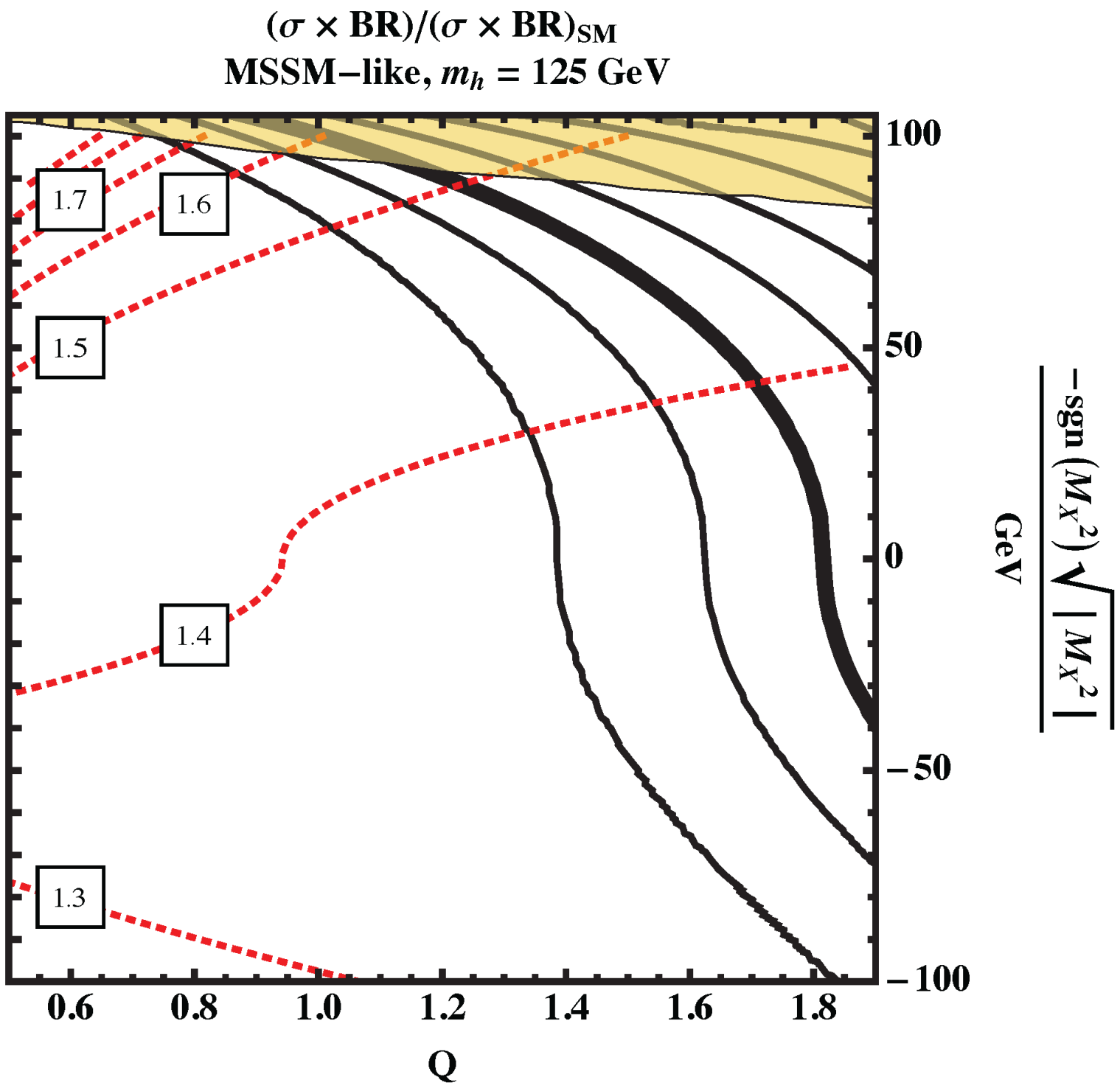

**Figure 8.2.** Contours of the ratio $\phi_c/T_c$ of the Higgs field value to the temperature of a first-order electroweak phase transition, for a new boson with the quantum numbers of $\tilde{t}_R$. The bold line denotes $\phi_c/T_c = 0.9$, and the adjacent solid lines delineate steps of $\Delta(\phi_c/T_c) = 0.2$. The yellow shaded region is excluded by the existence of a charge-color minimum. The red dotted lines show contours of the rate for $gg \to h \to \gamma\gamma$ from the Standard Model value, from [11]. The parameters $M_X$ and $Q$ are defined in (8.3).

## 8.1.2 Electroweak baryogenesis in composite Higgs models

The idea of Higgs compositeness has received a revival of interest in the last few years [17, 18], boosted by the dual description in terms of warped extra dimensional models. In composite Higgs models, the hierarchy between the Planck and TeV scale is due to the slow logarithmic running of an asymtotically free hypercolor interaction that becomes strong and confines close to the EW scale. In analogy with QCD, as the strong interaction confines, the global symmetry acting on the hyperquarks is broken down to a subgroup, delivering Goldstone bosons which are the analogs of the pions in QCD and may be identified as the degrees of freedom belonging to the Higgs doublet. The spectrum of composite Higgs bosons is determined by the structure of this symmetry breaking. The bosons are organized according to a coset space $G/H$, where $G$ is the symmetry group of the unbroken model and $H$ is the residual symmetry unbroken by the action of the new strong interactions. In these models, the top quark is also composite, since the hierarchy of Yukawa couplings is explained by partial fermion compositeness.

To preserve the custodial symmetry required in the electroweak theory, $G$ should contain an $SO(4)$ subgroup, with the Higgs multiplet transforming in the $(2, 2)$ representation. This restricts the possible choices of $G$ and $H$. In the minimal composite Higgs model, $G$ is the group $SO(5)$, spontaneously broken to $SO(4)$. The full symmetry $G$ is broken by loops of fermions or gauge bosons, which generate mass for the bosons and eventually generate the potential responsible for electroweak symmetry breaking.

To preserve the custodial $SO(4)$ symmetry of the SM, the Higgs should transform as a $(2, 2)$ of $SU(2)_L \times SU(2)_R \sim SO(4)$. In the minimal composite Higgs model SO(5) breaks to SO(4), delivering 4 goldstone bosons which are identified as the Higgs degrees of freedom. The $SO(5)$ symmetry is broken explicitly both by the fermions which do not come in complete representations of SO(5) and by the gauging of $SU(2)_L \in SO(5)$. A catalog of possible choices for $G$ and $H$ is presented in [19]. Loops of SM fermions or gauge bosons communicate the explicit breaking to the Goldstone bosons and generate a potential for the Higgs.

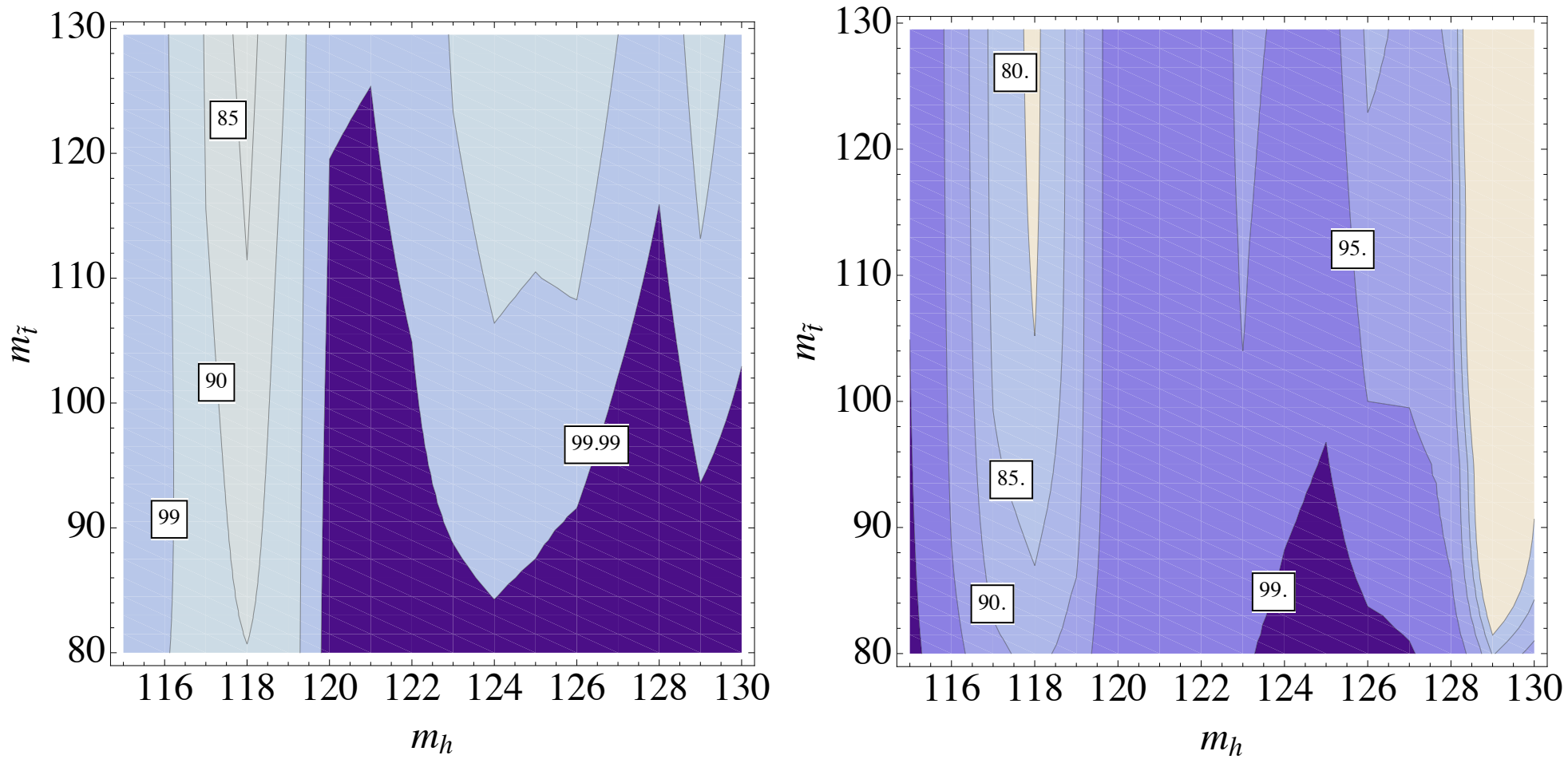

**Figure 8.3.** Confidence levels of exclusion of a general Light Stop scenario in the $(m_h, m_{\tilde{t}_R})$ plane. $\tilde{t_L}$ is taken very heavy while $m_A$ and $\tan\beta$ are varied in the range (1500, 2000) GeV and (5,15). From [15].

**Figure 8.4**
Diagram illustrating the largest contribution to the electron EDM due to the Higgs-singlet mixing where the new singlet $s$ couples only to the top quark, as needed for EW baryogenesis and as motivated by the scenario of partial compositeness. From [21].

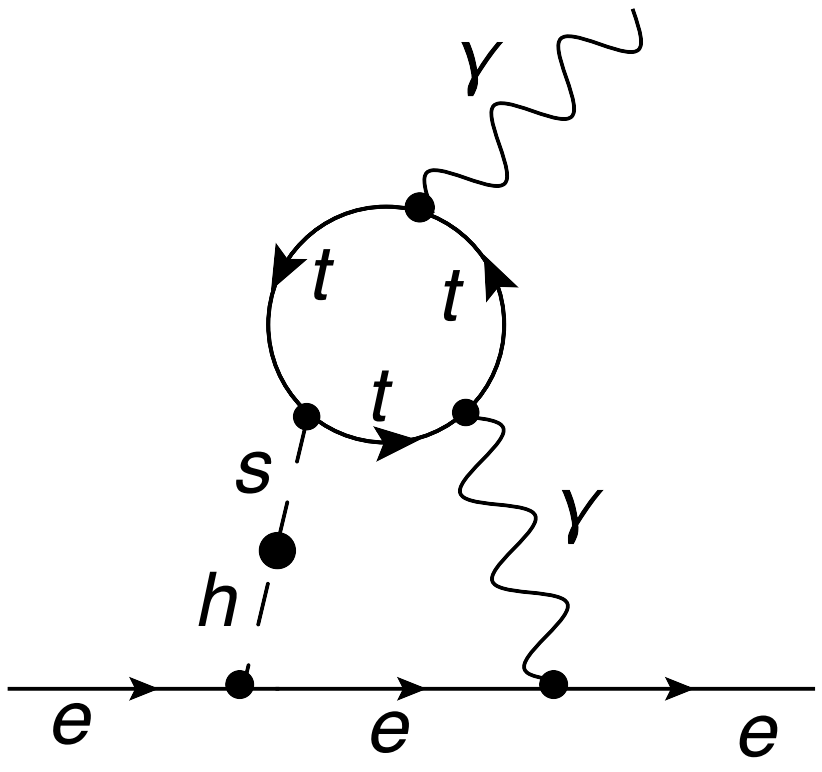

These composite Higgs models offer new possibilities for EW baryogenesis. Naturalness in these scenario implies modifications in the Higgs and top sectors, which are precisely the ones believed to be responsible for EW baryogenesis. Specific choices of $G$ and $H$ imply the presence of additional light scalars that can make the electroweak transition first-order. For instance, if the coset is $SO(6)/SO(5)$, we expect an additional singlet [20]. Another possibility is $SO(6)/SO(4) \times SO(2)$, which gives two Higgs doublets.

For the choice $SO(6)/SO(5)$, the extra singlet has a dimension-five pseudoscalar couplings to the top quarks that can break CP. EW baryogenesis in this context has been studied in [21]. The extra singlet is responsible for making the EW phase transition first order. Secondly, if that scalar couples to the top quark it can lead to a non-trivial CP-violating phase along the bubbles of the EW phase transition creating the seed for the sphaleron to generate a non-zero baryon asymmetry. It was shown that the correct amount of asymmetry can be produced in a large region of parameter space. The new complex phases and the mixing between the Higgs and the singlet lead to new contributions to the EDMs of neutron and electron not far from the reach of current and future experiments (see Fig. 8.4). The new singlet and the new CP-violating top couplings will be visible at the ILC as direct

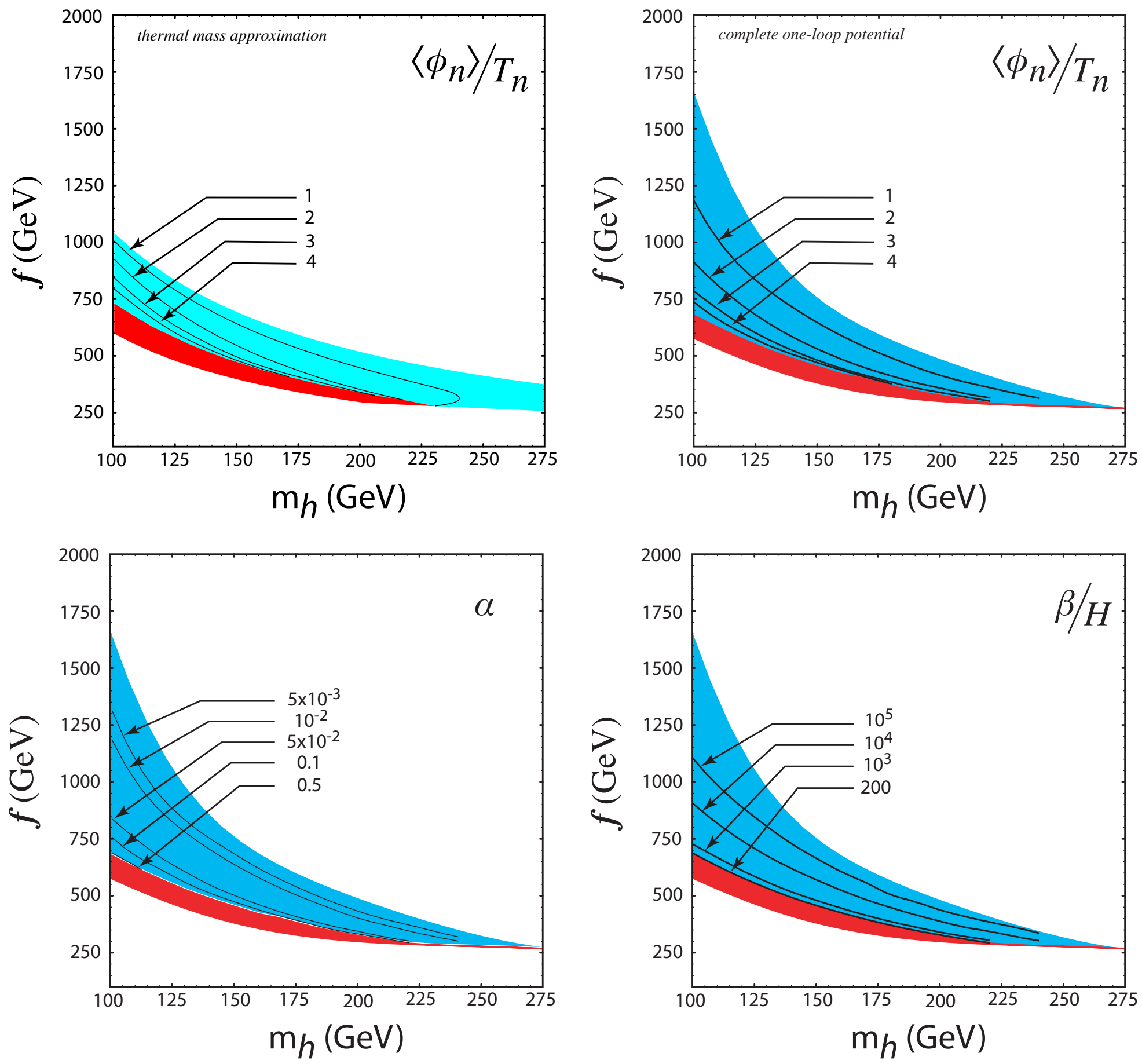

**Figure 8.5.** Phenomenology of the electroweak phase transition in the effective description (8.4). Upper panel: Contours of the ratio $\langle\phi\rangle/T$ evaluated at the nucleation temperature in the blue region that allows for a first-order EW phase transition. The left plot uses the thermal mass approximation [30] while the right plot uses the full one-loop potential [31]. Below the red lower bound, the EW symmetry remains intact in the vacuum while above the blue upper one, the phase transition is second order or not even occurs. Within the red band, the universe is trapped in a metastable vacuum and the transition never proceeds. The lower panel shows contours of $\alpha$, the ratio of latent heat to thermal energy density, and $\beta/H = T_n d(S_3/T)/dT$, approximately equal to the number of bubbles per horizon volume, from [31]. These quantities measure the amount of supercooling.

tests of this scenario.

The nature of the EW phase transition has also been studied in a number of contexts that give more specific models of the new strong interactions associated with composite Higgs bosons. These include models of technicolor [22], models with flat extra dimensions [23], and Randall-Sundrum models [24–28]. However, no full calculation of the baryon asymmetry has been carried out in these contexts. In some of these constructions, the EW phase transition can be too strongly first-order, leading to supersonic bubble growth which suppresses diffusion of CP violating densities in front of the bubble walls, thus preventing the mechanism of EW baryogenesis [29].

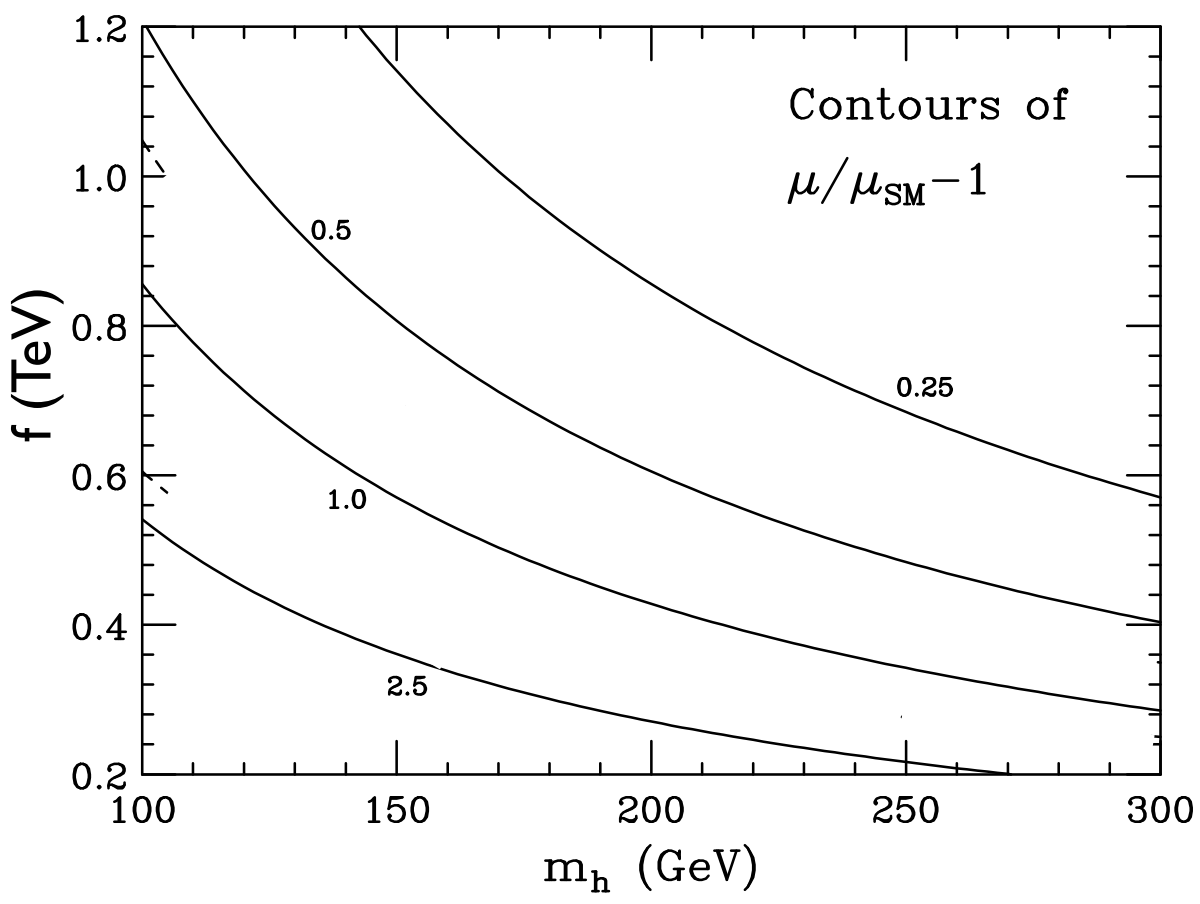

**Figure 8.6**
Contours of $\mu/\mu_{SM} - 1$ in the $(m_h, f)$ plane.

## 8.1.3 Effective field theory approach to the EW phase transition

The influence of tree-level modifications of the Higgs potential in making the EW phase transition strongly first order can be analyzed more generally using an effective field theory approach. For example, one can add dimension-6 operators to the Higgs potential, allowing a negative value for the quartic coupling [30, 31]:

$$V(\phi) = \mu^2|\phi|^2 - \lambda|\phi|^4 + \frac{|\phi|^6}{f^2} \tag{8.4}$$

The phenomenology of this effective theory is illustrated in Fig. 8.5, which shows contours of quantities characterizing the strength of the phase transition and the amount of supercooling in the $(m_h, f)$ plane. From these plots, it is clear that a phase transition that is strong enough for EW baryogenesis arises in a sizable region of parameter space.

In the parameter region of interest, a potential such as (8.4) leads to deviations of order 1 in the Higgs self-couplings. We can write the potential for the Higgs field $h$, the fluctuation from the vacuum expectation value, as

$$\mathcal{L} = m_H^2 h^2/2 + \mu h^3/3! + \eta h^4/4! + \cdots \tag{8.5}$$

when, from the effective theory (8.4),

$$\mu = 3\frac{m_H^2}{v} + 6\frac{v^3}{f^2} \qquad \eta = 3\frac{m_H^2}{v^2} + 36\frac{v^2}{f^2}. \tag{8.6}$$

The SM couplings are recovered as $f \to \infty$ [30]. Figure 8.6 shows contours of $\mu/\mu_{\rm SM} - 1$ in the $f$ vs. $m_H$ plane. Therefore, non-trivial probes of the Higgs potential may be obtained from precise measurements of the trilinear Higgs coupling. See [32] for other examples. As we have emphasized in Section 2.6.3, this is a difficult quantity to measure at any collider, but it is expected to be accessible at the ILC with an accuracy that clearly distinguishes the curves in the figure.

The bubble wall velocity is a key quantity entering the calculation of the baryon asymmetry. A model-independent and unified description of the different regimes (detonation, deflagration, hybrid, runaway) characterizing bubble growth was presented in [29]. The results are summarized in Fig. 8.7,, which shows contours for the bubble wall velocity in the plane $(\eta, \alpha_N)$ where $\eta$ and $\alpha_N$ are dimensionless parameters characterizing the strength of the phase transition (roughly the ratio of latent heat to thermal energy density) and the amount of friction. In the SM, $\eta \sim 1/1000$, while in the MSSM, $\eta \sim 1/30$. Eventually, one would have to calculate these quantities from measured parameters of the Higgs potential and the new particle spectrum for a reliable computation of the baryon asymmetry.

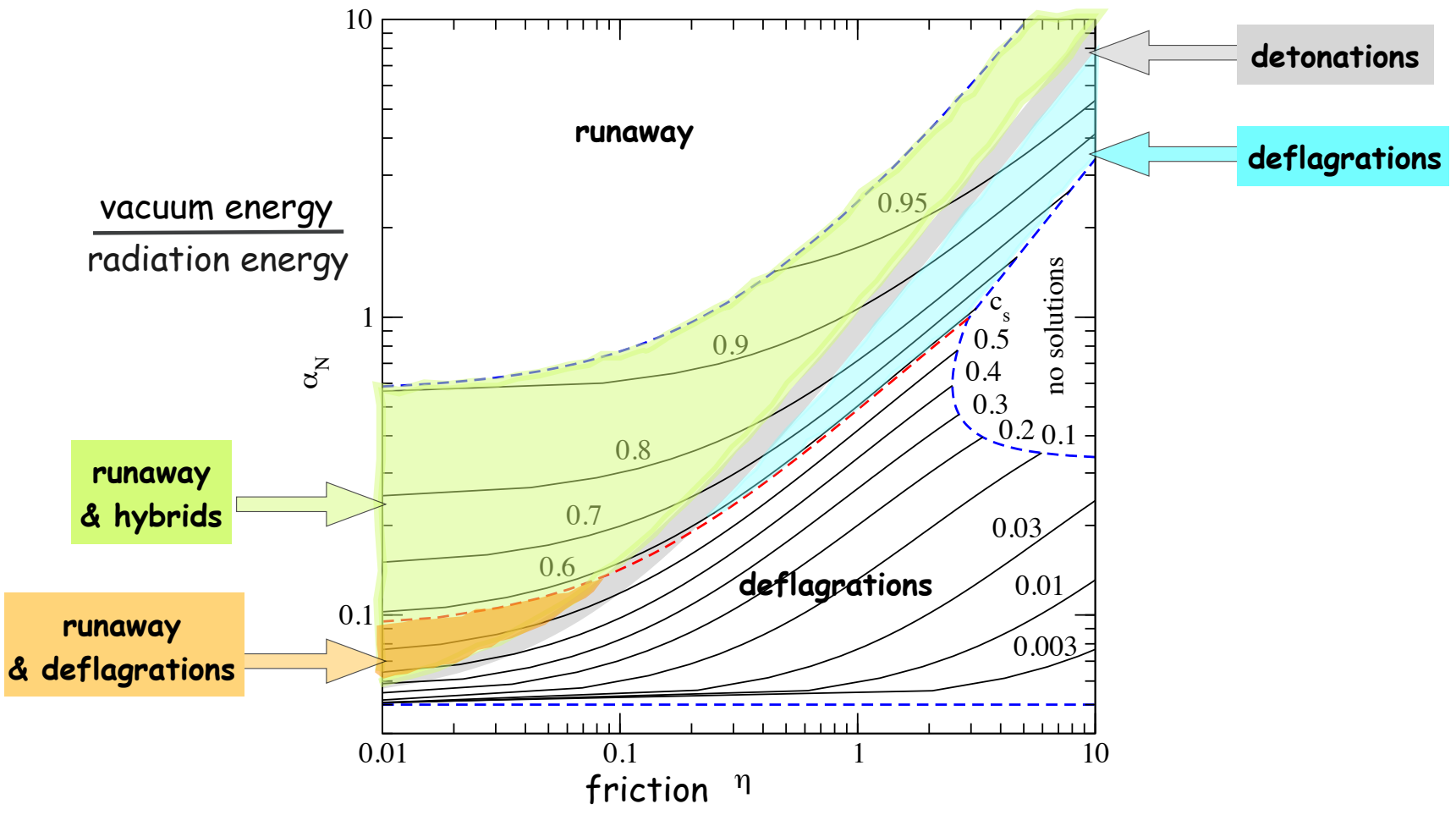

**Figure 8.7** Contours of the bubble wall velocity in a first-order cosmic phase transition in the $(\eta, \alpha_N)$ plane, from [29].

## 8.2 Dark matter

The existence and central role of dark matter is now one of the pillars of the standard model of cosmology. There are many pieces of evidence, from galactic length scales, cluster lengths scales, and the largest observable scales in the universe, that roughly 20% of the energy and 80% of the mass in the Universe is in the form of massive, non-baryonic particles with relatively weak interactions with ordinary matter [33, 34]. There are many proposals for the nature of this dark matter. The proposed particles span an enormous range in mass, from $10^{-5}$ eV to macroscopic and even planetary-scale masses. However, the most attractive proposal, and the one that we will concentrate on here, is that the particle that makes up dark matter is a 'weakly-interacting massive particle' (WIMP).

### 8.2.1 Dark matter and the WIMP paradigm

A WIMP is defined as a weakly interacting neutral particle that is stable over the lifetime of the universe. WIMPs can be created or destroyed only in pairs. The WIMP model further assumes that the WIMPs were in thermal equilibrium with the hot plasma of Standard Model particles early in the history of the universe. This initial condition allows us to predict the current density of WIMPs. In the model, when the temperature of the universe decreased below the WIMP mass, WIMPs began to annihilate, but, because the anninhilation requires a pair of WIMPs, the annihilation cut off when the density of WIMPs reached a well-defined small value. The density of WIMPs decreased further due to the expansion of the universe. However, as the Universe cooled, this small density of massive WIMPs eventually came to dominate the energy in radiation. By this logic, it is possible to derive an expression for $\Omega$, the current energy density of the universe in WIMPs, in the form

$$\Omega \sim \frac{x_F T_0^3}{\rho_c M_{Pl}} \frac{1}{\langle \sigma_{ann} v \rangle} . \tag{8.7}$$

In this expression, $x_F = m/T_F$, with $m$ the WIMP mass and $T_F$ is the freeze-out temperature at which annihilation turns off, $T_0$ is the temperature of photons today, $\rho_c$ is the critical energy density, $M_{Pl}$ is the Planck scale, and $\langle \sigma_{ann} v \rangle$ is the inclusive cross section for WIMP pair annihilation into SM particles, averaged over the WIMP thermal velocity distribution at freeze-out. Typically $x_F \approx 25$, with weak dependence on the WIMP mass, and the other parameters in the equation, including $\Omega$, are well measured. The expression (8.7) then determines the value of the annihilation cross section needed for the entire dark matter relic density to be composed of a single WIMP species. The result

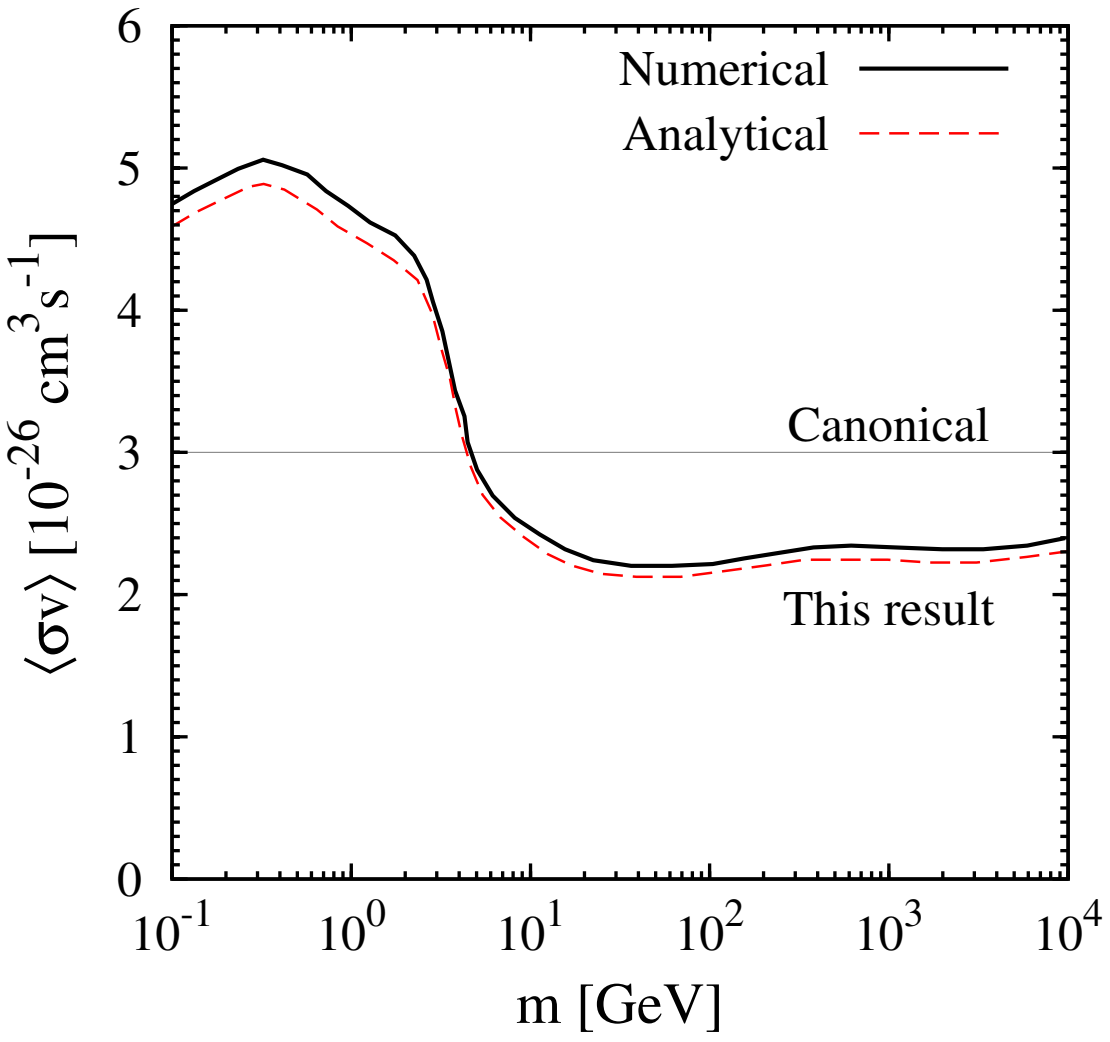

**Figure 8.8** Desired annihilation cross section $\langle\sigma v\rangle$ to obtain the measured thermal relic density, as a function of the WIMP mass $m$ (from [35]). The line marked "canonical" shows the often-quoted value $3 \times 10^{26}$ cm$^3$/s.

is shown in Fig. 8.8 [35]. The required value is roughly

$$\langle\sigma_{ann} v\rangle \approx (1 \;\; \text{pb}) \cdot c \; , \tag{8.8}$$

indicating that a WIMP with mass and interactions at the electroweak scale naturally leads to the required density of dark matter.

This observation motivates searches for WIMPs with masses of the order of 100 GeV, making use of techniques from particle physics. The three pillars of WIMP searches are: indirect detection of residual annihilation of WIMPs in the galactic neighborhood, direct detection of ambient WIMPs scattering in sensitive detectors on Earth, and artificial production of of WIMPs at high energy accelerators.

If a candidate particle for WIMP dark matter can be produced at the ILC, the precision study of its mass and properties available through the ILC measurements might make it possible to predict its pair annihilation cross section and thus its thermal relic density. This prediction could then be compared to the density of dark matter measured by astrophysical observations. This possibility of a direct connection between physics at the smallest and largest length scales is extremely enticing. Later in this section, we will discuss a number of scenarios in which the ILC makes such a comparison possible.

### 8.2.2 Theories of WIMPs

By far the most popular vision of WIMP dark matter is the neutralino found in supersymmetric theories. Supersymmetric theories are particularly amenable to searches at the LHC, because they contain a wealth of new colored states (squarks and gluons) with large hadroproduction cross sections. Such particles can decay into the dark matter plus jets of hadrons, leading to events characterized by hadronic activity together with a large imbalance of transverse momentum. As of this writing, the absence of a signal places limits on the masses of squarks and gluons to be substantially in excess of 1 TeV, depending on the fine details of the mass spectrum [36, 37]. The null results of these searches, especially when combined with the identification of the resonance near 125 GeV as the Higgs boson, have led some to propose that, if supersymmetry is realized in nature, it may not be minimal [38]. Nonetheless, viable points with modest fine-tuning still exist [39], and for the purposes of this discussion we will stay within the minimal supersymmetric extension of the SM. We have given a more detailed overview of the possibilities for supersymmetry consistent with the LHC constraints in Chapter 7.

Searches for supersymmetry based on the 2011 LHC data have focused on searches for the colored superpartners [40]. Such searches are important in terms of characterizing the overall scale of superpartner masses, but offer only limited information on the properties of supersymmetric dark matter. As the LHC collects more data and at higher energies, it becomes more sensitive to direct production of electroweak superpartners, and thus has more directly to say about the properties of dark matter. However, as we have stressed in Section 7.3, some spectra for electroweak SUSY spectra will continue to be very difficult to explore at the LHC.

Beyond supersymmetric theories, the most studied candidates for WIMP dark matter include the lightest Kaluza-Klein particle in 5-dimensional [41, 42] or 6-dimensional [43, 44] theories with Universal Extra Dimensions [45], and a light neutral vector boson in little Higgs theories [46, 47] incorporating $T$-parity [48]. All of these theories are primarily distinguished from supersymmetric theories in that the WIMP is a boson rather than a Majorana fermion. One other nonsupersymmetric theory which affords some contrast is based on a warped extra dimension [49] and has a dark matter particle which is a Dirac fermion [50–53].

Recently, there has also been activity aimed at capturing features of WIMP dark matter in cases where the particles mediating the interactions are heavy compared to the energy transfer of the processes of interest, by making use of effective field theory (EFT) descriptions of WIMPs [54–57] Such effective field theories allow for one to capture the low energy properties of any theory which is amenable to an EFT description, and facilitates comparisons between the different types of searches for dark matter. The picture which emerges from such studies is that there is a large degree of complementarity between direct, indirect, and collider searches. Direct and indirect detection constraints are typically stronger than collider bounds, but also subject to relatively large astrophysical uncertainties, and only apply to interactions which do not vanish in the limit in which WIMPs are non-relativistic. Instead, collider bounds apply roughly uniformly to any type of interaction involving the particles available in the initial state, but are limited for heavy WIMP masses by the finite energy available in the collision.

Another feature which is easily discerned from effective theory descriptions is that bounds from the Tevatron and LHC typically apply to WIMP couplings to quarks and gluons, whereas the couplings most relevant at a high energy $e^+e^-$ collider are the couplings to electrons and photons. While the most popular models of dark matter predict that couplings to quarks and leptons are comparable, it is possible to construct leptophilic models [58–60], motivated in part by the observation of an anomalous positron flux by the PAMELA and Fermi LAT collaborations [61, 62].

Beyond the straightforward freeze-out paradigm, there are other models of dark matter for which dark matter particles at the electroweak scale are relevant. The universe energy density stored in WIMPs may exhibit an explicit dependence on extra parameters, in particular the dark matter mass, for instance in models of asymmetric dark matter, *e.g.* [63]. Dark matter may also be produced by 'freeze-in' scenarios such as that in [64] or in scenarios where DM is is produced through decays [65].

### 8.2.3 Determination of dark matter parameters

Once dark matter is detected through a non-gravitational interaction, and is thus confirmed to be some kind of weakly interacting particle, the primary question will be whether or not its annihilation cross section is of the correct size for it to explain the cosmic dark matter as a thermal relic. If the annihilation cross section reconstructed from measurements on the particle is consistent with the determinations of the dark matter density, it will provide evidence that the thermal history of the Universe was (at least approximately) standard back to the time that the dark matter froze out—about 1 nsec after the Big Bang. This would parallel the argument the successful predictions of Big Bang nucleosynthesis based on measurements in nuclear physics lead to a compelling picture of the history of the Universe back to temperatures of order MeV [66] and times of order 1 second.

In principle, the most direct determination of the dark matter annihilation cross section would come from an observation by indirect detection experiments which look for annihilation of WIMPs in the galaxy. In practice, this is a daunting task, because of large uncertainties in astrophysical backgrounds, which can mask or pollute the signal, and in the distribution of dark matter itself, which enters into the observed photon flux as the density squared integrated along the line of sight of the observation. In addition, a relatively few final states are expected to be observable on the Earth, necessarily leading to an incomplete picture. In addition, the annihilation cross section observed in indirect detection might be very different from the one that determined the dark matter cross section in the early universe. If the cross section is strongly velocity-dependent, as happens, for example, in some SUSY models, annihilation channels which were important at the time of freeze-out ($v \sim 0.1$) may be subdominant in the galaxy today ($v \sim 10^{-3}$).

Direct detection experimentscan be used to estimate the annihilation cross section only if analyzed in an effective-interaction picture. In this context, they are sensitive only to couplings of dark matter to colored SM particles, which could turn out to represent a relatively unimportant fraction of the totality of WIMP annihilation. Direct detection also loses track of some types of interactions which may be important for WIMP annihilation, but are suppressed in the non-relativistic limit of elastic scattering.

Because of these limitations, colliders are likely to play the central role in providing the data from which to compute a WIMP relic density that can be compared with cosmological observations. We emphasize that this requires a complete picture of dark matter interactions with all SM species. Hadron colliders such as the LHC have large rates of production for exotic colored particles (and also typically higher energies, allowing searches for more massive particles), but also larger backgrounds that can hide many possible decay channels. In a typical theory of WIMPs such as the MSSM or UED models, the relic density is controlled by a a complicated interplay between annihilations into colored and uncolored states. For all of these reasons, input from an $e^+e^-$ collider such as ILC is likely to be essential.

### 8.2.4 ILC studies of dark matter parameter determination

In this section, we will review studies that have been done on the determination of dark matter parameters from collider data. Our discussion is based mainly on a few of the most detailed studies of the MSSM [67, 68]. These studies assume LHC running at $\sqrt{s} = 14$ TeV with data sets of hundreds of fb$^{-1}$. Under such conditions, many of the measurements will be systematics limited and thus the precise assumptions for collected data sample are less important than the assumed collision energy. The specific models analyzed in these papers are now excluded by LHC searches; however, as we have discussed in Section 7.4, very similar models with heavier squarks and gluinos are still viable and even attractive. Other examples of dark matter density determination are given in Section 7.5 and in [70–79].

In [68, 69], two mSUGRA-inspired models are investigated in terms of the ability of the LHC and 500 GeV ILC to reconstruct the spectrum and couplings of the neutralino. Model B′ is characterized by low sparticle masses and large mass splittings, resulting in a model that is particularly amenable to reconstruction using LHC measurements alone. In Figure 8.9, we show the sparticle spectrum and the range of reconstructed relic densities for model B′. The color-singlet sector of this model is similar to that of the benchmark model presented in Section 7.4.2. The derived relic density indicates that for this case, LHC data alone can predict the WIMP relic density to order 1 in $\Omega$. Adding data from the ILC, which would be very rich given the low values of the superparticle masses, the prediction for $\Omega$ is given to 20% accuracy. This model is very similar to model LCC1 studied in [67]. In that study, including information from a wider range of ILC observables, it is possible to predict the relic density to lie within a few percent of the underlying value.

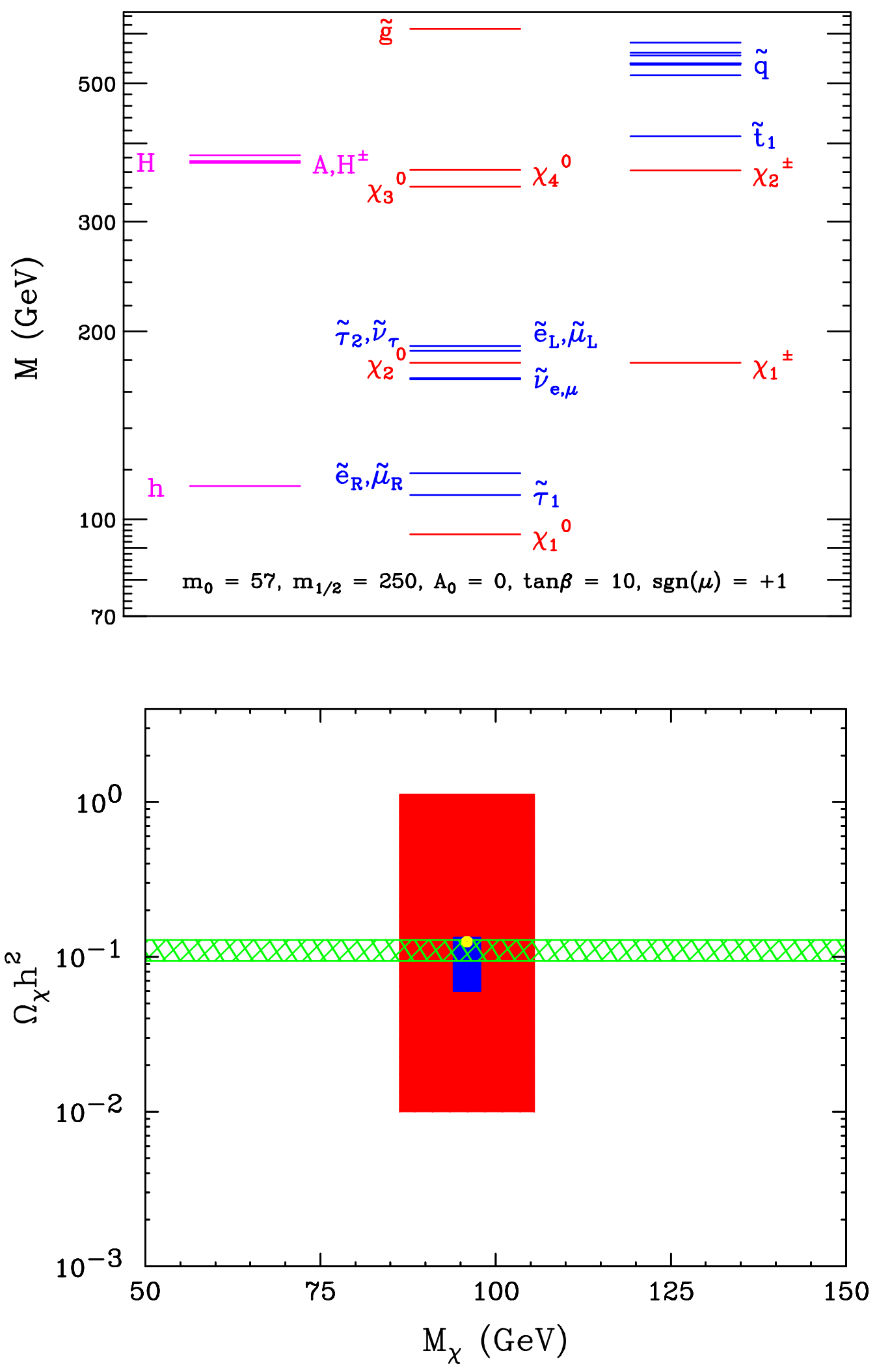

**Figure 8.9**
Projections for the estimation of the dark matter relic density from colliders in the supersymmetric B′ model described in [68, 69]. Top: Spectrum of the model. Bottom: Projections for determination of the WIMP mass and inferred relic density based on measurements at the LHC (red rectangle) and ILC (blue rectangle). The measurement of the relic density from cosmology is indicated by the green hatched region. The actual model prediction is shown as the yellow dot.

In [67], three additional MSSM parameter choices (LCC2-4) are investigated from the point of view of indirect and direct searches for dark matter, LHC searches, and an ILC at $\sqrt{s} = 500$ GeV and 1000 GeV, in order to see how many relevant dark matter properties can be reconstructed. In Model LCC3, the relic density is largely controlled by late coannihilation of the lightest neutralino with a stau. The small mass splitting renders the stau particularly challenging to reconstruct at the LHC. In Fig. 8.10, we show the sparticle spectrum and the range of reconstructed relic densities for model LCC3. As shown, the LHC has essentially no ability to reconstruct the relic density, because it is unable to obtain precise enough measurements of the neutralino and stau masses and the important parameter $\tan\beta$. In addition, and the neutralino and tau compositions leave large uncertainties in the coannhilation cross section. At the 500 GeV ILC, the situation clarifies, but remains rather uncertain, because while the neutralino and stau masses become much better measured, the neutralino composition remains uncertain. A 1 TeV ILC can fill in this remaining information, and results in a reasonably precise measurement of $\Omega h^2$ to within a factor of two.

In LCC4, the relic density is driven by neutralinos which annihilate through a heavy Higgs resonance that is approximately on-shell because the SUSY Higgses have masses $\sim 2m_{\chi_1^0}$. The colored sparticles are heavy (roughly at the current LHC exclusion limits for the gluino and first two generations of squarks and well above the current limits on third generation squarks). This point is a particular challenge for the LHC (despite the fact that it is able to observe much of the spectrum of particles) to reconstruct, because it requires very high precision measurements of the mass of the lightest neutralino and the mass and width of the pseudo-scalar Higgs boson $A^0$, as well as reasonably precise knowledge of the lightest neutralino composition; see Fig. 8.11. The resulting relic density prediction is peaked at very low values, with a substantial tail that extends past the WMAP measurement. At the 500 GeV ILC, the situation remains somewhat fuzzy, because the pseudo-scalar Higgs remains out of kinematic reach, though the composition of the neutralino becomes much-better understood. At the 1000 GeV ILC, the pair-production process $e^+e^- \to HA$ opens up, and the picture becomes reasonably clear.

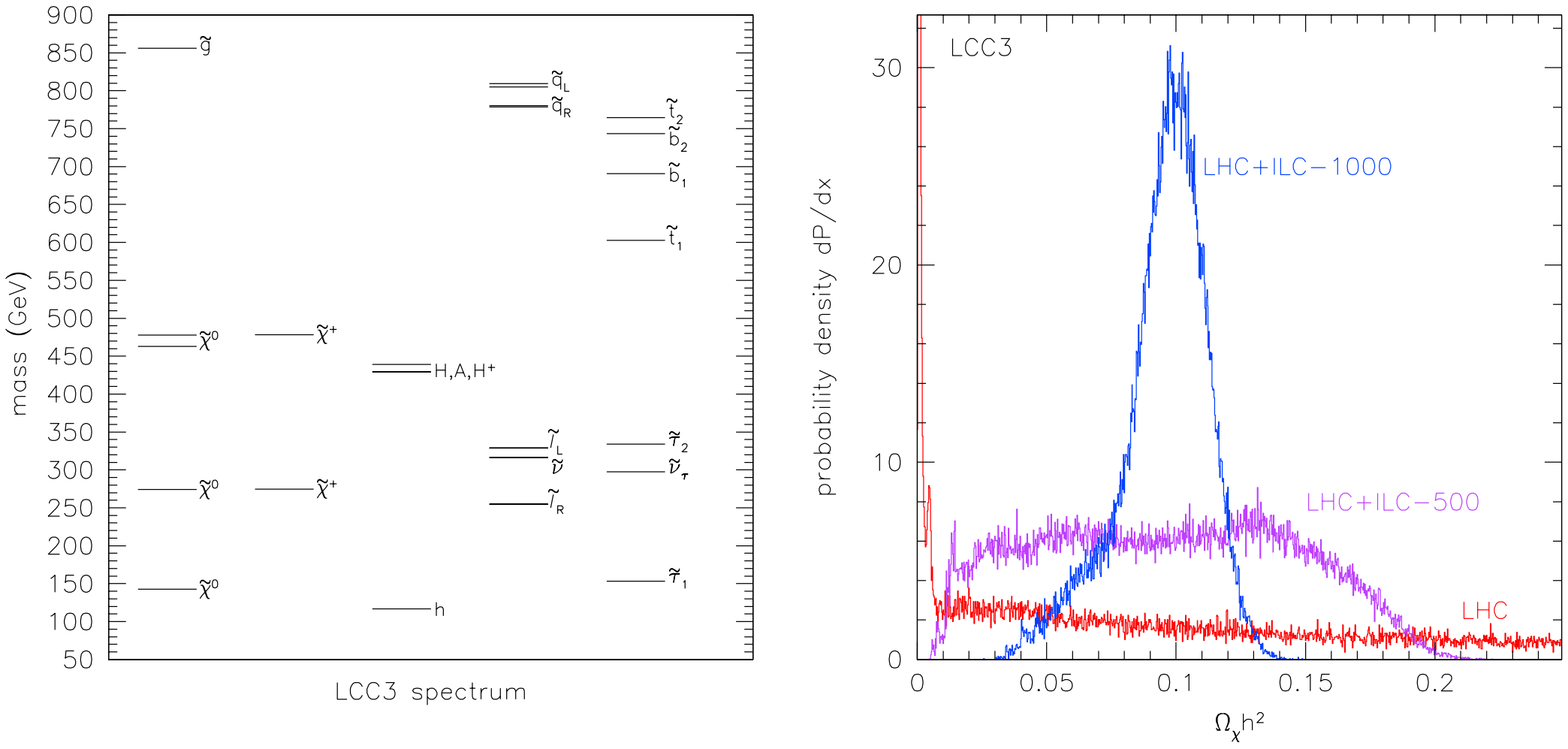

**Figure 8.10.** Projetions of the estimation of the dark matter relic density from colliders, at the MSSM point LCC3, from [67]. Left: Mass spectrum of the model. Right: Probability distribution of the predicted relic density based on measurements at the LHC alone (red histogram), LHC + a 500 GeV ILC (magenta histogram) and LHC + a 1000 GeV ILC (blue histogram).

Over-all, the picture that emerges is one in which the ILC is often necessary to provide the crucial information allowing one to reconstruct the relic density of neutralinos. Whether it is effective in accomplishing this goal is largely dependent on whether or not it has enough energy to access the important states. In the case studies shown here, the LHC data will be able to identify the relevant mass scales for new particles, but after the LHC program it still remains unclear which particles exactly are crucial to determining the neutralino annihilation rate and the relic density. That can be determined only by more detailed studies of the neutralino which are made possible at the ILC.

As a final example, we consider a leptophilic model of dark matter. If interactions between a generic Dirac WIMP $\chi$ and the SM leptons are mediated by a heavy vector particle, they may be described by the effective vertex,

$$\frac{1}{M_*^2}\,\overline{\chi}\gamma^\nu\chi \sum_{\ell=e,\mu,\tau} \bar{\ell}\gamma_\nu\ell \qquad (8.9)$$

We assume that there are no couplings to quarks at tree level. The parameter $M_*$ is a dimensionful coupling constant which maps on to the description of $Z'$ exchange through $1/M_*^2 \leftrightarrow g_\ell g_\chi/M_{Z'}^2$. If this interaction is the only way dark matter can interact with the SM, the observed relic density will be obtained for $M_* \sim 1$ TeV for a WIMP mass around 100 GeV [58]. A dark matter model of this type is constrained by LEP II through the L3 [80] and DELPHI [81] measurements of the process $e^+e^- \to \nu\overline{\nu}\gamma$ to $M_* \geq 480$ GeV [82]. While in principle the LHC could hope to observe processes such as $pp \to e^+e^-\chi\overline{\chi}$, these processes are very rare and unlikely to provide better bounds than the LEP searches. A recent 500 GeV ILC study of the process $e^+e^- \to \chi\overline{\chi}\gamma$ reveals the ability to place much more stringent limits on the cross section, particularly if the beams are polarized, which allows one to reduce the SM background [83]. The limits on the cross section translate into limits on $M_*$ of about 1.7 TeV for 100 GeV mass WIMPs. Then the ILC will be able to discover or rule out this class of leptophobic dark matter, and confirm its nature as a thermal relic.

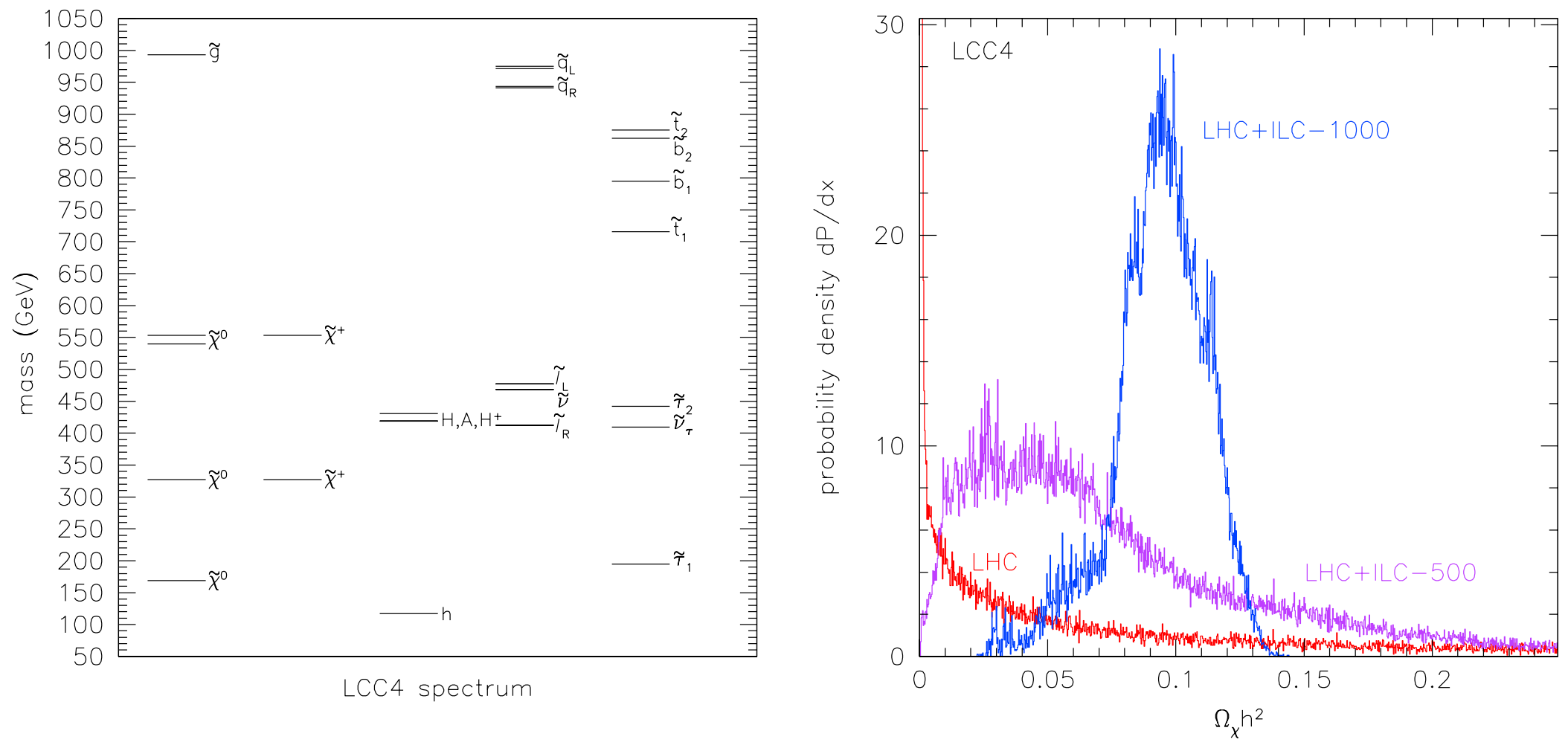

**Figure 8.11.** Projetions of the estimation of the dark matter relic density from colliders, at the MSSM point LCC4, from [67]. Left: Mass spectrum of the model. Right: Probability distribution of the predicted relic density based on measurements at the LHC alone (red histogram), LHC + a 500 GeV ILC (magenta histogram) and LHC + a 1000 GeV ILC (blue histogram).

## 8.3 Conclusions

In this section, we have reviewed in some detail models of baryogenesis and dark matter associated with new physics at the TeV energy scale. The discussion of models rapidly becomes complex and technical, because the predictions of the models for the baryon asymmetry and dark matter depend on detailed properties of the model. The most crucial aspects of the models come in the Higgs sector and in the superpartners or more general partners of Higgs and gauge bosons. At hadron colliders, it is very difficult even to discover these particles. In all but the simplest models, reaching the level of detail that is required to make predictions relevant to cosmology is quite beyond the capabilities of hadron collider experiments.

Experiments at the ILC also must be lucky. The relevant new particles—extended Higgs bosons, neutralinos, sleptons—must be light enough to be observed at the ILC in pair production. But, given this possibility, the ILC experiments will have the power to test theories of the type that we have discussed. Thus, the ILC offers unique opportunities to connect detailed aspects of particle physics to grand questions about the composition of the universe.

# Chapter 9 Conclusion

In this report, we have surveyed the range of physics topics that will be addressed by the ILC.

Our primary emphasis has been on the study of a Standard Model-like Higgs boson. The discovery of a new boson by the ATLAS and CMS experiments has vaulted the question of its properties of the top of the list of questions in high energy physics. We have argued that the ILC is perfectly matched to this problem. The ILC will be able to deliver a precise description of the properties of this new particle.

The ability of the ILC to operate at several different energies plays an important role in its ability to study the Higgs boson. We have described three phases of the Higgs boson program. First, at $\sqrt{s} = 250$ GeV, one may expect the precision measurement of the Higgs mass and its major branching fractions and the search for invisible and exotic modes. Second, at $\sqrt{s} = 500$ GeV, we anticipate precision measurements of the Higgs coupling to the $W$ boson and the higher statistics study of modes with small branching fractions. Finally, at $\sqrt{s} = 1$ TeV, for the measurement of the Higgs couplings to the top quark and the muon, and the Higgs self-coupling can be made. The suite of measurements at these three energies combines to provide a complete picture of the interactions of the Higgs particle and an incisive test of its role in the generation of mass for all elementary particles.

We have also emphasized the ability of the ILC to carry out precision measurements of the properties of the $W$ and $Z$ bosons and the top quark, and of elementary $e^+e^- \to 2$ fermion reactions. In addition, we have shown that the ILC has excellent capabilities to study new color-singlet particles that might be present in the mass range of a few hundred GeV.

The nature of the Higgs boson and the origin of electroweak symmetry breaking remains a central and puzzling problem. The traditional approaches to this problem either involve strong coupling in the Higgs sector, building the Higgs boson as a composite state, or weak coupling in the Higgs sector, realizing the Higgs as one member of a new multiplet of particles. Both types of models have been reshaped by the discoveries and exclusions from the LHC.

If the Higgs sector is strongly coupled, the model must be one with a light composite Higgs boson and additional vectorlike particles at the TeV scale. We have shown how the precision measurement capabilities of the ILC will give important clues to the properties of these models that will not be available from the LHC.

If the Higgs sector is weakly coupled, it is very likely that there are new color-singlet particles that are extremely difficult to study at the LHC. We have argued, in particular, that the LHC results motivate models of supersymmetry that have a spectrum of this type. The colored states of the supersymmetry spectrum may well be discovered in the 14 TeV program of the LHC. The lightest particles of supersymmetry, with their possible connection to the dark matter of the universe, will require the ILC for their proper understanding. For the highly motivated case of natural supersymmetry, the ILC could make the definitive test of this class of models, since charged higgsinos are expected to be present with mass below about 200 GeV. If these light higgsinos do indeed exist, then ILC would be a higgsino factory in addition to a Higgs factory.

For both types of models, the precision study of the Higgs boson will provide essential clues. To obtain these clues, we have shown that it will be necessary to measure the couplings of the Higgs boson at the few percent level. The ILC will give us that capability.

For all of these reasons, the physics questions that are before us now call for the ILC as the next major facility in high energy physics.

# List of Signatories

The following list of signatories represents a comprehensive list of those people who have contributed to the R&D and design work, for both the accelerator and the detectors, which is summarised in this report. The list also includes those people who wish to indicate their support for the next phases of the worldwide ILC effort.

It should be noted that inclusion in this list does not indicate any formal commitment by the signatories. It does not indicate commitment to the specific detector designs presented, nor exclusive support for ILC over other collider programs.

A. Abada[171], T. Abe[24], T. Abe[236], J. M. Abernathy[379], H. Abramowicz[265], A. Abusleme[231], S. Aderhold[47], O. Adeyemi[333], E. Adli[357,251], C. Adloff[164], C. Adolphsen[251], K. Afanaciev[209], M. Aguilar[31], S. Ahmad[93], A. Ahmed[382], H. Aihara[375], R. Ainsworth[237,139], S. Airi[154], M. Aizatskyi[208], T. Akagi[73], M. Akemoto[71], A. Akeroyd[367], J. Alabau-Gonzalvo[108], C. Albajar[46], J. E. Albert[379], C. Albertus[281], J. Alcaraz Maestre[31], D. Alesini[174], B. Alessandro[128], G. Alexander[265], J. P. Alexander[43], A. Alhaidari[243], N. Alipour Tehrani[33], B. C. Allanach[323], O. Alonso[311], J. Alwall[210], J. W. Amann[251], Y. Amhis[167], M. S. Amjad[167], B. Ananthanarayan[83], A. Andreazza[386,122], N. Andreev[58], L. Andricek[186], M. Anduze[172], D. Angal-Kalinin[258], N. Anh Ky[106,394], K. A. Aniol[18], K. I. Aoki[148], M. Aoki[148], H. Aoyagi[137], S. Aoyama[250], S. J. Aplin[47], R. B. Appleby[343,40], J. Arafune[96], Y. Arai[71], S. Araki[71], L. Arazi[404], A. Arbey[50], D. Ariza[47], T. Arkan[58], N. D. Arnold[7], D. Arogancia[194], F. Arteche[113], A. Aryshev[71], S. Asai[375], T. Asaka[137], T. Asaka[212], E. Asakawa[220], M. Asano[333], F. B. Asiri[58], D. Asner[225], M. Asorey[286], D. Attié[21], J. E. Augustin[169], D. B. Augustine[58], C. S. Aulakh[226], E. Avetisyan[47], V. Ayvazyan[47], N. Azaryan[142], F. Azfar[358], T. Azuma[246], O. Bachynska[47], H. Baer[355], J. Bagger[141], A. Baghdasaryan[407], S. Bai[102], Y. Bai[384], I. Bailey[40,176], V. Balagura[172,107], R. D. Ball[329], C. Baltay[405], K. Bamba[198], P. S. Bambade[167], Y. Ban[9], E. Banas[268], H. Band[384], K. Bane[251], M. Barbi[362], V. Barger[384], B. Barish[17,65], T. Barklow[251], R. J. Barlow[392], M. Barone[58,65], I. Bars[368], S. Barsuk[167], P. Bartalini[210], R. Bartoldus[251], R. Bates[332], M. Battaglia[322,33], J. Baudot[94], M. Baylac[170], P. Bechtle[295], U. Becker[185,33], M. Beckmann[47], F. Bedeschi[126], C. F. Bedoya[31], S. Behari[58], O. Behnke[47], T. Behnke[47], G. Belanger[165], S. Belforte[129], I. Belikov[94], K. Belkadhi[172], A. Bellerive[19], C. Belver Aguilar[108], A. Belyaev[367,259], D. Benchekroun[184], M. Beneke[57], M. Benoit[33], A. Benot-Morell[33,108], S. Bentvelsen[213], L. Benucci[331], J. Berenguer[31], T. Bergauer[224], S. Berge[138], E. Berger[7], J. Berger[251], C. M. U. Berggren[47], Z. Bern[318], J. Bernabeu[108], N. Bernal[295], G. Bernardi[169], W. Bernreuther[235], M. Bertucci[119], M. Besancon[21], M. Bessner[47], A. Besson[94,306], D. R. Bett[358,140], A. J. Bevan[234], A. Bhardwaj[326], A. Bharucha[333], G. Bhattacharyya[241], B. Bhattacherjee[150], B. Bhuyan[86], M. E. Biagini[174], L. Bian[102], F. Bianchi[128], O. Biebel[182], T. R. Bieler[190], C. Biino[128], B. Bilki[7,337], S. S. Biswal[221], V. Blackmore[358,140], J. J. Blaising[164], N. Blaskovic Kraljevic[358,140], G. Blazey[217], I. Bloch[48], J. Bluemlein[48], B. Bobchenko[107], T. Boccali[126], J. R. Bogart[251], V. Boisvert[237], M. Bonesini[121], R. Boni[174], J. Bonnard[168], G. Bonneaud[169], S. T. Boogert[237,139], L. Boon[233,7],

G. Boorman[237,139], E. Boos[180], M. Boronat[108], K. Borras[47], L. Bortko[48], F. Borzumati[272],
M. Bosman[294], A. Bosotti[119], F. J. Botella[108], S. Bou Habib[401], P. Boucaud[171], J. Boudagov[142],
G. Boudoul[89], V. Boudry[172], D. Boumediene[168], C. Bourgeois[167], A. Boveia[52], A. Brachmann[251],
J. Bracinik[315], J. Branlard[47,58], B. Brau[346], J. E. Brau[356], R. Breedon[320], M. Breidenbach[251],
A. Breskin[404], S. Bressler[404], V. Breton[168], H. Breuker[33], C. Brezina[295], C. Briegel[58],
J. C. Brient[172], T. M. Bristow[329], D. Britton[332], I. C. Brock[295], S. J. Brodsky[251], F. Broggi[119],
G. Brooijmans[42], J. Brooke[316], E. Brost[356], T. E. Browder[334], E. Brücken[69], G. Buchalla[182],
P. Buchholz[301], W. Buchmuller[47], P. Bueno[111], V. Buescher[138], K. Buesser[47], E. Bulyak[208],
D. L. Burke[251], C. Burkhart[251], P. N. Burrows[358,140], G. Burt[40], E. Busato[168], L. Butkowski[47],
S. Cabrera[108], E. Cabruja[32], M. Caccia[389,122], Y. Cai[251], S. S. Caiazza[47,333], O. Cakir[6],
P. Calabria[89], C. Calancha[71], G. Calderini[169], A. Calderon Tazon[110], S. Callier[93], L. Calligaris[47],
D. Calvet[168], E. Calvo Alamillo[31], A. Campbell[47], G. I. E. Cancelo[58], J. Cao[102], L. Caponetto[89],
R. Carcagno[58], M. Cardaci[201], C. Carloganu[168], S. Caron[97,213], C. A. Carrillo Montoya[123],
K. Carvalho Akiba[291], J. Carwardine[7], R. Casanova Mohr[311], M. V. Castillo Gimenez[108],
N. Castro[175], A. Cattai[33], M. Cavalli-Sforza[294], D. G. Cerdeno[111], L. Cerrito[234], G. Chachamis[108],
M. Chadeeva[107], J. S. Chai[260], D. Chakraborty[217], M. Champion[254], C. P. Chang[201], A. Chao[251],
Y. Chao[210], J. Charles[28], M. Charles[358], B. E. Chase[58], U. Chattopadhyay[81], J. Chauveau[169],
M. Chefdeville[164], R. Chehab[89], A. Chen[201], C. H. Chen[203], J. Chen[102], J. W. Chen[210],
K. F. Chen[210], M. Chen[330,102], S. Chen[199], Y. Chen[1], Y. Chen[102], J. Cheng[102], T. P. Cheng[351],
B. Cheon[66], M. Chera[47], Y. Chi[102], P. Chiappetta[28], M. Chiba[276], T. Chikamatsu[193],
I. H. Chiu[210,210], G. C. Cho[220], V. Chobanova[186], J. B. Choi[36,36], K. Choi[157], S. Y. Choi[37],
W. Choi[375,248], Y. I. Choi[260], S. Choroba[47], D. Choudhury[326], D. Chowdhury[83], G. Christian[358,140],
M. Church[58], J. Chyla[105], W. Cichalewski[263,47], R. Cimino[174], D. Cinca[337], J. Clark[58],
J. Clarke[258,40], G. Claus[94], E. Clement[316,259], C. Clerc[172], J. Cline[187], C. Coca[206], T. Cohen[251],
P. Colas[21], A. Colijn[213], N. Colino[31], C. Collard[94], C. Colledani[94], N. Collomb[258], J. Collot[170],
C. Combaret[89], B. Constance[33], C. A. Cooper[58], W. E. Cooper[58], G. Corcella[174], E. Cormier[29],
R. Cornat[172], P. Cornebise[167], F. Cornet[281], G. Corrado[123], F. Corriveau[187], J. Cortes[286],
E. Cortina Gil[303], S. Costa[308], F. Couchot[167], F. Couderc[21], L. Cousin[94], R. Cowan[185],
W. Craddock[251], A. C. Crawford[58], J. A. Crittenden[43], J. Cuevas[283], D. Cuisy[167], F. Cullinan[237],
B. Cure[33], E. Currás Rivera[110], D. Cussans[316], J. Cvach[105], M. Czakon[235], K. Czuba[401],
H. Czyz[365], J. D'Hondt[399], W. Da Silva[169], O. Dadoun[167], M. Dahiya[327], J. Dai[102],
C. Dallapiccola[346], C. Damerell[259], M. Danilov[107], D. Dannheim[33], N. Dascenzo[47,238], S. Dasu[384],
A. K. Datta[67], T. S. Datta[115], P. Dauncey[80], T. Davenne[259], J. David[169], M. Davier[167],
W. De Boer[90], S. De Cecco[169], S. De Curtis[120], N. De Groot[97,213], P. De Jong[213],
S. De Jong[97,213], C. De La Taille[93], G. De Lentdecker[307], S. De Santis[177],
J. B. De Vivie De Regie[167], A. Deandrea[89], P. P. Dechant[49], D. Decotigny[172], K. Dehmelt[257],
J. P. Delahaye[251,33], N. Delerue[167], O. Delferriere[21], F. Deliot[21], G. Della Ricca[388],
P. A. Delsart[170], M. Demarteau[7], D. Demin[142], R. Dermisek[87], F. Derue[169], K. Desch[295],
S. Descotes-Genon[171], A. Deshpande[252], A. Dexter[40], A. Dey[81], S. Dhawan[405], N. Dhingra[226],
V. Di Benedetto[58,123], B. Di Girolamo[33], M. A. Diaz[231], A. Dieguez[311], M. Diehl[47], R. Diener[47],
S. Dildick[331], M. O. Dima[206], P. Dinaucourt[167,93], M. S. Dixit[19], T. Dixit[252], L. Dixon[251],
A. Djouadi[171], S. Doebert[33], M. Dohlus[47], Z. Dolezal[34], H. Dong[102], L. Dong[102], A. Dorokhov[94],
A. Dosil[112], A. Dovbnya[208], T. Doyle[332], G. Doziere[94], M. Dragicevic[224], A. Drago[174],
A. J. Dragt[345], Z. Drasal[34], I. Dremin[179], V. Drugakov[209], J. Duarte Campderros[110],
F. Duarte Ramos[33], A. Dubey[272], A. Dudarev[142], E. Dudas[171,171], L. Dudko[180], C. Duerig[47],
G. Dugan[43], W. Dulinski[94], F. Dulucq[93], L. Dumitru[206], P. J. Dunne[80], A. Duperrin[27],
M. Düren[147], D. Dzahini[170], H. Eberl[224], G. Eckerlin[47], P. Eckert[297], N. R. Eddy[58],

W. Ehrenfeld[295], G. Eigen[314], S. Eisenhardt[329], L. Eklund[332], L. Elementi[58], U. Ellwanger[171], E. Elsen[47], I. Emeliantchik[209], L. Emery[7], K. Enami[71], K. Endo[71], M. Endo[375], J. Engels[47], C. Englert[49], S. Eno[345], A. Enomoto[71], S. Enomoto[197], F. Eozenou[21], R. Erbacher[320], G. Eremeev[269], J. Erler[287], R. Escribano[294], D. Esperante Pereira[108], D. Espriu[311], E. Etzion[265,33], S. Eucker[47], A. Evdokimov[336,107], E. Ezura[71], B. Faatz[47], G. Faisel[201], L. Fano[125], A. Faraggi[340], A. Fasso[251], A. Faus-Golfe[108], L. Favart[307], N. Feege[257], J. L. Feng[321], T. Ferber[47], J. Ferguson[33], J. Fernández[283], P. Fernández Martínez[108], E. Fernandez[294,293], M. Fernandez Garcia[110], J. L. Fernandez-Hernando[130], P. Fernandez-Martinez[32], J. Fernandez-Melgarejo[33], A. Ferrer[108], F. Ferri[21], S. Fichet[117], T. Fifield[347], L. Filkov[179], F. Filthaut[97,213], A. Finch[176], H. E. Fisk[58], T. Fiutowski[2], H. Flaecher[316], J. W. Flanagan[71], I. Fleck[301], M. Fleischer[47], C. Fleta[32], J. Fleury[93], D. Flores[32], M. Foley[58], M. Fontannaz[171], K. Foraz[33], N. Fornengo[128], F. Forti[126,391], B. Foster[47,140], M. C. Fouz[31], P. H. Frampton[353], K. Francis[7], S. Frank[224], A. Freitas[361], A. Frey[64], R. Frey[356], M. Friedl[224], C. Friedrich[48], M. Frigerio[163], T. Frisson[167], M. Frotin[172], R. Frühwirth[224], R. Fuchi[378], E. Fuchs[47], K. Fujii[71], J. Fujimoto[71], H. Fuke[134], B. Fuks[94,33], M. Fukuda[71], S. Fukuda[71], K. Furukawa[71], T. Furuya[71], T. Fusayasu[195], J. Fuster[108], N. Fuster[108], Y. Fuwa[95,159], A. Gaddi[33], K. Gadow[47], F. Gaede[47], R. Gaglione[164], S. Galeotti[126], C. Gallagher[356], A. A. Gallas Torreira[112], L. Gallin-Martel[170], A. Gallo[174], D. Gamba[140,33], D. Gamba[128], J. Gao[102], Y. Gao[24], P. H. Garbincius[58], F. Garcia[69], C. Garcia Canal[288], J. E. Garcia Navarro[108], P. Garcia-Abia[31], J. J. Garcia-Garrigos[108], L. Garcia-Tabares[31], C. García[108], J. V. García Esteve[286], I. García García[108], S. K. Garg[408], L. Garrido[311], E. Garutti[333], T. Garvey[167,261], M. Gastal[33], F. Gastaldi[172], C. Gatto[58,123], N. Gaur[326], D. Gavela Pérez[31], P. Gay[168], M. B. Gay Ducati[109], L. Ge[251], R. Ge[102], A. Geiser[47], A. Gektin[100], A. Gellrich[47], M. H. Genest[170], R. L. Geng[269], S. Gentile[387,127], A. Gerbershagen[33,358], R. Gerig[7], S. German[111], H. Gerwig[33], S. Ghazaryan[47], P. Ghislain[169], D. K. Ghosh[81], S. Ghosh[115], S. Giagu[387,127], L. Gibbons[43], S. Gibson[139,33], V. Gilewsky[143], A. Gillespie[328], F. Gilman[20], B. Gimeno Martínez[108], D. M. Gingrich[312,279], C. M. Ginsburg[58], D. Girard[164], J. Giraud[170], G. F. Giudice[33], L. Gladilin[180], P. Gladkikh[208], C. J. Glasman[343,40], R. Glattauer[224], N. Glover[49], J. Gluza[365], K. Gnidzinska[263], R. Godbole[83], S. Godfrey[19], F. Goertz[54], M. Goffe[94], N. Gogitidze[179,47], J. Goldstein[316], B. Golob[144,341], G. Gomez[110], V. Goncalves[290], R. J. Gonsalves[256], I. González[283], S. González De La Hoz[108], F. J. González Sánchez[110], G. Gonzalez Parra[294], S. Gopalakrishna[103], I. Gorelov[352], D. Goswami[86], S. Goswami[229], T. Goto[71], K. Gotow[398], P. Göttlicher[47], M. Götze[385], A. Goudelis[165], P. Goudket[258], S. Gowdy[33], O. A. Grachov[309], N. A. Graf[251], M. Graham[251], A. Gramolin[15], R. Granier De Cassagnac[172], P. Grannis[257], P. Gras[21], M. Grecki[47], T. Greenshaw[339], D. Greenwood[181], C. Grefe[33], M. Grefe[111], I. M. Gregor[47], D. Grellscheid[49], G. Grenier[89], M. Grimes[316], C. Grimm[58], O. Grimm[53], B. Grinyov[100], B. Gripaios[323], K. Grizzard[141], A. Grohsjean[47], C. Grojean[294,33], J. Gronberg[178], D. Grondin[170], S. Groote[370], P. Gros[240], M. Grunewald[310], B. Grzadkowski[382], J. Gu[102], M. Guchait[262], S. Guiducci[174], E. Guliyev[172], J. Gunion[320], C. Günter[47], C. Gwilliam[339], N. Haba[75], H. Haber[322], M. Hachimine[197], Y. Haddad[172], L. Hagge[47], M. Hagihara[378], K. Hagiwara[71,158], J. Haley[216], G. Haller[251], J. Haller[333], K. Hamaguchi[375], R. Hamatsu[276], G. Hamel De Monchenault[21], L. L. Hammond[58], P. Hamnett[47], L. Han[364], T. Han[361], K. Hanagaki[223], J. D. Hansen[211], K. Hansen[47], P. H. Hansen[211], X. Q. Hao[70], K. Hara[71], K. Hara[378], T. Hara[71], D. Harada[83], K. Harada[161], K. Harder[259], T. Harion[297], R. V. Harlander[385], E. Harms[58], M. Harrison[13], O. Hartbrich[47,385], A. Hartin[47], T. Hartmann[300], J. Harz[47], S. Hasegawa[197], T. Hasegawa[71], Y. Hasegawa[248], M. Hashimoto[38], T. Hashimoto[62], C. Hast[251], S. Hatakeyama[135], J. M. Hauptman[118], M. Hauschild[33], M. Havranek[105], C. Hawkes[315], T. Hayakawa[197], H. Hayano[71], K. Hayasaka[198], M. Hazumi[71,253], H. J. He[24], C. Hearty[317,104], H. F. Heath[316],

T. Hebbeker[235], M. Heck[90], V. Hedberg[183], D. Hedin[217], S. M. Heindl[90], S. Heinemeyer[110], I. Heinze[47], A. Hektor[205], S. Henrot-Versille[167], O. Hensler[47], A. Heo[23], J. Herbert[258], G. Herdoiza[138], B. Hermberg[47], J. J. Hernández-Rey[108], M. J. Herrero[111], B. Herrmann[165], A. Hervé[384], J. Hewett[251], S. Hidalgo[32], B. Hidding[333,318], N. Higashi[375], N. Higashi[71], T. Higo[71], E. Higón Rodríguez[108], T. Higuchi[150], M. Hildreth[354], C. T. Hill[58], S. Hillert[295], S. Hillier[315], T. Himel[251], A. Himmi[94], S. Himori[272], Z. Hioki[374], B. Hippolyte[94], T. Hiraki[159], K. Hirano[136], S. Hirano[197], K. Hirata[71], T. Hirose[276], M. Hirsch[108], J. Hisano[197], P. M. Ho[210], A. Hoang[302], A. Hocker[58], A. Hoecker[33], M. Hoeferkamp[352], M. Hoffmann[47], W. Hollik[186], K. Homma[72], Y. Homma[154], S. Honda[378], T. Honda[71], Y. Honda[71], N. T. Hong Van[106], K. Honkavaara[47], T. Honma[71], T. Hori[236], T. Horiguchi[272], Y. Horii[197], A. Horio[196], R. Hosaka[377], Y. Hoshi[271], H. Hoshino[197], K. Hosoyama[71], J. Y. Hostachy[170], G. W. Hou[210], M. Hou[102], A. Hoummada[184], M. S. Hronek[58], T. Hu[102], C. Hu-Guo[94], M. Huang[24], T. Huang[102], E. Huedem[58], F. Hügging[295], J. L. Hugon[330], C. Hugonie[173], K. Huitu[335], P. Q. Hung[380], C. Hunt[80], U. Husemann[90], G. Hussain[24], D. Hutchcroft[339], Y. Hyakutake[79], J. C. Ianigro[89], L. E. Ibanez[111], M. Ibe[96], M. Idzik[2], H. Igarashi[78], Y. Igarashi[71], K. Igi[236], A. Ignatenko[209], O. Igonkina[213], T. Iijima[198,197], M. Iinuma[73], Y. Iiyama[20], H. Ikeda[134], K. Ikeda[71], K. Ikematsu[301], J. I. Illana[281], V. A. Ilyin[207,180], A. Imhof[333], T. Inagaki[236], T. Inagaki[197], K. Inami[197], S. Inayoshi[248], K. Inoue[161], A. Irles[108], S. Isagawa[71], N. Ishibashi[378], A. Ishida[375], K. Ishida[212], N. Ishihara[71], S. Ishihara[77], K. Ishii[71], A. Ishikawa[272], K. Ishikawa[375], K. I. Ishikawa[72], K. Ishikawa[75], T. Ishikawa[71], M. Ishitsuka[275], K. Ishiwata[17], G. Isidori[174], A. Ismail[251], S. Iso[71], T. Isogai[197], C. Issever[358], K. Itagaki[272], T. Itahashi[223], A. Ito[275], F. Ito[378], S. Ito[272], R. Itoh[71], E. Itou[71], M. I. Ivanyan[26], G. Iwai[71], S. Iwamoto[375], T. Iwamoto[116], H. Iwasaki[71], M. Iwasaki[71], Y. Iwashita[95], S. Iwata[71], S. Iwata[276], T. Izubuchi[13,236], Y. Izumiya[272], S. Jablonski[401], F. Jackson[258], J. A. Jacob[316], M. Jacquet[167], P. Jain[40], P. Jaiswal[59], W. Jalmuzna[263], E. Janas[401], R. Jaramillo Echeverría[110], J. Jaros[251], D. Jatkar[67], D. Jeans[375], R. Jedziniak[58], M. J. Jenkins[176,40], K. Jensch[47], C. P. Jessop[354], T. Jezynski[47], M. Jimbo[35], S. Jin[102], O. Jinnouchi[275], M. D. Joergensen[211], A. S. Johnson[251], S. Jolly[309], D. T. Jones[340], J. Jones[258,40], R. M. Jones[343,40], L. Jönsson[183], N. Joshi[237], C. K. Jung[257,150], N. Juntong[343,40], A. Juste[88,294], W. Kaabi[167], M. Kadastik[205], M. Kado[167,33], K. Kadota[197], E. Kajfasz[27], R. Kajikawa[197], Y. Kajiura[197], M. Kakizaki[377], E. Kako[71], H. Kakuhata[377], H. Kakuno[276], A. Kalinin[258], J. Kalinowski[382], G. E. Kalmus[259], K. Kamada[47], J. Kaminski[295], T. Kamitani[71], Y. Kamiya[116], Y. Kamiya[71], R. Kammering[47], T. Kamon[266], J. I. Kamoshita[60], T. Kanai[275], S. Kananov[265], K. Kanaya[378], M. Kaneda[33], T. Kaneko[71,253], S. Kanemura[377], K. Kaneta[75], W. Kang[102], D. Kanjilal[115], K. Kannike[205], F. Kapusta[169], D. Kar[332], P. Karataev[237,139], P. E. Karchin[403], D. Karlen[379,279], S. Karstensen[47], Y. Karyotakis[164], M. Kasemann[47], V. S. Kashikhin[58], S. Kashiwagi[273], A. Kataev[98], V. Katalev[48], Y. Kataoka[116], N. Katayama[150], R. Katayama[375], E. Kato[272], K. Kato[155], S. Kato[71], Y. Kato[153], T. Katoh[71], A. Kaukher[47], S. Kawabata[71], S. I. Kawada[73], K. Kawagoe[161], M. Kawai[71], T. Kawamoto[116], H. Kawamura[71], M. Kawamura[71], Y. Kawamura[248], S. Kawasaki[71], T. Kawasaki[212], H. Kay[47], S. Kazama[375], L. Keegan[111], J. Kehayias[150], L. Keller[251], M. A. Kemp[251], J. J. Kempster[237], C. Kenney[251], I. Kenyon[315], R. Kephart[58], J. Kerby[7], K. Kershaw[33], J. Kersten[333], K. Kezzar[151], V. G. Khachatryan[26], M. A. Khan[23], S. A. Khan[242], Y. Khoulaki[184], V. Khoze[49], H. Kichimi[71], R. Kieffer[33], C. Kiesling[186], M. Kikuchi[377], Y. Kikuta[71], M. Killenberg[47], C. S. Kim[408], D. W. Kim[63], D. Kim[23], E. J. Kim[36,36], E. S. Kim[23], G. Kim[23], H. S. Kim[23], H. D. Kim[245], J. Kim[63], S. H. Kim[378], S. K. Kim[245], S. G. Kim[87], Y. I. Kim[358,140], Y. Kimura[71], R. E. Kirby[251], F. Kircher[21], Y. Kishimoto[96], L. Kisslinger[20], T. Kitahara[375], R. Kitano[272], Y. Kiyo[146,71], C. Kleinwort[47], W. Klempt[33], P. M. Kluit[213], V. Klyukhin[180,33], M. Knecht[28], J. L. Kneur[163], B. A. Kniehl[333], K. Ko[251], P. Ko[158], D. Kobayashi[275], M. Kobayashi[71], N. Kobayashi[71],

T. Kobayashi[116], M. Koch[295], P. Kodys[34], U. Koetz[47], E. N. Koffeman[213], M. Kohda[210],
S. Koike[71], Y. Kojima[71], K. Kolodziej[365], Y. Kolomensky[319,177], S. Komamiya[375], T. Kon[244],
P. Konar[229], Y. Kondou[71], D. Kong[23], K. Kong[338], O. C. Kong[201], T. Konno[275,275], V. Korbel[47],
J. G. Körner[138], S. Korpar[344,144], S. R. Koscielniak[279], D. Kostin[47], K. Kotera[248], W. Kotlarski[382],
J. Kotula[268], E. Kou[167], V. Kovalenko[333], S. V. H. Kox[170], K. Koyama[75], M. Krämer[235],
S. Kraml[170], M. Krammer[224], M. W. Krasny[169], F. Krauss[49], T. Krautscheid[295], M. Krawczyk[382],
K. Krempetz[58], P. Križan[341,144], B. E. Krikler[80], A. Kronfeld[58], K. Kruchinin[237,139], D. Krücker[47],
K. Krüger[47], B. Krupa[268], Y. P. Kuang[24], K. Kubo[71], T. Kubo[71], T. Kubota[347], T. Kubota[275],
Y. Kubyshin[298,180], V. Kuchler[58], I. M. Kudla[202], D. Kuehn[47], J. H. Kuehn[92], C. Kuhn[94], S. Kulis[2],
S. Kulkarni[170], A. Kumar[10], S. Kumar[86], T. Kumita[276], A. Kundu[16], Y. Kuno[223], C. M. Kuo[201],
M. Kurachi[198], A. Kuramoto[253], M. Kurata[375], Y. Kurihara[71], M. Kuriki[73,71], T. Kurimoto[377],
S. Kuroda[71], K. Kurokawa[71], S. I. Kurokawa[71], H. Kuwabara[276], M. Kuze[275], J. Kvasnicka[105],
P. Kvasnicka[34], Y. Kwon[408], L. Labun[210], C. Lacasta[108], T. Lackowski[58], D. Lacour[169],
V. Lacuesta[108], R. Lafaye[164], G. Lafferty[343], B. Laforge[169], I. Laktineh[89], R. L. Lander[320],
K. Landsteiner[111], S. Laplace[169], K. J. Larsen[213], R. S. Larsen[251], T. Lastovicka[105],
J. I. Latorre[311], S. Laurien[333], L. Lavergne[169], S. Lavignac[22], R. E. Laxdal[279], A. C. Le Bihan[94],
F. R. Le Diberder[167], A. Le-Yaouanc[171], A. Lebedev[13], P. Lebrun[33], T. Lecompte[7], T. Leddig[300],
F. Ledroit[170], B. Lee[25], K. Lee[158], M. Lee[177], S. H. Lee[260], S. W. Lee[267], Y. H. Lee[210],
J. Leibfritz[58], K. Lekomtsev[71], L. Lellouch[28], M. Lemke[47], F. R. Lenkszus[7], A. Lenz[49,33],
O. Leroy[27], C. Lester[323], L. Levchuk[208], J. Leveque[164], E. Levichev[15], A. Levy[265], I. Levy[265],
J. R. Lewandowski[251], B. Li[24], C. Li[364], C. Li[102], D. Li[102], H. Li[380], L. Li[195], L. Li[247], L. Li[364],
S. Li[102], W. Li[102], X. Li[102], Y. Li[24], Y. Li[24], Y. Li[24], Z. Li[251], Z. Li[102], J. J. Liau[210], V. Libov[47],
L. Lilje[47], J. G. Lima[217], C. J. D. Lin[204], C. M. Lin[154], C. Y. Lin[201], H. Lin[102], H. H. Lin[210],
F. L. Linde[213], R. A. Lineros[108], L. Linssen[33], R. Lipton[58], M. Lisovyi[47], B. List[47], J. List[47],
B. Liu[24], J. Liu[364], R. Liu[102], S. Liu[167], S. Liu[247], W. Liu[7], Y. Liu[102], Y. Liu[337,58], Z. Liu[361],
Z. Liu[102], Z. Liu[102], A. Lleres[170], N. S. Lockyer[279,317], W. Lohmann[48,12], E. Lohrmann[333],
T. Lohse[76], F. Long[102], D. Lontkovskyi[47], M. A. Lopez Virto[110], X. Lou[102,372], A. Lounis[167],
M. Lozano Fantoba[32], J. Lozano-Bahilo[281], C. Lu[232], R. S. Lu[210], S. Lu[47], A. Lucotte[170],
F. Ludwig[47], S. Lukic[396], O. Lukina[180], N. Lumb[89], B. Lundberg[183], A. Lunin[58], M. Lupberger[295],
B. Lutz[47], P. Lutz[21], T. Lux[294], K. Lv[102], M. Lyablin[142], A. Lyapin[237,139], J. Lykken[58],
A. T. Lytle[262], L. Ma[258], Q. Ma[102], R. Ma[312], X. Ma[102], F. Machefert[167], N. Machida[377],
J. Maeda[276], Y. Maeda[159], K. Maeshima[58], F. Magniette[172], N. Mahajan[229], F. Mahmoudi[168,33],
S. H. Mai[201], C. Maiano[119], H. Mainaud Durand[33], S. Majewski[356], S. K. Majhi[81],
N. Majumdar[241], G. Majumder[262], I. Makarenko[47], V. Makarenko[209], A. Maki[71], Y. Makida[71],
D. Makowski[263], B. Malaescu[169], J. Malcles[21], U. Mallik[337], S. Malvezzi[121], O. B. Malyshev[258,40],
Y. Mambrini[171], A. Manabe[71], G. Mancinelli[27], S. K. Mandal[150], S. Mandry[309,186], S. Manen[168],
R. Mankel[47], S. Manly[363], S. Mannai[303], Y. Maravin[149], G. Marchiori[169], M. Marcisovsky[105,45],
J. Marco[110], D. Marfatia[338], J. Marin[31], E. Marin Lacoma[251], C. Marinas[295], T. W. Markiewicz[251],
O. Markin[107], J. Marshall[323], S. Martí-García[108], A. D. Martin[49], V. J. Martin[329],
G. Martin-Chassard[93], T. Martinez De Alvaro[31], C. Martinez Rivero[110], F. Martinez-Vidal[108],
H. U. Martyn[235,47], T. Maruyama[251], A. Masaike[159], T. Mashimo[116], T. Masubuchi[116],
T. Masuda[159], M. Masuzawa[71], Z. Mateusz[401], A. Matheisen[47], H. Mathez[89], J. Matias[293],
H. Matis[177], T. Matsubara[276], T. Matsuda[71], T. Matsui[377], S. Matsumoto[161], S. Matsumoto[150],
Y. Matsumoto[220], H. Matsunaga[71], T. Matsushita[154], T. S. Mattison[317], V. A. Matveev[142],
U. Mavric[47], G. Mavromanolakis[33], K. Mawatari[399], S. J. Maxfield[339], K. Mazumdar[262],
A. Mazzacane[58,123], R. L. Mccarthy[257], D. J. Mccormick[251], J. Mccormick[251], K. T. Mcdonald[232],
R. Mcduffee[324], P. Mcintosh[258], B. Mckee[251], M. Medinnis[47], S. Mehlhase[211], T. Mehrling[47,333],

A. Mehta[339], B. Mele[127], R. E. Meller[43], I. A. Melzer-Pellmann[47], L. Men[102], G. Mendiratta[83], Z. Meng[316], M. H. Merk[213,400], M. Merkin[180], A. Merlos[32], L. Merminga[279], A. B. Meyer[47], A. Meyer[235], N. Meyners[47], Z. Mi[102], P. Michelato[119], S. Michizono[71], S. Mihara[71], A. Mikhailichenko[43], D. J. Miller[309], C. Milstene[403], Y. Mimura[210], D. Minic[398], L. Mirabito[89], S. Mishima[387], T. Misumi[13], W. A. Mitaroff[224], T. Mitsuhashi[71], S. Mitsuru[71], K. Miuchi[154], K. Miyabayashi[200], A. Miyamoto[71], H. Miyata[212], Y. Miyazaki[161], T. Miyoshi[71], R. Mizuk[107], K. Mizuno[3], U. Mjörnmark[183], J. Mnich[47], G. Moeller[47], W. D. Moeller[47], K. Moenig[48], K. C. Moffeit[251], P. Mohanmurthy[269], G. Mohanty[262], L. Monaco[119], S. Mondal[81], C. Monini[170], H. Monjushiro[71], G. Montagna[359,124], S. Monteil[168], G. Montoro[298], I. Montvay[47], F. Moortgat[53], G. Moortgat-Pick[333,47], P. Mora De Freitas[172], C. Mora Herrera[30], G. Moreau[171], F. Morel[94], A. Morelos-Pineda[280], M. Moreno Llacer[108], S. Moretti[367,259], V. Morgunov[47,107], T. Mori[71], T. Mori[272], T. Mori[116], Y. Morita[71], S. Moriyama[96,150], L. Moroni[121], Y. Morozumi[71], H. G. Moser[186], A. Moszczynski[268], K. Motohashi[275], T. Moulik[249], G. Moultaka[163], D. Moya Martin[110], S. K. Mtingwa[215], G. S. Muanza[27], M. Mühlleitner[91], A. Mukherjee[85], S. Mukhopadhyay[241], M. Mulders[33], D. Muller[251], F. Müller[47], T. Müller[90], V. S. Mummidi[83], A. Münnich[47], C. Munoz[111], F. J. Muñoz Sánchez[110], H. Murayama[375,319], R. Murphy[7], G. Musat[172], A. Mussgiller[47], R. Muto[71], T. Nabeshima[377], K. Nagai[71], K. Nagai[378], S. Nagaitsev[58], T. Nagamine[272], K. Nagano[71], K. Nagao[71], Y. Nagashima[223], S. C. Nahn[185], P. Naik[316], D. Naito[159], T. Naito[71], H. Nakai[71], K. Nakai[71,223], Y. Nakai[161], Y. Nakajima[177], E. Nakamura[71], H. Nakamura[71], I. Nakamura[71], S. Nakamura[159], S. Nakamura[71], T. Nakamura[116], K. Nakanishi[71], E. Nakano[222], H. Nakano[212], H. Nakano[272], Y. Namito[71], W. Namkung[230], H. Nanjo[159], C. D. Nantista[251], O. Napoly[21], Y. Nara[4], T. Narazaki[272], S. Narita[131], U. Nauenberg[324], T. Naumann[48], S. Naumann-Emme[47], J. Navarro[108], A. Navitski[47], H. Neal[251], K. Negishi[272], K. Neichi[270], C. A. Nelson[255], T. K. Nelson[251], S. Nemecek[105], M. Neubert[138], R. Neuhaus[218], L. J. Nevay[237], D. M. Newbold[316,259], O. Nezhevenko[58], F. Nguyen[175], M. Nguyen[172], M. N. Nguyen[251], T. T. Nguyen[106], R. B. Nickerson[358], O. Nicrosini[124], C. Niebuhr[47], J. Niehoff[58], M. Niemack[43], U. Nierste[92], H. Niinomi[159], I. Nikolic[169], H. P. Nilles[295], S. Nishida[71], H. Nishiguchi[71], K. Nishiwaki[67], O. Nitoh[277], L. Niu[24], R. Noble[251], M. Noji[71], M. Nojiri[71,150], S. Nojiri[197,198], D. Nölle[47], A. Nomerotski[358], M. Nomura[135], T. Nomura[201], Y. Nomura[319,177], C. Nonaka[198], J. Noonan[7], E. Norbeck[337], Y. Nosochkov[251], D. Notz[47], O. Novgorodova[48,12], A. Novokhatski[251], J. A. Nowak[349], M. Nozaki[71], K. Ocalan[192], J. Ocariz[169], S. Oda[161], A. Ogata[71], T. Ogawa[248], T. Ogura[248], A. Oh[343], S. K. Oh[156], Y. Oh[23], K. Ohkuma[61], T. Ohl[145], Y. Ohnishi[71], K. Ohta[189], M. Ohta[71,253], S. Ohta[71,253], N. Ohuchi[71], K. Oishi[161], R. Okada[71], Y. Okada[71,253], T. Okamura[71], H. Okawa[13], T. Okugi[71], T. Okui[59], K. I. Okumura[161], Y. Okumura[52,58], L. Okun[107], H. Okuno[236], C. Oleari[390], C. Oliver[31], B. Olivier[186], S. L. Olsen[245], M. Omet[253,71], T. Omori[71], Y. Onel[337], H. Ono[214], Y. Ono[272], D. Onoprienko[251], Y. Onuki[116,150], P. Onyisi[371], T. Oogoe[71], Y. Ookouchi[159], W. Ootani[116], M. Oreglia[52], M. Oriunno[251], M. C. Orlandea[206], J. Orloff[168], M. Oroku[375,71], R. S. Orr[376], J. Osborne[33], A. Oskarsson[183], P. Osland[314], A. Osorio Oliveros[282], L. Österman[183], H. Otono[223], M. Owen[343], Y. Oyama[71], A. Oyanguren[108], K. Ozawa[71,375], J. P. Ozelis[58,191], D. Ozerov[47], G. Pásztor[304,99], H. Padamsee[43], C. Padilla[294], C. Pagani[119,386], R. Page[316], R. Pain[169], S. Paktinat Mehdiabadi[101], A. D. Palczewski[269], S. Palestini[33], F. Palla[126], M. Palmer[58], F. Palomo Pinto[285], W. Pan[102], G. Pancheri[174], M. Pandurovic[396], O. Panella[125], A. Pankov[227], Y. Papaphilippou[33], R. Paparella[119], A. Paramonov[7], E. K. Park[75], H. Park[23], S. I. Park[23], S. Park[260], S. Park[373], W. Park[23], A. Parker[323], B. Parker[13], C. Parkes[343], V. Parma[33], Z. Parsa[13], R. Partridge[251], S. Pastor[108], E. Paterson[251], M. Patra[83], J. R. Patterson[43], M. Paulini[20], N. Paver[129], S. Pavy-Bernard[167],

B. Pawlik[268], A. Pérez Vega-Leal[285], B. Pearson[355], J. S. Pedersen[211], A. Pedicini[397], S. Pedraza López[108], G. Pei[102], S. Pei[102], G. Pellegrini[32], A. Pellegrino[213], S. Penaranda[286], H. Peng[364], X. Peng[102], M. Perelstein[43], E. Perez[112], M. A. Perez-Garcia[284,114], M. Perez-Victoria[281], S. Peris[293], D. Perret-Gallix[164], H. Perrey[47], T. M. Perry[384], M. E. Peskin[251], P. Petagna[33], R. Y. Peters[64,47], T. C. Petersen[211], D. P. Peterson[43], T. Peterson[58], E. Petrakou[210], A. A. Petrov[403], A. Petrukhin[89,107], S. Pfeiffer[47], H. Pham[94], K. H. Phan[395,71], N. Phinney[251], F. Piccinini[124], A. Pich[108], R. Pichai[85], J. Piedra[283], J. Piekarski[401], A. Pierce[348], P. Pierini[119], N. Pietsch[333,47], A. Pineda[293], J. Pinfold[312,152], A. Piotrowski[263], Y. Pischalnikov[58], R. Pittau[281], M. Pivi[251], W. Placzek[132], T. Plehn[296], M. A. Pleier[13], M. Poelker[269], L. Poggioli[167], I. Pogorelsky[13], V. Poireau[164], M. E. Pol[30], I. Polak[105], F. Polci[169], M. Polikarpov[107], T. Poll[316,259], M. W. Poole[258,40], W. Porod[145], F. C. Porter[17], S. Porto[333], J. Portolés[108], R. Pöschl[167], S. Poss[33], C. T. Potter[356], P. Poulose[86], K. T. Pozniak[401], V. Prahl[47], R. Prepost[384], C. Prescott[251], D. Price[87], T. Price[315], P. S. Prieto[58], D. Protopopescu[332], D. Przyborowski[2], K. Przygoda[263], H. Przysiezniak[164], F. Ptochos[325], J. Puerta-Pelayo[31], C. Pulvermacher[90], M. Purohit[366], Q. Qin[102], F. Qiu[102], H. Qu[102], A. Quadt[64], G. Quast[90], D. Quirion[32], M. Quiros[88], J. Rademacker[316], R. Rahmat[350], S. Rai[67], M. Raidal[205], S. Rakshit[84], M. Ramilli[333], F. Rarbi[170], P. Ratoff[176], T. Raubenheimer[251], M. Rauch[91], L. Raux[93], G. Raven[400,213], P. Razis[325], V. Re[124], S. Redford[33], C. E. Reece[269], I. Reichel[177], A. Reichold[358,140], P. Reimer[105], M. Reinecke[47], A. Rekalo[100], J. Repond[7], J. Resta-Lopez[108], J. Reuter[47], J. T. Rhee[156], P. M. Ribeiro Cipriano[47], A. Ribon[33], G. Ricciardi[292,123], F. Richard[167], E. Richter-Was[132], G. Riddone[33], S. Riemann[48], T. Riemann[48], M. Rijssenbeek[257], K. Riles[348], C. Rimbault[167], R. Rimmer[269], S. D. Rindani[229], A. Ringwald[47], L. Rinolfi[33], I. Ripp-Baudot[94], I. Riu[294], T. G. Rizzo[251], P. Robbe[167], J. Roberts[140,33], A. Robson[332], G. Rodrigo[108], P. Rodriguez[251], P. Rodriguez Perez[112], K. Rolbiecki[111], P. Roloff[33], R. S. Romaniuk[401], E. Romero Adam[108], A. Ronzhin[58], L. Roos[169], E. Ros[108], A. Rosca[47], C. Rosemann[47], J. Rosiek[382], M. C. Ross[251], R. Rossmanith[90], S. Roth[235], J. Rouëné[167], A. Rowe[58], P. Rowson[251], A. Roy[115], L. Royer[168], P. Royole-Degieux[93], C. Royon[21], A. Rozanov[27], M. Ruan[172], D. L. Rubin[43], I. Rubinskiy[47], R. Rückl[145], R. Ruiz[361], R. Ruiz De Austri[108], P. Ruiz Valls[108], P. Ruiz-Femenía[108], A. Ruiz-Jimeno[110], R. Ruland[251], V. Rusinov[107], J. J. Russell[251], I. Rutkowski[401], V. Rybnikov[47], A. Ryd[43], A. Sabio Vera[111], B. Sabirov[142], J. J. Saborido Silva[112], H. F. W. Sadrozinski[322], T. Saeki[71], B. Safarzadeh[101], P. Saha[305], H. Sahoo[7], A. Sailer[33], N. Saito[71], T. Saito[272], T. Sakaguchi[13], H. Sakai[71], K. Sakai[71], K. Sakaue[402], K. Sakurai[47], R. Salerno[172], J. Salfeld-Nebgen[47], J. Salt[108], L. Sanchez[31], M. A. Sanchis Lozano[108], J. Sandweiss[405], A. Santa[377], A. Santagata[286], A. Santamaria[108], P. Santorelli[292], T. Sanuki[272], A. A. Sapronov[142], M. Sasaki[96], H. Sato[248], N. Sato[71], Y. Sato[272], M. Satoh[71], E. Sauvan[164], V. Saveliev[238,47], A. Savoy-Navarro[166,126], M. Sawabe[71], R. Sawada[116], H. Sawamura[402], L. Sawyer[181], O. Schäfer[300,47], R. Schäfer[47], J. Schaffran[47], T. Schalk[322,251], R. D. Schamberger[257], J. Scheirich[34], G. Schierholz[47], F. P. Schilling[90], F. Schirra[89], F. Schlander[47], H. Schlarb[47], D. Schlatter[33], P. Schleper[333], J. L. Schlereth[7], R. D. Schlueter[177], C. Schmidt[47], U. Schneekloth[47], S. Schnetzer[239], T. Schoerner-Sadenius[47], M. Schram[225], H. J. Schreiber[48], S. Schreiber[47], K. P. Schüler[47], D. Schulte[33], H. C. Schultz-Coulon[297], M. Schumacher[5], S. Schumann[64], B. A. Schumm[322], M. H. Schune[167], S. Schuwalow[333,48], C. Schwanda[224], C. Schwanenberger[343], F. Schwartzkopff[295], D. J. Scott[258,58], F. Sefkow[47,33], A. Segui[286], N. Seguin-Moreau[93], S. Seidel[352], Y. Seiya[222], J. Sekaric[338], K. Seki[197], S. Sekmen[33], S. Seletskiy[13], S. Sen[337], E. Senaha[158], K. Senyo[406], S. Senyukov[94], I. Serenkova[227], D. A. Sergatskov[58], H. Sert[47,333], D. Sertore[119], A. Seryi[358,140], O. Seto[74], R. Settles[186], P. Sha[102], S. Shahid[301], A. Sharma[33], G. Shelkov[142], W. Shen[297], J. C. Sheppard[251], M. Sher[41], C. Shi[102], H. Shi[102], T. Shidara[71], W. Shields[237,139], M. Shimada[71], H. Shimizu[71], Y. Shimizu[272],

M. Shimojima[195], S. Shimojima[276], T. Shindou[155], N. Shinoda[272], Y. Shinzaki[272], M. Shioden[78,71], I. Shipsey[233], S. Shirabe[161], M. J. Shirakata[71], T. Shirakata[71], G. Shirkov[142], T. Shishido[71], T. Shishido[71], J. G. Shiu[210], R. Shivpuri[326], R. Shrock[257], T. Shuji[71], N. Shumeiko[209], B. Shuve[228,188], P. Sicho[105], A. M. Siddiqui[133], P. Sievers[33], D. Sikora[401], D. A. Sil[86], F. Simon[186], N. B. Sinev[356], W. Singer[47], X. Singer[47], B. K. Singh[10], R. K. Singh[82], N. Sinha[103], R. Sinha[103], K. Sinram[47], T. Sinthuprasith[14], P. Skubic[355], R. Sliwa[167], I. Smiljanic[396], J. R. Smith[373,7], J. C. Smith[251,43], S. R. Smith[251], J. Smolík[105,45], J. Snuverink[237,139], B. Sobloher[47], J. Sola[311], C. Soldner[186,57], S. Soldner-Rembold[343], D. Son[23], H. S. Song[260], N. Sonmez[51], A. Sopczak[44], D. E. Soper[356], P. Spagnolo[126], S. Spannagel[47], M. Spannowsky[49], A. Sparkes[329], C. M. Spencer[251], H. Spiesberger[138], M. Spira[261], M. Stahlhofen[47], M. Stanescu-Bellu[48], M. Stanitzki[47], S. Stapnes[33], P. Starovoitov[209], F. Staufenbiel[48], L. Steder[47], M. Steder[47], A. Steen[89], G. Steinbrueck[333], M. Steinhauser[92], F. Stephan[48], W. Stephen[237], S. Stevenson[358], I. Stewart[185], D. Stöckinger[264], H. Stoeck[369], M. Strauss[355], S. Striganov[58], D. M. Strom[356], R. Stromhagen[47], J. Strube[33], A. Strumia[205], G. Stupakov[251], N. Styles[47], D. Su[251], F. Su[102], S. Su[313], J. Suarez Gonzalez[209], Y. Sudo[161], T. Suehara[116], F. Suekane[274], Y. Suetsugu[71], R. Sugahara[71], A. Sugamoto[220], H. Sugawara[71], Y. Sugimoto[71], A. Sugiyama[240], H. Sugiyama[377], M. K. Sullivan[251], Y. Sumino[272], T. Sumiyoshi[276,71], H. Sun[102], M. Sun[20], X. Sun[170], Y. Sun[102], Y. Susaki[197], T. Suwada[71], A. Suzuki[71], S. Suzuki[240], Y. Suzuki[71], Y. Suzuki[73], Z. Suzuki[272], K. Swientek[2], C. Swinson[13], Z. M. Szalata[251], B. Szczezepanski[47], M. Szelezniak[94], J. Szewinski[202], A. Sznajder[289], L. Szymanowski[202], H. Tabassam[329], K. Tackmann[47], M. Taira[71], H. Tajima[197,251], F. Takahashi[272], R. Takahashi[71], R. Takahashi[75], T. Takahashi[73], Y. Takahashi[197], K. Takata[71], F. Takayama[160], Y. Takayasu[75], H. Takeda[154], S. Takeda[71], T. Takeshita[248], A. Taketani[236], Y. Takeuchi[378], T. Takimi[262], Y. Takubo[71], Y. Tamashevich[47], M. Tamsett[181], M. Tanabashi[198,197], T. Tanabe[116], G. Tanaka[161], M. M. Tanaka[71], M. Tanaka[223], R. Tanaka[73], H. Taniuchi[377], S. Tapprogge[138], E. Tarkovsky[107], M. A. Tartaglia[58], X. R. Tata[334], T. Tauchi[71], M. Tawada[71], G. Taylor[347], A. M. Teixeira[168], V. I. Telnov[15,219], P. Tenenbaum[251], E. Teodorescu[206], S. Terada[71], Y. Teramoto[222], H. Terao[200], A. Terashima[71], S. Terui[71], N. Terunuma[71], M. Terwort[47], M. Tesar[186], F. Teubert[33], T. Teubner[340], R. Teuscher[376], T. Theveneaux-Pelzer[168], D. Thienpont[93,172], J. Thom-Levy[43], M. Thomson[323], J. Tian[71], X. Tian[366], M. Tigner[43], J. Timmermans[213], V. Tisserand[164], M. Titov[21], S. Tjampens[164], K. Tobe[197], K. Tobioka[150,319], K. Toda[278], M. Toda[71], N. Toge[71], J. Tojo[161], K. Tokushuku[71], T. Toma[49], R. Tomas[33], T. Tomita[161], A. Tomiya[223], M. Tomoto[197,198], K. Toms[352], M. Tonini[47], F. Toral[31], E. Torrence[356], E. Torrente-Lujan[33], N. Toumbas[325], C. Touramanis[339], F. Toyoda[161], K. Toyomura[71], G. Trahern[55], T. H. Tran[172], W. Treberspurg[224], J. Trenado[311], M. Trimpl[58], S. Trincaz-Duvoid[169], M. Tripathi[320], W. Trischuk[376], M. Trodden[360], G. Trubnikov[142], H. C. Tsai[39], J. F. Tsai[210], K. H. Tsao[336], R. Tschirhart[58], E. Tsedenbaljir[210], S. Y. Tseng[201], T. Tsuboyama[71], A. Tsuchiya[250], K. Tsuchiya[71], T. Tsukamoto[71], K. Tsumura[197], S. Tsuno[71], T. Tsurugai[189], T. Tsuyuki[96], B. Tuchming[21], P. V. Tyagi[68,254], I. Tyapkin[142], M. Tytgat[331], K. Uchida[295], F. Uchiyama[71], Y. Uchiyama[116], S. Uehara[71], H. Ueno[161], K. Ueno[71], K. Ueno[71], K. Ueshima[274], Y. Uesugi[73], N. Ujiie[71], F. Ukegawa[378], N. Ukita[378], M. Ullán[32], H. Umeeda[72], K. Umemori[71], Y. Unno[66], S. Uozumi[23], J. Urakawa[71], A. M. Uranga[111], J. Urresti[32], A. Ushakov[333], I. Ushiki[272], Y. Ushiroda[71], A. V[83], P. Vázquez Regueiro[112], L. Vacavant[27], G. Valencia[118], L. Valery[168], I. Valin[94], J. W. Valle[108], C. Vallee[27], N. Van Bakel[213], H. Van Der Graaf[213], N. Van Der Kolk[167], E. Van Der Kraaij[33], B. Van Doren[338], B. Van Eijk[213], R. Van Kooten[87], W. T. Van Oers[279], D. Vanegas[108], P. Vanhoefer[186], P. Vankov[47], P. Varghese[58], A. Variola[167], R. Varma[85], G. Varner[334], G. Vasileiadis[162], A. Vauth[47], J. Velthuis[316], S. K. Vempati[83], V. Vento[108], M. Venturini[177], M. Verderi[172], P. Verdier[89], A. Verdugo[31],

A. Vicente[171], J. Vidal-Perona[108], H. L. R. Videau[172], I. Vila[110], X. Vilasis-Cardona[299], E. Vilella[311], A. Villamor[32], E. G. Villani[259], J. A. Villar[286], M. A. Villarejo Bermúdez[108], D. Vincent[169], P. Vincent[169], J. M. Virey[28], A. Vivoli[58], V. Vogel[47], R. Volkenborn[47], O. Volynets[47], F. Von Der Pahlen[110], E. Von Toerne[295], B. Vormwald[47], A. Voronin[180], M. Vos[108], J. H. Vossebeld[339], G. Vouters[164], Y. Voutsinas[94,47], V. Vrba[105,45], M. Vysotsky[107], D. Wackeroth[256], A. Wagner[47], C. E. Wagner[7,52], R. Wagner[7], S. R. Wagner[324], W. Wagner[385], J. Wagner-Kuhr[90], A. P. Waite[251], M. Wakayama[197], Y. Wakimoto[276], R. Walczak[358,140], R. Waldi[300], D. G. E. Walker[251], N. J. Walker[47], M. Walla[47], C. J. Wallace[49], S. Wallon[171,393], D. Walsh[328], S. Walston[178], W. A. T. Wan Abdullah[342], D. Wang[102], G. Wang[102], J. Wang[251], L. Wang[251], L. Wang[52], M. H. Wang[251], M. Z. Wang[210], Q. Wang[102], Y. Wang[102], Z. Wang[24], R. Wanke[138], C. Wanotayaroj[356], B. Ward[8], D. Ward[323], B. Warmbein[47], M. Washio[402], K. Watanabe[71], M. Watanabe[212], N. Watanabe[71], T. Watanabe[155], Y. Watanabe[71], S. Watanuki[272], Y. Watase[71], N. K. Watson[315], G. Watts[383], M. M. Weber[90], H. C. Weddig[47], H. Weerts[7], A. W. Weidemann[251], G. Weiglein[47], A. Weiler[47], S. Weinzierl[138], H. Weise[47], A. Welker[138], N. Welle[47], J. D. Wells[33,348], M. Wendt[58,33], M. Wenskat[47], H. Wenzel[58], N. Wermes[295], U. Werthenbach[301], W. Wester[58], L. Weuste[186,57], A. White[373], G. White[251], K. H. Wichmann[47], M. Wielers[183], R. Wielgos[58], W. Wierba[202], T. Wilksen[47], S. Willocq[346], F. F. Wilson[259], G. W. Wilson[338], P. B. Wilson[251], M. Wing[309], M. Winter[94], K. Wittenburg[47], P. Wittich[43], M. Wobisch[181], A. Wolski[339,40], M. D. Woodley[251], M. B. Woods[251], M. Worek[385], S. Worm[33,259], G. Wormser[167], D. Wright[178], Z. Wu[251], C. E. Wulz[224], S. Xella[211], G. Xia[40,343], L. Xia[7], A. Xiao[7], L. Xiao[251], M. Xiao[102], Q. Xiao[102], J. Xie[7], C. Xu[102], F. Xu[210], G. Xu[102], K. Yagyu[201], U. A. Yajnik[85], V. Yakimenko[251], S. Yamada[71,116], S. Yamada[71], Y. Yamada[272], Y. Yamada[402], A. Yamaguchi[274], D. Yamaguchi[275], M. Yamaguchi[272], S. Yamaguchi[272], Y. Yamaguchi[375], Y. Yamaguchi[75], A. Yamamoto[71,375], H. Yamamoto[272], K. Yamamoto[222], K. Yamamoto[118], M. Yamamoto[71], N. Yamamoto[197], N. Yamamoto[71], Y. Yamamoto[71], Y. Yamamoto[375], T. Yamamura[375], T. Yamanaka[116], S. Yamashita[116], T. Yamashita[3], Y. Yamashita[214], K. Yamauchi[197], M. Yamauchi[71], T. Yamazaki[375], Y. Yamazaki[154], J. Yan[375,71], W. Yan[364], C. Yanagisawa[257,11], H. Yang[247], J. Yang[56], U. K. Yang[245,343], Z. Yang[24], W. Yao[177], S. Yashiro[71], F. Yasuda[375], O. Yasuda[276], I. Yavin[188,228], E. Yazgan[331], H. Yokoya[377], K. Yokoya[71], H. Yokoyama[375], S. Yokoyama[275], R. Yonamine[71], H. Yoneyama[240], M. Yoshida[71], T. Yoshida[62], K. Yoshihara[116,33], S. Yoshihara[116,33], M. Yoshioka[71,272], T. Yoshioka[161], H. Yoshitama[73], C. C. Young[251], H. B. Yu[348], J. Yu[373], C. Z. Yuan[102], F. Yuasa[71], J. Yue[102], A. Zabi[172], W. Zabolotny[401], J. Zacek[34], I. Zagorodnov[47], J. Zalesak[105,58], A. F. Zarnecki[381], L. Zawiejski[268], M. Zeinali[101], C. Zeitnitz[385], L. Zembala[401], K. Zenker[47], D. Zeppenfeld[91], D. Zerwas[167], P. Zerwas[47], M. Zeyrek[192], A. Zghiche[164], J. Zhai[102], C. Zhang[102], J. Zhang[102], J. Zhang[7], Y. Zhang[24,33], Z. Zhang[167], F. Zhao[102], F. Zhao[102], T. Zhao[102], Y. Zhao[251], H. Zheng[102], Z. Zhengguo[364], L. Zhong[24], F. Zhou[251], X. Zhou[364,102], Z. Zhou[102], R. Y. Zhu[17], X. Zhu[24], X. Zhu[102], M. Zimmer[47], F. Zomer[167], T. Zoufal[47], R. Zwicky[329] 1 Academia Sinica - 128 Sec. 2, Institute of Physics, Academia Rd., Nankang, Taipei 11529, Taiwan, R.O.C.

2 AGH University of Science and Technology, Akademia Gorniczo-Hutnicza im. Stanislawa Staszica w Krakowie, Al. Mickiewicza 30 PL-30-059 Cracow, Poland

3 Aichi Medical University, Nagakute, Aichi, 480-1195, Japan

4 Akita International University, Yuwa, Akita City, 010-1292, Japan

5 Albert-Ludwigs Universität Freiburg, Physikalisches Institut, Hermann-Herder Str. 3, D-79104 Freiburg, Germany

6 Ankara Üniversitesi Fen Fakültesi, Fizik Bölümü, Dögol Caddesi, 06100 Tandoĝan Ankara, Turkey

7 Argonne National Laboratory (ANL), 9700 S. Cass Avenue, Argonne, IL 60439, USA

8 Baylor University, Department of Physics, 101 Bagby Avenue, Waco, TX 76706, USA

9 Beijing University, Department of Physics, Beijing, China 100871

10 Benares Hindu University, Benares, Varanasi 221005, India

11 Borough of Manhattan Community College, The City University of New York, Department of Science, 199 Chambers Street, New York, NY 10007, USA

12 Brandenburg University of Technology, Postfach 101344, D-03013 Cottbus, Germany

13 Brookhaven National Laboratory (BNL), P.O.Box 5000, Upton, NY 11973-5000, USA

14 Brown University, Department of Physics, Box 1843, Providence, RI 02912, USA

15 Budker Institute for Nuclear Physics (BINP), 630090 Novosibirsk, Russia

16 Calcutta University, Department of Physics, 92 A.P.C. Road, Kolkata 700009, India

17 California Institute of Technology, Physics, Mathematics and Astronomy (PMA), 1200 East California Blvd, Pasadena, CA 91125, USA

18 California State University, Los Angeles, Dept. of Physics and Astronomy, 5151 State University Dr., Los Angeles, CA 90032, USA

19 Carleton University, Department of Physics, 1125 Colonel By Drive, Ottawa, Ontario, Canada K1S 5B6

20 Carnegie Mellon University, Department of Physics, Wean Hall 7235, Pittsburgh, PA 15213, USA

21 CEA Saclay, IRFU, F-91191 Gif-sur-Yvette, France

22 CEA Saclay, Service de Physique Théorique, CEA/DSM/SPhT, F-91191 Gif-sur-Yvette Cedex, France

23 Center for High Energy Physics (CHEP) / Kyungpook National University, 1370 Sankyuk-dong, Buk-gu, Daegu 702-701, Republic of Korea

24 Center for High Energy Physics (TUHEP), Tsinghua University, Beijing, China 100084

25 Center For Quantum Spacetime (CQUeST), Sogang University, 35 Baekbeom-ro, Mapo-gu, Seoul 121-742, Republic of Korea

26 Center for the Advancement of Natural Discoveries using Light Emission (CANDLE), Acharyan 31, 0040, Yerevan, Armenia

27 Centre de Physique des Particules de Marseille (CPPM), Aix-Marseille Université, CNRS/IN2P3, 163, Avenue de Luminy, Case 902, 13288 Marseille Cedex 09, France

28 Centre de Physique Theorique, CNRS - Luminy, Universiti d''Aix - Marseille II, Campus of Luminy, Case 907, 13288 Marseille Cedex 9, France

29 Centre Lasers Intenses et Applications (CELIA), Université Bordeaux 1 - CNRS - CEA, 351 Cours de la Libération, 33405 Talence Cedex, France

30 Centro Brasileiro de Pesquisas Físicas (CBPF), Rua Dr. Xavier Sigaud, n.150 22290-180, Urca - Rio de Janeiro, RJ, Brazil

31 Centro de Investigaciones Energéticas, Medioambientales y Tecnológicas, CIEMAT, Avenida Complutense 22, E-28040 Madrid, Spain

32 Centro Nacional de Microelectrónica (CNM), Instituto de Microelectrónica de Barcelona (IMB), Campus UAB, 08193 Cerdanyola del Vallès (Bellaterra), Barcelona, Spain

33 CERN, CH-1211 Genève 23, Switzerland

34 Charles University, Institute of Particle & Nuclear Physics, Faculty of Mathematics and Physics, V Holesovickach 2, CZ-18000 Prague 8, Czech Republic

35 Chiba University of Commerce, 1-3-1 Konodai, Ichikawa-shi, Chiba, 272-8512, Japan

36 Chonbuk National University, Division of Science Education, Jeonju 561-756, Republic of Korea

37 Chonbuk National University, Physics Department, Jeonju 561-756, Republic of Korea

38 Chubu University, 1200 Matsumoto-cho, Kasugai-shi, Aichi, 487-8501, Japan

39 Chung Yuan Christian University, Department of Physics, 200 Chung Pei Rd., Chung Li 32023 Taiwan, R.O.C

40 Cockcroft Institute, Daresbury, Warrington WA4 4AD, UK

41 College of William and Mary, Department of Physics, Williamsburg, VA, 23187, USA

42 Columbia University, Department of Physics, New York, NY 10027-6902, USA

43 Cornell University, Laboratory for Elementary-Particle Physics (LEPP), Ithaca, NY 14853, USA

44 Czech Technical University in Prague, Institute of Experimental and Applied Physics (IEAP), Horska 3a/22, 12800 Prague 2, Czech Republic

45 Czech Technical University, Faculty of Nuclear Science and Physical Engineering, Brehova 7, CZ-11519 Prague 1, Czech Republic

46 Departamento de Física Teórica, Facultad de Ciencias, Módulo 15 (antiguo C-XI) y Módulo 8, Universidad Autónoma de Madrid, Campus de Cantoblanco, 28049 Madrid, Spain

47 Deutsches Elektronen-Synchrotron DESY, A Research Centre of the Helmholtz Association, Notkestrasse 85, 22607 Hamburg, Germany (Hamburg site)

48 Deutsches Elektronen-Synchrotron DESY, A Research Centre of the Helmholtz Association, Platanenallee 6, 15738 Zeuthen, Germany (Zeuthen site)

49 Durham University, Department of Physics, Ogen Center for Fundamental Physics, South Rd., Durham DH1 3LE, UK

50 École Normale Supérieure de Lyon, 46 allée d'Italie, 69364 Lyon Cedex 07, France

51 Ege University, Department of Physics, Faculty of Science, 35100 Izmir, Turkey

52 Enrico Fermi Institute, University of Chicago, 5640 S. Ellis Avenue, RI-183, Chicago, IL 60637, USA
53 ETH Zürich, Institute for Particle Physics (IPP), Schafmattstrasse 20, CH-8093 Zürich, Switzerland
54 ETH Zürich, Institute for Theoretical Physics (ITP), Wolfgang-Pauli-Str. 27, Zürich, Switzerland
55 European Spallation Source ESS AB, Box 176, 221 00 Lund, Sweden
56 Ewha Womans University, 11-1 Daehyun-Dong, Seodaemun-Gu, Seoul, 120-750, Republic of Korea
57 Excellence Cluster Universe, Technische Universität München, Boltzmannstr. 2, 85748 Garching, Germany
58 Fermi National Accelerator Laboratory (FNAL), P.O.Box 500, Batavia, IL 60510-0500, USA
59 Florida State University, Department of Physics, 77 Chieftan Way, Tallahassee, FL 32306-4350, USA
60 Fujita Gakuen Health University, Department of Physics, Toyoake, Aichi 470-1192, Japan
61 Fukui University of Technology, 3-6-1 Gakuen, Fukui-shi, Fukui 910-8505, Japan
62 Fukui University, Department of Physics, 3-9-1 Bunkyo, Fukui-shi, Fukui 910-8507, Japan
63 Gangneung-Wonju National University, 210-702 Gangneung Daehangno, Gangneung City, Gangwon Province, Republic of Korea
64 Georg-August-Universität Göttingen, II. Physikalisches Institut, Friedrich-Hund-Platz 1, 37077 Göttingen, Germany
65 Global Design Effort
66 Hanyang University, Department of Physics, Seoul 133-791, Republic of Korea
67 Harish-Chandra Research Institute, Chhatnag Road, Jhusi, Allahabad 211019, India
68 Helmholtz-Zentrum Berlin für Materialien und Energie (HZB), Wilhelm-Conrad-Röntgen Campus, BESSY II, Albert-Einstein-Str. 15, 12489 Berlin, Germany
69 Helsinki Institute of Physics (HIP), P.O. Box 64, FIN-00014 University of Helsinki, Finland
70 Henan Normal University, College of Physics and Information Engineering, Xinxiang, China 453007
71 High Energy Accelerator Research Organization, KEK, 1-1 Oho, Tsukuba, Ibaraki 305-0801, Japan
72 Hiroshima University, Department of Physics, 1-3-1 Kagamiyama, Higashi-Hiroshima, Hiroshima 739-8526, Japan
73 Hiroshima University, Graduate School of Advanced Sciences of Matter, 1-3-1 Kagamiyama, Higashi-Hiroshima, Hiroshima 739-8530, Japan
74 Hokkai-Gakuen University, 4-1-40 Asahimachi, Toyohira-ku, Sapporo 062-8605, Japan
75 Hokkaido University, Department of Physics, Faculty of Science, Kita, Kita-ku, Sapporo-shi, Hokkaido 060-0810, Japan
76 Humboldt Universität zu Berlin, Fachbereich Physik, Institut für Elementarteilchenphysik, Newtonstr. 15, D-12489 Berlin, Germany
77 Hyogo University of Teacher Education, 942-1 Shimokume, Kato-city, Hyogo 673-1494, Japan
78 Ibaraki National College of Technology, 866 Nakane, Hitachinaka, Ibaraki 312-8508, Japan
79 Ibaraki University, College of Technology, Department of Physics, Nakanarusawa 4-12-1, Hitachi, Ibaraki 316-8511, Japan
80 Imperial College, Blackett Laboratory, Department of Physics, Prince Consort Road, London, SW7 2BW, UK
81 Indian Association for the Cultivation of Science, Department of Theoretical Physics and Centre for Theoretical Sciences, Kolkata 700032, India
82 Indian Institute of Science Education and Research (IISER) Kolkata, Department of Physical Sciences, Mohanpur Campus, PO Krishi Viswavidyalaya, Mohanpur 741252, Nadia, West Bengal, India
83 Indian Institute of Science, Centre for High Energy Physics, Bangalore 560012, Karnataka, India
84 Indian Institute of Technology Indore, IET Campus, M-Block, Institute of Engineering and Technology (IET), Devi Ahilya Vishwavidyalaya Campus, Khandwa Road, Indore - 452017, Madhya Pradesh, India
85 Indian Institute of Technology, Bombay, Powai, Mumbai 400076, India
86 Indian Institute of Technology, Guwahati, Guwahati, Assam 781039, India
87 Indiana University, Department of Physics, Swain Hall West 117, 727 E. 3rd St., Bloomington, IN 47405-7105, USA
88 Institucio Catalana de Recerca i Estudis, ICREA, Passeig Lluis Companys, 23, Barcelona 08010, Spain
89 Institut de Physique Nucléaire de Lyon (IPNL), Domaine scientifique de la Doua, Bâtiment Paul Dirac 4, rue Enrico Fermi, 69622 Villeurbanne, Cedex, France
90 Institut für Experimentelle Kernphysik, KIT,Universität Karlsruhe (TH), Wolfgang-Gaede-Str. 1, Postfach 6980, 76128 Karlsruhe, Germany
91 Institut für Theoretische Physik (ITP), Karlsruher Institut für Technologie (KIT), Fakultät für Physik, Postfach 6980, 76049 Karlsruhe, Germany
92 Institut für Theoretische Teilchenphysik, Campus Süd, Karlsruher Institut für Technologie (KIT), 76128 Karlsruhe, Germany
93 Institut National de Physique Nucleaire et de Physique des Particules, 3, Rue Michel- Ange, 75794 Paris Cedex 16, France
94 Institut Pluridisciplinaire Hubert Curien, 23 Rue du Loess - BP28, 67037 Strasbourg Cedex 2, France
95 Institute for Chemical Research, Kyoto University, Gokasho, Uji, Kyoto 611-0011, Japan
96 Institute for Cosmic Ray Research, University of Tokyo, 5-1-5 Kashiwa-no-Ha, Kashiwa, Chiba 277-8582, Japan
97 Institute for Mathematics, Astrophysics and Particle Physics (IMAPP), P.O. Box 9010, 6500 GL Nijmegen, Netherlands
98 Institute for Nuclear Research, Russian Academy of Sciences (INR RAS), 60-th October Anniversary Prospect 7a, 117312, Moscow, Russia
99 Institute for Particle and Nuclear Physics, Wigner Research Centre for Physics, Hungarian Academy of Sciences, P.O. Box 49, 1525 Budapest, Hungary
100 Institute for Scintillation Materials (ISMA), 60 Lenina Ave, 61001, Kharkiv, Ukraine
101 Institute for studies in fundamental sciences (IPM), Niavaran Square, P.O. Box 19395-5746, Tehran, Iran
102 Institute of High Energy Physics - IHEP, Chinese Academy of Sciences, P.O. Box 918, Beijing, China 100049
103 Institute of Mathematical Sciences, Taramani, C.I.T. Campus, Chennai 600113, India
104 Institute of Particle Physics, Canada

105 Institute of Physics, ASCR, Academy of Science of the Czech Republic, Division of Elementary Particle Physics, Na Slovance 2, CZ-18221 Prague 8, Czech Republic
106 Institute of Physics, Vietnam Academy of Science and Technology (VAST), 10 Dao-Tan, Ba-Dinh, Hanoi 10000, Vietnam
107 Institute of Theoretical and Experimetal Physics, B. Cheremushkinskawa, 25, RU-117259, Moscow, Russia
108 Instituto de Fisica Corpuscular (IFIC), Centro Mixto CSIC-UVEG, Edificio Investigacion Paterna, Apartado 22085, 46071 Valencia, Spain
109 Instituto de Física da Universidade Federal do Rio Grande do Sul (UFRGS), Av. Bento Gonçalves 9500, Caixa Postal 15051, CEP 91501-970, Porto Alegre, RS, Brazil
110 Instituto de Fisica de Cantabria, (IFCA, CSIC-UC), Facultad de Ciencias, Avda. Los Castros s/n, 39005 Santander, Spain
111 Instituto de Física Teórica UAM/CSIC, C/ Nicolás Cabrera 13-15, Universidad Autónoma de Madrid, Cantoblanco, 28049 Madrid, Spain
112 Instituto Galego de Fisica de Altas Enerxias (IGFAE,USC) Facultad de Fisica, Campus Sur E-15782 Santiago de Compostela, Spain
113 Instituto Tecnológico de Aragón (ITA), C/ María de Luna 7-8, 50018 Zaragoza, Spain
114 Instituto Universitario de Física Fundamental y Matemáticas de la Universidad de Salamanca (IUFFyM), Casas del Parque, 37008 Salamanca, Spain
115 Inter-University Accelerator Centre, Aruna Asaf Ali Marg, Post Box 10502, New Delhi 110067, India
116 International Center for Elementary Particle Physics, University of Tokyo, Hongo 7-3-1, Bunkyo District, Tokyo 113-0033, Japan
117 International Institute of Physics, Federal University of Rio Grande do Norte, Av. Odilon Gomes de Lima, 1722 - Capim Macio - 59078-400 - Natal-RN, Brazil
118 Iowa State University, Department of Physics, High Energy Physics Group, Ames, IA 50011, USA
119 Istituto Nazionale di Fisica Nucleare (INFN), Laboratorio LASA, Via Fratelli Cervi 201, 20090 Segrate, Italy
120 Istituto Nazionale di Fisica Nucleare (INFN), Sezione di Firenze, Via G. Sansone 1, I-50019 Sesto Fiorentino (Firenze), Italy
121 Istituto Nazionale di Fisica Nucleare (INFN), Sezione di Milano Bicocca, Piazza della Scienza 3, I-20126 Milano, Italy
122 Istituto Nazionale di Fisica Nucleare (INFN), Sezione di Milano, Via Celoria 16, I-20133 Milano, Italy
123 Istituto Nazionale di Fisica Nucleare (INFN), Sezione di Napoli, Complesso Universitá di Monte Sant'Angelo,via, I-80126 Naples, Italy
124 Istituto Nazionale di Fisica Nucleare (INFN), Sezione di Pavia, Via Bassi 6, I-27100 Pavia, Italy
125 Istituto Nazionale di Fisica Nucleare (INFN), Sezione di Perugia, Via A. Pascoli, 06123 Perugia, Italy
126 Istituto Nazionale di Fisica Nucleare (INFN), Sezione di Pisa, Edificio C - Polo Fibonacci Largo B. Pontecorvo, 3, I-56127 Pisa, Italy
127 Istituto Nazionale di Fisica Nucleare (INFN), Sezione di Roma, c/o Dipartimento di Fisica - Università degli Studi di Roma "La Sapienza", P.le Aldo Moro 2, I-00185 Roma, Italy
128 Istituto Nazionale di Fisica Nucleare (INFN), Sezione di Torino, c/o Universitá di Torino, facoltá di Fisica, via P Giuria 1, 10125 Torino, Italy
129 Istituto Nazionale di Fisica Nucleare (INFN), Sezione di Trieste, Padriciano 99, I-34012 Trieste (Padriciano), Italy
130 ITER Organization, Route de Vinon-sur-Verdon, 13115 St. Paul-lez-Durance, France
131 Iwate University, 4-3-5 Ueda, Morioka, Iwate, 020-8551, Japan
132 Jagiellonian University, Institute of Physics, Ul. Reymonta 4, PL-30-059 Cracow, Poland
133 Jamia Millia Islamia, Department of Physics, Jamia Nagar, New Delhi 110025, India
134 Japan Aerospace Exploration Agency, Sagamihara Campus, 3-1-1 Yoshinodai, Sagamihara, Kanagawa 220-8510 , Japan
135 Japan Atomic Energy Agency, 4-49 Muramatsu, Tokai-mura, Naka-gun, Ibaraki 319-1184, Japan
136 Japan Atomic Energy Agency, Tokai Research and Development Center, 2-4 Shirane Shirakata, Tokai-mura, Naka-gun, Ibaraki 319-1195, Japan
137 Japan Synchrotron Radiation Research Institute (JASRI), 1-1-1, Kouto, Sayo-cho, Sayo-gun, Hyogo 679-5198, Japan
138 Johannes Gutenberg Universität Mainz, Institut für Physik, 55099 Mainz, Germany
139 John Adams Institute for Accelerator Science at Royal Holloway University of London, Egham Hill, Egham, Surrey TW20 0EX, UK
140 John Adams Institute for Accelerator Science at University of Oxford, Denys Wilkinson Building, Keble Road, Oxford OX1 3RH, UK
141 Johns Hopkins University - Henry A. Rowland Department of Physics & Astronomy 3701 San Martin Drive, Baltimore, Maryland (MD) 21218, USA
142 Joint Institute for Nuclear Research (JINR), Joliot-Curie 6, 141980, Dubna, Moscow Region, Russia
143 Joint Institute for Power and Nuclear Research "Sosny" at National Academy of Sciences of Belarus, 99 Academician A.K.Krasin Str., Minsk BY-220109, Belarus
144 Jozef Stefan Institute, Jamova cesta 39, 1000 Ljubljana, Slovenia
145 Julius-Maximilians-Universität Würzburg, Fakultät für Physik und Astronomie, Am Hubland, 97074 Würzburg, Germany
146 Juntendo University, School of Medicine, Dept. of Physics, Hiraga-gakuendai 1-1, Inzai-shi, Chiba 270-1695, Japan
147 Justus-Liebig-Universität Gießen, II. Physikalisches Institut, Heinrich-Buff-Ring 16, 35392 Gießen, Germany
148 Kanazawa University, Institute for Theoretical Physics (KITP), School of Mathematics and Physics, College of Science and Engineering, Kakuma-machi, Kanazawa city, Ishikawa 920-1192, Japan
149 Kansas State University, Department of Physics, 116 Cardwell Hall, Manhattan, KS 66506, USA
150 Kavli Institute for the Physics and Mathematics of the Universe (Kavli IPMU), University of Tokyo, 5-1-5 Kashiwanoha, Kashiwa, 277-8583, Japan
151 King Saud University (KSU), Dept. of Physics, P.O. Box 2454, Riyadh 11451, Saudi Arabia
152 King's College London - Department of physics, Strand, London WC2R 2LS, London, UK
153 Kinki University, Department of Physics, 3-4-1 Kowakae, Higashi-Osaka, Osaka 577-8502, Japan
154 Kobe University, Department of Physics, 1-1 Rokkodai-cho, Nada-ku, Kobe, Hyogo 657-8501, Japan

155 Kogakuin University, Department of Physics, Shinjuku Campus, 1-24-2 Nishi-Shinjuku, Shinjuku-ku, Tokyo 163-8677, Japan
156 Konkuk University, 93-1 Mojin-dong, Kwanglin-gu, Seoul 143-701, Republic of Korea
157 Korea Advanced Institute of Science & Technology, Department of Physics, 373-1 Kusong-dong, Yusong-gu, Taejon 305-701, Republic of Korea
158 Korea Institute for Advanced Study (KIAS), School of Physics, 207-43 Cheongryangri-dong, Dongdaemun-gu, Seoul 130-012, Republic of Korea
159 Kyoto University, Department of Physics, Kitashirakawa-Oiwakecho, Sakyo-ku, Kyoto 606-8502, Japan
160 Kyoto University, Yukawa Institute for Theoretical Physics, Kitashirakawa-Oiwakecho, Sakyo-Ku, Kyoto 606-8502, Japan
161 Kyushu University, Department of Physics, 6-10-1 Hakozaki, Higashi-ku, Fukuoka 812-8581, Japan
162 L.P.T.A., UMR 5207 CNRS-UM2, Université Montpellier II, Case Courrier 070, Bât. 13, place Eugène Bataillon, 34095 Montpellier Cedex 5, France
163 Laboratoire Charles Coulomb UMR 5221 CNRS-UM2, Université Montpellier 2, Place Eugène Bataillon - CC069, 34095 Montpellier Cedex 5, France
164 Laboratoire d'Annecy-le-Vieux de Physique des Particules (LAPP) , Université de Savoie, CNRS/IN2P3, 9 Chemin de Bellevue, BP 110, F-74941 Annecy-Le-Vieux Cedex, France
165 Laboratoire d'Annecy-le-Vieux de Physique Theorique (LAPTH), Chemin de Bellevue, BP 110, F-74941 Annecy-le-Vieux Cedex, France
166 Laboratoire d'AstroParticules et Cosmologie (APC), Université Paris Diderot-Paris 7 - CNRS/IN2P3, Bâtiment Condorcet, Case 7020, 75205 Paris Cedex 13, France
167 Laboratoire de l'Accélérateur Linéaire (LAL), Université Paris-Sud 11, Bâtiment 200, 91898 Orsay, France
168 Laboratoire de Physique Corpusculaire de Clermont-Ferrand (LPC), Université Blaise Pascal, I.N.2.P.3./C.N.R.S., 24 avenue des Landais, 63177 Aubière Cedex, France
169 Laboratoire de Physique Nucléaire et des Hautes Energies (LPNHE), UPMC, UPD, IN2P3/CNRS, 4 Place Jussieu, 75005, Paris Cedex 05, France
170 Laboratoire de Physique Subatomique et de Cosmologie (LPSC), Université Joseph Fourier (Grenoble 1), CNRS/IN2P3, Institut Polytechnique de Grenoble, 53 rue des Martyrs, F-38026 Grenoble Cedex, France
171 Laboratoire de Physique Theorique, Université de Paris-Sud XI, Batiment 210, F-91405 Orsay Cedex, France
172 Laboratoire Leprince-Ringuet (LLR), École polytechnique – CNRS/IN2P3, Route de Saclay, F-91128 Palaiseau Cedex, France
173 Laboratoire Univers et Particules de Montpellier (LUPM) - UMR5299, Université de Montpellier II, Place Eugène Bataillon - Case courrier 72, 34095 Montpellier Cedex 05, France
174 Laboratori Nazionali di Frascati, via E. Fermi, 40, C.P. 13, I-00044 Frascati, Italy
175 Laboratório de Instrumentação e Física Experimental de Partículas (LIP LISBOA), Av. Elias Garcia 14 - 1°, 1000-149 Lisbon, Portugal
176 Lancaster University, Physics Department, Lancaster LA1 4YB, UK
177 Lawrence Berkeley National Laboratory (LBNL), 1 Cyclotron Rd, Berkeley, CA 94720, USA
178 Lawrence Livermore National Laboratory (LLNL), Livermore, CA 94551, USA
179 Lebedev Physical Institute, Leninsky Prospect 53, RU-117924 Moscow, Russia
180 Lomonosov Moscow State University, Skobeltsyn Institute of Nuclear Physics (MSU SINP), 1(2), Leninskie gory, GSP-1, Moscow 119991, Russia
181 Louisiana Tech University, Department of Physics, Ruston, LA 71272, USA
182 Ludwig-Maximilians-Universität München, Fakultät für Physik, Am Coulombwall 1, D - 85748 Garching, Germany
183 Lunds Universitet, Fysiska Institutionen, Avdelningen för Experimentell Högenergifysik, Box 118, 221 00 Lund, Sweden
184 L'Université Hassan II, Aïn Chock, "Réseau Universitaire de Physique des Hautes Energies" (RUPHE), Département de Physique, Faculté des Sciences Aïn Chock, B.P 5366 Maarif, Casablanca 20100, Morocco
185 Massachusetts Institute of Technology (MIT), Laboratory for Nuclear Science, 77 Massachusetts Avenue, Cambridge, MA 02139, USA
186 Max-Planck-Institut für Physik (Werner-Heisenberg-Institut), Föhringer Ring 6, 80805 München, Germany
187 McGill University, Department of Physics, Ernest Rutherford Physics Bldg., 3600 University Street, Montreal, Quebec, H3A 2T8 Canada
188 McMaster University, Department of Physics & Astronomy, 1280 Main Street West, Hamilton, ON, L8S 4M1, Canada
189 Meiji Gakuin University, Department of Physics, 2-37 Shirokanedai 1-chome, Minato-ku, Tokyo 244-8539, Japan
190 Michigan State University, Department of Chemical Engineering & Materials Science, 2527 Engineering Building East Lansing, MI 48824-1226, USA
191 Michigan State University, Department of Physics and Astronomy, East Lansing, MI 48824, USA
192 Middle East Technical University, Department of Physics, TR-06531 Ankara, Turkey
193 Miyagi Gakuin Women's University, Faculty of Liberal Arts, 9-1-1 Sakuragaoka, Aoba District, Sendai, Miyagi 981-8557, Japan
194 MSU-Iligan Institute of Technology, Department of Physics, Andres Bonifacio Avenue, 9200 Iligan City, Phillipines
195 Nagasaki Institute of Applied Science, 536 Abamachi, Nagasaki-Shi, Nagasaki 851-0193, Japan
196 Nagoya University, Department of Materials Science and Engineering, Furo-cho, Chikusa-ku, Nagoya, 464-8603, Japan
197 Nagoya University, Department of Physics, School of Science, Furo-cho, Chikusa-ku, Nagoya, Aichi 464-8602, Japan
198 Nagoya University, Kobayashi-Maskawa Institute for the Origin of Particles and the Universe (KMI), Furo-cho, Chikusa-ku, Nagoya Aichi 464-8602, Japan
199 Nanjing University, Department of Physics, Nanjing, China 210093
200 Nara Women's University, High Energy Physics Group, Kitauoya-Nishimachi, Nara 630-8506, Japan
201 National Central University, High Energy Group, Department of Physics, Chung-li, Taiwan 32001, R.O.C
202 National Centre of Nuclear Research (NCBJ), ul. Andrzeja Soltana 7, 05-400 Otwock-Swierk, Poland
203 National Cheng Kung University, Physics Department, 1 Ta-Hsueh Road, Tainan, Taiwan 70101, R.O.C

204 National Chiao-Tung University, Institute of Physics, 1001 Ta Hsueh Rd, Hsinchu, Taiwan 300, R.O.C.
205 National Institute of Chemical Physics and Biophysics (NICPB), Ravala pst 10, 10143 Tallinn, Estonia
206 National Institute of Physics and Nuclear Engineering “Horia Hulubei” (IFIN-HH), Str. Reactorului no.30, P.O. Box MG-6, R-76900 Bucharest - Magurele, Romania
207 National Research Centre “Kurchatov Institute”, 1 Akademika Kurchatova pl., Moscow, 123182, Russia
208 National Science Center - Kharkov Institute of Physics and Technology (NSC KIPT), Akademicheskaya St. 1, Kharkov, 61108, Ukraine
209 National Scientific & Educational Centre of Particle & High Energy Physics (NCPHEP), Belarusian State University, M.Bogdanovich street 153, 220040 Minsk, Belarus
210 National Taiwan University, Physics Department, Taipei, Taiwan 106, R.O.C
211 Niels Bohr Institute (NBI), University of Copenhagen, Blegdamsvej 17, DK-2100 Copenhagen, Denmark
212 Niigata University, Department of Physics, Ikarashi, Niigata 950-218, Japan
213 Nikhef, National Institute for Subatomic Physics, P.O. Box 41882, 1009 DB Amsterdam, Netherlands
214 Nippon Dental University School of Life Dentistry at Niigata, 1-8 Hamaura-cho, Chuo-ku, Niigata 951-1500, Japan
215 North Carolina A&T State University, 1601 E. Market Street, Greensboro, NC 27411, USA
216 Northeastern University, Physics Department, 360 Huntington Ave, 111 Dana Research Center, Boston, MA 02115, USA
217 Northern Illinois University, Department of Physics, DeKalb, Illinois 60115-2825, USA
218 Northwestern University, Department of Physics and Astronomy, 2145 Sheridan Road., Evanston, IL 60208, USA
219 Novosibirsk State University (NGU), Department of Physics, Pirogov st. 2, 630090 Novosibirsk, Russia
220 Ochanomizu University, Department of Physics, Faculty of Science, 1-1 Otsuka 2, Bunkyo-ku, Tokyo 112-8610, Japan
221 Orissa University of Agriculture & Technology, Bhubaneswar 751003, Orissa, India
222 Osaka City University, Department of Physics, Faculty of Science, 3-3-138 Sugimoto, Sumiyoshi-ku, Osaka 558-8585, Japan
223 Osaka University, Department of Physics, 1-1 Machikaneyama, Toyonaka, Osaka 560-0043, Japan
224 Österreichische Akademie der Wissenschaften, Institut für Hochenergiephysik, Nikolsdorfergasse 18, A-1050 Vienna, Austria
225 Pacific Northwest National Laboratory, (PNNL), PO Box 999, Richland, WA 99352, USA
226 Panjab University, Chandigarh 160014, India
227 Pavel Sukhoi Gomel State Technical University, ICTP Affiliated Centre & Laboratory for Physical Studies, October Avenue, 48, 246746, Gomel, Belarus
228 Perimeter Institute for Theoretical Physics, 31 Caroline Street North, Waterloo, Ontario N2L 2Y5, Canada
229 Physical Research Laboratory, Navrangpura, Ahmedabad 380 009, Gujarat, India
230 Pohang Accelerator Laboratory (PAL), San-31 Hyoja-dong, Nam-gu, Pohang, Gyeongbuk 790-784, Republic of Korea
231 Pontificia Universidad Católica de Chile, Avda. Libertador Bernardo OHiggins 340, Santiago, Chile
232 Princeton University, Department of Physics, P.O. Box 708, Princeton, NJ 08542-0708, USA
233 Purdue University, Department of Physics, West Lafayette, IN 47907, USA
234 Queen Mary, University of London, Mile End Road, London, E1 4NS, United Kingdom
235 Rheinisch-Westfälische Technische Hochschule (RWTH), Physikalisches Institut, Physikzentrum, Otto-Blumenthal-Straße, 52056 Aachen
236 RIKEN, 2-1 Hirosawa, Wako, Saitama 351-0198, Japan
237 Royal Holloway, University of London (RHUL), Department of Physics, Egham, Surrey TW20 0EX, UK
238 Russian Academy of Science, Keldysh Institute of Applied Mathematics, Muiskaya pl. 4, 125047 Moscow, Russia
239 Rutgers, The State University of New Jersey, Department of Physics & Astronomy, 136 Frelinghuysen Rd, Piscataway, NJ 08854, USA
240 Saga University, Department of Physics, 1 Honjo-machi, Saga-shi, Saga 840-8502, Japan
241 Saha Institute of Nuclear Physics, 1/AF Bidhan Nagar, Kolkata 700064, India
242 Salalah College of Technology (SCOT), Engineering Department, Post Box No. 608, Postal Code 211, Salalah, Sultanate of Oman
243 Saudi Center for Theoretical Physics, King Fahd University of Petroleum and Minerals (KFUPM), Dhahran 31261, Saudi Arabia
244 Seikei University, Faculty of Science and Technology, 3-3-1 Kichijoji-Kitamachi, Musashino-shi, Tokyo 180-8633, Japan
245 Seoul National University, San 56-1, Shinrim-dong, Kwanak-gu, Seoul 151-742, Republic of Korea
246 Setsunan University, Institute for Fundamental Sciences, 17-8 Ikeda Nakamachi, Neyagawa, Osaka, 572-8508, Japan
247 Shanghai Jiao Tong University, Department of Physics, 800 Dongchuan Road, Shanghai, China 200240
248 Shinshu University, 3-1-1, Asahi, Matsumoto, Nagano 390-8621, Japan
249 Shiv Nadar University, Village Chithera, Tehsil Dadri, District Gautam Budh Nagar, 203207 Uttar Pradesh, India
250 Shizuoka University, Department of Physics, 836 Ohya, Suruga-ku, Shizuoka 422-8529, Japan
251 SLAC National Accelerator Laboratory, 2575 Sand Hill Road, Menlo Park, CA 94025, USA
252 Society for Applied Microwave Electronics Engineering and Research (SAMEER), I.I.T. Campus, Powai, Post Box 8448, Mumbai 400076, India
253 Sokendai, The Graduate University for Advanced Studies, Shonan Village, Hayama, Kanagawa 240-0193, Japan
254 Spallation Neutron Source (SNS), Oak Ridge National Laboratory (ORNL), P.O. Box 2008 MS-6477, Oak Ridge, TN 37831-6477, USA
255 State University of New York at Binghamton, Department of Physics, PO Box 6016, Binghamton, NY 13902, USA
256 State University of New York at Buffalo, Department of Physics & Astronomy, 239 Franczak Hall, Buffalo, NY 14260, USA
257 State University of New York at Stony Brook, Department of Physics and Astronomy, Stony Brook, NY 11794-3800, USA
258 STFC Daresbury Laboratory, Daresbury, Warrington, Cheshire WA4 4AD, UK

259 STFC Rutherford Appleton Laboratory, Chilton, Didcot, Oxon OX11 0QX, UK
260 Sungkyunkwan University (SKKU), Natural Science Campus 300, Physics Research Division, Chunchun-dong, Jangan-gu, Suwon, Kyunggi-do 440-746, Republic of Korea
261 Swiss Light Source (SLS), Paul Scherrer Institut (PSI), PSI West, CH-5232 Villigen PSI, Switzerland
262 Tata Institute of Fundamental Research, School of Natural Sciences, Homi Bhabha Rd., Mumbai 400005, India
263 Technical University of Lodz, Department of Microelectronics and Computer Science, al. Politechniki 11, 90-924 Lodz, Poland
264 Technische Universität Dresden, Institut für Kern- und Teilchenphysik, D-01069 Dresden, Germany
265 Tel-Aviv University, School of Physics and Astronomy, Ramat Aviv, Tel Aviv 69978, Israel
266 Texas A&M University, Physics Department, College Station, 77843-4242 TX, USA
267 Texas Tech University, Department of Physics, Campus Box 41051, Lubbock, TX 79409-1051, USA
268 The Henryk Niewodniczanski Institute of Nuclear Physics, Polish Academy of Sciences (IFJ PAN), ul. Radzikowskiego 152, PL-31342 Cracow, Poland
269 Thomas Jefferson National Accelerator Facility (TJNAF), 12000 Jefferson Avenue, Newport News, VA 23606, USA
270 Tohoku Gakuin University, Department of Business Administration, 1-3-1 Tsuchitoi, Aoba-ku Sendai, Miyagi 980-8511, Japan
271 Tohoku Gakuin University, Faculty of Technology, 1-13-1 Chuo, Tagajo, Miyagi 985-8537, Japan
272 Tohoku University, Department of Physics, Aoba District, Sendai, Miyagi 980-8578, Japan
273 Tohoku University, Research Center for Electron Photon Science, Taihaku District, Sendai, Miyagi 982-0826, Japan
274 Tohoku University, Research Center for Neutrino Science, Aoba District, Sendai, Miyagi 980-8578, Japan
275 Tokyo Institute of Technology, Department of Physics, 2-12-1 O-Okayama, Meguro, Tokyo 152-8551, Japan
276 Tokyo Metropolitan University, Faculty of Science and Engineering, Department of Physics, 1-1 Minami-Osawa, Hachioji-shi, Tokyo 192-0397, Japan
277 Tokyo University of Agriculture Technology, Department of Applied Physics, Naka-machi, Koganei, Tokyo 183-8488, Japan
278 Toyama Prefectural University, Department of Mathematical Physics, 5180 Kurokawa Imizu-shi, Toyama, 939-0398, Japan
279 TRIUMF, 4004 Wesbrook Mall, Vancouver, BC V6T 2A3, Canada
280 Universidad Autónoma de San Luis Potosí, Alvaro Obregon 64, Col. Centro, San Luis Potosí, S.L.P. 78000, México
281 Universidad de Granada, Departamento de Física Teórica y del Cosmos, Campus de Fuentenueva, E-18071 Granada, Spain
282 Universidad de los Andes, Faculty of Science, Department of Physics, Carrera 1 18A-10, Bloque Ip. Bogotá, Colombia
283 Universidad de Oviedo, Departamento de Física, Campus de Llamaquique. C/ Calvo Sotelo, s/n 33005 Oviedo, Spain
284 Universidad de Salamanca, Departamento de Física Fundamental, Plaza de la Merced, s/n., 37008 Salamanca, Spain
285 Universidad de Sevilla, Escuela Técnica Superior de Ingeniería, Departamento Ingeniería Electrónica, Camino de los Descubrimientos s/n, 41092 Sevilla, Spain
286 Universidad de Zaragoza - Departamento de Física Teórica, Pedro Cerbuna 12, E-50009 Zaragoza, Spain
287 Universidad Nacional Autónoma de México, Instituto de Física, Circuito de la Investigación Cientifica s/n, Ciudad Universitaria, CP 04510 México D.F., Mexico
288 Universidad Nacional de La Plata, Departamento de Física, Facultad de Ciencias Exactas, C.C. N 67, 1900 La Plata, Argentina
289 Universidade do Estado do Rio de Janeiro (UERJ), Rio de Janeiro, RJ - Brasil 20550-900, Brazil
290 Universidade Federal de Pelotas, Instituto de Física e Matemática, Campus Universitário, Caixa Postal 354, 96010-900 Pelotas, RS, Brazil
291 Universidade Federal do Rio de Janeiro (UFRJ), Instituto de Física, Av. Athos da Silveira Ramos 149, Centro de Tecnologia - Bloco A, Cidade Universitária, Ilha do Fundão, Rio de Janeiro, RJ, Brazil
292 Universitá degli Studi di Napoli "Federico II", Dipartimento di Fisica, Via Cintia, 80126 Napoli, Italy
293 Universitat Autònoma de Barcelona, Departament de Física, Edifici C, 08193 Bellaterra, Barcelona, Spain
294 Universitat Autònoma de Barcelona, Institut de Fisica d'Altes Energies (IFAE), Campus UAB, Edifici Cn, E-08193 Bellaterra, Barcelona, Spain
295 Universität Bonn, Physikalisches Institut, Nußallee 12, 53115 Bonn, Germany
296 Universität Heidelberg, Institut für Theoretische Physik, Philosophenweg 16, 69120 Heidelberg, Germany
297 Universität Heidelberg, Kirchhoff-Institut für Physik, Im Neuenheimer Feld 227, 69120 Heidelberg, Germany
298 Universitat Politècnica de Catalunya, Institut de Tècniques Energètiques, Campus Diagonal Sud, Edifici PC (Pavelló C). Av. Diagonal, 647 08028 Barcelona, Spain
299 Universitat Ramon Llull, La Salle, C/ Quatre Camins 2, 08022 Barcelona, Spain
300 Universität Rostock, 18051 Rostock, Germany
301 Universität Siegen, Naturwissenschaftlich-Technische Fakultät, Department Physik, Emmy Noether Campus, Walter-Flex-Str.3, 57068 Siegen, Germany
302 Universität Wien - Theoretische Physik Boltzmanngasse 5, A-1090 Vienna, Austria
303 Université catholique de Louvain, Centre for Cosmology, Particle Physics and Phenomenology (CP3), Institute of Mathematics and Physics, 2 Chemin du Cyclotron, 1348 Louvain-la-Neuve, Belgium
304 Université de Genève, Section de Physique, 24, quai E. Ansermet, 1211 Genève 4, Switzerland
305 Université de Montréal, Département de Physique, Groupe de Physique des Particules, C.P. 6128, Succ. Centre-ville, Montréal, Qc H3C 3J7, Canada
306 Université de Strasbourg, UFR de Sciences Physiques, 3-5 Rue de l'Université, F-67084 Strasbourg Cedex, France
307 Université Libre de Bruxelles, Boulevard du Triomphe, 1050 Bruxelles, Belgium
308 Universittà di Catania, Dipartimento di Fisica e Astronomia, Via Santa Sofia 64, 95123 Catania, Italy
309 University College London (UCL), High Energy Physics Group, Physics and Astronomy Department, Gower Street, London WC1E 6BT, UK

310 University College, National University of Ireland (Dublin), Department of Experimental Physics, Science Buildings, Belfield, Dublin 4, Ireland
311 University de Barcelona, Facultat de Física, Av. Diagonal, 647, Barcelona 08028, Spain
312 University of Alberta - Faculty of Science, Department of Physics, 4-181 CCIS, Edmonton AB T6G 2E1, Canada
313 University of Arizona, Department of Physics, 1118 E. Fourth Street, PO Box 210081, Tucson, AZ 85721, USA
314 University of Bergen, Institute of Physics, Allegaten 55, N-5007 Bergen, Norway
315 University of Birmingham, School of Physics and Astronomy, Particle Physics Group, Edgbaston, Birmingham B15 2TT, UK
316 University of Bristol, H. H. Wills Physics Lab, Tyndall Ave., Bristol BS8 1TL, UK
317 University of British Columbia, Department of Physics and Astronomy, 6224 Agricultural Rd., Vancouver, BC V6T 1Z1, Canada
318 University of California (UCLA), Los Angleles, CA 90095, US
319 University of California Berkeley, Department of Physics, 366 Le Conte Hall, #7300, Berkeley, CA 94720, USA
320 University of California Davis, Department of Physics, One Shields Avenue, Davis, CA 95616-8677, USA
321 University of California Irvine, Department of Physics and Astronomy, High Energy Group, 4129 Frederick Reines Hall, Irvine, CA 92697-4575 USA
322 University of California Santa Cruz, Institute for Particle Physics, 1156 High Street, Santa Cruz, CA 95064, USA
323 University of Cambridge, Cavendish Laboratory, J J Thomson Avenue, Cambridge CB3 0HE, UK
324 University of Colorado at Boulder, Department of Physics, 390 UCB, University of Colorado, Boulder, CO 80309-0390, USA
325 University of Cyprus, Department of Physics, P.O.Box 20537, 1678 Nicosia, Cyprus
326 University of Delhi, Department of Physics and Astrophysics, Delhi 110007, India
327 University of Delhi, S.G.T.B. Khalsa College, Delhi 110007, India
328 University of Dundee, Department of Physics, Nethergate, Dundee, DD1 4HN, Scotland, UK
329 University of Edinburgh, School of Physics, James Clerk Maxwell Building, The King's Buildings, Mayfield Road, Edinburgh EH9 3JZ, UK
330 University of Florida, Department of Physics, Gainesville, FL 32611, USA
331 University of Ghent, Department of Subatomic and Radiation Physics, Proeftuinstraat 86, 9000 Gent, Belgium
332 University of Glasgow, SUPA, School of Physics & Astronomy, University Avenue, Glasgow G12 8QQ, Scotland, UK
333 University of Hamburg, Physics Department, Luruper Chaussee 149, 22761 Hamburg, Germany
334 University of Hawaii, Department of Physics and Astronomy, HEP, 2505 Correa Rd., WAT 232, Honolulu, HI 96822-2219, USA
335 University of Helsinki, Department of Physical Sciences, P.O. Box 64 (Vaino Auerin katu 11), FIN-00014, Helsinki, Finland
336 University of Illinois at Chicago, Department Of Physics, 845 W Taylor St., Chicago IL 60607, USA
337 University of Iowa, Department of Physics and Astronomy, 203 Van Allen Hall, Iowa City, IA 52242-1479, USA
338 University of Kansas, Department of Physics and Astronomy, Malott Hall, 1251 Wescoe Hall Drive, Room 1082, Lawrence, KS 66045-7582, USA
339 University of Liverpool, Department of Physics, Oliver Lodge Lab, Oxford St., Liverpool L69 7ZE, UK
340 University of Liverpool, Division of Theoretical Physics, Department of Mathematical Sciences, Chadwick Building, Liverpool L69 3BX, UK
341 University of Ljubljana, Faculty of Mathematics and Physics, Jadranska ulica 19, 1000 Ljubljana, Slovenia
342 University of Malaya, Faculty of Science, Department of Physics, 50603 Kuala Lumpur, Malaysia
343 University of Manchester, School of Physics and Astronomy, Schuster Lab, Manchester M13 9PL, UK
344 University of Maribor, Faculty of Chemistry and Chemical Engineering (FKKT), Smetanova ulica 17, 2000 Maribor, Slovenia
345 University of Maryland, Department of Physics and Astronomy, Physics Building (Bldg. 082), College Park, MD 20742, USA
346 University of Massachusetts - Amherst, Department of Physics, 1126 Lederle Graduate Research Tower (LGRT), Amherst, MA 01003-9337, USA
347 University of Melbourne, School of Physics, Victoria 3010, Australia
348 University of Michigan, Department of Physics, 500 E. University Ave., Ann Arbor, MI 48109-1120, USA
349 University of Minnesota, 148 Tate Laboratory Of Physics, 116 Church St. S.E., Minneapolis, MN 55455, USA
350 University of Mississippi, Department of Physics and Astronomy, 108 Lewis Hall, PO Box 1848, Oxford, Mississippi 38677-1848, USA
351 University of Missouri – St. Louis, Department of Physics and Astronomy, 503 Benton Hall One University Blvd., St. Louis Mo 63121, USA
352 University of New Mexico, New Mexico Center for Particle Physics, Department of Physics and Astronomy, 800 Yale Boulevard N.E., Albuquerque, NM 87131, USA
353 University of North Carolina at Chapel Hill, Department of Physics and Astronomy, Phillips Hall, CB #3255, 120 E. Cameron Ave., Chapel Hill, NC 27599-3255, USA
354 University of Notre Dame, Department of Physics, 225 Nieuwland Science Hall, Notre Dame, IN 46556, USA
355 University of Oklahoma, Department of Physics and Astronomy, Norman, OK 73071, USA
356 University of Oregon, Department of Physics, 1371 E. 13th Ave., Eugene, OR 97403, USA
357 University of Oslo, Department of Physics, P.O box 1048, Blindern, 0316 Oslo, Norway
358 University of Oxford, Particle Physics Department, Denys Wilkinson Bldg., Keble Road, Oxford OX1 3RH England, UK
359 University of Pavia, Department of Physics, via Bassi 6, I-27100 Pavia, Italy
360 University of Pennsylvania, Department of Physics and Astronomy, 209 South 33rd Street, Philadelphia, PA 19104-6396, USA
361 University of Pittsburgh, Department of Physics and Astronomy, 100 Allen Hall, 3941 O'Hara St, Pittsburgh PA 15260, USA
362 University of Regina, Department of Physics, Regina, Saskatchewan, S4S 0A2 Canada

363 University of Rochester, Department of Physics and Astronomy, Bausch & Lomb Hall, P.O. Box 270171, 600 Wilson Boulevard, Rochester, NY 14627-0171 USA
364 University of Science and Technology of China, Department of Modern Physics (DMP), Jin Zhai Road 96, Hefei, China 230026
365 University of Silesia, Institute of Physics, Ul. Uniwersytecka 4, PL-40007 Katowice, Poland
366 University of South Carolina, Department of Physics and Astronomy, 712 Main Street, Columbia, SC 29208, USA
367 University of Southampton, School of Physics and Astronomy, Highfield, Southampton S017 1BJ, England, UK
368 University of Southern California, Department of Physics & Astronomy, 3620 McClintock Ave., SGM 408, Los Angeles, CA 90089-0484, USA
369 University of Sydney, Falkiner High Energy Physics Group, School of Physics, A28, Sydney, NSW 2006, Australia
370 University of Tartu, Institute of Physics, Riia 142, 51014 Tartu, Estonia
371 University of Texas at Austin, Department of Physics, 1 University Station C1600, Austin, Texas 78712, USA
372 University of Texas at Dallas, Department of Physics, 800 West Campbell Road, Richardson, Texas 75080, USA
373 University of Texas, Center for Accelerator Science and Technology, Arlington, TX 76019, USA
374 University of Tokushima, Institute of Theoretical Physics, Tokushima-shi 770-8502, Japan
375 University of Tokyo, Department of Physics, 7-3-1 Hongo, Bunkyo District, Tokyo 113-0033, Japan
376 University of Toronto, Department of Physics, 60 St. George St., Toronto M5S 1A7, Ontario, Canada
377 University of Toyama, Department of Physics, 3190 Gofuku, Toyama 930-8555, Japan
378 University of Tsukuba, Faculty of Pure and Applied Sciences, 1-1-1 Ten'nodai, Tsukuba, Ibaraki 305-8571, Japan
379 University of Victoria, Department of Physics and Astronomy, P.O.Box 3055 Stn Csc, Victoria, BC V8W 3P6, Canada
380 University of Virginia, Department of Physics, 382 McCormick Rd., PO Box 400714, Charlottesville, VA
381 University of Warsaw, Institute of Experimental Physics, Ul. Hoza 69, PL-00 681 Warsaw, Poland
382 University of Warsaw, Institute of Theoretical Physics, Ul. Hoza 69, PL-00 681 Warsaw, Poland
383 University of Washington, Department of Physics, PO Box 351560, Seattle, WA 98195-1560, USA
384 University of Wisconsin, Physics Department, Madison, WI 53706-1390, USA
385 University of Wuppertal, Gaußstraße 20, D-42119 Wuppertal, Germany
386 Università degli Studi di Milano, Dipartimento di Fisica, Via Celoria 16, 20133 Milano, Italy
387 Università degli Studi di Roma "La Sapienza", Dipartimento di Fisica, Piazzale Aldo Moro 5, 00185 Roma, Italy
388 Università degli Studi di Trieste, Dipartimento di Fisica, via A. Valerio 2, I-34127 Trieste, Italy
389 Università dell'Insubria in Como, Dipartimento di Scienze CC.FF.MM., via Vallegio 11, I-22100 Como, Italy
390 Università di Milano-Bicocca, Dipartimento di Fisica"G. Occhialin", Piazza della Scienza 3, 20126 Milano, Italy
391 Università di Pisa, Departimento di Fisica "Enrico Fermi", Largo Bruno Pontecorvo 3, I-56127 Pisa, Italy
392 Universiy of Huddersfield, International Institute for Accelerator Applications, Queensgate Campus, Huddersfield HD1 3DH, UK
393 UPMC Univ. Paris 06, Faculté de Physique (UFR 925), 4 Place Jussieu, 75252 Paris Cedex 05, France
394 Vietnam National University, Laboratory of High Energy Physics and Cosmology, Faculty of Physics, College of Science, 334 Nguyen Trai, Hanoi, Vietnam
395 Vietnam National University, University of Natural Sciences, 227 Nguyen Van Cu street, District 5, Ho Chi Minh City, Vietnam
396 VINCA Institute of Nuclear Sciences, Laboratory of Physics, PO Box 522, YU-11001 Belgrade, Serbia
397 Virginia Commonwealth University, Department of Physics, P.O. Box 842000, 701 W. Grace St.,Richmond, VA. 23284-2000, USA
398 Virginia Polytechnic Institute and State University, Physics Department, Blacksburg, VA 2406, USA
399 Vrije Universiteit Brussel, Pleinlaan 2, 1050 Brussels, Belgium
400 Vrije Universiteit, Department of Physics, Faculty of Sciences, De Boelelaan 1081, 1081 HV Amsterdam, Netherlands
401 Warsaw University of Technology, The Faculty of Electronics and Information Technology, ul. Nowowiejska 15-19, 00-665 Warsaw, Poland
402 Waseda University, Advanced Research Institute for Science and Engineering, Shinjuku, Tokyo 169-8555, Japan
403 Wayne State University, Department of Physics, Detroit, MI 48202, USA
404 Weizmann Institute of Science, Department of Particle Physics, P.O. Box 26, Rehovot 76100, Israel
405 Yale University, Department of Physics, New Haven, CT 06520, USA
406 Yamagata University, 1-4-12 Kojirakawa-cho, Yamagata-shi, Yamagata, 990-8560, Japan
407 Yerevan Physics Institute, 2 Alikhanyan Brothers St., Yerevan 375036, Armenia
408 Yonsei University, Department of Physics, 134 Sinchon-dong, Sudaemoon-gu, Seoul 120-749, Republic of Korea

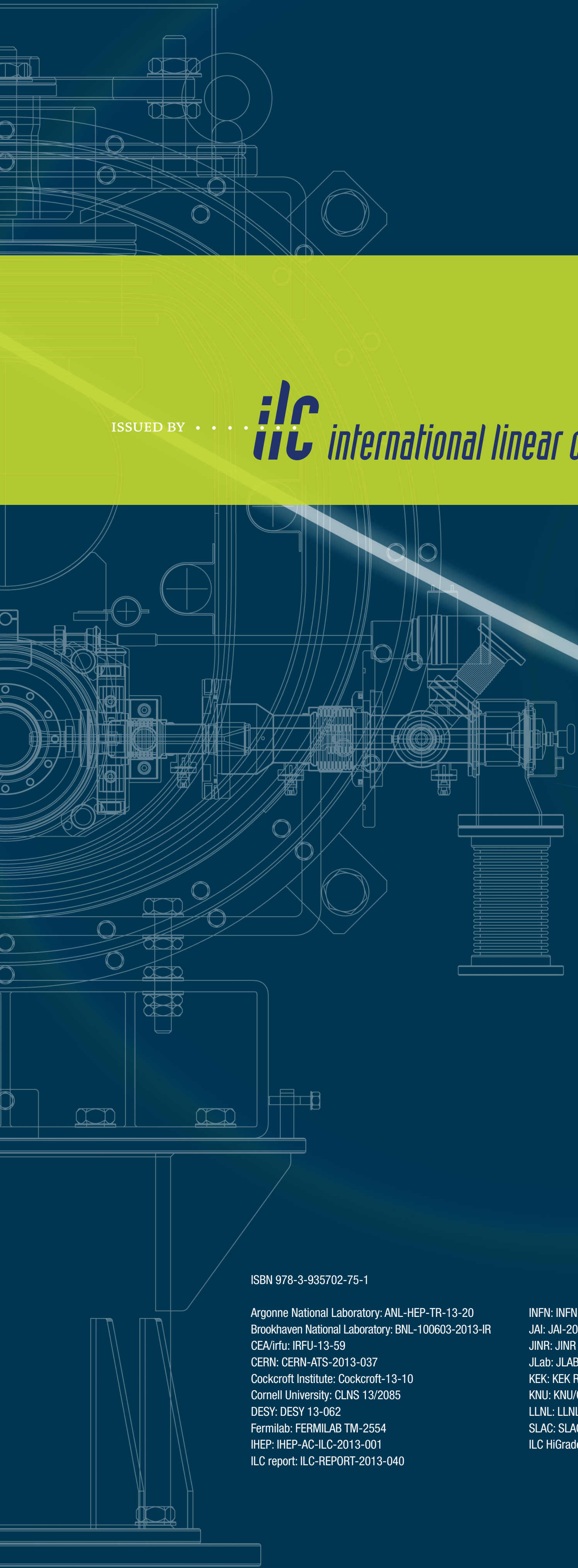

ISSUED BY ilc international linear collider

WWW.LINEARCOLLIDER.ORG